\documentclass[a4paper,11pt]{book}
\usepackage[nottoc,numbib]{tocbibind}
\usepackage{geometry} 
\usepackage{hyperref}
\usepackage{cite}
\usepackage{subcaption}  
\usepackage{graphicx}
\usepackage[utf8]{inputenc}
\usepackage{amssymb}
\usepackage{amsmath}
\usepackage{makeidx}
\usepackage{appendix}
\usepackage{mmacells}
\usepackage{afterpage}
\usepackage{geometry}

\title{Entanglement Entropy from Numerical Holography}
\author{Christian Ecker}
\makeindex
\date{\today}                                           

\begin{document}
\begin{titlepage}
\begin{figure}[ht]
\begin{center}
\vspace{-2cm}
\includegraphics[scale=0.3]{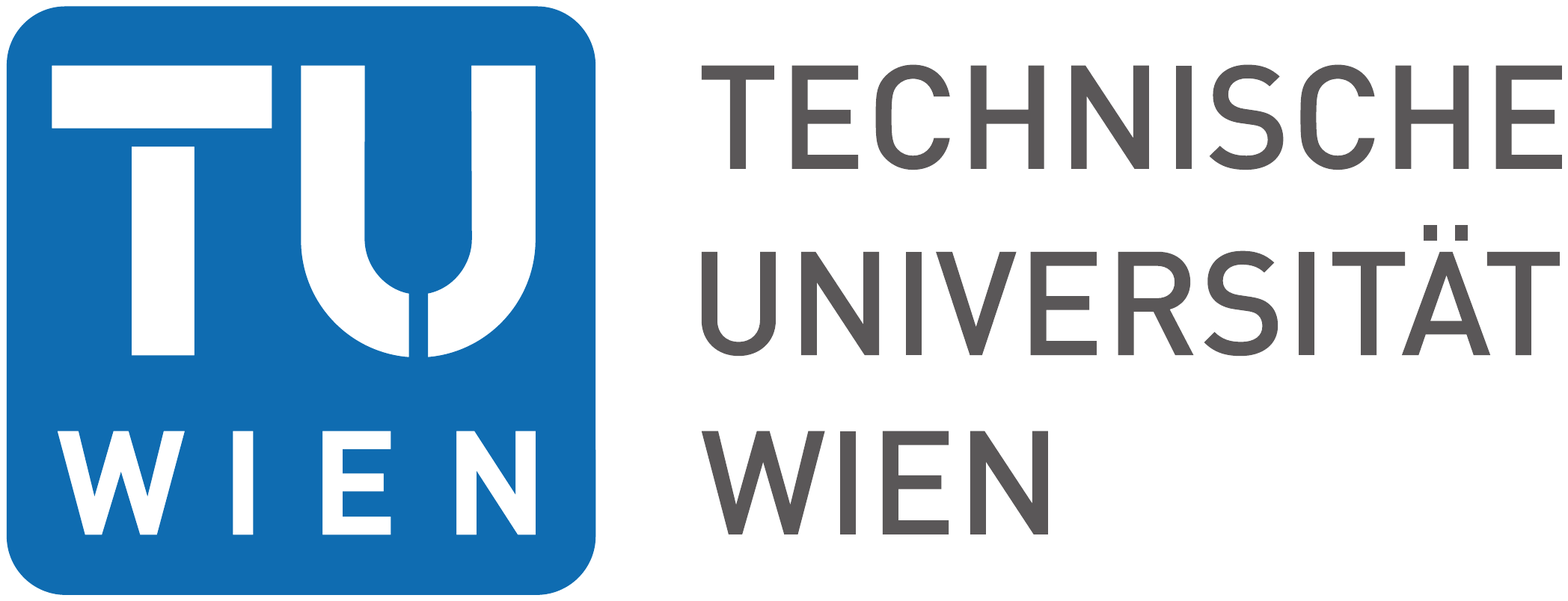}\\[1.0cm]
\end{center}
\end{figure}
\vspace{-3cm}
\begin{center}
{\LARGE Dissertation\\[1.0cm]}
{\LARGE\textbf{Entanglement Entropy from Numerical Holography}\\[1.0cm]}
{\normalsize Ausgeführt zum Zwecke der Erlangung des Akademischen Grades \\
 eines Doktors der technischen Wissenschaften}\\[1cm]
{\normalsize unter der Leitung von\\
Assoc.-Prof. Priv.-Doz. Dr. Daniel Grumiller\\
Institut für theoretische Physik\\
Technische Universität Wien, AT\\[1cm]}

{\normalsize eingereicht an der Technischen Universität Wien\\
Fakultät für Physik\\[1cm]}
{\normalsize von\\

Dipl.-Ing. Christian Ecker\\

Blütengasse 9/2/22\\

1030 Wien}\\[2cm]

\noindent\begin{tabular}{ll}

\makebox[2cm]{Wien, am 07.09.2018}\hspace*{1cm} & \makebox[4cm]{\hrulefill}\\

& \\[1ex]

\end{tabular}
\vspace*{2.5cm}

\begin{tabular}{p{5cm}@{}p{5cm}@{}p{5cm}@{}}
 \makebox[4cm]{\hrulefill} & \makebox[4cm]{\hrulefill} & \makebox[4cm]{\hrulefill} \\
 Prof. Daniel Grumiller    & Prof. Javier Mas Sol\'{e} & Prof. Paul Romatschke     \\
 (Betreuer)                & (Gutachter)               & (Gutachter)                 \\
\end{tabular}
\end{center}
\end{titlepage}
\newpage \vspace*{8cm}
\thispagestyle{empty}
\begin{center}
  \Large \emph{}
\end{center}

\newpage \vspace*{8cm}
\thispagestyle{empty}
\begin{center}
  \Large \emph{To my lovely wife Daniela \\and \\my wonderful daughter Mia.}
\end{center}

\newpage \vspace*{8cm}
\thispagestyle{empty}
\begin{center}
  \Large \emph{}
\end{center}
\frontmatter

\chapter{Abstract}\label{Chap:Abstract}
In this thesis I present numerical studies of entanglement entropy and the quantum null energy condition in strongly coupled far-from-equilibrium quantum states using holography.
The holographic prescription for entanglement entropy requires to determine the area of extremal surfaces in asymptoticaly Anti-de Sitter spacetimes which I do both numerically and, when possible, analytically.
I give a careful introduction into the numerical methods used and provide the computer codes to compute entanglement entropy and the quantum null energy condition.
These methods are then applied to systems of various degrees of complexity, including homogeneous and isotropic far-from-equilibrium quantum quenches dual to Vaidya spacetimes, to homogeneous and anisotropic finite temperature states dual to anisotropic black brane geometries, and to inhomogeneous and anisotropic states of colliding walls of energy dual to gravitational shock wave collisions in Anti-de Sitter space.   
For all these scenarios I compute the fully non-linear dynamics of the dual geometry, which requires to numerically solve five-dimensional Einstein equations with negative cosmological constant and asymptotic Anti-de Sitter boundary conditions.
The numerical solutions for the geometries allow to extract the time evolution of the holographic energy momentum tensor and provides the background for computing two-point functions, entanglement entropy and the quantum null energy condition.
From the anisotropic system one learns that the near-equilibrium dynamic of entanglement entropy has an accurate description in terms of quasinormal modes. 
In the shock wave system I identify characteristic features of entanglement entropy that allow to discriminate between thick and narrow shocks.
All my numerical studies confirm the quantum null energy condition, also the shock wave system, which itself can violate the classical null energy condition for sufficiently narrow shocks. My results also show that the quantum null energy condition can saturate in the far-from-equilibrium regime.

\chapter{Zusammenfassung}\label{Chap:Zusammenfassung}
In dieser Dissertation beschreibe ich numerische Studien der Verschränkungsentropie und der Quanten-Null-Energie-Bedingung in stark gekoppelten Nicht-Gleichgewichts-Quantenzuständen unter Verwendung des holographischen Prinzips.
Die holographische Beschreibung der Verschränkungsentropie erfordert die Bestimmung von extremalen Flächen in asymptotischen Anti-de Sitter Raumzeiten, welche ich numerisch mache.
Ich geben eine detaillierte Einführung in die verwendeten numerischen Methoden und stellen die Computercodes zur Verfügung, um die Verschränkungsentropie und die Quanten-Null-Energie-Bedingung zu berechnen. 
Diese Methoden werden dann auf Systeme verschiedener Komplexität angewendet, einschließlich homogener und isotroper Nicht-Gleichgewichts-Quantenzustände, die dual zu Vaidya Raumzeiten sind, auf homogene aber anisotrope Zustände bei endlicher Temperatur, welche dual zu anisotropen Black-Brane Geometrien sind, und auf inhomogene und anisotrope Zustände kollidierender Energiewände, die sich mit kollidierenden Gravitationsschockwellen im dualen Anti-de Sitter Raum beschreiben lassen. 
Für alle diese Systeme berechne ich vollständig die nichtlineare Dynamik der dualen Geometrie, indem ich die zeitabhängigen Einstein Gleichungen mit negativer kosmologischer Konstante und asymptotischen Anti-de Sitter Randbedingungen in fünf Dimensionen numerisch löse. 
Die numerische Lösung für die Geometrien erlaubt es die zeitliche Entwicklung des holographischen Energie-Impuls-Tensors zu extrahieren und liefert den notwendigen Hintergrund zur Berechnung von Zweipunktfunktionen, der Verschränkungsentropie und der Quanten-Null-Energie-Bedingung.
Aus dem anisotropen System sieht man, dass sich die Dynamik der Verschränkungsentropie nahe am Gleichgewicht präzise mit Quasinormalmoden beschreiben lässt.
Im Schockwellensystem identifiziere ich charakteristische Merkmale der Verschränkungsentropie, die es erlauben zwischen dicken und dünnen Schockwellen zu unterscheiden.
Alle meine numerischen Ergebnisse bestätigen die Quanten-Null-Energie-Bedingung, auch die des Schockwellensystems, in welchem die klassische Null-Energie-Bedingung für hinreichend dünne Schocks verletzt sein kann. Meine Ergebnisse zeigen weiters, dass die Quanten-Null-Energie-Bedingung in diesem System auch in weit aus dem Gleichgewicht gebrachten Zuständen saturiert werden kann.

\chapter{Publications}\label{Chap:Publications}
The research results obtained during my PhD studies have been published in peer reviewed articles in different scientific journals and are freely available on \href{http://arXiv.org}{{arXiv.org}}.
This thesis is based on the following articles which will be reviewed in Chapter \ref{chap:Aniso}, Chapter \ref{Chap:Shocks} and Chapter \ref{Chap:QNEC} 

\begin{itemize}
\item[1] C.~Ecker, D.~Grumiller and S.~A. Stricker, {\it {Evolution of holographic
  entanglement entropy in an anisotropic system}},  {\em JHEP} {\bf 07} (2015)
  146 [\href{http://arXiv.org/abs/1506.02658}{{\tt 1506.02658}}].
\item[2] C.~Ecker, {\it {Holographic Entanglement Entropy from Numerical Relativity}},
  {\em PoS} {\bf CORFU2015} (2016) 066 [\href{http://inspirehep.net/record/1498004/files/PoS%28CORFU2015%29066.pdf}{{\tt PoS(CORFU2015)066.pdf}}]. 
\item[3] C.~Ecker, D.~Grumiller, P.~Stanzer, S.~A. Stricker and W.~van~der Schee, {\it
  {Exploring nonlocal observables in shock wave collisions}},  {\em JHEP} {\bf
  11} (2016) 054 [\href{http://arXiv.org/abs/1609.03676}{{\tt 1609.03676}}].
\item[4] C.~Ecker, D.~Grumiller, W.~van~der Schee and P.~Stanzer, {\it {Saturation of
  the Quantum Null Energy Condition in Far-From-Equilibrium Systems}},  {\em
  Phys. Rev.} {\bf D97} (2018), no.~12 126016
  [\href{http://arXiv.org/abs/1710.09837}{{\tt 1710.09837}}].
\end{itemize}
Besides the work mentioned above, several wonderful collaborations were formed and resulted in a list of articles which are not directly related to the main topic of this thesis.
\begin{itemize}
\item[5] B.~Doucot, C.~Ecker, A.~Mukhopadhyay and G.~Policastro, {\it {Density response
  and collective modes of semiholographic non-Fermi liquids}},  {\em Phys.
  Rev.} {\bf D96} (2017), no.~10 106011
  [\href{http://arXiv.org/abs/1706.04975}{{\tt 1706.04975}}].
\item[6] C.~Ecker, C.~Hoyos, N.~Jokela, D.~Rodríguez~Fernández and A.~Vuorinen, {\it
  {Stiff phases in strongly coupled gauge theories with holographic duals}},
  {\em JHEP} {\bf 11} (2017) 031 [\href{http://arXiv.org/abs/1707.00521}{{\tt 1707.00521}}].
\item[7] E.~Annala, C.~Ecker, C.~Hoyos, N.~Jokela, D.~Rodríguez~Fernández and
  A.~Vuorinen, {\it {Holographic compact stars meet gravitational wave
  constraints}},  [\href{http://arXiv.org/abs/1711.06244}{{\tt 1711.06244}}].
\item[8] C.~Ecker, A.~Mukhopadhyay, F.~Preis, A.~Rebhan and A.~Soloviev, {\it {Time
  evolution of a toy semiholographic glasma}},
  [\href{http://arXiv.org/abs/1806.01850}{{\tt 1806.01850}}].
\item[9] C.~Ecker, D.~Grumiller, M.M.~Sheikh-Jabbari, P.~Stanzer and W.~van~der Schee, {\it {On QNEC and its saturation in 2d CFTs}},
  TUW-18-07, in preparation.
\end{itemize}
A considerable amount of time was spent on the development of computer code of which selected examples are available on my homepage \href{http://christianecker.com/}{{christianecker.com}}.

\chapter{Acknowledgements}\label{Chap:Acknowledgements}
Most of the scientific achievements published in the articles listed earlier would not have been possible without the contribution of a number of people. 
First and foremost I would like to thank my supervisor Daniel Grumiller for hiring me as his PhD student.
I am grateful for the many insightful discussions and the excellent collaboration, especially towards the end of my PhD studies, and for his valuable remarks on the manuscript.
I am deeply impressed by Daniels technical brilliance and profited a lot from his deep understanding of gravity and his broad physics knowledge. 

I would like to thank Wilke van der Schee for the excellent collaboration, the many enlightening discussions and for sharing his extraordinary numerics know-how.
I am also grateful to Esperanza L\'{o}pez and Paul Romatschke for sharing their relaxation and spectral codes, which provided the basis for developing my own implementations.
Furthermore, I thank my PhD colleague Philipp Stanzer who made invaluable contributions to developing the codes used to compute the entanglement entropy and the quantum null energy condition and for carefully proofreading the manuscript.
Without Esperanza, Paul, Philipp and Wilke most of the numerical results presented in this thesis would not have been achieved.

I profited a lot from several excellent lectures during my time as (under)graduate student.
Most notably, I like to thank Herbert Balasin for his excellent lecture series on General Relativity and Anton Rebhan for his Quantum Field Theory lectures. These courses ignited my passion for theoretical physics and provided the necessary foundation for learning string theory and holography.
Furthermore, I like to thank Peter van Nieuwenhuizen for his brilliant lecture series on supergravity, which I enjoyed a lot and will always remember.
I am particularly thankful to Stefan Stricker, especially for his strong support during the start of my PhD and the excellent collaboration on our articles about entanglement entropy. 
I am also grateful to Florian Preis for countless discussions, for sharing his deep physics knowledge and the excellent collaboration on our article about semiholography.
Florian and Stefan were not only appreciated collaborators, but became also close friends of mine.
I like to thank Ayan Mukhopadhyay, Benoit Doucot and Giuseppe Policastro for the excellent collaboration on our paper on semiholographic non-Fermi liquids and in particular Benoit for his pedagogical explanations about Fermi liquids.

I am extremely thankful to Aleksi Vuorinen for hosting me during my three months visit in winter 2016 at the University of Helsinki and in particular for his strong support and his invaluable advises for my postdoc applications. I am grateful for the wonderful personal and scientific experience in the Helsinki group and especially like to thank Aleksi, Eemeli Annala, Niko Jokela, Keijo Kajantie, Ville Keränen, and Jere Remes who provided a warm and welcoming environment during the cold and dark season in Finland. 
Furthermore, I thank Aleksi, Niko, Eemeli, Carlos Hoyos, and David Rodriguez Fernández for the excellent collaboration on our neutron star related projects. I profited a lot from technical discussions with Carlos.

I thank David Mateos for hosting me during my three months visit in spring 2017 at the University of Barcelona and for sharing his knowledge on non-conformal holographic plasmas. During my stay in Barcelona I profited from discussions with Maximilian Attems, Jorge Casalderrey-Solana, Yago Bea, Miquel Triana, and especially Helvi Witek who gave me an invaluable introduction to the Einstein toolkit.

I also like to thank my office mates at TU Wien, Frederic Brünner, Alexander Haber, David Müller and Alexander Soloviev for the many interesting and entertaining discussions. It was a great pleasure sharing an office with you guys.

I am very thankful to my referees Javier Mas Sol\'{e} and Paul Romatschke for reading the manuscript and preparing their reports.

Most importantly, I want to thank my wife Daniela and my daughter Mia.
I am deeply grateful to Daniela for her permanent support, for baring all the inconvenience resulting from the duties of a scientist husband, and for carefully reading the manuscript and spotting a shere endless number of typos.
Finally, I like to thank my beloved daughter Mia for her understanding for all the hours I was either travelling or working at home, and for her interest in my work.
\tableofcontents

\mainmatter
\chapter{Introduction}\label{Chap:Introduction}
The gauge/gravity duality \cite{Maldacena:1997re, Witten:1998qj, Gubser:1998bc} has established itself as a valuable tool in the quest for a better understanding of strongly coupled systems. 
In particular it is used  to gain insights into the thermalization process of non-Abelian plasmas by studying the gravitational dual of $\mathcal{N}=4$ Super--Yang--Mills (SYM) theory, a maximally supersymmetric conformal field theory (CFT) in four spacetime dimensions. 
Thermal equilibration of the field theory is then mapped to black hole formation on the gravity side.

Important examples are the experiments at RHIC and the LHC which revealed that the quark gluon plasma created in heavy ion collisions behaves as a strongly coupled viscous liquid that thermalizes extremely fast \cite{Teaney:2000cw,Huovinen:2001cy,Hirano:2002ds,Romatschke:2007mq}. 
In the last decade there has been considerable progress in setting up collisions of SYM matter in various scenarios and studying its evolution.
One of the  starting points was the study of perfect fluid dynamics in a boost invariant setting \cite{Janik:2005zt,Janik:2006gp}.
In \cite{Chesler:2008hg} it was possible to study far-from-equilibrium dynamics by numerically solving the full Einstein equations in an anisotropic but otherwise completely homogeneous system. 
Trying to come closer to mimic a  heavy ion collision led to the idea \cite{Janik:2005zt} of colliding delta like gravitational shock waves \cite{Grumiller:2008va,Albacete:2009ji}, which are dual to lumps of energy in the SYM theory moving at the speed of light.
The next step was to make the system anisotropic and inhomogeneous by the collision of  gravitational shock waves which are homogeneous in the transverse direction and have finite width in the longitudinal direction \cite{Chesler:2010bi}. It was found that a hydrodynamic description of the plasma is valid even when the anisotropy is still large \cite{Heller:2011ju}. This onset of  hydrodynamic behavior is now termed \textit{hydrodynamization}\index{hydrodynamization}. 

Further advances include radial flow \cite{vanderSchee:2012qj}, the effect of different initial conditions \cite{Casalderrey-Solana:2013aba}, the collision of two black holes \cite{Bantilan:2014sra}, and more \cite{Chesler:2015fpa,Keegan:2015avk,Chesler:2016ceu}.
Now it is even possible to simulate  the collision of two localized lumps of matter to mimic off-central nucleus-nucleus \cite{Chesler:2015wra,vanderSchee:2015rta} and proton-nucleus collisions \cite{Chesler:2015bba}.
More recently, these studies were extended to collisions including electromagnetic fields \cite{Casalderrey-Solana:2016xfq}, to non-conformal systems including scalar fields in the bulk \cite{Attems:2016tby,Attems:2017ezz,Attems:2017zam,Attems:2018gou}, and to collisions including finite coupling corrections using Gauss-Bonnet gravity \cite{Grozdanov:2016zjj}. An excellent review of the state-of-the-art of such calculations, including a discussion of remaining challenges, can be found in \cite{Busza:2018rrf}.

Despite all the advances one has to keep in mind that in heavy ion collisions there are many energy scales involved. To get an accurate understanding of all the relevant thermalization mechanisms, strong and weak coupling phenomena must be combined. 
One step towards this direction is the combination of different effective descriptions \cite{vanderSchee:2013pia} or by constructing a semi-holographic framework where the weakly and strongly coupled sector can interact with each other \cite{Iancu:2014ava,Mukhopadhyay:2015smb,Ecker:2018ucc}.

So far, in most colliding shock wave studies the quantities of interest are local quantities, i.e.\ the components of the energy momentum tensor, such as the energy density or the pressures. This allows to determine if local equilibrium is reached, here understood as the local applicability of hydrodynamics.
In order to gain further insights into the thermalization process the time evolution of nonlocal quantities, such as various correlation functions (e.g. Wightman function or Feynman propagator), in coordinate space needs to be considered.
 This is still a complicated task but two such nonlocal quantities can be obtained relatively easy with the help of the gauge/gravity duality, namely the equal time two-point function for scalar operators of large conformal weight and entanglement entropy. 

The equal time two-point function can be obtained from the length of spacelike geodesics which are anchored to the boundary of Anti-de~Sitter (AdS) and extending into the bulk. Although the geodesic approximation is only valid for operators of large conformal weight,
a comparison of the Feynman propagator for a scalar field with conformal dimension $\Delta=3/2$ with the geodesic approximation revealed that they show the same qualitative behavior \cite{Keranen:2014lna}.
Similarly the holographic entanglement entropy can be obtained from the area of extremal surfaces \cite{Ryu:2006bv,Hubeny:2007xt}.
A motivation to study entanglement entropy in $\mathcal{N}=4$ SYM theory comes from the question how to measure entropy production in holographic models for heavy ion collisions \cite{Muller:2011ra}. Within the gauge/gravity duality the entropy of the (stationary) black hole corresponds to the entropy of the field theory. However, in time dependent backgrounds entropy as defined from the area of the apparent horizon is ambiguous because it depends on the choice of time slicing. By contrast, the definition of the holographic entanglement entropy is unique and therefore may serve as an alternative measure for entropy production.  

Another set of examples for thermal equilibration in strongly interacting time-dependent physics systems are condensed matter experiments, where it is now possible to drive an isolated system to a far-from-equilibrium state by a quantum quench, i.e.~a control parameter of the system is varied rapidly \cite{Polkovnikov:2010yn}. 
In a seminal work, Calabrese and Cardy \cite{Calabrese:2005in} were able to compute the time evolution of the entanglement entropy after a quench in a two-dimensional conformal field theory and in the Ising spin chain model in a transverse magnetic field. In both cases they find that for an entangling interval of length $l$, the entanglement entropy increases linearly with time until $t\sim l/2$ after which it saturates. The linear scaling with time and the crossover at $t\sim l/2$ can be understood in terms of entangled quasiparticle pairs emitted from the initial state and is therefore  expected to hold for a wider class of systems. 

The simplest example where one can study the analog of quenches in the holographic setup are spacetimes where thin shells collapse to form a black hole.
This was first worked out in the Vaidya spacetime  \cite{AbajoArrastia:2010yt,Balasubramanian:2010ce, Liu:2013iza, Liu:2013qca} where the shell is composed out of null dust and later generalized to matter with  arbitrary equation of states \cite{Keranen:2014zoa, Keranen:2015fqa}. The linear  scaling is even present in geometries with  Lifshitz scaling and hyperscaling violation \cite{Fonda:2014ula}. 
The behavior of the entanglement entropy in these setups has been studied extensively \cite{AbajoArrastia:2010yt, Balasubramanian:2011ur, Albash:2010mv, Baron:2012fv,Galante:2012pv,Keranen:2011xs, Liu:2013qca} and indeed shows universal behavior consistent with the findings of Cardy and Calabrese. 

However, this universal behavior disappears in more complicated setups \cite{Buchel:2014gta,Rangamani:2015agy, Abajo-Arrastia:2014fma,daSilva:2014zva,Ecker:2015kna,Bellantuono:2016tkh}.
For example, a radially collapsing scalar field in global AdS can have many bounces between the boundary and the center of AdS before a black hole forms \cite{Bizon:2011gg}, resulting in a periodic behavior of the entanglement entropy \cite{Abajo-Arrastia:2014fma,daSilva:2014zva}.
In \cite{Buchel:2014gta}  a massive scalar field dual to a massive fermionic operator was turned on, treating the quench as a perturbation on the static spacetime. In this setup the entanglement entropy is also not monotonic and in some cases approaches the equilibrium value from above. This reveals a qualitative difference of entanglement entropy to the thermal entropy which, on general grounds, must be monotonically growing in a closed system.
Note that these quenches are thermal quenches because the end state is a thermal state.
In anisotropic $\mathcal{N}=4$ SYM the entanglement entropy and equal time two-point functions show oscillatory behavior with exponential damping at late times which is given by the lowest quasinormal mode \cite{Ecker:2015kna}.
Analytic progress has been made in \cite{Pedraza:2014moa} where the late-time behavior of two-point functions, Wilson loops and entanglement entropy has been studied perturbatively in a boost-invariant system.

Another motivation to study entanglement entropy comes from the recently proposed quantum null energy condition (QNEC) \cite{Bousso:2015mna,Bousso:2015wca}, which is currently the only known local energy condition that is supposed to hold in any relativistic QFT.
The QNEC inequality states that the null projection of the expectation values of the energy momentum tensor for any quantum state is larger than, or equal to, the second variation of entanglement entropy with respect to lightlike deformations of the entangling region evaluated for the same state.
Energy conditions rose to prominence in the 1960s as requisites for proofs of singularity theorems or Hawking's area theorem \cite{Hawking:1969sw, Hawking:1973uf}. While the specific energy condition needed depends on details of the particular theorem, all local classical ones are violated by quantum effects \cite{Epstein1965}.
Instead, QFTs typically obey nonlocal conditions such as the Averaged Null Energy Condition (ANEC, \cite{Klinkhammer:1991ki, Ford:1994bj}), which is the statement that negative energy density along a complete null geodesic is compensated by positive energy density (with ``quantum interest'' \cite{Ford:1999qv}).
These averaged energy conditions can sometimes be proven for QFTs (see \cite{Faulkner:2016mzt,Hartman:2016lgu} for ANEC) and hence provide non-trivial consistency conditions for general QFTs. 
A better understanding of quantum energy conditions can then even lead to bounds on inflationary parameters, such as conjectured in \cite{Cordova:2017zej}.
In general relativity the null energy condition (NEC), which states that the null projection of the energy momentum tensor is strictly positive or zero, together with the Raychaudhuri equation implies the focusing of geodesics.
This in turn implies that classically the total area of black hole horizons does not decrease. When including quantum effects this needs to be generalized by adding area and (entanglement) entropy together, after which the generalized entropy cannot decrease. This quantum focusing conjecture \cite{Bousso:2015mna} then implies QNEC.
QNEC has been proven for free and superrenormalizable theories \cite{Bousso:2015wca}, theories with holographic duals \cite{Koeller:2015qmn}, and interacting quantum field theories in $d\geq3$ \cite{Balakrishnan:2017bjg}.
Quantum energy conditions are particularly relevant for systems that violate the classical ones. 
This is for instance the case in holographic models of heavy ion collisions, where sufficiently narrow shocks can produce regions of negative energy density in the forward light cone. As a consequence the classical null energy condition is violated in these regions \cite{Arnold:2014jva}, but it can be shown that QNEC still holds \cite{Ecker:2017jdw}.

The outline of this thesis is as follows:
Chapter \ref{Chap:Theoretical Background} provides the necessary background material, starting with a basic introduction to entanglement entropy, followed by a review of relevant concepts in string theory and a motivation of the AdS/CFT correspondence. With these basic concepts at our disposal we introduce the holographic prescription for entanglement entropy and close the chapter with a section on the quantum null energy condition. 
In Chapter \ref{Chap:Numerics} we explain the numerical methods used in this thesis, starting with the method of characteristics which we use to solve the Einstein equations, followed by the shooting method and the relaxation method which we use to compute entanglement entropy, and ending with our numerical method to compute QNEC.
In Chapter \ref{chap:Aniso} we present our analysis of two-point functions and entanglement entropy in an anisotropic far-from-equilibrium system. 
In Chapter \ref{Chap:Shocks} we study two-point functions and entanglement entropy in shock wave collisions and in Chapter \ref{Chap:QNEC} we present our QNEC computations in this system. 
We summarize, conclude and suggest possible extensions to our work in Chapter \ref{Chap:ClosingRemarks}.
In Appendices \ref{App:AreaFunct}--\ref{App:MathematicaQNEC} we collect basic derivations and provide selected code listings of our numerical procedures.

\chapter{Theoretical Background}\label{Chap:Theoretical Background}
The aim of this chapter is to introduce the theoretical concepts, and to make basic definitions used in this thesis. 
We start in Section \ref{Sec:Entanglement} with the definition and important properties of entanglement entropy in quantum mechanics and quantum field theory. After that we briefly discuss in Section \ref{Sec:Holography} the holographic principle, followed by a lightning review of string theory before we motivate and define the AdS/CFT correspondence. In Section \ref{Sec:HolographicEE} we discuss the Ryu-Takayanagi prescription for the holographic entanglement entropy including some basic calculations of extremal surfaces which will be used in later chapters. Finally, in Section \ref{Sec:QNEC} we introduce the quantum null energy condition.
\section{Entanglement Entropy}\label{Sec:Entanglement}
Entanglement is a genuine feature of quantum states which is absent in systems that follow the laws of classical mechanics.
It is highly desirable to formulate measures that quantify how much ``quantum" a given system is. Several such entanglement measures have been introduced, including R\'{e}nyi entropies, mutual information, entanglement entropy and many more (see \cite{Plenio:2007zz} for a more complete list). Mainly because of its computational accessibility, the entanglement entropy plays a particularly important role in quantifying entanglement.

Currently, entanglement entropy is studied in many different research areas, including information theory, condensed matter physics, high energy physics and quantum gravity. Originating in quantum information theory \cite{nielsen_chuang_2010}, entanglement entropy provides a connection between seemingly unrelated fields and initiated many ideas leading to interesting new insights.
In condensed matter theory the entanglement entropy provides a useful order parameter for quantum-phase transitions which are characterized by their entanglement pattern and not by the usual Ginzburg-Landau paradigm of spontaneous symmetry breaking \cite{Vidal:2002rm,Kitaev:2005dm,Fendley:2006gr}. 
In high energy physics, entanglement entropy is important because it characterizes the number of degrees of freedom under renormalization group flow in QFTs \cite{Ryu:2006ef}.
The AdS/CFT correspondence led to an interesting geometric construction of entanglement entropy, establishing a tight connection between spacetime geometry and entanglement.
This led to the idea of formulating quantum gravity from quantum entanglement \cite{VanRaamsdonk:2009ar} and resulted in the famous ER=EPR slogan \cite{Maldacena:2013xja}.
As we will discuss in some detail in later chapters, entanglement entropy plays an important role in formulating a local energy bound in quantum field theories, called quantum null energy condition \cite{Bousso:2015mna,Bousso:2015wca}.
In the context of this thesis we will study entanglement entropy in holographic models relevant for the quark gluon plasma produced in relativistic heavy ion collisions.

An overview on recent developments in (holographic) entanglement entropy, including an extensive list of references, can be found for instance in the textbook \cite{Rangamani:2016dms} and the review article \cite{Nishioka:2018khk} on which some of the discussion in this section is based. 

\subsection{Definition and Important Properties}
\begin{figure}[htb]
\center
\includegraphics[width=0.40\linewidth]{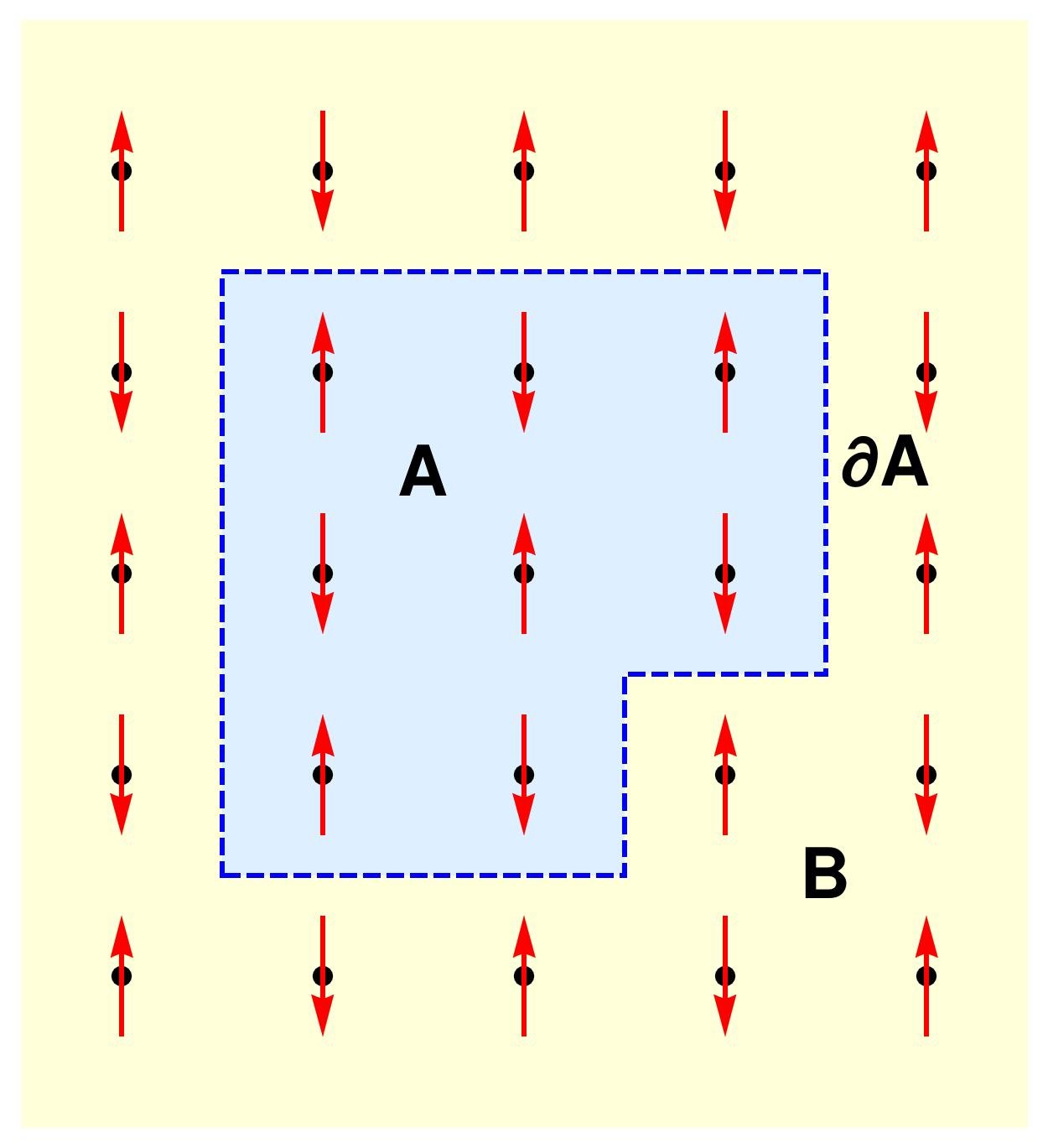}\quad\quad \includegraphics[width=0.40\linewidth]{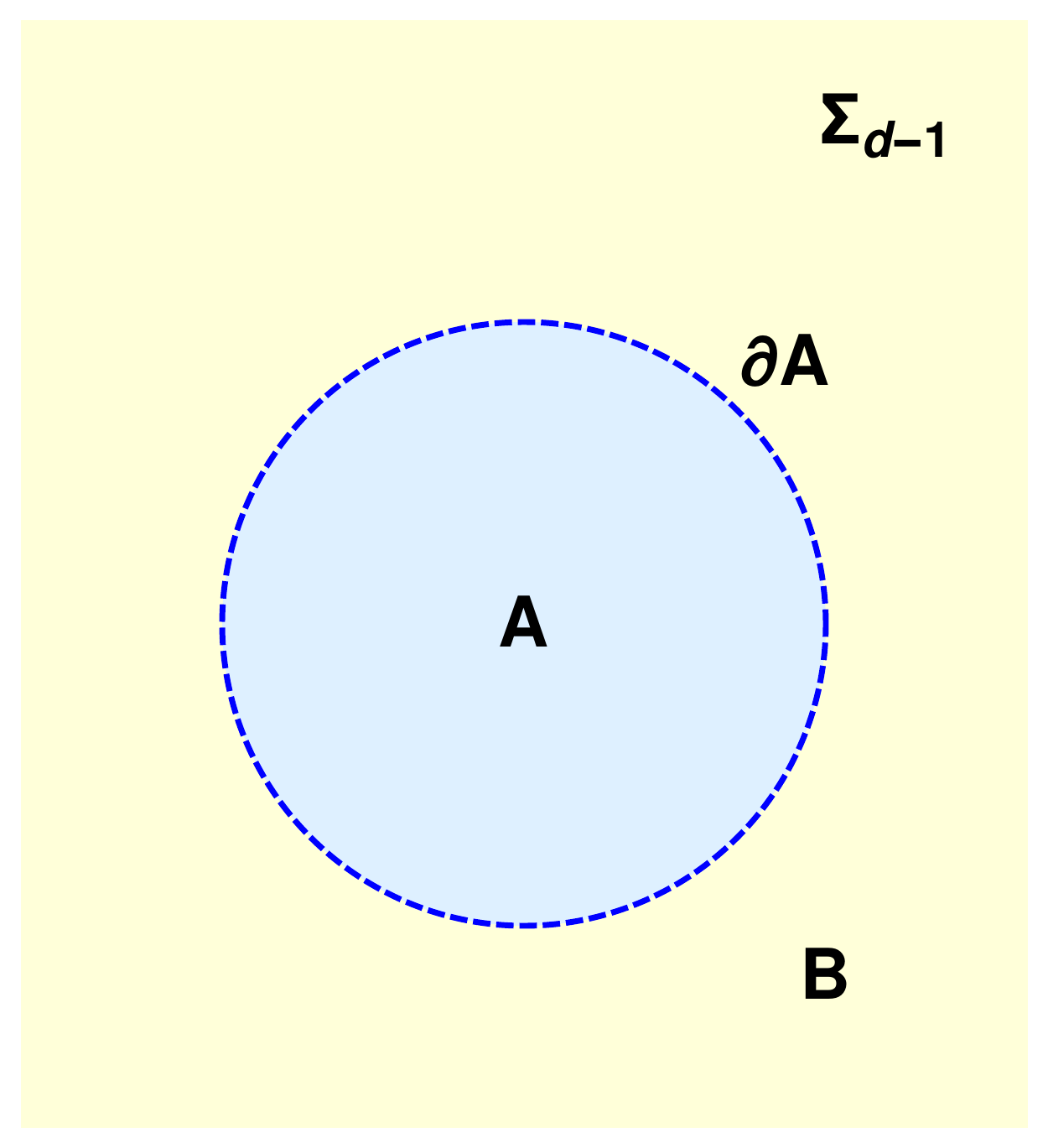}
\caption{%
Left: Example for bipartitioning a discrete quantum system by grouping lattice sites into two sets $A$ and $B$. 
Right: In a continuous QFT on a $d$-dimensional Lorentzian manifold a Cauchy slice $\Sigma_{d-1}$ is split into two spatial regions $A$ and $B$ such that $A\cup B=\Sigma_{d-1}$. }
\label{Fig:regionAB}
\end{figure}

We start with the general definition of entanglement entropy.
Let us assume a quantum system in the ground state $|\psi\rangle$ which is pure in the Hilbert space $\mathcal{H}$ and normalized ($\langle\psi|\psi\rangle=1$) such that the corresponding density matrix $\rho=|\psi\rangle\langle\psi|$ has unit trace $\mathrm{Tr}(\rho)=1$.
In order to define entanglement entropy it is necessary to divide the whole system into two parts $A$ and $B=\bar{A}$ such that the total Hilbert space takes a direct product form $\mathcal{H}=\mathcal{H}_A\otimes\mathcal{H}_B$\footnote{We want to stress that for theories without gauge symmetries the total Hilbert space can always be  written in a tensor product form. However, for pure gauge theories without any matter degrees of freedom no such decomposition exists and the prescription remains ambiguous. For a recent review on this issue see \cite{Pretko:2018nsz}.}. The shared boundary $\partial A$ is called \textit{entangling surface}\index{entangling surface}.
We want to stress that this dividing procedure only assigns the degrees of freedom in the system to two disjoint sets and does not correspond to a physical change of the system. 
This situation is illustrated in Figure \ref{Fig:regionAB} for the case of a discrete lattice system, where the lattice points are grouped into two sets, and a continuous quantum field theory in which the bipartitioning is realized by splitting a Cauchy slice into two spatial regions.
As a next step let us define the \textit{reduced density matrix}\index{reduced density matrix} $\rho_A$ of the subsystem $A$ by taking the \textit{partial trace}\index{partial trace} over the degrees of freedom in $B$
\begin{equation}\label{Eq:ReducedDensityMatrix}
\rho_A=\mathrm{Tr}_B(\rho)\,.
\end{equation}
Intuitively, $\rho_A$ corresponds to the density matrix for an observer who is causally connected to the degrees of freedom in $A$ only.
The exact definition of the trace operation in \eqref{Eq:ReducedDensityMatrix} depends on the theory. For instance in a discrete quantum system for which an orthonormal basis $\{|i_B\rangle,i=1,\ldots, \mathrm{dim}(\mathcal{H}_B)\}$ of the Hilbert space $\mathcal{H}_B$ is available the partial trace is defined as
\begin{equation}\label{Eq:ReducedDensityMatrixQM}
\mathrm{Tr}_B(\rho)=\sum_{i} \langle i_B | \rho| i_B \rangle\,.
\end{equation}
The entanglement entropy associated to the subregion $A$ is then defined as the \textit{von Neumann entropy}\index{von Neumann entropy} of the reduced density matrix
\begin{equation}\label{Eq:DefEE}
S_A=-\mathrm{Tr}_{A}(\rho_A \log(\rho_A))\,.
\end{equation}
It quantifies how much entanglement is in the state $|\psi\rangle$ when bipartitioning the system.
In general a state is called \textit{separable}\index{separable} when it can be written in product form $|\psi\rangle=\sum_{ij} c_i c_j | i_A\rangle | j_B \rangle=|\psi_A\rangle \otimes |\psi_B\rangle$, such that the reduced density matrix becomes pure $\rho_A=|\psi_A\rangle\langle\psi_A|$ in which case the entanglement entropy vanishes $S_A=0$. A state which is not separable $|\psi\rangle=\sum_{ij} c_{ij} | i_A\rangle | j_B \rangle\neq|\psi_A\rangle \otimes |\psi_B\rangle$, where $c_{ij}\neq c_ic_j$, is called \textit{entangled}\index{entangled state} or \textit{mixed}\index{mixed state} and the entanglement entropy takes a finite value $S_A>0$.
Entanglement entropy is a measure for how much a given quantum state differs from a pure state. 
We want to stress that the amount of entanglement in a state depends on the Hilbert space in which it is defined. It is always possible to purify a state that is mixed in the original Hilbert space by extending the Hilbert space such that it is pure in the extended Hilbert space. This procedure is called \textit{entanglement purification}\index{entanglement purification}.

\begin{figure}[htb]
\center
\includegraphics[width=0.6\linewidth]{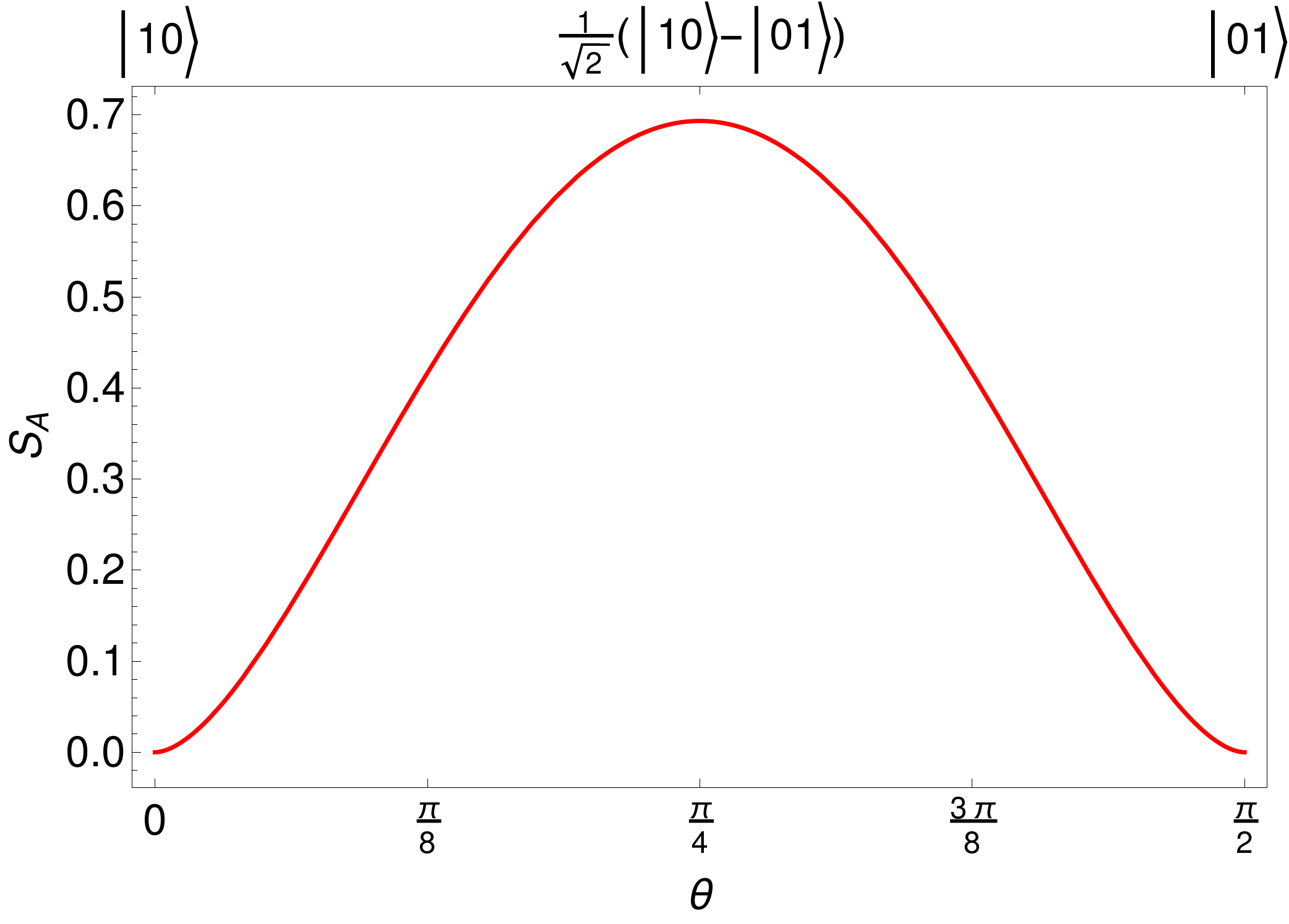}
\caption{%
Entanglement entropy for the one-parameter family of states given in \eqref{Eq:BellState}. For $\theta=0,\pi/2$ the state is pure and the entanglement entropy vanishes, where for $\theta=\pi/4$ one obtains a Bell state for which the entanglement entropy takes its maximum value $\log\mathrm{dim}(\mathcal{H}_A)=\log 2$.}
\label{Fig:Squbit}
\end{figure}

Until now we were not explicit about the quantum system under consideration. 
In order to build an understanding of the concept of entanglement and to see why the definition \eqref{Eq:DefEE} provides a useful measure for it, let us have a look at a simple discrete quantum mechanical system consisting of two interacting spin $1/2$ degrees of freedom.
The total Hilbert space has by construction a tensor product form $\mathcal{H}=\mathcal{H}_A\otimes\mathcal{H}_B$ with basis $\{ | 00 \rangle,| 01 \rangle,| 10 \rangle,| 11 \rangle \}$ where $| ij\rangle=|i_A\rangle\otimes|j_B\rangle$ for $i,j=0,1$.
Let us further consider a general state of the form
\begin{equation}\label{Eq:BellState}
|\psi\rangle=\cos(\theta) |01\rangle-\sin(\theta)|10\rangle\,,\quad\theta\in [0,\pi/2 ]\,,
\end{equation}
for which the entanglement entropy evaluates to
\begin{equation}S_A(\theta)=-\cos^2(\theta)\log(\cos^2(\theta))-\sin^2(\theta)\log(\sin^2(\theta))\,.
\end{equation}
The angle $\theta$ parametrizes a linear superposition of the pure states $|01\rangle=|0_A\rangle\otimes|1_B\rangle$, obtained for $\theta=0$, and $|10\rangle=|1_A\rangle\otimes|0_B\rangle$, obtained for $\theta=\pi/2$, which both satisfy individually $S_A=S_B=0$. For $0<\theta<\pi/2$ the state is not pure. In particular for $\theta=\pi/4$ one obtains a \textit{maximally entangled state}\index{maximally entangled state} $\frac{1}{2}(|01\rangle-|10\rangle)$, i.e. a superposition of all possible states with equal weight, which satisfies $S_A=\log \mathrm{dim}(\mathcal{H}_A)=\log(2)$. Such maximally entangled states are called \textit{Bell states}\index{Bell state} or \textit{Einstein-Podolski-Rosen pairs}\index{Einstein-Podolski-Rosen pair}. This is shown in Figure \ref{Fig:Squbit} for the simple example given in \eqref{Eq:BellState}.

Entanglement entropy satisfies a number of interesting inequalities \cite{nielsen2000quantum}.
\begin{itemize}
\item In bipartite systems $\mathcal{H}=\mathcal{H}_A\otimes\mathcal{H}_B$ the entanglement entropies of the respective subsystems satisfy \textit{subadditivity}\index{subadditivity}
\begin{equation}
 S_A+S_B\geq S_{A\cup B}\,.
\end{equation}
Subadditivity motivates the definition of a strictly positive quantity called \textit{mutual information}\index{mutual information}
\begin{equation}
I_{AB}= S_A+S_B-S_{A\cup B}\geq0\,.
\end{equation}
It is particularly useful in QFTs, where entanglement entropy is UV-divergent, because the divergent contributions cancel each other out and mutual information realizes a divergence free measure for entanglement.

\item \textit{Araki-Lieb inequality}\index{Araki-Lieb inequality} \cite{Araki1970}
\begin{equation}\label{Eq:Araki}
|S_A-S_B|\leq S_{A\cup B}\,.
\end{equation}

\item Tripartite systems $\mathcal{H}=\mathcal{H}_A\otimes\mathcal{H}_B\otimes\mathcal{H}_C$ satisfy \textit{strong subadditivity}\index{strong subadditivity}
\begin{equation}\label{Eq:SSA}
S_{A\cup B} +S_{B\cup C}\geq S_A+S_C\,.
\end{equation}
This inequality relies on unitarity of the underlying quantum systems and plays an important role in proving c- and F-theorems for renormalization group flows in QFTs.
Strong subadditivity is regarded as the most fundamental one because both, the Araki-Lieb inequality and subadditivity, are derivable from it \cite{Araki1970}.
\end{itemize}

There is another set of entropies called R\'{e}nyi entropies \cite{Renyi1961} which is defined in terms of the $n$-th moments of the reduced density matrix
\begin{equation}\label{Eq:Renyi}
S_A^{(n)}=\frac{1}{1-n}\log\mathrm{Tr}_{A}(\rho_A^n),\quad  n\in\mathbb{Z}_+  \,.
\end{equation}
After analytically continuing the definition to $n\in \mathbb{R}_+$ and taking the $n\to 1$ limit the R\'{e}nyi entropy converges to the entanglement entropy
\begin{equation}\label{Eq:replica}
S_A=\lim\limits_{n\to 1} S_A^{(n)}=-\lim\limits_{n\to 1}\partial_n\log\mathrm{Tr}_{A}(\rho_A^n)\,.
\end{equation}
This way of expressing the entanglement entropy is particularly useful for calculations in two-dimensional QFTs by the so called replica method as we will discuss in the next section. We can readily apply \eqref{Eq:replica} to the two spin system.
For simplicity we consider the maximally entangled state ($\theta=\frac{\pi}{4}$) for which the $n$-th power of the reduced density matrix is given by
\begin{equation}
\mathrm{Tr}_A(\rho_A^n)=2^{1-n}\,.
\end{equation}
Using this result in \eqref{Eq:replica} gives the expected result
\begin{equation}
S_A=-\lim\limits_{n\to 1}\partial_n\log 2^{1-n}=\log 2\,.
\end{equation}

\subsection{Entanglement Entropy in Quantum Field Theories}
Let us assume a relativistic local QFT on a Lorentzian $d$-dimensional spacetime $\mathcal{M}$ which is globally hyperbolic such that the notion of a Cauchy slice ($\Sigma_{d-1}$) is available. 
The specific structure of the spacetime is not essential for the following discussion so we can, for simplicity, assume it to be Minkowski $\mathcal{M}=\mathbb{R}^{1,d-1}$. 
One can then describe an instantaneous state of the system by its density matrix $\rho_{\Sigma}$ on $\Sigma_{d-1}$.
After bipartitioning the system according to $A\cup B=\Sigma_{d-1}$ (see right panel of Figure \ref{Fig:regionAB}), the reduced density matrix $\rho_A$ can be constructed by integrating $\rho_{\Sigma}$ over all field configurations in region B. 
The desired expression for the entanglement entropy then follows from \eqref{Eq:Renyi} and \eqref{Eq:replica}.

\begin{figure}[htb]
\center
\includegraphics[width=0.6\linewidth]{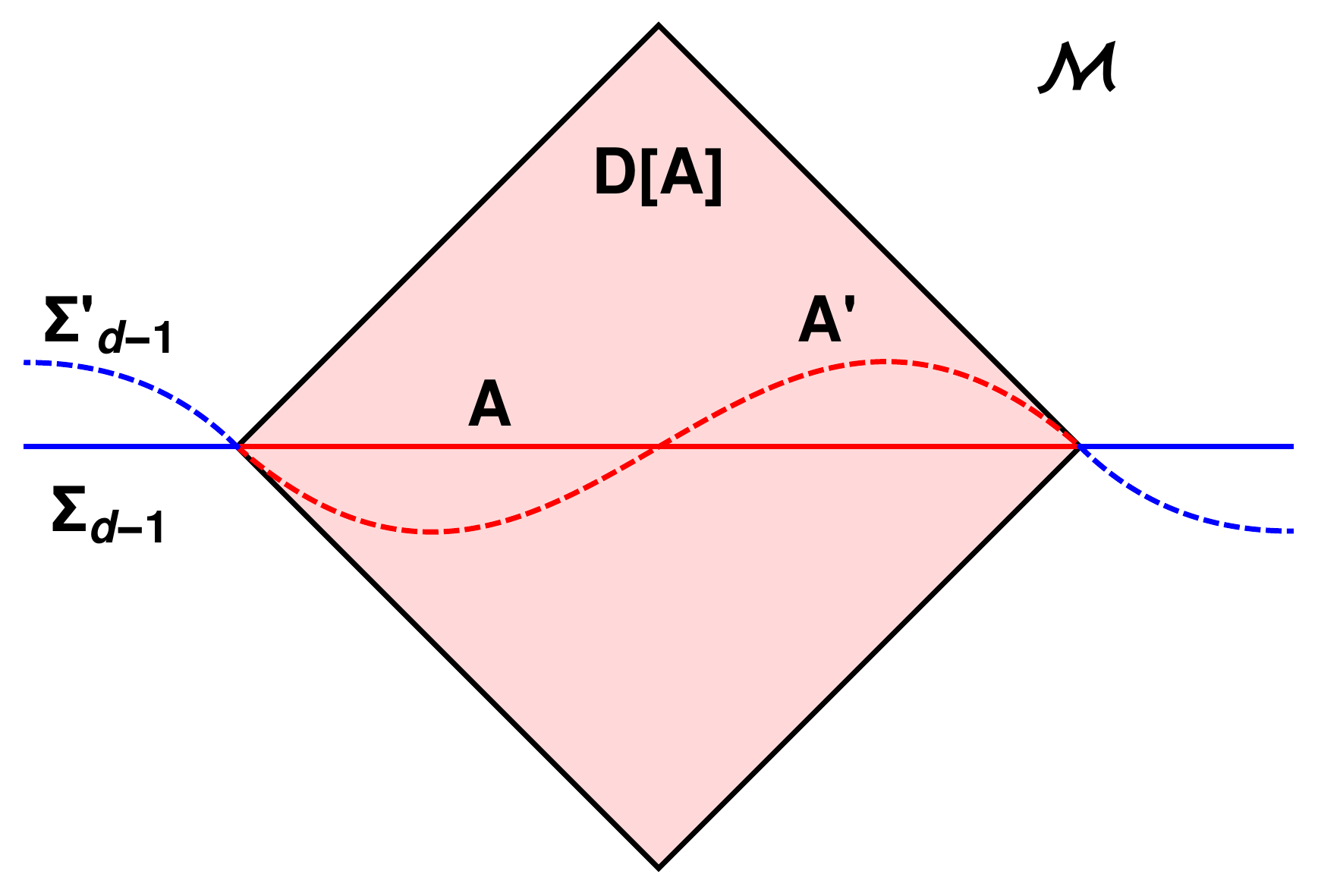}
\caption{Illustration of the causal domain $D[A]$ of the subregion $A$ which is a subset of the Cauchy slice $\Sigma_{d-1}$ of $\mathcal{M}$. The entanglement entropy associated to $A$ remains invariant under deformations $A\to A'$ that leave the causal domain unchanged $D[A]=D[A']$, and for which $A'$ is part of another Cauchy slice $\Sigma'_{d-1}$. }
\label{Fig:wedge}
\end{figure}

Causality puts some important constraints on entanglement entropy in any local relativistic QFT \cite{Headrick:2014cta}. Given a state $\rho_A$ on $A\subset \Sigma_{d-1}$ there exists a unitary transformation (localized on $A$) which allows to evolve it within its domain of dependence $D[A]$. The entanglement entropy remains invariant under such transformations. Consider now a deformation of the spatial domain $A\to A'$, which is part of another Cauchy slice $\Sigma_{d-1}'$, such that $D[A']=D[A]$. This situation is illustrated in Figure \ref{Fig:wedge}. The state $\rho_{A'}$ on the new slice is related to $\rho_A$ by a unitary transformation. Therefore entanglement entropy does not depend on the particular choice of Cauchy slice. It only depends on $D[A]$. This is why entanglement entropy is called a \textit{wedge observable}\index{wedge observable}. 

Since we are in the continuum we anticipate UV-singularities coming from contributions in the neighborhood of the entangling surface $\partial A$.
Physically, the major contributions to the entanglement entropy originate from Einstein-Podolski-Rosen pairs across $\partial A$ which suggests that the most divergent contribution scales like the area of the entangling surface. 
This is known as the \textit{area law of entanglement entropy}\index{area law of entanglement entropy}\cite{Srednicki:1993im,Bombelli:1986rw}.
The analogy in the scaling behavior to the entropy of black holes initiated the interest in entanglement entropy \cite{Bombelli:1986rw}.

The divergences in entanglement entropy are usually regulated using a UV-cutoff $\epsilon$, which effectively introduces a minimal length scale or, equivalently, a maximal energy scale in the theory. 
In general one obtains the following scaling behavior for entanglement entropy in the UV
\begin{equation}\label{Eq:UV-scaling}
S_A=\begin{cases}
 s_{d-2}(\frac{L}{\epsilon})^{d-2}+s_{d-4}(\frac{L}{\epsilon})^{d-4}+\ldots+s_1 \frac{L}{\epsilon}+(-1)^{\frac{d-1}{2}}s_{0}+\mathcal{O}(\epsilon)\,,\quad &\mbox{if } d\, \mathrm{odd}\,, \\
 s_{d-2}(\frac{L}{\epsilon})^{d-2}+s_{d-4}(\frac{L}{\epsilon})^{d-4}+\ldots+(-1)^{\frac{d-2}{2}}s_{0}\log(\frac{L}{\epsilon})+\mathcal{O}(\epsilon^0)\,,\quad &\mbox{if } d\, \mathrm{even}\,,
\end{cases} 
\end{equation}
where $L$ denotes a characteristic length scale of the entangling region (like the radius in case of a spherical entangling region). The coefficient $s_0$ turns out to be universal, where the coefficients $s_{i>0}$ depend on the particular regularization and the shape of the entangling surface. The appearance of the logarithmic contribution originates in conformal anomalies which only exist in even dimensions.
In the limiting case of $d=2$, where $\partial A$ degenerates to two disconnected points, the area law fails in the sense that one obtains to leading order a $\log$-contribution $S_A\!\propto\! \log(L/\epsilon)$ instead of $S_A\!\propto\! (L/\epsilon)^{d-2}$.

The IR behavior of entanglement entropy can be analyzed by assuming a fixed value for the cutoff $\epsilon$ and studying the scaling with the characteristic length scale $L$.
Ground states typically exhibit an area law behavior $S_A\propto L^{d-2}$, where excited states, like thermal ensembles at high temperature, show a volume law $S_A\propto L^{d-1}$.
This characteristic scaling will become particularly clear once we have the holographic picture of entanglement entropy available. 
There are important exceptions to the general rule, namely systems with Fermi surfaces, which are known to violate the area law behavior, due to the presence of an IR-scale given by the Fermi momentum $k_F$, and one obtains instead $S_A\propto L^{(d-1)}\log(k_F L)$ \cite{Swingle:2011np}. 

Explicit calculations of entanglement entropy within QFT can be done using the so-called \textit{replica method}\index{replica method}\cite{Calabrese:2004eu}.
In this approach the trace of the $n$-th power of the reduced density matrix is expressed in terms of a path integral $\mathcal{Z}$ over a $n$-branch cover $\mathcal{M}_n$ of the original spacetime manifold $\mathcal{M}$
\begin{equation}
\mathrm{Tr}(\rho_A^n)=\mathcal{Z}[\mathcal{M}_n]\,.
\end{equation}
This allows to write the R\'{e}nyi, by using \eqref{Eq:Renyi}, in the following way
\begin{equation}
S_A^{(n)}=\frac{1}{1-n}\log\left(\frac{\mathcal{Z}[\mathcal{M}_n]}{\mathcal{Z}[\mathcal{M}]^n}\right)\,.
\end{equation}
The entanglement entropy follows then from \eqref{Eq:replica}, by analytically continuing to $n\in\mathbb{R}_+$, and sending $n\to1$. Such calculations turn out to be tractable only for few examples, like free field theories \cite{Srednicki:1993im} in $d>2$ and for interacting CFTs in 1+1 dimensions \cite{Holzhey:1994we}, where the conformal symmetry algebra $so(2,d)$ gets enhanced and becomes of Virasoro type, which simplifies the calculation significantly.
A technical discussion of path integral methods to compute entanglement entropy in $2d$ CFTs is out of the scope of this thesis; we will rather state here some important results which have been obtained in closed form in the literature.
These results will also provide important consistency checks for the holographic calculations of entanglement entropy we present later.

For a single interval $A$ for the vacuum state in a CFT$_2$ on $\mathbb{R}^{1,1}$ one obtains
\begin{equation}
S_A=\frac{c}{3}\log\left(\frac{l}{\epsilon}\right)\,,
\end{equation}
where $c$ denotes the central charge in the CFT$_2$, and $l$ parametrizes the spatial size of the entangling region $A=\{x\in \mathbb{R}|-l/2\leq x \leq l/2\}$.
For a thermal system with temperature $T=1/\beta$ one finds
\begin{equation}
S_A=\frac{c}{3}\log\left(\frac{\beta}{\pi \epsilon}\mathrm{sinh}\left(\frac{\pi l}{\beta}\right)\right)\,.
\end{equation}

Interestingly, the method is also applicable to dynamic systems like quantum-quenches giving the result \cite{Calabrese:2005in}
\begin{equation}\label{Eq:CFTquench}
S_A(t)\propto\begin{cases}
 \frac{\pi c}{6\epsilon} t &\mbox{if } t<\frac{l}{2}\,, \\
 \frac{\pi c}{12\epsilon} l  &\mbox{if } t\geq \frac{l}{2}\,.
\end{cases} 
\end{equation} 
which means that $S_A(t)$ increases linearly until it saturates at $t=l/2$. Later in Chapter \ref{Chap:Numerics}  we will obtain exactly the same scaling in a holographic calculation where the quench is realized by the so-called Vaidya geometry.

We close this section by mentioning one major disadvantage of the pure field theoretic approach of computing entanglement entropy.
While the path integral method is a powerful tool to compute entanglement entropy in CFTs, its feasibility heavily relies on the enhanced symmetry in two dimensions.
In spacetime dimensions larger than two, such calculations are practically impossible, except in non-interacting theories for highly symmetric entangling regions (half spaces, balls and infinite strips).
As we will see in later sections, for theories with holographic dual, one can overcome this limitation.

\section{Holography and the AdS/CFT Correspondence}\label{Sec:Holography}
The goal of this chapter is to introduce and motivate the AdS/CFT correspondence, which is the main theoretical tool in this thesis.
We start with a discussion of the holographic principle, followed by a short review of basic concepts in string theory which are necessary to understand the arguments which led to the correspondence before we finally state the AdS/CFT correspondence.

It should be emphasized that the AdS/CFT correspondence has the status of a conjecture which currently lacks a proof. Nevertheless, as we will discuss below, in certain limits exist a number of non-trivial checks of the correspondence.
In that sense, we assume the validity of the AdS/CFT correspondence as a working hypothesis.

\subsection{The Holographic Principle}

The \textit{holographic principle}\index{holographic principle} goes back to an idea by 't Hooft \cite{tHooft:1993dmi} and Susskind \cite{Susskind:1994vu}, who conclude, from unitarity and counting arguments, that the physics involving gravity in a given number of dimensions can be completely captured by a lower-dimensional description without gravity. The holographic principle is motivated by the behavior of black holes which obey the famous \textit{Bekenstein-Hawking formula}\index{Bekenstein-Hawking formula} \cite{Bekenstein:1972tm,Bekenstein:1973ur,Bekenstein:1974ax}
\begin{equation}\label{BekensteinHawking}
S_{BH}=\frac{A}{4G_N \hbar}\,,
\end{equation}
which relates the entropy of a black hole $S_{BH}$ to the area of its event horizon $A$ (measured in Planck units) and $G_N$ denotes Newtons constant. The coefficient $\frac{1}{4}$ was originally fixed by the first law of thermodynamics $dE=TdS$ in Hawking's calculation of the black hole temperature \cite{Hawking:1974rv,Hawking:1974sw}.

The argument that leads to the holographic principle goes as follows:
any attempt to probe the structure of a theory down to the shortest distances, in order to learn about the microscopic degrees of freedom, will eventually probe energy densities which will dynamically form a black hole for which \eqref{BekensteinHawking} holds. Hence, the largest obtainable entropy for a given volume is that held by a black hole filling that volume. The black hole entropy puts an upper limit on the number of degrees of freedom inside a volume which is then given by the surrounding area.
This leads to the conclusion that in any theory of quantum gravity the number of degrees of freedom in any volume are of order one per unit area of the surface surrounding that volume.

The holographic principle is not specific about the theories it relates, only that they are formulated in different dimensions and that one of them includes gravity, and the other not. For that reason, the holographic principle is assumed to hold generally. For the same reason, specific examples can not be constructed from the principle itself, but rather need to be discovered.
The probably most famous example discovered so far is the AdS/CFT correspondence \cite{Maldacena:1997re}, which is a concrete realization of the holographic principle in string theory. It relates gravity in ($d+1$)-dimensional Anti-de Sitter space to a $d$-dimensional gauge theory.
Before we give a precise statement of the correspondence we will introduce in the next section some basic concepts in string theory which are necessary to understand the AdS/CFT correspondence.

\subsection{Strings and D-Branes}
The basic idea of string theory is to use one-dimensional objects, called \textit{strings}\index{string}, rather than point like objects, as fundamental building blocks of the theory.
Historically, string theory arose in the late 1960s as an attempt to describe the physics of strong interactions. The vibration modes of a string provided a way to explain the relation between mass and angular momentum $M^2\propto J$, the so-called \textit{Regge behavior}\index{Regge behavior}, of particles discovered in the first collider experiments. The formulation of Quantum Chromo Dynamics (QCD) and the discovery of confinement made string theory obsolete as a theory for the strong interaction.
Today string theory is seen as a framework that consistently unifies quantum theory and gravity, thus making it a promising candidate for a theory of quantum gravity.
Remarkably, the AdS/CFT correspondence in some sense closes this cycle by relating string theory again with gauge theories that are similar to QCD. 
\begin{figure}[htb]
\center
\includegraphics[width=0.25\linewidth]{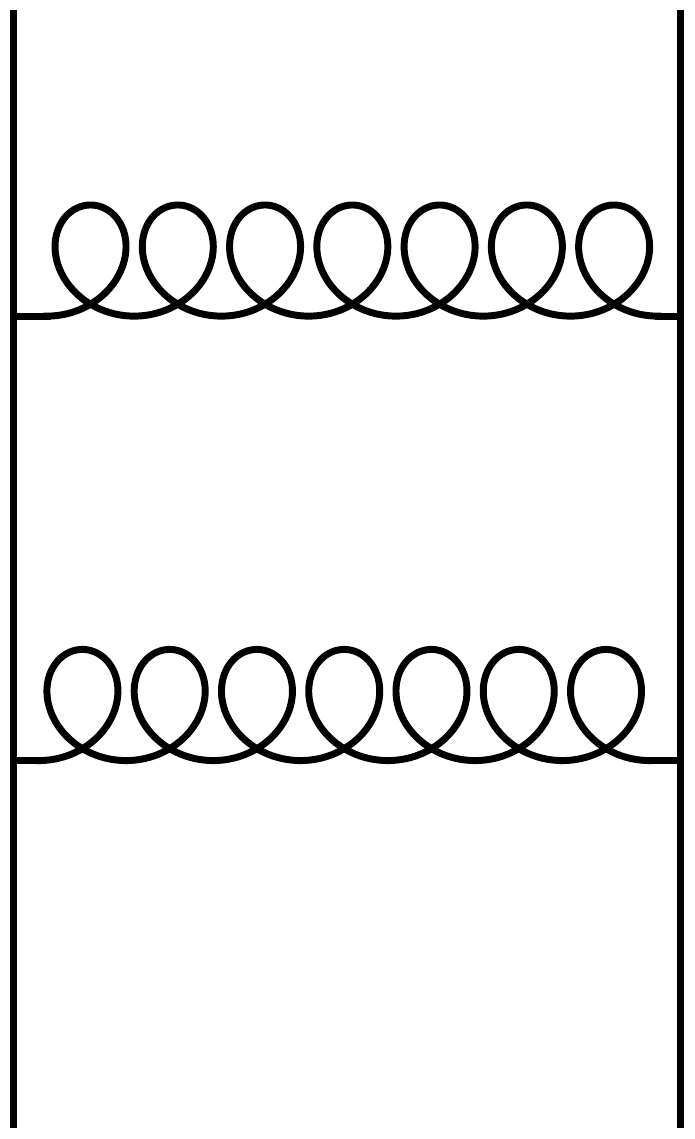}$\quad\quad\quad\quad\quad\quad$ \includegraphics[width=0.4\linewidth]{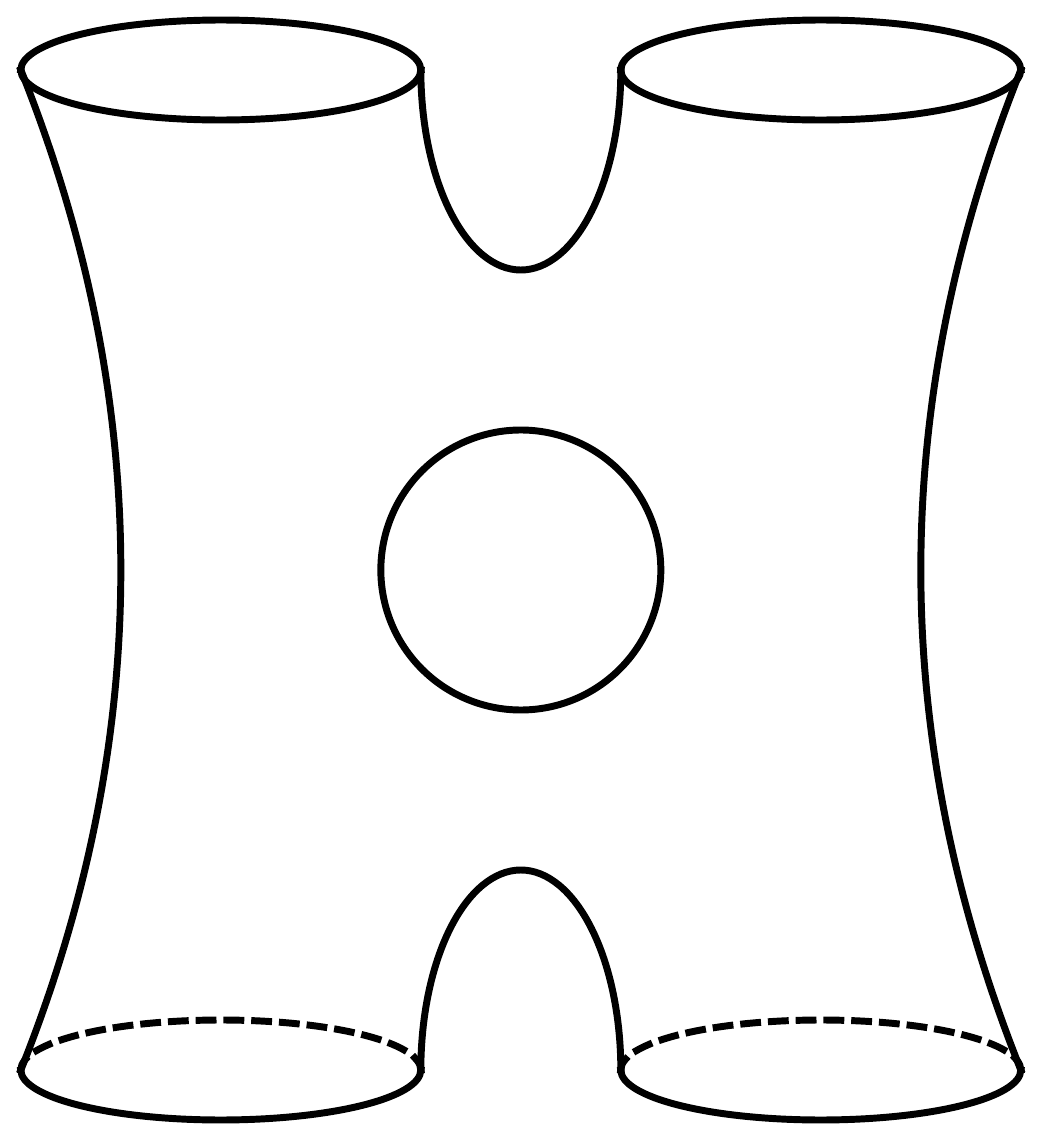}
\caption{%
Left: Feynman diagram of two propagating particles (straight lines) corrected by two-graviton (curly lines) exchange. 
Right: World sheet of two closed strings giving rise to a similar process such as shown left. Here the interaction is not point-like, thus avoiding the short range singularities that appear when quantizing general relativity. Figures are adopted from \cite{polchinski1998string}.}
\label{Fig:worldSheets}
\end{figure}

The strings of string theory are characterized by the \textit{string length}\index{string length} $l_s$ and by the dimensionless \textit{string coupling}\index{string coupling} $g_s$ which is not a free parameter but is determined dynamically by the expectation value of the dilaton field.
Being extended objects, strings can come in two topologically different realizations, namely in the form of open strings and closed strings. The vibration modes of open strings can be associated to excitations of gauge fields, whereas those of closed strings represent gravitational degrees of freedom.

As mentioned above, string theory provides a consistent framework for quantum gravity which elegantly resolves the issues of short-distance divergences resulting in non-renormalizability when quantizing Einsteins theory of general relativity. 
The argument \cite{polchinski1998string} goes as follows: consider, for example, the process of two propagating particles corrected by the exchange of two gravitons such as depicted in Figure \ref{Fig:worldSheets}.
In a quantized version of Einstein gravity the scattering amplitude of this process at the energy scale $E$ is proportional to $G_N^2 E^2\int dE' E'$,  which diverges for large internal energy transfer $E'$. The corresponding Feynman diagram is shown on the left side of Figure \ref{Fig:worldSheets}. In position space the divergences appear in the limit where the graviton vertices coincide. Moreover they grow worse with each additional graviton, resulting in non-renormalizability of the theory. In string theory the strings sweep out a world sheet and interactions are represented by joining and separating of the sheets such as shown on the right side of Figure \ref{Fig:worldSheets}. The interaction is not localized but rather smeared out in a way that Lorentz invariance is maintained and the divergences are avoided.

Quantization imposes strict constraints on the spacetime dimensions in which string theories can be formulated consistently. The consistency condition is the requirement that the Lorentz group remains free of anomalies on the quantum level.
Bosonic string theories are anomaly free only in 26 spacetime dimensions. However, string theories containing only bosonic degrees of freedom are, from the physics point of view, incomplete because they do not include fermions. Furthermore, they have a tachyonic ground state, i.e. the lowest energy state has negative mass squared, making the theory unstable. It is unknown if stable vacua for bosonic string theories exist.
A natural way to include fermions is to impose supersymmetry which requires to project out\footnote{The so-called GSO projection was originally introduced in \cite{1977NuPhB.122..253G}.} the tachyonic states from the spectrum which also leads to a stable vacuum.
The resulting theories are called \textit{superstring theories}\index{superstring theory} and they are anomaly free in ten spacetime dimensions only.
An active field of research, called string phenomenology, aims to construct the standard model of particle physics in four spacetime dimensions from such theories by compactifiying the superfluous six dimensions on so-called \textit{Calabi-Yau manifolds}\index{Calabi-Yau manifold} \cite{ibanez_uranga_2012}.  

\begin{figure}[htb]
\center
\includegraphics[width=0.7\linewidth]{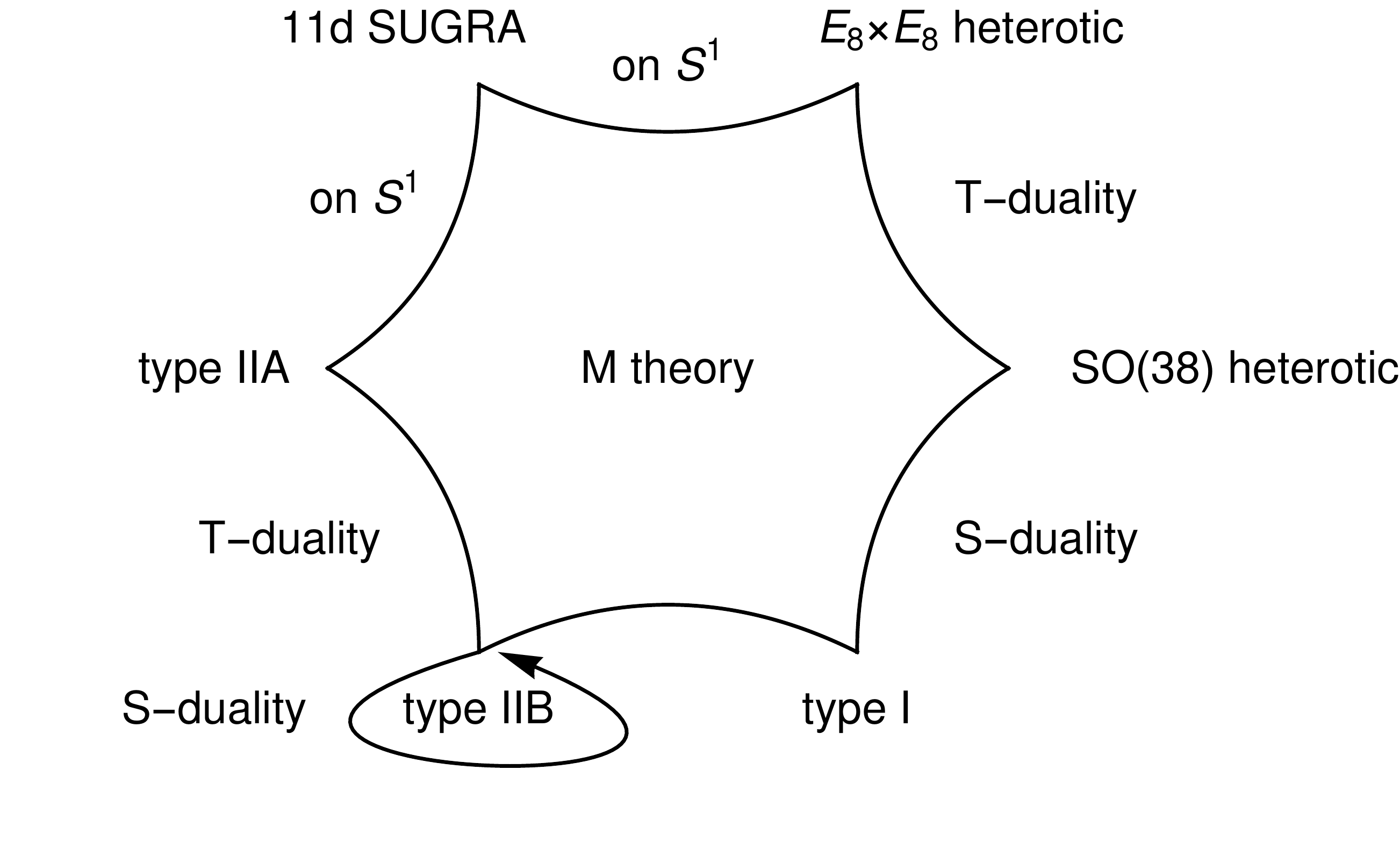} 
\caption{Web of dualities relating basic string theories and M-theory.}
\label{Fig:dualityWeb}
\end{figure}

It is natural to ask how many different ten-dimensional superstring theories exist.
As shown in the seminal paper by Michael Green and John Schwarz \cite{GREEN1984117}, only a few choices of gauge groups for the fields in the low energy limit eliminate quantum anomalies, which is an important consistency condition. The so-called \textit{Green-Schwarz anomaly cancellation mechanism} allows to restrict the gauge groups to $SO(32)$ and $E_8\times E_8$, resulting in only five different consistent superstring theories. Furthermore, these five superstring theories plus 11d supergravity (SUGRA) were found to be linked by a chain of dualities, known as the \textit{duality web}\index{duality web}\cite{Hull:1994ys,Hull:1995xh,Witten:1995ex}, which we illustrate in Figure \ref{Fig:dualityWeb}. All of them can be interpreted as effective theories obtained in certain limits of so-called \textit{M-theory}\index{M-theory}\cite{Duff:1996aw,Sen:1998kr}.

The canonical example of the AdS/CFT correspondence is realized in type IIB string theory, a theory of closed strings.
On the perturbative level string theory is a theory of one-dimensional strings only. However, on the non-perturbative level there exist also higher-dimensional solitonic objects, extending along $(p+1)$ directions $\{x_0,x_1,\ldots,x_p\}$, called Dirichlet-branes or D-branes for short \cite{Polchinski:1995mt}.
It can be shown that stable\footnote{In these cases the corresponding Dp-branes are the lightest states that carry a conserved charge and preserve part of the supersymmetry of the underlying theory.} Dp-branes exist in IIA theory for $p=0,2,4,6,8$ and in type IIB theory for $p=1,3,5,7$ \cite{Polchinski:1995mt}. 
Including D-branes adds open string sectors to type II theories which originally contain closed strings only.
The endpoints of the open strings in this sector have to obey Dirichlet boundary conditions in directions of the Dp-branes, i.e. the endpoints are allowed to move only along the $(p+1)$ directions of the Dp-brane and not in the transverse directions $\{ x_{p+1},x_{p+2},\ldots,x_9\}$.
The massless excitations of these open string sectors correspond to an Abelian gauge field $A_\mu(x_0,x_1,\ldots,x_p)$, describing longitudinal excitations of the brane, plus $(9-p)$ scalar fields $\phi^i(x_0,x_1,\ldots,x_p)$, accounting for transverse excitations, and their superpartners. Moreover, a stack of $N$ coincident Dp-branes gives rise to a non-Abelian $U(N)$ gauge theory living in the $(p+1)$-dimensions of the brane plus extra fields, all in the adjoined representation of the gauge group.

In 1995 Joseph Polchinski made the important discovery that D-branes are the same as extremal (mass = electric charge) p-brane solutions of supergravity which started the second superstring revolution \cite{Polchinski:1995mt}.
As we will explain in the next section, the two different effective descriptions of D3-branes in type IIB superstring theory in terms of a supersymmetric gauge theory and in terms of p-brane solutions of type IIB supergravity play a crucial role in the formulation of the AdS/CFT correspondence.

Let us close this section by summarizing the most important string theoretic concepts relevant for the AdS/CFT correspondence.
A consistent formulation of superstring theory requires ten spacetime dimensions. One of the five basic superstring theories is type IIB superstring theory which on the perturbative level consists only of closed strings and whose low energy spectrum contains the graviton.
Adding a stack of $N$ coincident D-branes to the theory includes an open string sector whose low energy spectrum includes a non-Abelian $U(N)$ gauge field depending only on the directions of the D-branes. D-branes have an alternative description in terms of extremal p-brane solutions of supergravity giving rise to a curved background geometry. In the next section we will use these statements to motivate the AdS/CFT correspondence.

\subsection{Motivation and Definition of the AdS/CFT Correspondence}

We are now ready to motivate and define the AdS/CFT correspondence in the way it was originally proposed by Juan Maldacena in 1997 \cite{Maldacena:1997re}. As mentioned in the beginning of this chapter, there exists currently no proof of the correspondence, rather a convincing chain of arguments which is supported by a number of non-trivial consistency checks.
The correspondence can be stated as follows \cite{ammon_erdmenger_2015}:
\newline

\fbox{
\begin{minipage}{0.9\linewidth}
$\mathcal{N}=4$ Super Yang-Mills (SYM) theory with gauge group $SU(N)$ and Yang-Mills coupling $g_{YM}$ is equivalent to type IIB string theory with string length $l_s$ and string coupling $g_s$ on AdS$_5\times$S$^5$ with curvature radius $L$ related by \eqref{Eq:AdSCFTparam}.
\end{minipage}
}
\\
\newline
\newline
The free parameters on the field theory side, the Yang-Mills coupling $g_{YM}$ and the rank of the gauge group $N$, are mapped by the correspondence to the free parameters on the string theory side, the string coupling $g_s$ and $L/l_s$ in the following way
\begin{equation}\label{Eq:AdSCFTparam}
g_{YM}^2=2\pi g_s,\quad\quad 2g_{YM}^2 N=L^4/l_s^4\,.
\end{equation}
Note that the relevant combination that characterizes finite-size effects of the strings is the dimensionless ratio $L/l_s$, i.e. the size of the strings relative to the curvature scale $L$ of the background geometry, and not $l_s$ by itself. The coupling strength of the field theory dual comes in form of the \textit{'t Hooft coupling}\index{'t Hooft coupling} $\lambda=g_{YM}^2N$ which was originally introduced in the 1970s as parameter in the large-$N$ expansion of non-Abelian gauge theories \cite{tHooft:1973alw}. Already then it was realized that the Feynman diagrams in the large-$N$ expansion organize in a similar way as the world sheets in the perturbative expansion of closed strings, thereby providing a first hint to the gauge/gravity correspondence.

After having stated the correspondence, let us now review how it was originally motivated.
The relevant  setup is type IIB superstring theory with a stack of $N$ coincident D3-branes embedded into (9+1)-dimensional Minkowski space. The D3-branes extend along the spacetime directions $\{x_0,x_1,x_2,x_3\}$ and are transversal to the remaining six directions (see Table \ref{D3braneEmbedding}). 
\begin{table}[t]  
 \caption[D3 brane embedding in AdS/CFT.]{Embedding of $N$ coincident D3-branes in ten-dimensional Minkowski space such as relevant for the AdS/CFT correspondence. Directions parallel to the D3-brane are indicated by $\bullet$, directions orthogonal to the branes by --. Endpoints of open strings are restricted to the four $\bullet$-directions, whereas closed strings can propagate in all ten spacetime directions. }\label{D3braneEmbedding}
 \centering
 \begin{tabular}{lccccccccccc} 
 \hline\hline
   &      & 0         & 1         & 2         & 3         & 4  & 5  & 6  & 7  & 8  & 9 \\ 
  \hline
   & N D3 & $\bullet$ & $\bullet$ & $\bullet$ & $\bullet$ & -- & -- & -- & -- & -- & -- \\
 \hline
\end{tabular}
\end{table}
As already mentioned, the stack of D3-branes can be viewed from two perspectives, namely from an open and a closed string point of view.
Which perspective is more appropriate depends on the value of $g_s$ which controls the coupling strength between open and closed strings.

\begin{figure}[htb]
\center
\includegraphics[width=0.45\linewidth]{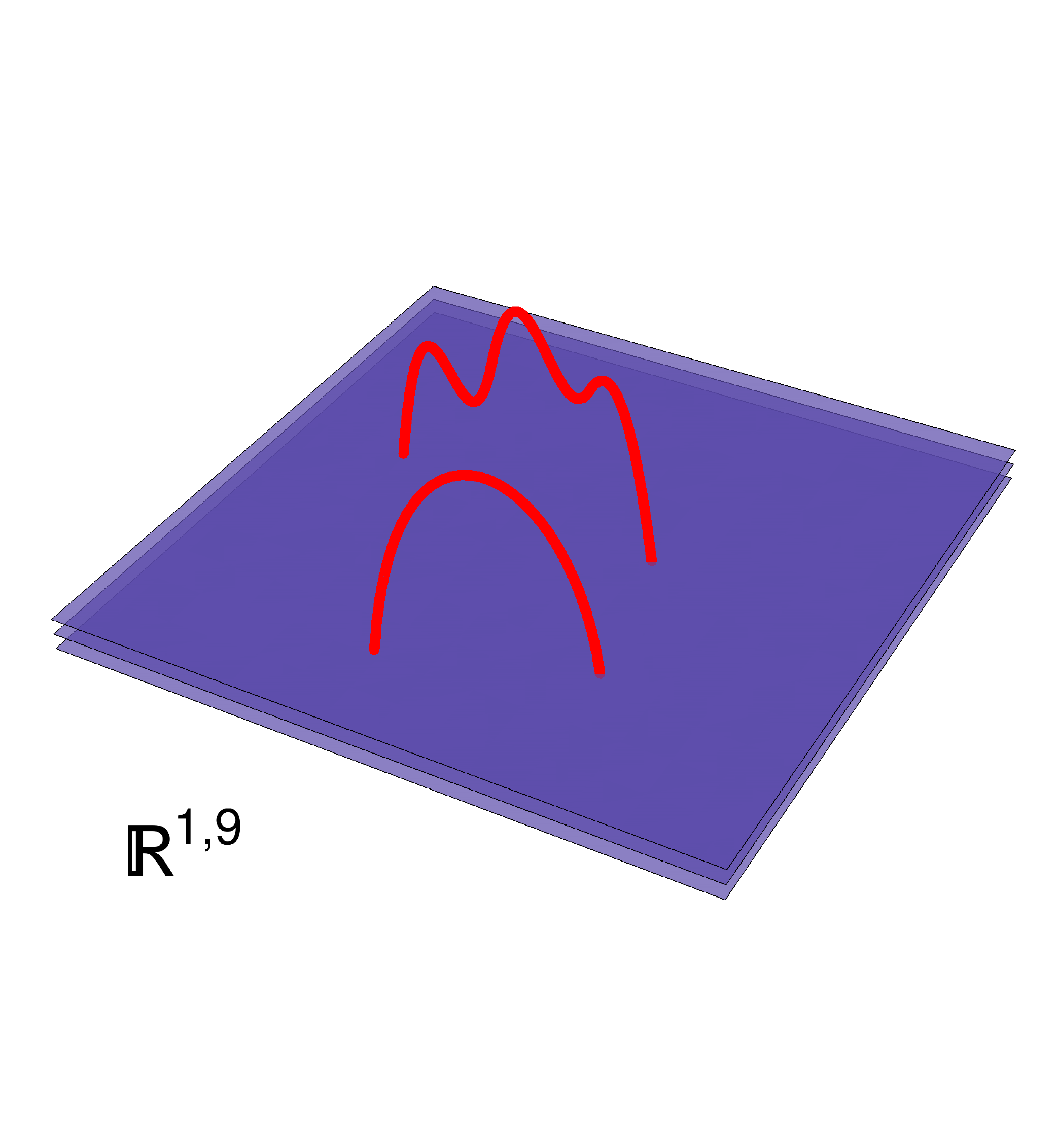}\quad\quad\quad \includegraphics[width=0.45\linewidth]{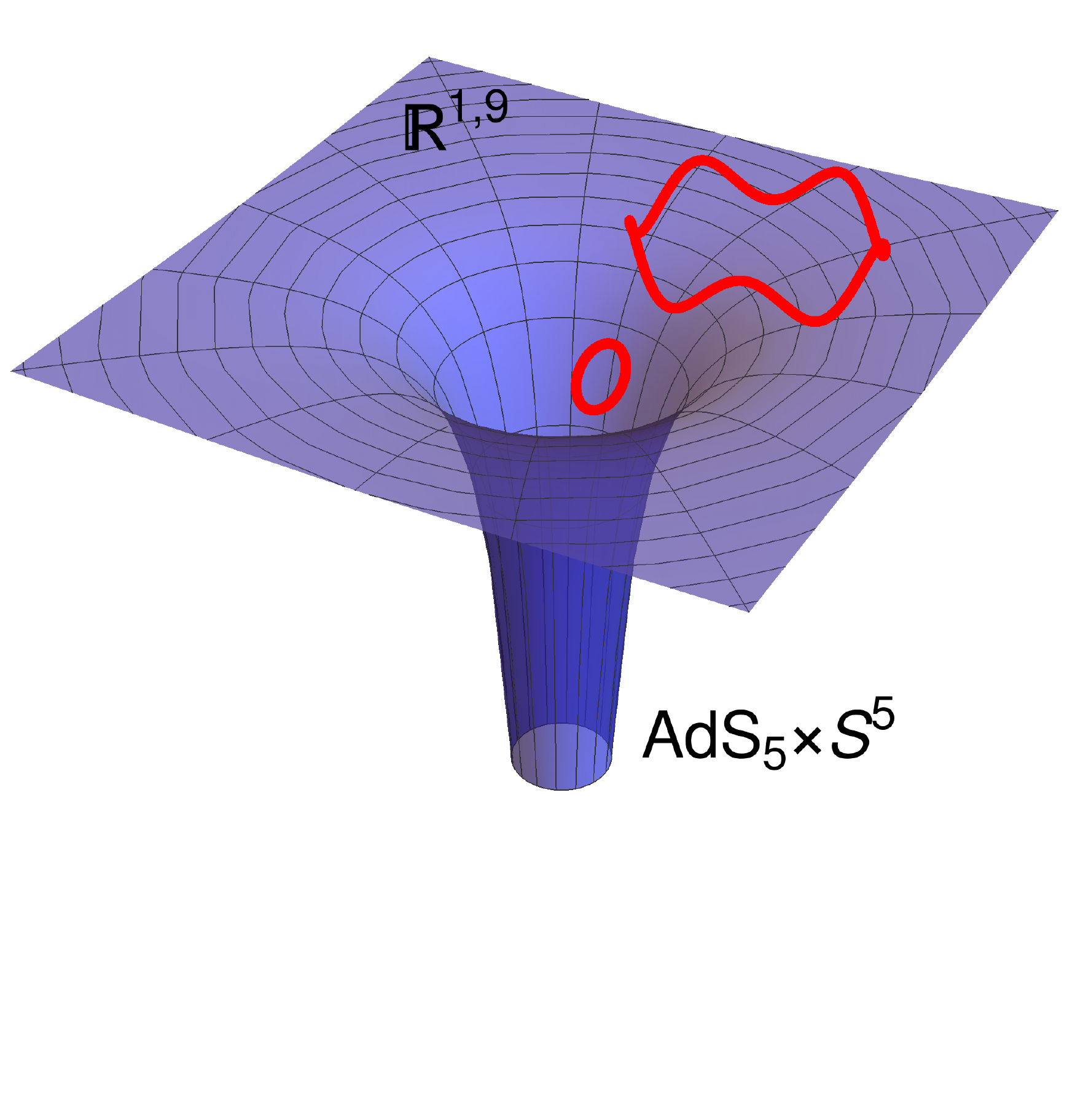}
\caption{%
Left: Open strings (red lines) stretched between a stack of $N$ coincident D3-branes (blue surfaces) giving rise to $(3+1)$-dimensional $\mathcal{N}=4$ SU(N) SYM theory in the open string picture. Right: Throat geometry induced by the D-branes in the closed string picture. }
\label{Fig:Dbarens}
\end{figure}

In the open string picture, which is valid for $g_s N\ll 1$, D-branes can be viewed as higher-dimensional objects where open strings end, such as depicted on the left side of Figure \ref{Fig:Dbarens}.
The dynamics of open strings in the low energy limit ($E\ll 1/l_s$) can then be described by a supersymmetric gauge theory living on the ($3+1$)-dimensional world volume of the D-branes. The fields of this theory are all in the adjoint representation of the $U(N)$ gauge group and can be organized into a $\mathcal{N}=4$ supermultiplet. The $U(1)$ part, describing simultaneous center of mass motions of the branes, decouples from the remaining $SU(N)\subset U(N)$.
The associated $SU(N)$ \textit{Super-Yang-Mills}\index{Super-Yang-Mills theory} (SYM) \textit{theory} is very special because it is the most general renormalizable $SU(N)$ gauge theory in four dimensions with global $\mathcal{N}=4$ supersymmetry. Furthermore, its $\beta$-function vanishes exactly,
 hence the coupling constant is independent of the energy scale which makes the theory \textit{superconformal}\index{superconformal}.  
The full system also contains closed strings which are allowed to propagate in all ten spacetime dimensions and correspond to excitations of the (9+1)-dimensional Minkowski background. The effective action for these massless string modes is given by ten-dimensional supergravity plus higher derivative terms proportional to powers of $\alpha'=l_s^2$.
In the so-called \textit{decoupling limit}\index{decoupling limit} ($\alpha'\to 0$) the effective interaction contribution between closed and open string sector vanishes.
We conclude that in the open string picture ($g_s N\ll 1$) in the decoupling limit the dynamics of open strings is governed by $\mathcal{N}=4$ SYM theory, where the dynamics of closed strings is described by type IIB supergravity on $\mathbb{R}^{9,1}$.
This closes the discussion of the open string picture.

In the closed string perspective, which is appropriate in the strong coupling limit  $g_s N\gg 1$, the stack of D3-branes is viewed as a p-brane solution of \textit{type IIB supergravity}\index{type IIB supergravity},  the low-energy limit of type IIB superstring theory. 
The metric part of this solution is given by
\begin{equation}
ds^2=H(r)^{-1/2}\eta_{\mu\nu}dx^\mu dx^\nu + H(r)^{1/2}\delta_{ij}dx^i dx^j\,, \quad H(r)=1+\frac{L^4}{r^4}\,,
\end{equation}
with $\mu,\nu=0,1,2,3$ and $i,j=4,\ldots,9$ and the radial coordinate is defined as $r^2=\delta_{ij}x^ix^j$.
The geometry consists of two different regions. Firstly, for $r\gg L$ where $H(r)\approx 1$ one obtains the asymptotically flat region and the effective low energy theory for the closed strings becomes type IIB SUGRA on $\mathbb{R}^{9,1}$. Secondly, the case $r\ll L$, where $H(r)\approx L^4/r^4$, corresponds to the so-called \textit{near-horizon}\index{near-horizon region} or \textit{throat region}\index{throat region} in which the metric is given by
\begin{equation}\label{Eq:Metric_AdS5xS5}
ds^2=\frac{r^2}{L^2}\eta_{\mu\nu}dx^\mu dx^\nu+\frac{L^2}{r^2}dr^2+L^2ds^2_{S^5}\,.
\end{equation}
The first two parts in \eqref{Eq:Metric_AdS5xS5} correspond to the metric of five-dimensional \textit{Anti-de Sitter space}\index{Anti-de Sitter space} (AdS$_5$) which is a maximally symmetric solution of the Einstein equations with negative cosmological constant and curvature radius $L$. The last part ($ds^2_{S^5}$) is the metric of a five-dimensional sphere (S$^5$) with radius $L$.
In the decoupling limit ($\alpha'\to 0$) one can show that the dynamics of closed strings in the asymptotic and the throat region decouple.
We arrive at the conclusion that in the decoupling limit and for $g_s N\gg 1$ closed strings are described by type IIB SUGRA on $\mathbb{R}^{9,1}$ in the asymptotic region and on AdS$_5\times$S$^5$ in the throat region.

The open and the closed string picture gives two different ways of describing the same physical system. Since type IIB SUGRA on $\mathbb{R}^{9,1}$ appears in both perspectives, also the other two effective theories, namely $\mathcal{N}=4$ SYM theory in four dimensions and type IIB SUGRA on AdS$_5\times$S$^5$ are conjectured to be equivalent. This is the main assumption on which the AdS/CFT correspondence is based.

A first plausibility check of the conjecture is to see if the symmetries of $\mathcal{N}=4$ SYM theory in four dimensions and type IIB SUGRA on AdS$_5\times$S$^5$ match, which is indeed the case. It turns out that the bosonic and fermionic symmetries in the respective theories can be combined into the supergroup $SU(2,2|4)$.
The bosonic part of $SU(2,2|4)$, for instance, is $SO(2,4)\times SO(6)$ and corresponds on the string theory side to the isometries of AdS$_5 \times$S$^5$. In $\mathcal{N}=4$ SYM theory the $SO(2,4)$ part is realized in form of the conformal group of $(3+1)$-dimensional Minkowski space and the $SO(6)$ part corresponds to the R-symmetry group under which the scalar fields in the theory transform as vectors.

The AdS/CFT correspondence provides a one-to-one map, the so-called \textit{holographic dictionary}\index{holographic dictionary}, between gauge invariant operators $\mathcal{O}(x)$ in the field theory side and their dual fields $\Phi(r,x)$ on the string theory side.
This field-operator map is established by equating the generating functional for one-particle irreducible (1PI) correlation functions of the CFT side and the string theory partition function \cite{Gubser:1998bc,Witten:1998qj} 
\begin{equation}\label{Eq:Zstring}
 \big\langle e^{\int d^dx\mathcal{O}(x)\phi_{(0)}(x)} \big\rangle_{CFT}=Z_{string}[\Phi(r,x)]\,,
\end{equation}
where the boundary conditions for the bulk field $\Phi$ are identified with the source functions $\phi_{(0)}$ in the CFT via $\phi_{(0)}(x)=\lim\limits_{r\to\infty} r^{\alpha_\phi} \Phi(r,x)$ with $\alpha_\phi$ determined by the asymptotic behavior of $\Phi$.

The AdS/CFT correspondence, and therefore \eqref{Eq:Zstring}, is conjectured to hold for all values of the parameters $g_s$ and $l_s$, or equivalently, for all values of $g_{YM}$ and $N$. 
From the theoretical point of view this is truly remarkable, because it allows to express a consistent theory of quantum gravity in terms of a quantum field theory.
However, expressing the right hand side of \eqref{Eq:Zstring} for generic values for the string coupling is extremely difficult and in explicit calculations one usually has to consider the limit where $g_s\ll 1$, while the string length $l_s$ (in units of $L$) is kept fixed.

At leading order in $g_s$ only tree level (genus zero) diagrams are kept in $Z_{string}$, which physically means that quantum gravity effects are neglected. On the field theory side this gives the so-called \textit{'t Hooft limit}\index{'t Hooft limit}, characterized by $N\to \infty$ and fixed (but arbitrary) 't Hooft coupling $\lambda$.

By neglecting finite size effects of the strings as well, i.e. sending $L/l_s\to \infty$, one obtains the \textit{point like approximation}\index{point like approximation} for gravity excitations which is governed by classical supergravity. On the field theory side this corresponds, via \eqref{Eq:AdSCFTparam}, to the infinite coupling limit $\lambda\to \infty$. 
From this relation one can also see that the AdS/CFT correspondence is a \textit{strong-weak duality}\index{strong-weak duality}, meaning when the field theory side becomes strongly coupled the gravity side becomes weakly curved and vice versa.
In this so-called \textit{supergravity limit}\index{supergravity limit} the right hand side of \eqref{Eq:Zstring} reduces, by the \textit{saddle point approximation}\index{saddle point approximation}, to 
\begin{equation}
Z_{string}[\Phi]\approx e^{S_{ren}[\Phi_c]}\,,
\end{equation}
where $S_{ren}[\Phi_c]$ is the renormalized on-shell supergravity action evaluated at a solution $\Phi_c$ to the classical equations of motion.
The connected (1PI) correlation functions in the large-$N$ CFT can then be obtained from variations of the on-shell action of the classical gravity dual
\begin{equation}\label{Eq:expval}
\langle \mathcal{O}(x_1)\ldots \mathcal{O}(x_n)\rangle_{CFT}=\frac{\delta^n S_{ren}[\Phi_c]}{\delta\phi(x_1)\ldots \delta\phi(x_n)}\Big|_{\phi=0}\,,
\end{equation}
where the expectation value $\langle\ldots\rangle_{CFT}$ can either be in vacuum or in a thermal state.

All applications of the AdS/CFT correspondence presented in this thesis assume the supergravity limit in which the AdS side is represented by Einstein gravity and the CFT side by a large-$N$ $\mathcal{N}=4$ SYM theory at infinite 't Hooft coupling $\lambda$. This drastic restriction is extremely useful because it makes first principle studies in strongly coupled gauge theories possible by solving classical Einstein equations of the dual gravity problem.
In the next section we will use \eqref{Eq:expval} to compute the holographic energy momentum tensor for $\mathcal{N}=4$ SYM theory from the dual gravity action.

\subsection{The Holographic Energy-Momentum-Tensor}
In this section we derive the holographic energy momentum tensor \cite{Balasubramanian:1999re,Emparan:1999pm} using the method of \textit{holographic renormalization}\index{holographic renormalization}\cite{deHaro:2000xn,Skenderis:2002wp}.
For simplicity we restrict the discussion to the case where the gravity action (in Euclidean signature) is of the following form
\begin{equation}\label{Eq:action}
S=-\frac{1}{16 \pi G_N}\int d^{d+1}x \sqrt{g}\left(R+\frac{d(d-1)}{L^2}\right)-\frac{1}{8\pi G_N}\int d^dx\sqrt{\gamma}K\,.
\end{equation}
Cases including in addition to the metric also bulk matter fields will not be considered in this thesis.
Next we express the asymptotically AdS metric in terms of a so-called \textit{Fefferman-Graham expansion} \index{Fefferman-Graham expansion}
\begin{equation}
ds^2=G_{MN}dx^M dx^N=L^2\left(\frac{d\rho^2}{4\rho^2}+\frac{1}{\rho}g_{\mu\nu}(\rho,x)dx^\mu dx^\nu\right)\,.
\end{equation}
The \textit{Fefferman-Graham theorem}\index{Fefferman-Graham theorem} \cite{2007arXiv0710.0919F} states that if $G_{MN}$ satisfies Einstein equations, then the metric $g_{\mu\nu}(\rho,x)$ can be expanded in the following way
\begin{equation}\label{Eq:FGexpansion}
g_{\mu\nu}(\rho,x)=g_{(0)\mu\nu}(x)+\rho g_{(2)\mu\nu}(x)+\ldots+\rho^{d/2}\left(\log(\rho) h_{(d)\mu\nu}(x)+g_{(d)\mu\nu}(x)\right)+\ldots\,,
\end{equation}
For given boundary conditions $g_{(0)\mu\nu}(x)$ the Einstein equations, following from \eqref{Eq:action}, can be solved order by order in $\rho$. The coefficient matrices up to $\mathcal{O}(\rho^{d/2})$ are functions of $g_{(0)\mu\nu}(x)$ only and only integer powers of $\rho$ appear up to that order in the expansion. In $d=4$, for example, the matrix $g_{(2)\mu\nu}$ is given by
\begin{equation}
g_{(2)\mu\nu}(x)=\frac{L}{d-2}\left(R_{(0)\mu\nu}-\frac{1}{2(d-1)}R_{(0)}g_{(0)\mu\nu} \right)\,,
\end{equation}
where $R_{(0)\mu\nu}$ ($R_{(0)}$) is the Ricci tensor (scalar) associated to the boundary metric $g_{(0)\mu\nu}$.
Furthermore, the logarithmic terms are present only for even $d$ and are related to the conformal anomaly which vanishes for odd $d$ \cite{Henningson:1998gx}.
The higher order coefficients contain contributions which can only be extracted from a full bulk solution.

Putting the asymptotic expansion of the metric into the action and evaluate it at a cutoff $\epsilon$ gives
\begin{equation}
S_\epsilon=-\frac{1}{16\pi G_N}\int d^{d+1}x\sqrt{g_{(0)}}\left(a_{(0)}\epsilon^{-d/2}+ a_{(2)}\epsilon^{-d/2+1}+\ldots -\log a_{(d)}\epsilon\right)+S_{finite}\,,
\end{equation}
where $S_{finite}$ summarizes the finite contributions and the coefficients of the divergent parts read
\begin{equation}
a_{(0)}=\frac{2(d-1)}{L},\quad a_{(2)}=\frac{L R_{(0)}}{2(d-1)},\quad a_{(4)}=\frac{L^3}{2(d-2)^2}\left(R_{(0)}^{\mu\nu} R_{(0)\mu\nu}-\tfrac{1}{(d-1)}R_{(0)}^2 \right)\,.
\end{equation}
A renormalized version of the action is obtained by adding an appropriate counterterm $S_{ct}$ that cancels the divergences\footnote{Note that this way of renormalizing the action is ambiguous, because it is possible to include in $S_{ct}$ terms that are finite in the $\epsilon\to 0$ limit. In the following we neglect such terms.} in the limit $\epsilon\to 0$
\begin{equation}
S_{ren}=\lim\limits_{\epsilon\to 0}(S_\epsilon+S_{ct})\,.
\end{equation}
The holographic energy momentum tensor is then obtained by varying the renormalized action $S_{ren}$ with respect to the boundary metric $g_{(0)\mu\nu}$
\begin{equation}
\langle T_{\mu\nu}(x) \rangle =-\frac{2}{\sqrt{g_{(0)}}}\frac{\delta S_{ren}}{\delta g_{(0)}^{\mu\nu}(x)}\,.
\end{equation}
In $d=4$ this gives the following expression for the holographic energy momentum tensor \cite{deHaro:2000xn}
\begin{equation}\label{Eq:HolEMT}
\langle T_{\mu\nu} \rangle=\frac{4}{16\pi G_N}\left( g_{(4)\mu\nu}+\frac{1}{8}\left(\mathrm{Tr} g_{(2)}^2-(\mathrm{Tr}g_{(2)})^2\right)g_{(0)\mu\nu}-\frac{1}{2}(g_{(2)}^2)_{\mu\nu}+\frac{1}{4}g_{(2)\mu\nu}\mathrm{Tr}g_{(2)}\right)\,.
\end{equation}

\subsection{AdS/CFT and the Quark Gluon Plasma}
In heavy ion collisions at the Relativistic Heavy Ion Collider (RHIC) and the Large Hadron Collider (LHC) at CERN a deconfined state of quarks and gluons, called quark gluon plasma (QGP) is produced.
This plasma has several remarkable properties including a very small shear viscosity over entropy density ratio $\eta/s$ whose explanation provides a theoretical challenge to first principle QCD approaches.
The weak and strong coupling results for this dimensionless ratio turn out to differ parametrically and experimental data favor the strong coupling result $\eta/s=\frac{1}{4\pi}$ obtained from the AdS/CFT correspondence \cite{ Policastro:2001yc,Kovtun:2004de}. While strong coupling provides no challenge to first principle lattice QCD techniques, the inherent dynamical nature of the collision makes it not accessible to lattice simulations, which are best suited to study equilibrium properties. 
Today, the AdS/CFT correspondence can be seen as the only first principle approach in which real time calculations of strongly coupled non-Abelian gauge theories are feasible.

However, there are several subtleties one has to address when using the AdS/CFT approach.
It is a well-known fact that the vacua of QCD and $\mathcal{N}=4$ SYM theory have very different properties, but there are good arguments for why AdS/CFT can nevertheless teach important lessons about the QGP in heavy ion collisions.
For example, at $T> T_c$, where $T_c\approx 170 MeV$ is the temperature in QCD at which the hadron gas crosses over to a deconfined quark-gluon plasma phase, many of the differences become unimportant \cite{casalderrey-solana_liu_mateos_rajagopal_wiedemann_2014}. 
One might criticize that $\mathcal{N}=4$ SYM is superconformal, has no running coupling and therefore no confinement like QCD. However, at finite $T$ SUSY is explicitly broken and above $T_c$ both theories are non-confining. Therefore, the fact that QCD has a good quasiparticle description in terms of hadrons becomes irrelevant.
It is also true that QCD is significantly non-conformal just above $T_c$ but at higher temperatures the quark gluon plasma becomes more and more scale invariant.

Another fact is that AdS/CFT calculations are usually performed in the infinite coupling limit $(\lambda\to \infty)$, which is certainly not the case in experimentally realized QCD plasmas. 
Furthermore, QCD is asymptotically free, which means that high energy processes are weakly coupled. However, at temperatures slightly above $T_c$, which are accessible in heavy ion collision experiments, the QGP turns out to be strongly coupled.
It is an active field of research to include finite coupling corrections \cite{Grozdanov:2016zjj,Waeber:2018bea} in AdS/CFT simulations. As mentioned earlier, taking finite coupling corrections into account requires higher curvature terms in the classical gravity action, which can be technically involved.

A more severe challenge is to overcome the large $N$ limit. QCD has $N_c=3$ colors, but AdS/CFT calculations are typically tractable only in the $N_c\to \infty$ limit. Computing finite $N_c$ corrections is conceptually far more involved than computing finite coupling corrections, because it requires to take quantum corrections in the gravity theory into account.

QCD also has $N_f=N_c=3$ flavors in the fundamental representation, but in AdS/CFT it is usually necessary to work either in the \textit{quenched approximation}\index{quenched approximation}, where $0<N_f\ll N_c$, or in the \textit{Veneziano limit}\index{Veneziano limit} in which both $N_c$ and $N_f$ are taken to be large but having fixed ratio $N_f/N_c$ \cite{Jarvinen:2011qe}.

\section{Holographic Entanglement Entropy}\label{Sec:HolographicEE}
\begin{figure}[htb]
\center
\includegraphics[width=0.6\linewidth]{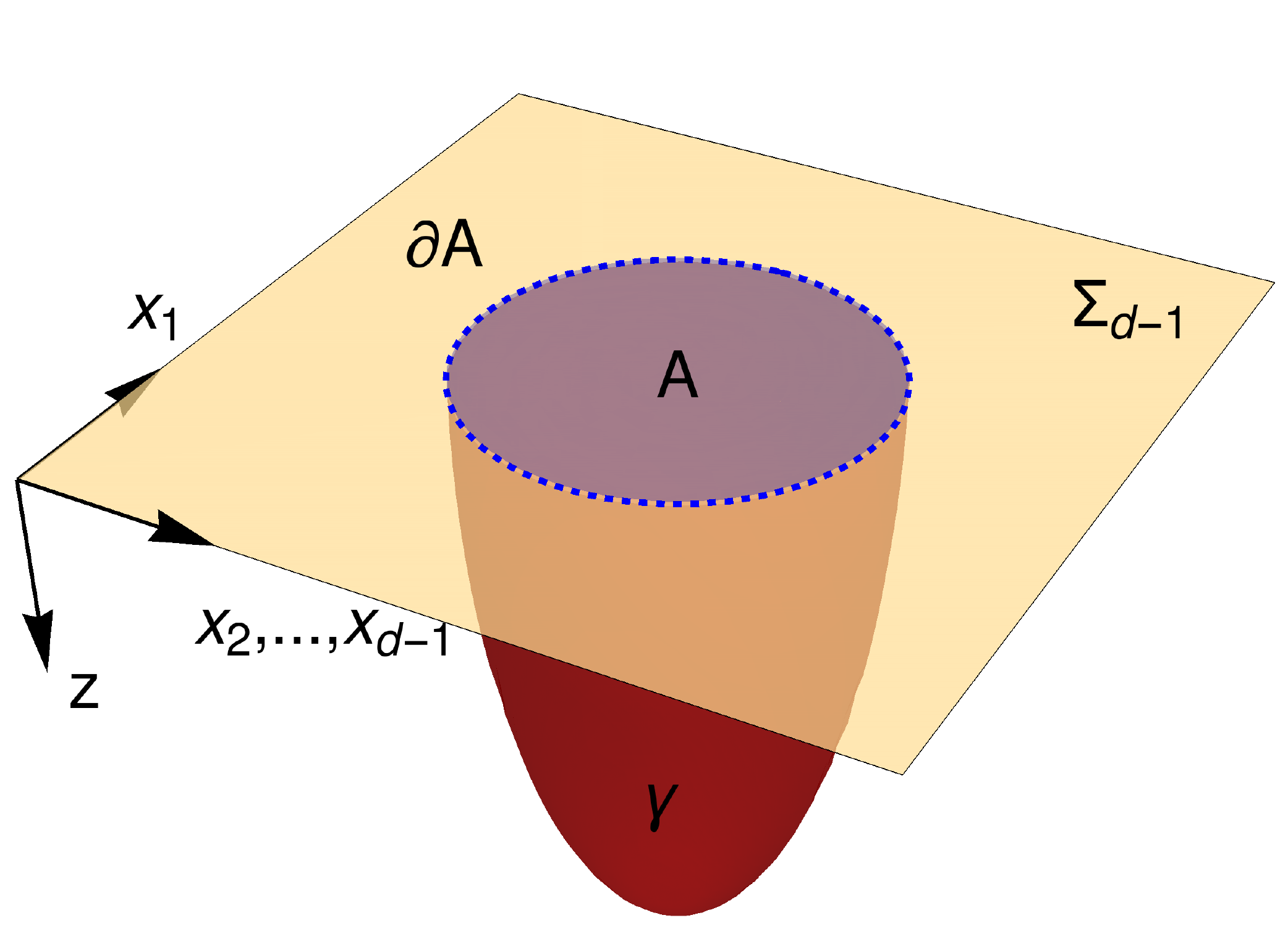}
\caption{Illustration of the (H)RT prescription for the holographic entanglement entropy. The red surface, anchored on the boundary $\partial A$ of the entangling region $A\in\Sigma_{d-1}$, is the extremal co-dimension two surface $\gamma$ whose surface area determines the entanglement entropy of region $A$.
}
\label{Fig:RTsurface}
\end{figure}

After having reviewed some basic aspects of entanglement entropy and the AdS/CFT correspondence we are now ready to introduce the holographic prescription for entanglement entropy.
The original proposal for the holographic entanglement entropy formula, valid for static states, was given in the seminal paper by Ryu and Takayanagi (RT) \cite{Ryu:2006bv} (see also \cite{Ryu:2006ef}) and is since then referred to as the RT-formula.
The proposal by Ryu and Takayanagi states that the entanglement entropy for a spatial region $A$, which is part of a Cauchy slice in a holographic CFT$_d$, is given by the minimal area $\mathcal{A}_{\partial A}$ of a co-dimension two surface $\gamma$ in the dual bulkspace time which is anchored ($\partial\gamma=\partial A$) on the boundary at $\partial A$ 
\begin{equation}\label{Eq:RTformula}
S_{A}=\frac{\mathcal{A}_{\partial A}}{4G_N}\,.
\end{equation}
This situation is illustrated in Figure \ref{Fig:RTsurface}.
Notice also the striking similarity to the Bekenstein-Hawking formula \eqref{BekensteinHawking}.
In addition, the bulk surface $\gamma$ has to satisfy a homology constraint which demands that the extremal surface must be smoothly retractable to the boundary region.
The latter is for instance relevant in global AdS black hole spacetimes where several topologically different surfaces can satisfy this constraint. If this is the case one has to pick the surface with the smallest area. 

Because entanglement entropy has a well defined prescription in terms of the time dependent density matrix in quantum field theory, it is expected that also a dual geometric prescription for the time dependent case exists. 
Such a prescription was given by Hubeny, Rangamani and Takayanagi (HRT) \cite{Hubeny:2007xt} who proposed the \textit{covariant holographic entanglement entropy}\index{covariant holographic entanglement entropy}. In their proposal the notion of minimal surfaces is replaced by the one of extremal surfaces. More precisely, in time dependent situations one has to find surfaces which extremize the corresponding area functional, subject to the boundary conditions $\partial \gamma=\partial A$ and the homology constraint mentioned above. 
We will not attempt to repeat existing derivations of the holographic entanglement entropy formulas, but rather refer the interested reader to \cite{Lewkowycz:2013nqa} where a proof of the original RT prescription is given, and to \cite{Dong:2016hjy}, where the HRT prescription was derived.

We also mention an alternative formulation of the extremal surface prescription in terms of so-called \textit{maximin surfaces}\index{maximin surface}, such as given by Wall \cite{Wall:2012uf}.
In this construction one picks a Cauchy slice in the bulk $\Sigma_{d}^{guess}$ such that $\partial \Sigma^{guess}_d=A\cup\bar{A}=\Sigma_{d-1}$. On this slice one finds a minimal surface $\gamma^{guess}$. One then varies the choice of the bulk Cauchy slice $\Sigma_{d}^{guess}$ and generates a family of minimal surfaces $\gamma^{guess}$ associated to those slices and computes their areas. The maximin surface is then defined as the minimal surface of maximal area in the entire family, which can be shown to be equivalent to the extremal surface associated to $A$ \cite{Wall:2012uf}.

Before closing this section we mention how general the holographic entanglement entropy formula stated in \eqref{Eq:RTformula} is.
The HRT proposal in terms of extremal surfaces in the classical bulk theory assumes the semiclassical limit provided by $N\to\infty$. Furthermore, it assumes classical Einstein gravity, i.e. only second derivative terms in the bulk action, which corresponds to the infinite coupling limit in the dual CFT. 
Including finite $N$ and finite coupling corrections requires to modify the holographic entanglement entropy prescription.
The case for higher curvature bulk theories, such as relevant for finite coupling corrections in the CFT, has been worked out in \cite{Dong:2013qoa,Bhattacharyya:2014yga}
\begin{equation}
S_A=\frac{1}{8G_N}\int d^{d-1}x\sqrt{h}\mathrm{D}_\mathcal{L}\,,
\end{equation}
where $h$ denotes the determinant of the induced metric on the relevant bulk surface and $\mathrm{D}_\mathcal{L}$ is given by a complicated functional, involving first and second variations of the Lagrangian, associated to the higher derivative theory, with respect to the Riemann tensor.
This expression can be seen as a generalization of the Iyer-Wald black hole entropy for higher derivative theories \cite{Iyer:1994ys}.

Quantum corrections in the bulk, such as necessary for finite $N$ corrections, can be organized in powers of the Newton constant 
\begin{equation}
S_A=S_{RT}+S_q+\mathcal{O}(G_N)\,,
\end{equation}
where the leading contribution $S_{RT}\propto 1/G_N$ is given by the RT-formula \eqref{Eq:RTformula}.
The sub-leading term $S_q$ has been worked out in \cite{Barrella:2013wja,Faulkner:2013ana} 
\begin{equation}\label{Eq:QantumCorrEE}
S_q= S_{bulk-ent}+\frac{\delta A}{4G_N}+\langle\Delta S_{W-like}  \rangle + S_{counterterms}\,,
\end{equation}
where the first term denotes the bulk entanglement, the second term is the change in area due to quantum backreactions of the classical background, the third term is the quantum expectation value of the Wald-like entropy \cite{Iyer:1994ys} and the last term is a collection of counter-terms necessary to render the expression finite.
In this thesis we stick to the simplest case of infinite $N$ and infinite coupling where the above mentioned corrections are not required.
In the next section we shall demonstrate the power of the geometric picture for entanglement entropy by reviewing the holographic proof of strong subadditivity.
\subsection{Holographic Proof of Strong Subadditivity}
\begin{figure}[htb]
\center
\includegraphics[width=0.33\linewidth]{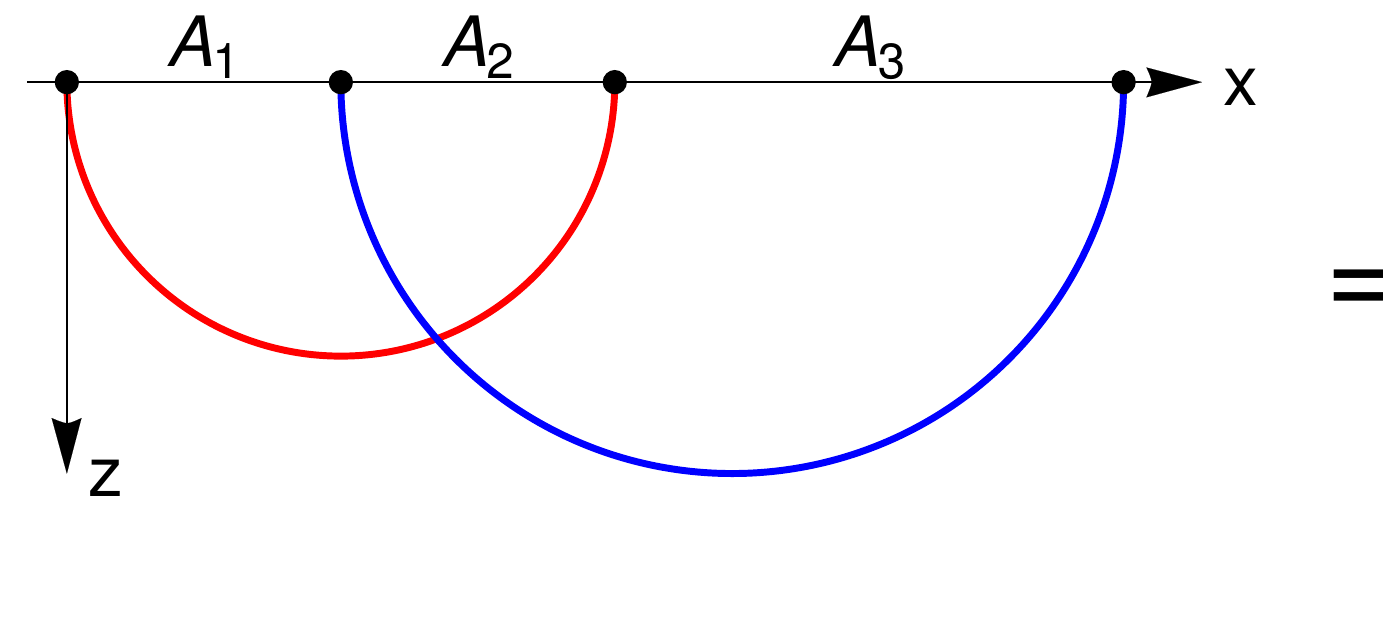}\includegraphics[width=0.33\linewidth]{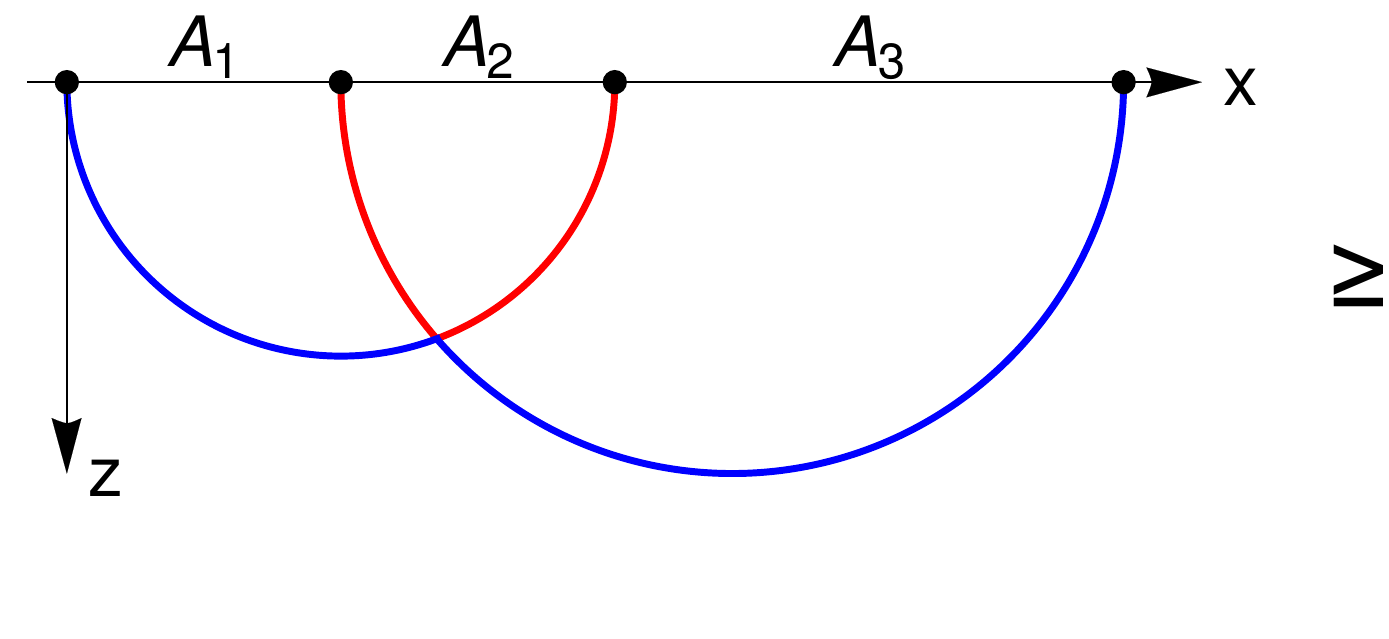}\includegraphics[width=0.33\linewidth]{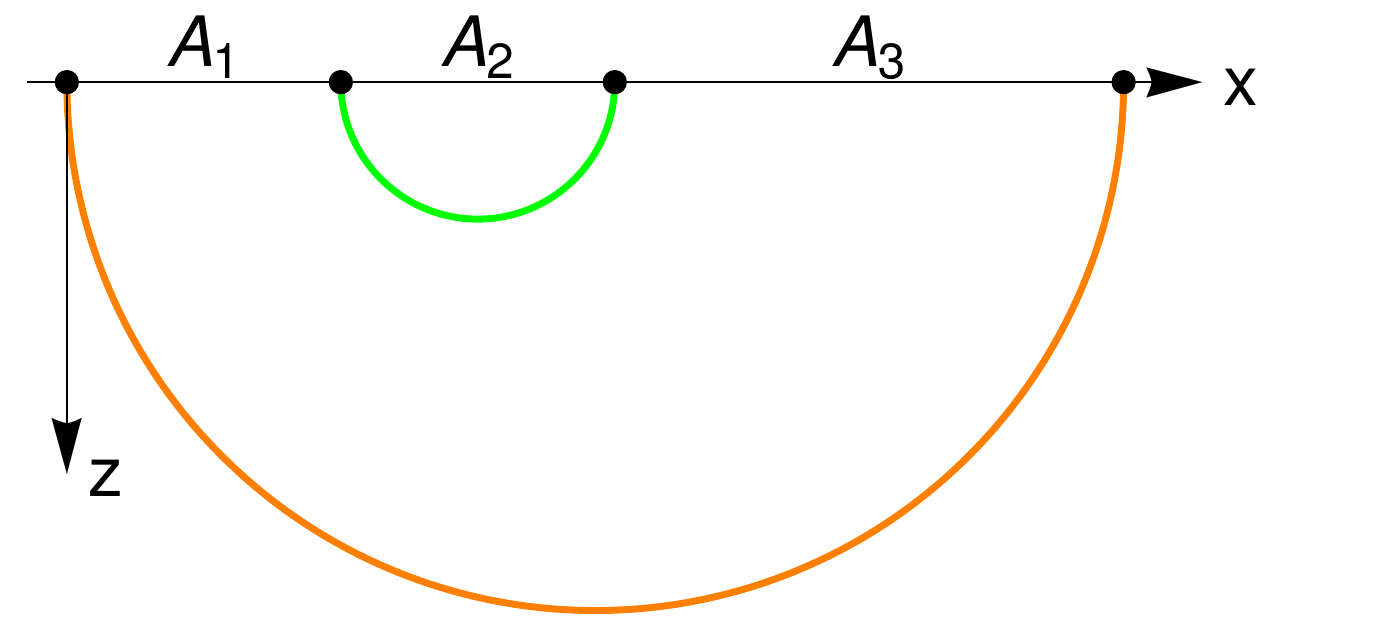}
\caption{Geometric proof of strong subadditivity in time independent spacetimes. 
}
\label{Fig:SSA}
\end{figure}

To appreciate the power of the holographic prescription let us revisit the strong subadditivity inequality introduced in \eqref{Eq:SSA}.
In the time independent case, where the entanglement entropy is given by the area of a minimal bulk surface, the holographic proof of strong subadditivity becomes extremely simple.
The argument goes as follows: using the RT-formula the strong subadditivity inequality \eqref{Eq:SSA} can be written as 
\begin{equation}\label{Eq:SAAhol}
\mathcal{A}_{1,2}+\mathcal{A}_{2,3}\geq\mathcal{A}_{1,2,3}+\mathcal{A}_{2}\,,
\end{equation}
where $\mathcal{A}_{i,j,\ldots}$ denotes the minimal surface areas associated to a combined region $A_i\cup A_j\cup \ldots$.
The key aspect of the geometric proof is to interpret the left hand side of \eqref{Eq:SAAhol} in two different ways. The first way is to consider the areas associated to regions $A_1\cup A_2$ and $A_2\cup A_3$ to be computed from genuine minimal surfaces such as illustrated by the red and blue curves in the first plot of Figure \ref{Fig:SSA}.
The second way is to intersect these surfaces, without changing the total area, and construct two new surfaces which are now associated to regions $A_2$ and $A_1\cup A_2\cup A_3$ such as shown in the center of Figure \ref{Fig:SSA}. In the latter picture the red and blue surfaces are in general not minimal because, by definition, already the green and orange surfaces shown on the right hand side are. Hence $(\mathcal{A}_{1,2}+\mathcal{A}_{2,3})$ must be larger than the sum of the areas of the true minimal surfaces $(\mathcal{A}_{1,2,3}+\mathcal{A}_{2})$. This completes the proof.


The proof for the time dependent case using the HRT prescription is more subtle, because the extremal surfaces associated to the different entangling regions in general do not reside on the same Cauchy slice. Nevertheless, an analogous proof was obtained in \cite{Wall:2012uf}, using the earlier mentioned maximin surface construction.

\subsection{Extremal Surfaces}
Computing entanglement entropy using the (H)RT prescription requires the computation of minimal/extremal co-dimension two surfaces in asymptotically AdS spacetimes.
In this section we provide the differential equations such extremal surfaces have to sa-tisfy.
We write the line element of a completely general asymptotic $\mathrm{AdS}_{d+1}$ spacetime as follows
\begin{equation}
ds^2=G_{\mu\nu}dx^\mu dx^\nu\,.
\end{equation}
A ($d-1$)-dimensional surface in the bulk can be written in terms of embedding functions $X^\mu=X^\mu(\sigma^a,z)$ which are parametrized with $d-2$ intrinsic coordinates $\sigma^a$ and the bulk coordinate $z$.
The induced metric on the surface is given by
\begin{equation}\label{inducedMetric}
H_{\alpha\beta}=\partial_\alpha X^\mu\partial_\beta X^\nu G_{\mu\nu}\,.
\end{equation}
The area functional can be written in terms of the induced metric
\begin{equation}
\mathcal{A}=\int dz d^{d-2}\sigma \sqrt{H[X]}.
\end{equation}
We are interested in stationary solutions $\delta\mathcal{A}=0$, which means we have to perform the variation of the surface functional with respect to the embedding functions
\begin{equation}\label{area}
\delta \mathcal{A}=\int dz d^{d-2}\sigma \delta\left(\sqrt{H[X]}\right).
\end{equation}
This variation is performed explicitly in Appendix \ref{App:AreaFunct}. Here we only state the final result which is the differential equation a surface extremizing the area functional has to satisfy
\begin{align}\label{Eq:extrSurface}
\frac{1}{\sqrt{H}}\partial_\alpha (\sqrt{H}H^{\alpha\beta}\partial_\beta X^\mu)+H^{\alpha\beta} \partial_\alpha X^\sigma \partial_\beta X^\nu \Gamma^{\mu}_{\sigma\nu}=0\,,
\end{align}
where $\Gamma^{\mu}_{\sigma\nu}$ denote the Christoffel symbols associated to the bulk metric $G_{\mu\nu}$.

We note that solving the above non-linear partial differential equation subject to boundary conditions describing the entangling region is a formidable task. Explicit solutions are only available for certain highly symmetric cases in which the shape of the entangling region in the boundary respects the symmetries of the bulk geometry. Some of these basic examples, where the extremal surface can be found either in closed form or using simple shooting methods will be discussed in the next section. For more complicated time dependent geometries and/or unregularly shaped entangling regions one typically has to resort to more sophisticated numerical methods like relaxation which we discuss in Chapter \ref{Chap:Numerics}.

\subsection{Basic Examples}\label{Sec:BasicExamples}
In this subsection we give some examples in which the geodesic or extremal surface equations can either be solved explicitly or by means of a simple shooting method. For simplicity we restrict ourselves to stripe-shaped entangling regions $A$ that preserve ($d-2$)-dimensional translational symmetry and are defined by 
\begin{equation}\label{Eq:stripe}
A=\{ x_1\in (-l/2,l/2),\, x_i\in \mathbb{R}\, \forall \, i=2,3,\ldots ,d-1\}\,.
\end{equation}
For ball-shaped entangling regions, preserving $SO(d-2)$ rotational symmetry, the calculations are completely analogous, but these will not be important for the later discussion.
From these basic examples we can already learn a lot about characteristic features of these surfaces. Furthermore, they play an important role in finding numerical solutions in more complicated situations. To unclutter the notation we set from here on the AdS-radius $L\equiv 1$.

\begin{figure}[htb]
\center
 \includegraphics[width=0.45\linewidth]{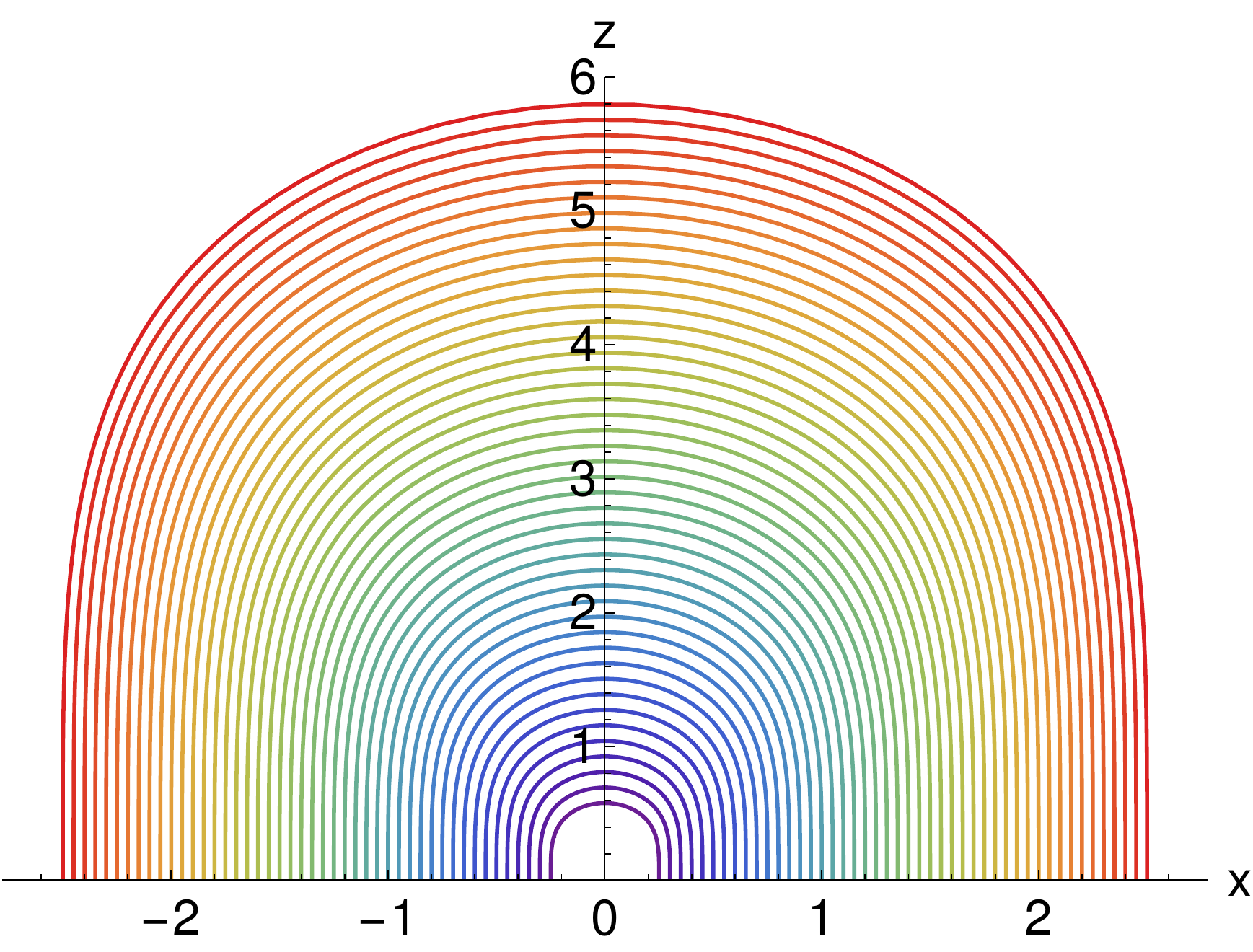} \quad\includegraphics[width=0.45\linewidth]{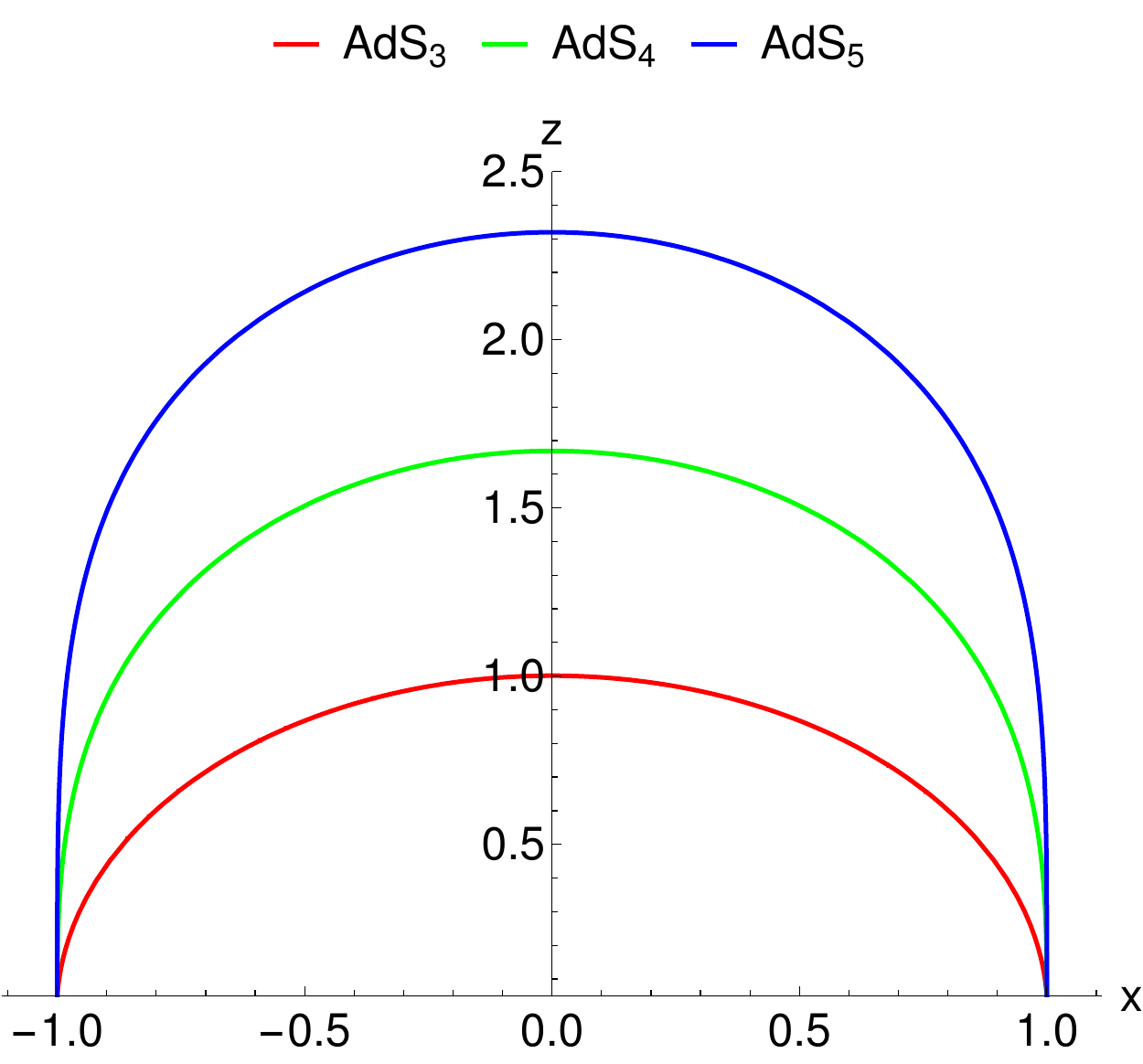}
\caption{Left: Minimal surfaces in Poincar\'e patch AdS$_5$ for different boundary separations. Right: Comparison of minimal surfaces in Poincar\'e patch AdS$_{d+1}$ of fixed boundary separation $l=2$.}
\label{Fig:VacSurf}
\end{figure}

Let us start with the simplest possible case, namely minimal surfaces in Poincar\'e patch AdS$_{d+1}$. The areas of these surfaces correspond to the entanglement entropy of vacuum states in $d$-dimensional Minkowski space $\mathbb{R}^{d-1,1}$.
For later convenience we choose \textit{Eddington-Finkelstein coordinates}\index{Eddington-Finkelstein coordinates} for this example in which the line element is given by
\begin{equation}
d s^2=\frac{1}{z^2}\left(-d v^2-2 d z d v + d \vec{x}^2\right)\,.
\end{equation}
Because it is beneficial for later numerical applications we use a non-affine parametrization of the embedding functions $X^\alpha(\sigma)=(Z(\sigma),V(\sigma),X(\sigma))$ defined by $Z(\sigma)=z_*\big(1-\sigma^2\big)$ and $V(\sigma)=v_0-Z(\sigma)$ with $\sigma \in [-1,1]$ and fixed boundary time $V(\pm 1)=v_0$.
In this parametrization the extremal surface equation \eqref{Eq:extrSurface} reduces to a non-affine geodesic equation\footnote{In Appendix \ref{App:AreaFunct} we show how to obtain the non-affine form of the geodesic equation from \eqref{Eq:extrSurface}.}
\begin{equation}\label{Eq:GeoNonAffine}
\ddot{X}^\alpha(\sigma)+\Gamma^{\alpha}_{\beta\gamma}(X^{\delta}(\sigma))\dot{X}^\beta(\sigma)\dot{X}^\gamma(\sigma)=J(\sigma)\dot{X}^\alpha(\sigma)\,,
\end{equation}
where $\Gamma^{\alpha}_{\beta\gamma}$ denote the Christoffel symbols associated to the auxiliary spacetime $\tilde{g}_{\alpha\beta}=\tfrac{1}{z^{2(d-2)}}G_{\alpha\beta}$ and $J(\sigma)=\tfrac{d^2\tau}{d\sigma^2}\Big/\tfrac{d\tau}{d\sigma}$ denotes the Jacobian for the transformation to the affine parameter $\tau$ defined by $\tfrac{dX^\alpha(\tau)}{d\tau}\tfrac{dX^\beta(\tau)}{d\tau}\tilde{g}_{\alpha\beta}\overset{!}{=}1$.
The solution for $X(\sigma)$, which satisfies the boundary conditions defined in \eqref{Eq:stripe}, can be expressed as
\begin{equation}\label{nonAffine2}
X(\sigma)=\mathrm{sgn}(\sigma)\Big(-\frac{l}{2} + \frac{Z(\sigma)^{d}}{d z_*^{d-1}}\, {}_2F_1 \left[ \tfrac{1}{2},\tfrac{d}{2(d-1)},\tfrac{3d-8}{2d-6};\left(\tfrac{Z(\sigma)}{z_*}\right)^{2(d-1)}\right]\Big)\,,
\end{equation}
where $z_*=\frac{2l}{\sqrt{\pi}}\Gamma(\tfrac{1}{2(d-1)})\Big/\Gamma(\tfrac{d}{2(d-1)})$ denotes the $z$-position at which the two branches join.
We realize a UV-cutoff at a given value $z_{cut}$ by truncating the non-affine parameter $\sigma\in[\sigma_-,\sigma_+]$ with $\sigma_\pm$ given by
\begin{equation}\label{Eq:sigmaCut}
\sigma_\pm=\pm\sqrt{1-\frac{z_{cut}}{z_{*}}}\,.
\end{equation}
For the parametrization given above the Jacobian in $d=2,3,4$ evaluates to
\begin{equation}\label{Eq:Jacobian}
J(\sigma)=\frac{d^2\tau}{d\sigma^2}\Big/\frac{d\tau}{d\sigma}= 
  \begin{cases}
    \frac{5\sigma-3\sigma^3}{2-3\sigma^2+\sigma^4}       & \quad d=2\,,\\
\\
    \frac{\sigma  \left(7 \sigma ^6-27 \sigma ^4+38 \sigma ^2-22\right)}{\sigma ^8-5 \sigma ^6+10 \sigma ^4-10 \sigma ^2+4}  & \quad d=3\,,\\
\\  
  \frac{-51\sigma+145\sigma^3-205\sigma^5+159\sigma^7-65\sigma^9+11\sigma^{11}}{(2-\sigma^2)(1-\sigma^2)(3-3\sigma^2+\sigma^4)(1-\sigma^2+\sigma^4)}  & \quad d=4\,.
  \end{cases}
\end{equation}
At this point it is interesting to note that the result for $d=2$ represents a genuine solution to the geodesic equation in Poincare patch AdS$_{d+1}$, irrespective of $d$ 
\begin{subequations}
\begin{eqnarray}
Z(\sigma)&=&\frac{l}{2}\big(1-\sigma^2\big)\,,\label{Eq:Ansatz2PF1}\\
V(\sigma)&=&v_0-Z(\sigma)\,,\label{Eq:Ansatz2PF2}\\
X(\sigma)&=&\frac{l}{2} \sigma \sqrt{2-\sigma^2}\,.\label{Eq:Ansatz2PF3}
\end{eqnarray}
\end{subequations}
These geodesics will become important in the holographic computation of two-point functions, which are determined by their geodesic length as we will discuss in Chapter \ref{chap:Aniso}.
As can be seen from Figure \ref{Fig:VacSurf} (right) for the same boundary separation the surfaces in higher dimensions reach further into the bulk. 

\begin{figure}[htb]
\center
\includegraphics[width=0.35\linewidth]{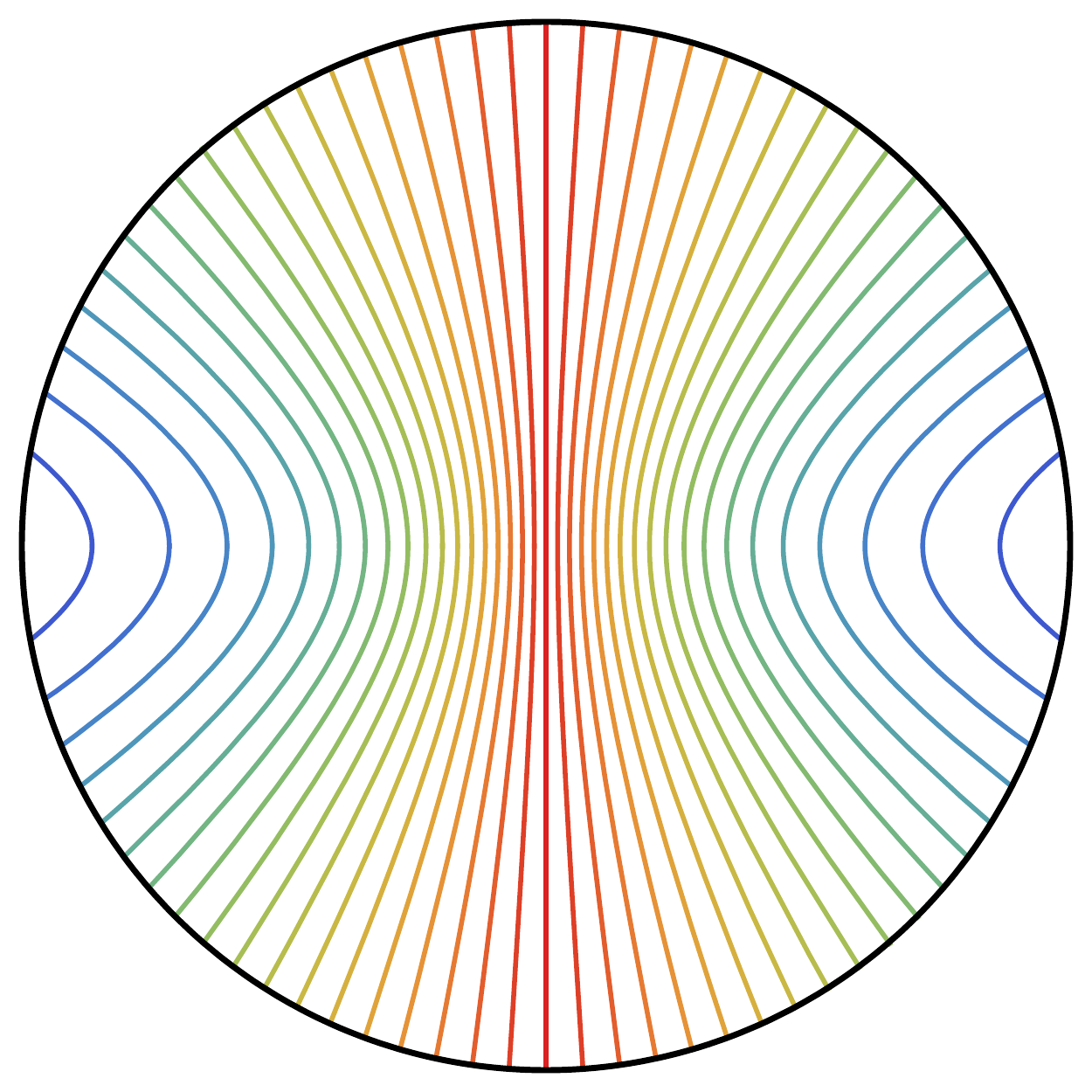}\quad\quad \quad\quad  \includegraphics[width=0.35\linewidth]{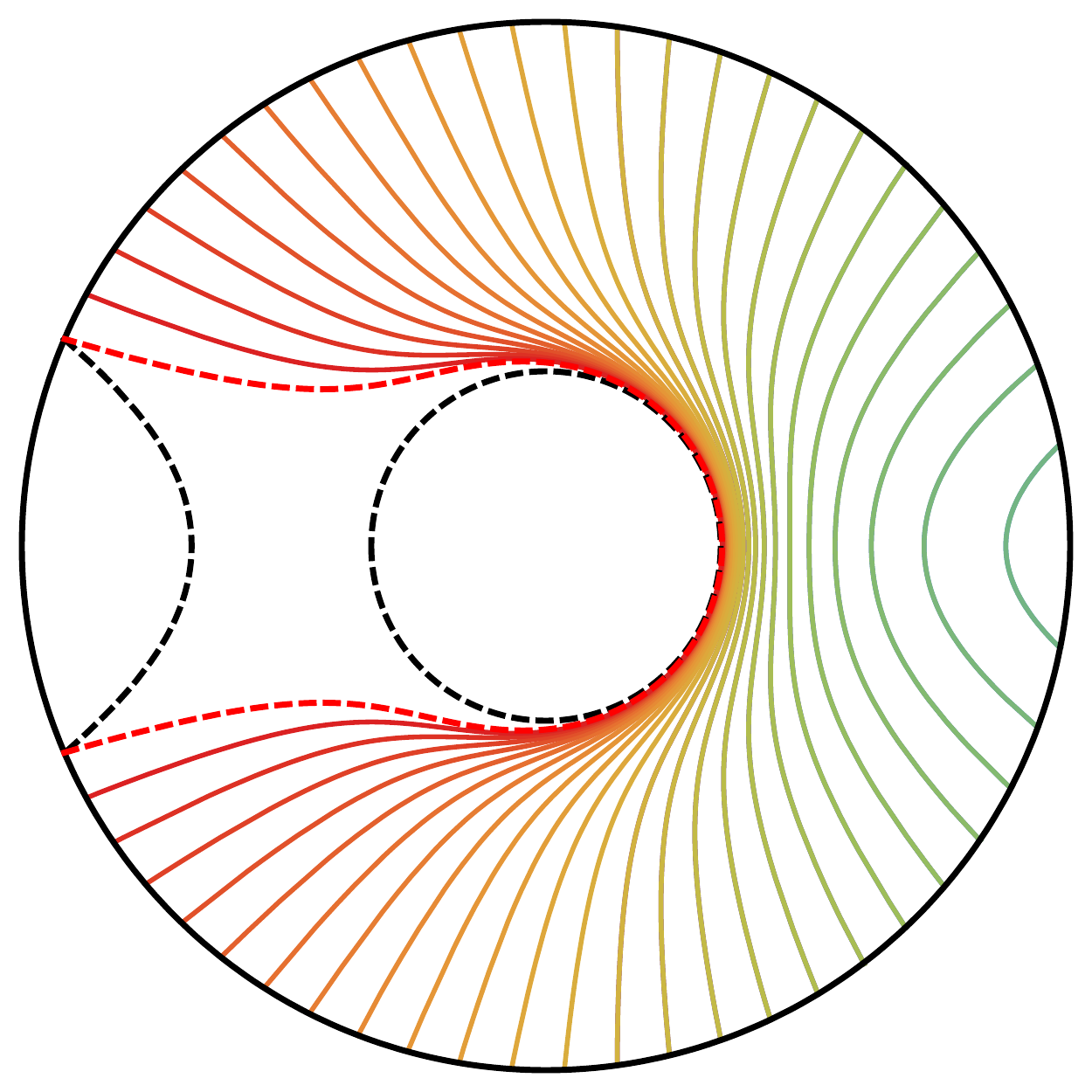}
\caption{%
Left: Minimal surfaces in global AdS$_3$ obtained from the limiting case $r_+=0$ of \eqref{Eq:surfBTZ}. The geometry is shown for $0\le r\le 3$. 
Right: Minimal surfaces in the BTZ geometry with black hole radius $r_+=1$ shown as the black dashed circle in the center. The red dashed line is the minimal surface for the limiting case $\varphi_0=\varphi_*(r_+)$ where the saddle points of the area functional exchange dominance.
}
\label{Fig:surfBTZ}
\end{figure}

As next example we discuss minimal surfaces in the non-rotating BTZ geometry which is holographically dual to thermal states in a CFT$_2$ on $S^1\times \mathbb{R}$.  
The corresponding line element can be written as follows
\begin{equation}
ds^2=-(r^2-r_+^2) dt^2+\frac{dr^2}{r^2-r_+^2}+r^2d\varphi^2\,, \quad \varphi\in[0,2\pi)\,,
\end{equation}
where $r_+=2\pi T$ is the radius of the black hole with Hawking temperature $T$, which can be identified via the duality with the temperature of the thermal state in the CFT$_2$.
As entangling region we assume a single connected segment of the spatial $S^1$ of the boundary geometry
\begin{equation}\label{Eq:GlobalAdS}
A=\{t=t_0,-\varphi_0<\varphi<\varphi_0\}\,.
\end{equation}
This case allows to determine the minimal surfaces $X^\alpha=(R(\varphi),T(\varphi))$ in closed form \cite{Hubeny:2012wa}
\begin{subequations}
\begin{eqnarray}\label{Eq:surfBTZ}
R(\varphi)&=&r_+\left(1-\cosh^2(r_+ \varphi)/\cosh^2(r_+\varphi_0)\right)^{-1/2}\,,\\
T(\varphi)&=&t_0\,,
\end{eqnarray}
\end{subequations}
where the solution for pure AdS$_3$, which is dual to the vacuum state that has $T=0$, is obtained for $r_+=0$.
In Figure~\ref{Fig:surfBTZ} we plot these surfaces for different sizes $\varphi_0$ of the entangling region.
Interestingly, in the finite temperature case, which has the black hole in the center, the area functional has two saddle points. For a given value of $r_+$, these saddle points exchange dominance for sufficiently large values of $\varphi_0$. The first saddle point, which is dominant for small $\varphi_0$, corresponds to surfaces that are homologous to the entangling region $A$ such as given by the differently colored lines in Figure~\ref{Fig:surfBTZ} (right). The second one is dominant for large values of $\varphi_0$ and corresponds to disconnected surfaces, where the first part wraps the horizon and the second part is homologous to $\bar{A}$, such as given by the black dashed lines in Figure~\ref{Fig:surfBTZ} right.
The RT prescription dictates to pick the solution which has minimal area.
The areas of the two possibilities are given by
\begin{equation}
\mathcal{A}= 
  \begin{cases}
    \log\left(\frac{2}{r_+ \epsilon}\sinh\left(\frac{R r_+}{2\pi}\varphi_0  \right) \right)       & \quad \varphi_0<\varphi_*\,,\\
     \pi r_+ + \log\left(\frac{2}{r_+ \epsilon}\sinh\left(\frac{R r_+}{2\pi}(\pi-\varphi_0)  \right) \right)       & \quad \varphi_0\ge\varphi_*\,,
  \end{cases}
\end{equation}
where $\epsilon$ is a UV-cutoff and $R$ is the size of the spatial cycle $S^{1}$.
There is a critical value $\varphi_*(r_+)$ at which the saddle points exchange dominance and where the areas of both solutions become equal
\begin{equation}
\varphi_*(r_+)=\frac{1}{r_+} \coth^{-1}(2\coth(\pi r_+)-1)\,.
\end{equation}
This situation is illustrated in Figure~\ref{Fig:surfBTZ} (right) where the surface shown in dashed red and the disconnected dashed black surface have the same area.
For large entangling regions $\varphi_0\gg\varphi_*$ the entanglement entropy approaches the thermal entropy 
\begin{equation}
\lim\limits_{\varphi_0\to\pi} S_A = \frac{c}{3}\pi r_+=\frac{\mathcal{A}}{4G_N^3}=S_{BH}\,,
\end{equation}
where we have expressed the central charge via $c=\frac{3}{2G_N^{(3)}}$ \cite{Brown:1986nw}.
The homology constraint implies that for $\varphi\ge\varphi_*$ the deviation of the entanglement entropy from its pure state value ($\delta S_A=S_A-S_{\bar{A}}$) becomes constant and equal to the thermal entropy $S_{BH}$. This is the so-called \textit{entanglement plateau}\index{entanglement plateau} phenomenon discussed in \cite{Hubeny:2013gta}.
Interestingly, the case $\varphi_0=\varphi_*$ saturates the Araki-Lieb inequality \cite{Araki1970}, as can easily be seen by setting $B=\bar{A}$ and using $S_{A\cup \bar{A}}=S_{BH}$ in \eqref{Eq:Araki}.

\begin{figure}[htb]
\center
\includegraphics[width=0.45\linewidth]{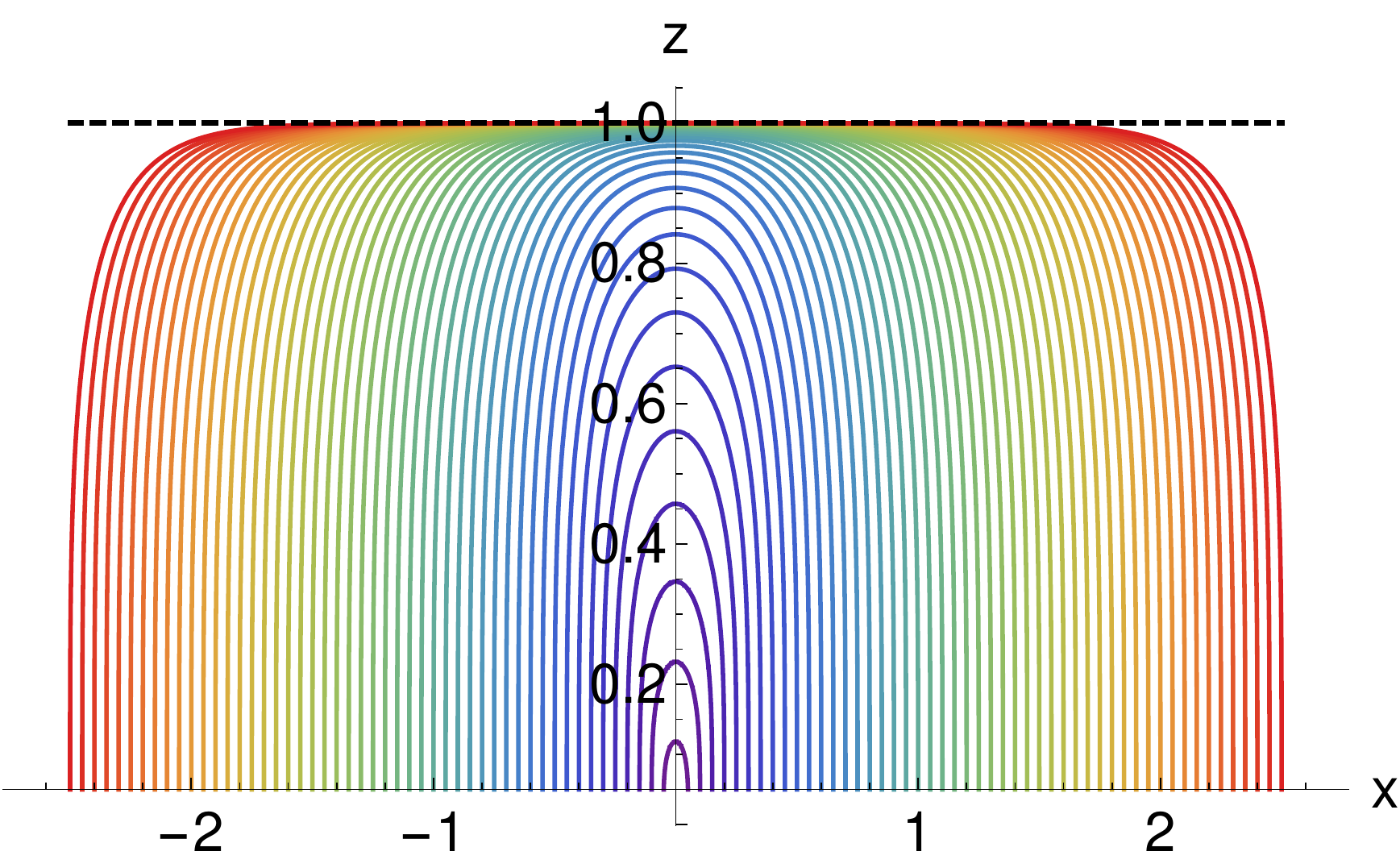}\quad\quad \includegraphics[width=0.45\linewidth]{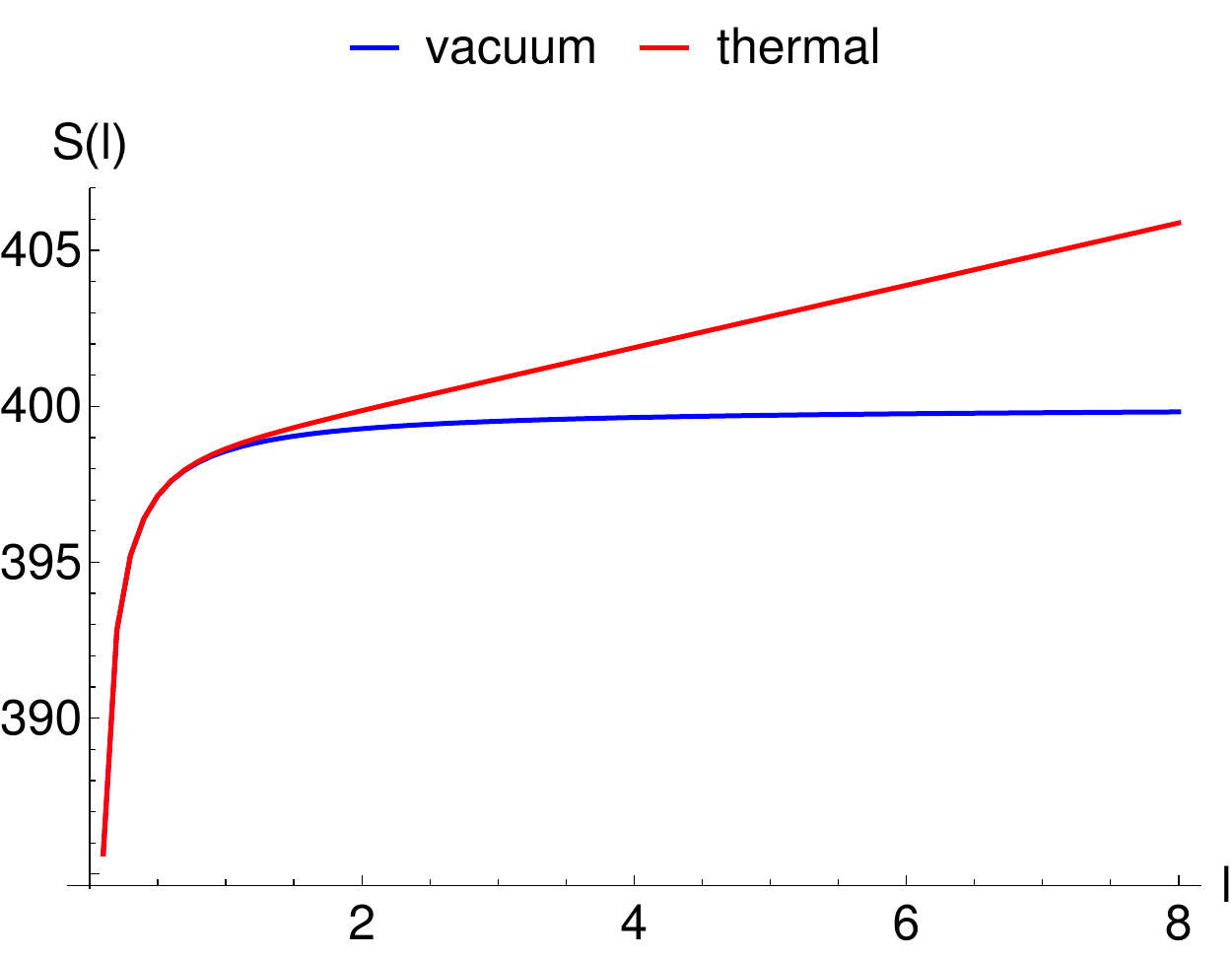}
\caption{%
Left: Minimal surfaces in Poincar\'e patch AdS$_5$ black brane geometry for different boundary separations. The black dashed line indicates the radial position of the horizon.
Right: Scaling of entanglement entropy with $l$ in the vacuum and a thermal state of temperature $T=1/\pi$ in a CFT$_4$ on $\mathbb{R}^{3,1}$. Both curves are for fixed UV-cutoff $z_{cut}=5*10^{-3}$.}
\label{Fig:BBSurf}
\end{figure}

\begin{figure}[htb]
\center
\includegraphics[width=0.40\linewidth]{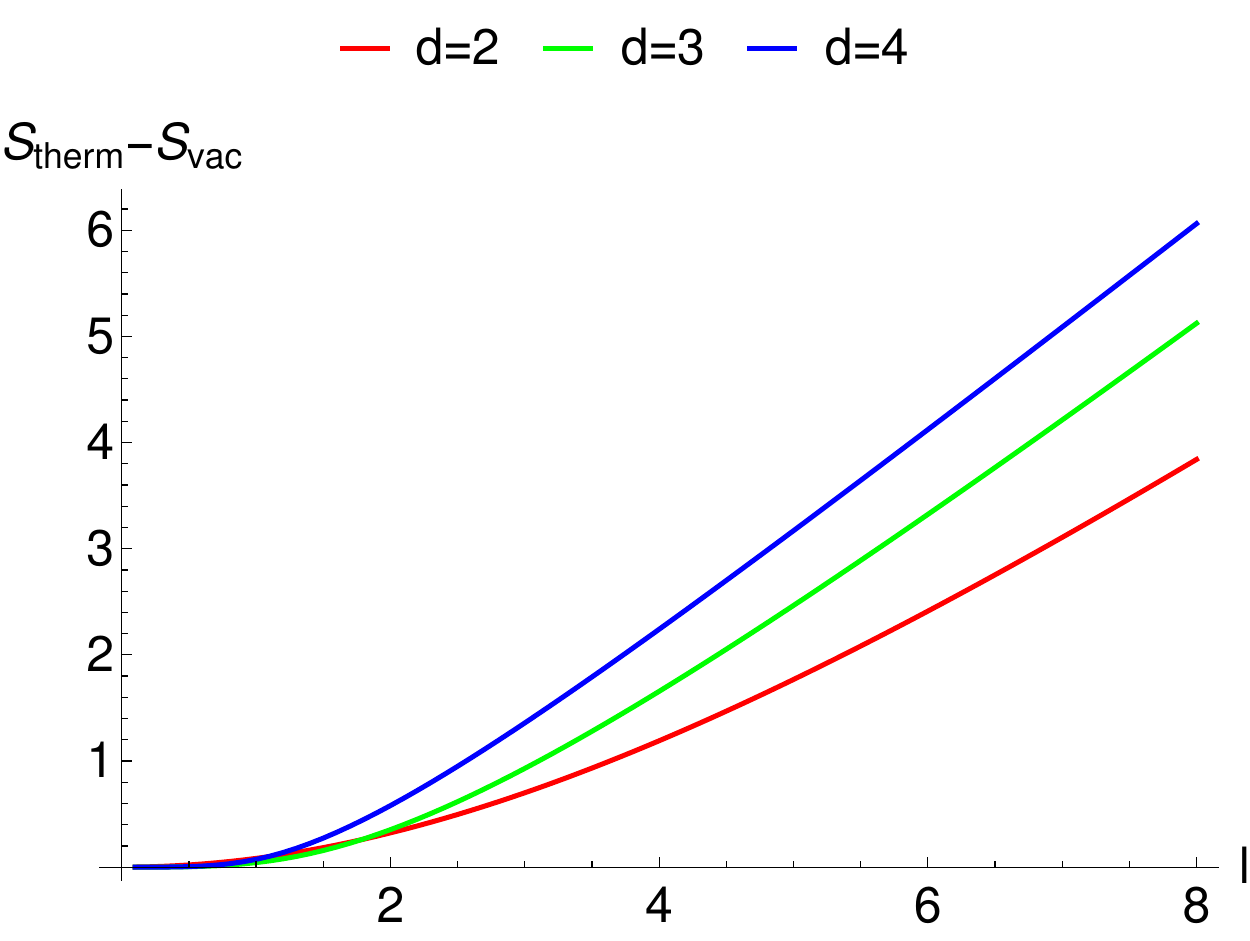}\quad\quad\includegraphics[width=0.40\linewidth]{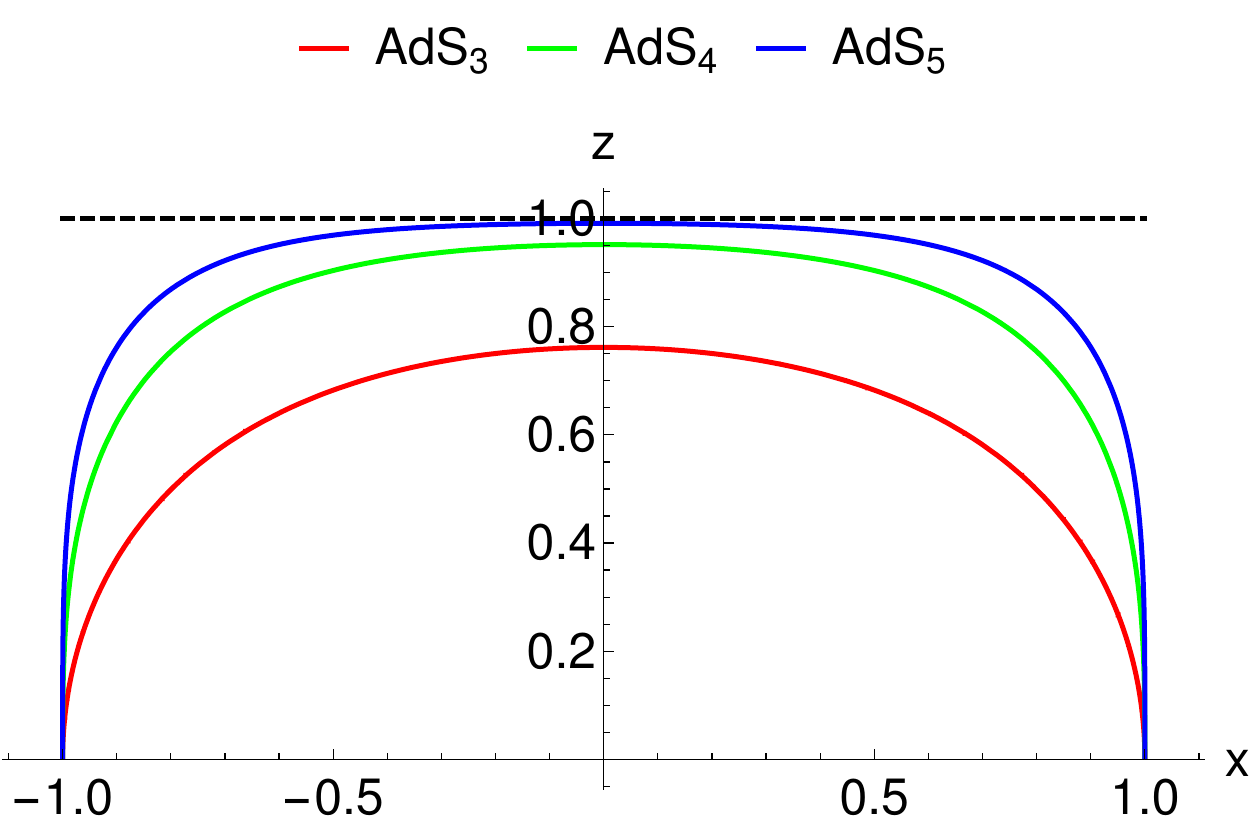}
\caption{
Left: Difference between thermal and vacuum entanglement entropy in $d=2,3,4$. 
Right: Minimal surfaces in AdS$_{d+1}$ black brane geometries for $d=2,3,4$ and fixed boundary separation. 
}
\label{Fig:SbbVac}
\end{figure}

As last example we consider homogeneous and isotropic finite temperature states in a CFT$_d$ for which the entanglement entropy is computed from extremal surfaces in a Poincar\'e patch AdS$_{d+1}$ Schwarzschild black brane geometry
\begin{equation}
d s^2=\frac{1}{z^2}\left(-(1-M z^{d})d v^2-2 d z d v + d \vec{x}^2\right)\,,
\end{equation}
where the temperature of the dual state is given by $T=\frac{d}{4\pi}\sqrt[d]{M}$.
Like in the pure AdS case we consider strip-shaped entangling regions \eqref{Eq:stripe} for which the area functional again reduces to a geodesic equation in the relevant three-dimensional auxiliary spacetime
\begin{equation}
d\tilde{s}^2=\frac{1}{z^{2(d-1)}}\left(-(1-M z^{d})d v^2-2 d z d v + d x^2 \right)\,.
\end{equation}
This time we solve the geodesic equation numerically with a simple shooting method that will be described in the next chapter. The corresponding Mathematica code is listed in Appendix \ref{App:Shoot}.
We show the solution for the minimal surfaces, together with the radial position of the black brane horizon, in Figure \ref{Fig:BBSurf} (left).
We find surfaces of increasing boundary separation approach the horizon closely but never cross it. In \cite{Hubeny:2012ry} it was shown that this is true in any static black hole spacetime, irrespective of the dimension and the shape of the boundary region. As we will see later, this is no longer the case once we consider time dependent systems, where one finds surfaces which can penetrate both, apparent and event horizons. 
Like in the pure AdS$_{d+1}$ case, we find that in higher dimensions surfaces of the same boundary separation reach deeper into the bulk (see right plot in Figure \ref{Fig:SbbVac}).
It is instructive to study the scaling of the entanglement entropy with the system size $l$ for both, vacuum and thermal states.
For small entangling regions we expect the universal UV-scaling given in \eqref{Eq:UV-scaling} for both kinds of states.
On the other hand, for large regions we expect different IR-scalings for vacuum and thermal states.
In the IR we expect an area law scaling ($\propto l^{d-2}$) for the vacuum and an extensive volume law scaling ($\propto l^{d-1}$) for the thermal state, similar to the thermal entropy.
The holographic entanglement entropy nicely satisfies these expectations\footnote{Note that we plot the entanglement entropy density in a (1+1)-dimensional subspace which gives the constant scaling for the vacuum and the linear scaling for the thermal state.} as can be seen in Figure \ref{Fig:BBSurf} (right), where we plot the entanglement entropy for the vacuum, computed from surfaces in pure AdS$_5$, and for the thermal state with temperature $T=1/\pi$, computed from surfaces in the AdS$_5$-Schwarzschild with $M=1$.
On the geometry side both, the UV- and the IR-scaling, can be nicely explained. The geometric reason for the universal UV-scaling comes from the fact that surfaces with small $l$ reside in the asymptotic region which, to leading order, is the same in both geometries.
The extensive IR-scaling behavior of the thermal state comes from the part of the surface that stretches along the horizon. This part gives a contribution which scales exactly like the horizon area that, via the Bekenstein-Hawking law \eqref{BekensteinHawking}, in holography is identified with the thermal entropy of the dual field theory state.

It is also instructive to study the scaling of entanglement entropy in different dimensions. 
In the Figure \ref{Fig:SbbVac} (left) we plot the difference between the entanglement entropy of the thermal state and the vacuum in different dimension. In higher dimensions the entanglement entropy enters the volume law scaling already for smaller $l$ than in lower dimensions. 
The geometric reason is that in higher dimensions the surfaces already for smaller $l$ get closer to the horizon where the thermal scaling takes over. This is illustrated in the Figure \ref{Fig:SbbVac} (right), where we show surfaces with fixed separation in different dimensions.

\section{The Quantum Null Energy Condition}\label{Sec:QNEC}
The quantum null energy condition (QNEC) is a local energy condition that provides a lower bound for expectation values of null projections of the energy momentum tensor in relativistic quantum field theories. 
QNEC connects the null projection of the energy momentum tensor to the second variation of the entanglement entropy with respect to a lightlike deformation of the entangling region
\begin{equation}\label{Eq:DefQNEC}
\langle T_{kk}\rangle \geq \frac{\hbar}{2\pi\sqrt{h}}S''\,,
\end{equation}
where $h$ is the determinant of the induced metric of the entangling surface and $\langle T_{kk}\rangle$ denotes the expectation value of the energy momentum tensor projected onto two lightlike vectors
\begin{equation}
\langle T_{kk}\rangle\equiv \langle T_{\mu\nu}(x)k^\mu(x) k^\nu(x)\rangle\,, \qquad  k^2(x)=0\,.
\end{equation}
The function $S''$ on the right hand side of \eqref{Eq:DefQNEC} is the diagonal part of the following expression \cite{Koeller:2015qmn}
\begin{equation}\label{Eq:S''}
\frac{\delta S}{\delta X^\mu(x)\delta X^\nu(x')}k^\mu(x)k^\nu(x')=S''(x)\delta(x-x')+(\mathrm{off-diagonal})\,,
\end{equation}
where $\delta X^\mu(x)$ is a local variation of the entangling region $A$ at point $x\in\partial A$.
Note that the off-diagonal contributions only exist in $d>2$. 

Originally, QNEC was discovered by studying the properties of the \textit{generalized entropy}\index{generalized entropy}
\begin{equation}
S_{gen}=\frac{A}{4G \hbar}+S_{out}\,,
\end{equation}
where the first contribution is the usual Bekenstein-Hawking entropy \eqref{BekensteinHawking} and $S_{out}$ is the matter entropy outside the horizon. 
The notion of generalized entropy allowed Bekenstein to formulate the \textit{Generalized Second Law}\index{Generalized Second Law} $dS_{gen}\geq0$ \cite{PhysRevD.9.3292} that corrects the ordinary Second Law \cite{PhysRevLett.26.1344}, which fails when matter entropy falls behind a black hole horizon. 

A generalized entropy can be associated not only to horizons but to any surface that splits a Cauchy slice into two disjoint regions.
The generalized entropy can be used to formulate a semiclassical version of the classical focussing theorem by Penrose \cite{PhysRevLett.14.57} which states that light rays are always focussed and never repelled by matter
\begin{equation}\label{Eq:FocTh}
\frac{d\theta}{d\lambda}=\frac{d}{d\lambda}\left( \frac{1}{\mathcal{A}}\frac{d\mathcal{A}}{d\lambda}\right)\leq 0\,,
\end{equation}
where $\mathcal{A}$ is an infinitesimal area element generated by a congruence of null geodesics parametrized by an affine parameter $\lambda$.
The classical focussing theorem in Einstein gravity follows from the null energy condition which states that the lightlike projection of the energy momentum tensor is positive 
\begin{equation}
T_{ab}k^ak^b\ge0 \quad \forall k^2=0\,.
\end{equation}
The null energy condition is sufficient to proof many theorems of classical general relativity, including area theorems and singularity theorems, without knowing the energy momentum tensor explicitly \cite{Hawking:1973uf,0264-9381-4-2-015}.
It is widely obeyed by classical fields and by coherent quantum states.

However, NEC does not hold in general and can be violated by reasonable quantum states \cite{Epstein1965}.
In QFTs the energy density at a given point can be made arbitrarily negative by an appropriate choice of the quantum state. However, for stability any negative energy density must be accompanied by a positive energy elsewhere which can be formalized in terms of the \textit{quantum interest conjecture}\index{quantum interest conjecture} \cite{Ford:1999qv}. Positive-definite expressions of the stress tensor that are bounded from below may exist, but must be nonlocal.
For instance the so-called \textit{averaged null energy condition}\index{averaged null energy condition} (ANEC) can be defined by integrating $\langle T_{kk} \rangle$ along a lightlike geodesic and can be proven in certain QFTs \cite{Faulkner:2016mzt,Hartman:2016lgu}. In Table \ref{Tab:EnergyCond} we give a (incomplete) list of energy conditions together with circumstances under which they are violated. 
\begin{table}[t]  
 \caption[D3 brane embedding in AdS/CFT.]{Summary of some energy conditions and when they are violated. 'EC' means 'energy condition' and 'S', 'W', 'N', 'AN' and 'QN' mean 'strong', 'weak', 'null', 'averaged null' and 'quantum null'. Timelike and null vectors are denoted as $t^\mu$ and $k^\mu$ respectively and $\int_\gamma$ means integration along a null geodesic $\gamma$.}\label{Tab:EnergyCond}
 \centering
 \begin{tabular}{lccccccccccc} 
 \hline\hline
   &   EC  &Inequality & Local?  & True?  & Violated by \\ 
  \hline
   & SEC   &              $T_{\mu\nu}t^\mu t^\nu\ge  T^\mu_\mu t^\nu t_\nu $   & yes & NO! & free scalar field \\
   & WEC   &              $T_{\mu\nu}t^\mu t^\nu\ge0$   & yes & NO  & negative cosmological constant \\
   & NEC   &              $T_{\mu\nu}k^\mu k^\nu\ge0$   & yes & no  & quantum effects \\
   & ANEC  & $\int_{\gamma}T_{\mu\nu}k^\mu k^\nu\ge 0 $& no  & yes & not violated in reasonable QFTs \\
   & QNEC  & $\langle T_{kk}\rangle \geq \tfrac{\hbar}{2\pi\sqrt{h}}S''$                             & yes & yes & not violated in reasonable QFTs \\
 \hline
\end{tabular}
\end{table}
\enlargethispage{1\baselineskip}
Extending the notion of generalized entropy to arbitrary surfaces allows to lift many classical theorems in general relativity to the semiclassical level. For instance, the classical focussing theorem fails for evaporating black holes because quantum fluctuation can have negative energy violating NEC.
Replacing the area in the classical expansion \eqref{Eq:FocTh} by the generalized entropy via $\mathcal{A}=4G_N S_{gen}$ allows to define a \textit{quantum expansion}\index{quantum expansion}   
\begin{equation}\label{Eq:Qexpansion}
\Theta[X(x);x_1]\equiv \frac{2G_N\hbar}{\sqrt{h(x_1)}}\frac{\delta S_{gen}[X(x)]}{\delta X(x_1)}\,.
\end{equation}
The quantum expansion $\Theta[X(x);x_1]$ measures the rate at which the generalized entropy changes when locally deforming the associated surface generated by $X(x)$ in one of its future orthogonal null directions $X(x_1)$.
In the limit $\hbar \to 0$ it becomes the classical expansion.

The concept of quantum expansion allows to lift the classical focussing theorem to the semiclassical level. This leads to the so-called \textit{quantum focussing conjecture}\index{quantum focussing conjecture} (QFC) \cite{Bousso:2015mna} 
\begin{equation}
\frac{\delta}{\delta X(x_2)} \Theta[X(x);x_1]  \leq 0\,,
\end{equation}
which states that any variation of the quantum expansion along the null direction $X(x_2)$ will not increase.
QNEC arises as a special case of the quantum focussing conjecture \cite{Bousso:2015mna}.
Specializing the QFC to the diagonal case ($x_1=x_2$) gives
\begin{eqnarray}\label{Eq:QFCtoQNEC}
0&\leq& \theta'+ \frac{4G_N\hbar}{\sqrt{h}}(S''_{out}-S'_{out}\theta)\nonumber\\
 &=& -\frac{1}{2}\theta^2-\sigma^2-8\pi G_N \langle T_{kk}\rangle+\frac{4G_N\hbar}{\sqrt{h}}(S''_{out}-S'_{out}\theta)\,,
\end{eqnarray}
where prime denotes derivatives with respect to the affine parameter of the null geodesic generating the deformation vector and in the last step we used the Raychaudhuri equation and NEC to express $\theta'$.
Evaluated for vanishing classical expansion $\theta=0$ and shear $\sigma=0$, the last line of \eqref{Eq:QFCtoQNEC} implies the QNEC inequality stated in \eqref{Eq:DefQNEC}.
It is interesting to note that the derivation of QNEC from the quantum focussing conjecture employs concepts of semiclassical gravity, but the QNEC inequality itself is a statement about the energy momentum tensor and entanglement entropy in quantum field theory only. Newtons constant nicely drops out in \eqref{Eq:QFCtoQNEC} for $\theta=\sigma=0$.

Proofs of QNEC exist for free and superrenormalizable theories \cite{Bousso:2015wca}, theories with holographic duals \cite{Koeller:2015qmn}, and interacting quantum field theories in $d\geq3$ \cite{Balakrishnan:2017bjg}.
These proofs assume that the quantum field theory is formulated on a flat background. QNEC in curved backgrounds is discussed in \cite{Fu:2017evt}.
In \cite{Leichenauer:2018obf} it was shown that in $d>2$ the diagonal part of QNEC always saturates, i.e. the inequality becomes an equality.

In explicit holographic calculations in higher dimensions it is challenging to isolate the diagonal part $S''$ defined in \eqref{Eq:S''}, because one would have to perform a local deformation of the RT-surface that breaks translational symmetry in all spatial directions. This would require to solve the full set of partial differential equations \eqref{Eq:extrSurface} and not only an auxiliary geodesic equation. We will not do that in the examples presented below, but rather homogeneously deform the infinitely extended bounding surface of our strip-shaped entangling region without braking translation symmetry in the infinite directions. This can be interpreted as the so-called \textit{global form of QNEC}\index{global form of QNEC}\cite{Koeller:2015qmn}.
But before we come to the more advanced numerical examples, we perform in the next section an explicit calculation of QNEC in a CFT$_2$ using holography.

\subsection{QNEC in Thermal CFT$_2$}\label{Sec:QNECCFT2}
In this subsection we compute the quantum null energy condition for finite temperature states in the grand canonical ensemble in CFT$_{2}$ with angular momentum $J$.
In two-dimensional QFTs the quantum null energy condition takes the stronger form
\begin{equation}\label{Eq:QNEC}
\langle T_{kk}\rangle \geq \frac{\hbar}{2\pi}\left(S''+\frac{6}{c}(S')^2\right)\,,
\end{equation}
where $c$ denotes the central charge of the CFT and we use units in which Planck's constant $\hbar=1$.
Ultimately we are interested in the right hand side of (\ref{Eq:QNEC}) as a function of the size $l$ of the entangling region. 
In order to do that we have to compute the first and second derivative of the entanglement entropy with respect to a lightlike deformation of one of the boundary points.
The dual gravity prescription for these finite temperature and angular momentum states is given by the BTZ geometry
\begin{equation}\label{ds2BTZ}
ds^2=-\frac{(r^2-r_+^2)(r^2-r_-^2)}{r^2}dt^2+\frac{r^2}{(r^2-r_+^2)(r^2-r_-^2)}dr^2+r^2(dx+\frac{r_+r_-}{r^2}dt)^2\,,
\end{equation}
where $r_+$ and $r_-$ denote the radial positions of the outer and inner horizon respectively and we have set the AdS-radius to unity. 
The mass $M$ and angular momentum $J$ are related to $r_+$ and $r_-$ in the following way
\begin{equation}
M=\frac{r_+^2+r_-^2}{8G_N^{(3)}}\,, \quad \quad\quad \quad J=\frac{r_+r_-}{4G_N^{(3)}}\,,
\end{equation}
where $G_N^{(3)}$ denotes the three-dimensional Newton constant. 
The non-vanishing components of the holographic energy momentum tensor in the dual field theory are given by \cite{Detournay:2014fva}
\begin{equation}
T_{tt}=T_{xx}=\frac{r_+^2+r_-^2}{16\pi G_N^{(3)}}\,, \qquad T_{tx}=\frac{r_+r_-}{8\pi G_N^{(3)}}\,.
\end{equation}
This allows us to evaluate the left hand side of \eqref{Eq:QNEC}.
There are two linearly independent lightlike projection vectors $k^a_\pm=(1,\pm 1)$ which give
\begin{equation}\label{Eq:Tkk}
T_{\pm \pm}=T_{ab}k_\pm^a k_\pm^b=\frac{1}{8\pi G_N^{(3)}}(r_+\pm r_-)^2=\frac{1}{2\pi}(M\pm J)\,.
\end{equation}
To compute the holographic entanglement entropy we use the fact that all BTZ black holes are locally equivalent to pure AdS$_3$.
The relevant coordinate transformation can be written as follows \cite{Carlip:1994gc}
\begin{subequations}
\begin{eqnarray}
w_\pm&=&\sqrt{\frac{r^2-r_+^2}{r^2-r_-^2}}e^{(x\pm t)(r_+\pm r_-)} \equiv X \pm T\,,\\
z &=&\sqrt{\frac{r_+^2-r_-^2}{r^2-r_-^2}}e^{x r_+ + t r_-}\label{Eq:z}\,.
\end{eqnarray}
\end{subequations}
In the new coordinates the line element is manifestly Poincar\'{e} patch AdS$_3$
\begin{equation}
ds^2=\frac{1}{z^2}(dw_+dw_-+dz^2)=\frac{1}{z^2}(-dT^2+dX^2+dz^2)\,.
\end{equation}
This is extremely useful because for this case the formula for the entanglement entropy is well known \cite{Ryu:2006bv}
\begin{equation}\label{Eq:EE}
S_{EE}=\frac{\mathcal{A}}{4G_{N}^{(3)}}=\frac{c}{3}\mathrm{log}\frac{l}{\epsilon}\,,
\end{equation}
where $\mathcal{A}$ is the area of the extremal co-dimension two bulk surface homologous to the entanglement region in the boundary, $c=\frac{3}{2G_N^{(3)}}$ is the central charge of the CFT \cite{Brown:1986nw}, $l$ is the length of the entangling region and $\epsilon$ is a UV-cutoff.
We take our entangling region to range from $(t_0,x_1)$ to $(t_0,x_2)$ such that $l=|x_2-x_1|$. 

Without loss of generality we choose to deform the second boundary point $(t_0,x_2)\to(t_0+\lambda_t,x_2+\lambda_x)$ with a deformation vector $(\lambda_t,\lambda_x)$.
Since the rotating BTZ geometry is anisotropic (the sign of $J$ singles out a preferred spatial direction), deformations with $(\lambda_t,\lambda_x)=(\lambda,\lambda)$ and $(\lambda_t,\lambda_x)=(\lambda,-\lambda)$ for $\lambda>0$ are qualitatively different and need to be treated separately.
Our strategy is to construct the proper distance and the cutoff (supplemented with the deformation $(\lambda_t,\lambda_x)$) in Poincar\'{e} coordinates and use them in (\ref{Eq:EE}).
This will give us a formula for the entanglement entropy $S_\pm(\lambda)$ as a function of $\lambda$ from which we can compute $S_\pm'=\frac{d}{d\lambda}S_\pm(\lambda)|_{\lambda=0}$ and $S_\pm''=\frac{d^2}{d\lambda^2}S_\pm(\lambda)|_{\lambda=0}$ and obtain the right side of (\ref{Eq:QNEC}).
To construct the proper distance squared it is useful to define the following quantities at the boundary ($r=\infty$)
\begin{eqnarray}
\Delta w_\pm&=&(X_1\pm T_1)-(X_2\pm T_2)\nonumber\\
            &=&e^{(x_1\pm t_1)(r_+\pm r_-)}-e^{((x_2+\lambda_x)\pm (t_2+\lambda_t))(r_+ \pm r_-)}\,,
\end{eqnarray}
where $(X_i,T_i)$ with $i=1,2$ denote the endpoints of the entangling region in Poincar\'{e} coordinates and we have used $\lim_{r\to\infty}\sqrt{\frac{r^2-r_+^2}{r^2-r_-^2}}=1$. 
The proper distance squared can now be expressed as
\begin{eqnarray}\label{Eq:dist}
(\Delta x)^2&=&\Delta w_+ \Delta w_-\nonumber\\
&=& [(X_1+T_1)-(X_2+T_2)][(X_1-T_1)-(X_2-T_2)]\nonumber\\
&=& -(T_1-T_2)^2+(X_1-X_2)^2\nonumber\\
&=&\left(e^{(r_++r_-)(t_0+x_1)}-e^{(r_++r_-)(t_0+\lambda_t+x_2+\lambda_x)}\right)\nonumber\\
&\times&\left(e^{(r_+-r_-)(t_0-x_1)}-e^{(r_+-r_-)(t_0+\lambda_t-x_2-\lambda_x)}\right)\,.
\end{eqnarray}
Next we use (\ref{Eq:z}) to express the cutoffs $\epsilon_1,\epsilon_2$ at the two endpoints of the extremal surface
\begin{eqnarray}\label{Eq:cutoff}
\epsilon_1=\frac{\sqrt{r_+^2-r_-^2}}{r_\infty}e^{r_+ x_1+r_-t_0}\,, \quad\quad \epsilon_2=\frac{\sqrt{r_+^2-r_-^2}}{r_\infty}e^{r_+ (x_2+\lambda_x)+r_-(t_0+\lambda_t)}\,,
\end{eqnarray}
where $r_\infty$ can be identified with the UV-cutoff in the field theory $\epsilon=\frac{1}{r_\infty}$.
We can now use (\ref{Eq:dist}) and (\ref{Eq:cutoff}) in a slightly generalized form of the entanglement entropy formula 
\begin{equation}
S_{EE}=\frac{c}{6}\mathrm{log}\frac{(\Delta x)^2}{\epsilon_1\epsilon_2}\,.
\end{equation}
Evaluating the previous formula for a deformation in positive ($S_+$) or negative ($S_-$) x-direction gives
\begin{equation}
S_{\pm}(\lambda)=\frac{c}{6}\mathrm{log}\frac{2\cosh(l \, r_++\lambda(r_+\pm r_-))-2\cosh(l \, r_-+\lambda(\pm r_++r_-))}{\epsilon^2(r_+^2-r_-^2)}\,,
\end{equation}
where we have used $x_2-x_1=l$.
The right hand side of (\ref{Eq:QNEC}) is then given by
\begin{eqnarray}\label{Eq:QNEC_BTZ}
\frac{1}{2\pi}\Big(S_\pm''+\frac{6}{c}(S_\pm')^2\Big)&=&-\frac{c}{12\pi}(r_+\pm r_-)^2\Big(\mathrm{csch}[\tfrac{1}{2}l(r_+\pm r_-)]^2+\mathrm{coth}[\tfrac{1}{2}l(r_-\pm r_+)]^2 \Big)\nonumber\\
&=& \frac{c}{12\pi}(r_+\pm r_-)^2\,,
\end{eqnarray}
which, after using $c=\frac{3}{2G_N^{(3)}}$, exactly agrees with the stress tensor null projections \eqref{Eq:Tkk}.
This means that the QNEC inequality for states dual to BTZ geometries is not only satisfied but also saturated, independently of the size of the entangling region.
In Appendix \ref{App:QNECsat} we give an alternative derivation using the bulk equations of motion of the RT-surfaces directly.
In \cite{Khandker:2018xls} it is argued using holography that for two-dimensional field theories QNEC is not saturated if the corresponding RT-surface passes through matter in the bulk.
We present an explicit example for this in Chapter \ref{Chap:Numerics}.

\chapter[Numerical Entanglement Entropy and QNEC]{Computing Entanglement Entropy and QNEC Numerically}\label{Chap:Numerics}

In this chapter we introduce the numerical methods used to compute solutions to Einstein equations with asymptotically AdS boundary conditions in Section \ref{Sec:NumericEinstein} and to find extremal surfaces and geodesics in these spacetimes in Section \ref{Sec:NumericSurface}.

\section{Solving Einstein Equations on AdS Numerically}\label{Sec:NumericEinstein}

In a general coordinate system the Einstein equations consist of a complicated set of coupled non-linear partial differential equations (PDE) which need to be solved for a given set of initial and boundary conditions.
In applications with asymptotically flat boundary conditions, like for astrophysical simulations of merging black holes or neutron stars, the resulting initial value problem is typically formulated in the so-called \textit{BSSN formulation}\index{BSSN formulation}\cite{PhysRevD.52.5428,Baumgarte:1998te}. State of the art simulations of this kind play an important role in the interpretation of gravitational wave signals recently detected by LIGO and VIRGO \cite{Abbott:2016blz,TheLIGOScientific:2017qsa} and usually require to run sophisticated numerical GR codes\footnote{For example, the Einstein toolkit \cite{Toolkit} is a publicly available numerical GR code that is able to perform such simulations.} on supercomputing facilities \footnote{All the simulations presented in this thesis were carried out on an ordinary desktop computer. However, see for example the numerical AdS/CFT simulations in \cite{Attems:2016tby} which were performed on a supercomputer.}.

For spacetimes with asymptotic AdS boundary conditions the so-called \textit{method of characteristics}\index{method of characteristics}\cite{Bondi21,Sachs103,Winicour:2008vpn}, in which the spacetime is foliated with lightlike slices, turns out to be particularly well suited.
In this formulation the Einstein equations on each characteristic slice decouple into a nested set of ordinary differential equations (ODE) which can be efficiently solved using \textit{spectral methods}\index{spectral methods} \cite{boyd01,trefethen2000spectral,Grandclement:2007sb}.
The stepping between slices is usually done with simple \textit{Runge-Kutta}\index{Runge-Kutta} or \textit{Adams-Bashforth}\index{Adams-Bashforth} time-stepping algorithms \cite{Press:2007:NRE:1403886} which are typically sufficient to obtain a stable time evolution.
The lightlike slicing is realized with generalized Eddington-Finkelstein coordinates which are regular across the black hole horizon.
In the context of numerical AdS/CFT, this method has been applied by several authors to various time evolution problems \cite{Chesler:2008hg,Chesler:2009cy,Chesler:2010bi,vanderSchee:2012qj,Heller:2012km,Casalderrey-Solana:2013aba,Chesler:2015bba,Chesler:2015fpa,Chesler:2015wra,Attems:2016tby}. A detailed explanation of this method can be found in \cite{Chesler:2013lia}. 

In the following we will illustrate this method for the example of a homogeneous but anisotropic AdS$_5$ black brane for which the initial value problem is solved with fixed boundary conditions and non-trivial initial conditions in the bulk.
We realize Eddinton-Finkelstein gauge in the line element $ds^2=g_{\mu\nu}dx^\mu dx^\nu$ by setting $g_{rv}=1,\,g_{r\mu}=0$ and we fix $g_{vy}=0$ by imposing parity invariance in the y-direction. Spatial homogeneity is realized by the Killing vectors $\xi_1=\partial_y,\xi_2=\partial_{x_1},\xi_3=\partial_{x_2},$ and SO(2) invariance in the $x_{1,2}$-plane by the Killing vector $\xi_4=x_2\partial_1-x_1\partial_2$.
The most general ansatz for the line element which satisfies these assumptions can be written as
\begin{equation}\label{Eq:metricAniso}
d s^2=-A(r,v)d v^2+2d vd r +S^2(r,v)\Big( e^{-2B(r,v)} d y^2 +e^{B(r,v)} d\vec{x}^2 \Big) \;,
\end{equation}
where $r$ is the holographic coordinate in which the asymptotic boundary is located at $r=\infty$, $v$ the (advanced) time, $y$ the longitudinal and $\vec{x}=(x_1,x_2)$ the two transversal coordinates of the five-dimensional spacetime.
The aim is to solve the vacuum Einstein equations with negative cosmological constant subject to fixed asymptotically AdS boundary conditions and non-trivial initial conditions for the field $B(r,v)$ in the bulk
\begin{equation}
R_{\mu\nu}-\frac{1}{2}g_{\mu\nu}R +\Lambda g_{\mu\nu}=0\,,
\end{equation}
where $R_{\mu\nu}$ and $R$ denote the Ricci tensor and Ricci scalar associated to the metric $g_{\mu\nu}$ respectively. The cosmological constant is given by $\Lambda=-\tfrac{(d-1)(d-2)}{2L^2}$, where $d=5$ and we set the AdS radius $L\equiv 1$.
In the coordinate system defined by \eqref{Eq:metricAniso} the Einstein equations read
\begin{subequations}
\label{Eq:Einstein}
\begin{align}
0&= 2S'' + (B')^2S \label{eq:E5}\,,\\
0&= S (\dot{S})'+2S' \dot{S}-2S^2\,, \label{eq:E1}\\
0&=2S(\dot{B})'+3(S'\dot{B}+B'\dot{S})\,, \label{eq:E2} \\
0&=A''+ 3B'\dot{B}-12 S' \dot{S}/S^2+4\,, \label{eq:E3}\\
0&= 2\ddot{S}-A' \dot{S}+\dot{B}^2S\,, \label{eq:E4}
\end{align}
\end{subequations}
where prime denotes radial derivative and dot means time derivative, via
\begin{equation}\label{Eq:dot}
h'\equiv \partial_r h\,,\qquad \dot{h}\equiv\partial_v h+\frac{1}{2}A \partial_r h\,,
\end{equation}
for any function $h(r,v)$.
Near the boundary ($r=\infty$) solutions to these equations can be expressed as generalized power series in $r$
\begin{subequations}
\begin{align}
A(r,v)&=r^2 \sum_{n=0}^{\infty}\left(a_{n}(v)+\alpha_{1,n}(v)\log(r)+\ldots+\alpha_{n,n}(v)\log(r)^n\right) r^{-n}\,, \\
S(r,v)&=r\sum_{n=0}^{\infty}\left(s_{n}(v)+\sigma_{1,n}(v)\log(r)+\ldots+\sigma_{n,n}(v)\log(r)^n\right) r^{-n}\,,\\
B(r,v)&=\sum_{n=0}^{\infty}\left(b_{n}(v)+\beta_{1,n}(v)\log(r)+\ldots+\beta_{n,n}(v)\log(r)^n\right) r^{-n}\,.
\end{align}
\end{subequations}
Fixing the conformal boundary metric to Minkowski $ds_{b}^2=r^2\eta_{\mu\nu}dx^\mu dx^\nu$ determines the leading coefficients $a_{0}=1$ and $s_{0}=1$, and the residual gauge freedom $r \rightarrow r + \xi(v)$ of the metric ansatz \eqref{Eq:metricAniso} is fixed by setting the subleading coefficient $a_{1}=0$.
In order to get a well defined initial value problem resulting in a stable time evolution it is necessary to choose a computational domain in the bulk direction that contains the apparent horizon $r_{ah}$, defined by $\dot{S}(r,v)|_{r_{ah}}=0$, on the initial slice.\footnote{Note that other authors \cite{Chesler:2010bi,Casalderrey-Solana:2013aba,Attems:2016tby} use the residual gauge freedom $r\rightarrow r+\xi(v)$ to fix the apparent horizon to a constant value in the radial direction which is then used to bound the computational domain.} 

Solving the Einstein equations order by order in $r$ gives
\begin{subequations}\label{asymptotic}
\begin{align}
A(r,v)&=r^2+\frac{a_4(v)}{r^2}+\frac{ a_4'(v)}{2r^3}+\mathcal{O}(r^{-4})\,,\\
S(r,v)&=r-\frac{ a_4'(v)}{20r^4}-\frac{ b_4(v)^2}{7r^7}+\mathcal{O}(r^{-8})\,,\\ 
B(r,v)&=\frac{b_4(v)}{r^4}+\frac{b_4'(v)}{r^5}+\mathcal{O}(r^{-6})\,,
\end{align}
\end{subequations}
where the normalizable modes $a_4(v)$ and $b_4(v)$ remain undetermined in this procedure. In the specific case, where the boundary metric is fixed by $b_0=0$, one obtains at $\mathcal{O}(r^{-3})$ the relation $a_4'(v)=0$ which means that $a_4=const.$. As we will see later, this relation implies the covariant conservation of the holographic stress tensor and $a_4$ is proportional to the energy of the dual field theory state.
The coefficient $a_4$ has to be provided as initial datum and the function $b_4(v)$ needs to be extracted from the full bulk solution. Furthermore, since the boundary metric is flat, all coefficients of the logarithmic terms vanish $\alpha_{i,j}(v)=\beta_{i,j}(v)=\sigma_{i,j}(v)=0$. These terms only appear for curved boundary metrics of even dimension at orders $\ge d$.

It is convenient to transform the near boundary solution from Eddington-Finkelstein to Fefferman-Graham form \eqref{Eq:FGexpansion} and use the expression \eqref{Eq:HolEMT} to determine the holographic stress tensor.
We first write a series ansatz for the Eddington-Finkelstein coordinates in powers of the radial Fefferman-Graham coordinate
\begin{subequations}\label{BoundarySeries}
\begin{align}
 r_{EF}(r_{FG},t_{FG})&=\sum_{n=1}^\infty \left[r_n(t_{FG})+\rho_n(t_{FG})\log(r_{FG})\right] (r_{FG})^n\,,\\
 t_{EF}(r_{FG},t_{FG})&=t_{FG}+\sum_{n=1}^\infty \left[t_n(t_{FG})+\tau_n(t_{FG})\log(r_{FG})\right] (r_{FG})^n\,.
\end{align}
\end{subequations}
The metric transforms as follows
\begin{equation}\label{CoordTrafo}
g^{FG}_{\mu\nu}=\frac{\partial x_{EF}^\alpha}{\partial x_{FG}^\mu} \frac{\partial x_{EF}^\beta}{\partial x_{FG}^\nu}g_{\alpha\beta}^{EF}\,,
\end{equation}
where $x_{EF}^\mu=(r_{EF},t_{EF})$ and $x_{FG}^\mu=(r_{FG},t_{FG})$.
Using the expressions for the metric in Eddington-Finkelstein and Fefferman-Graham coordinates\footnote{We only need to consider time and radial indices in the transformation, since the remaining components of the metric are already in Fefferman-Graham form.}
\begin{equation}
g^{EF}_{\mu\nu}=\begin{pmatrix} 0 & 1 \\ 1 & g^{EF}_{1,1} \end{pmatrix}\,, \qquad g^{FG}_{\mu\nu}=\begin{pmatrix} r_{FG}^2 & 0 \\ 0 & g^{FG}_{1,1} \end{pmatrix}\,,
\end{equation}
in the transformation law \eqref{CoordTrafo} leads to a set of two equations 
\begin{subequations}
\begin{align}
 0&=\frac{\partial{t_{EF}}}{\partial{r_{FG}}} \frac{\partial{t_{EF}}}{\partial{t_{FG}}}g^{EF}_{1,1}+\frac{\partial{r_{EF}}}{\partial{r_{FG}}}\frac{\partial{t_{EF}}}{\partial{t_{FG}}}+\frac{\partial{t_{EF}}}{\partial{r_{FG}}}\frac{\partial{r_{EF}}}{\partial{t_{FG}}}\,,\\
 0&=r_{FG}^2-2\frac{\partial{r_{EF}}}{\partial{r_{FG}}}\frac{\partial{t_{EF}}}{\partial{r_{FG}}}-\left(\frac{\partial{t_{EF}}}{\partial{r_{FG}}}\right)^2 g^{EF}_{1,1}\,,
\end{align}
\end{subequations}
which we can solve order by order in $r_{FG}$ and $r_{FG}\log(r_{FG})$.
The near boundary expansion of the metric in Fefferman-Graham coordinates is given by
\begin{subequations}\label{MetricFG}
\begin{align}
 g^{FG}_{1,1}&=-r^2-\frac{3 a_{4}}{4r^2}+\mathcal{O}(r^{-4})\,,\\
 g^{FG}_{2,2}&=r^2-\left(\frac{a_4}{4}-2b_4(t)\right)\frac{1}{r^2}+\mathcal{O}(r^{-4})\,,\\
 g^{FG}_{3,3}&=g^{FG}_{4,4}=r^2-\left(\frac{a_4}{4}+b_4(t)\right)\frac{1}{r^2}+\mathcal{O}(r^{-4})\,,
\end{align}
\end{subequations}
where we have used $r\equiv r_{FG}$ and $t \equiv t_{FG}$ to shorten the notation.
As expected, the near-boundary expansion in Fefferman-Graham coordinates contains only even powers of $r$ and all logarithmic terms vanish.
After replacing the radial coordinate by $\rho=\frac{1}{r^2}$ we obtain explicit expressions for the non-vanishing metric components in Fefferman-Graham coordinates (\ref{Eq:FGexpansion}) in terms of the Eddington-Finkelstein coefficients
\begin{subequations}
\begin{align}
g_{(0)tt}&=-1\,,\quad g_{(0)yy}=g_{(0)x_1x_1}=g_{(0)x_2x_2}=1\,,\\
g_{(2)tt}&=g_{(2)yy}=g_{(2)x_1x_1}=g_{(2)x_2x_2}=0\,,\\
g_{(4)tt}&=-\frac{3}{4}a_4\,,\quad g_{(4)yy}=-\frac{1}{4}a_4-2b_4(t)\,,\\
g_{(4)x_1x_1}&=g_{(4)x_2x_2}=-\frac{1}{4}a_4+b_4(t)\,.
\end{align}
\end{subequations}
The coefficients in the asymptotic expansion determine via \eqref{Eq:HolEMT} the expectation value of the stress energy tensor in the dual field theory
\begin{equation}
\langle T^{\mu\nu}\rangle =\frac{N_c^2}{2\pi^2}\,\mathrm{diag}\left[ \mathcal{E},~P_\parallel (t),~P_\perp (t),~P_\perp (t) \right]\,,
\end{equation}
where 
\begin{equation}
\mathcal{E} = -\frac{3}{4}a_4\,,\qquad P_\parallel(t) = -\frac{1}{4}a_4 -2 b_4(t)\,,\qquad P_\perp(t) = -\frac{1}{4}a_4 +b_4(t)\,.
\end{equation}

In the numerical procedure it is convenient to work with the inverse radial coordinate $z \equiv 1/r$ such that the boundary is located at $z=0$.
Due to asymptotic AdS boundary conditions, the metric functions $A$ and $S$ diverge as $z\to 0$.
It is numerically favorable to define new functions with the known divergent pieces removed and rescale them with appropriate powers of $z$ so that the resulting functions are finite or vanish as $z\to 0$ and the normalizable modes are easy to read of from the numerical solution.
This leads us to the following field redefinitions
\begin{subequations}
\begin{eqnarray}
       A(z,v) \!\!\!\!&\rightarrow \frac{1}{z^2}+z A(z,v)\,,\qquad                        B(z,v) \!\!\!\!& \to z^3 B(z,v)\,,\\
       S(z,v) \!\!\!\!&\rightarrow \frac{1}{z}+z^2S(z,v)\,,\qquad                  \dot{S}(z,v)  \!\!\!\!&\to \frac{1}{2z^2}+\frac{1}{2}z^2\dot{S}(z,v)\,,\\
 \ddot{S}(z,v) \!\!\!\!&\rightarrow \frac{1}{2z^3}+\frac{1}{4}\ddot{S}(z,v)\,,\qquad \dot{B}(z,v) \!\!\!\!& \to -2z^3\dot{B}(z,v)\,.
\end{eqnarray}
\end{subequations}
The redefined function $B(z,v)$ allows to extract $b_4(t)$ simply from the boundary value of $B'(z,v)$ 
\begin{equation}
b_4(t)=B'(z=0,v=t),
\end{equation}
where $B'=\partial_z B$. Note that at $z=0$ the advanced time is equal to the usual Minkowski time in the boundary theory $v|_{z=0}=t$.
In terms of the redefined fields the first four Einstein equations \eqref{Eq:Einstein} can be rewritten in the form
\begin{subequations}\label{redefEom}
\begin{align}
S''+\frac{6}{z}S'+\left(\frac{6}{z^2}+\frac{9}{2}z^4B^2+3z^5BB'+\frac{1}{2}z^6B'^2\right)S=&j_S\,,\label{redefEom1}\\
\dot{S}'+\frac{2z^2(3S+zS')}{1+z^3S}\dot{S}=&j_{\dot{S}}\,,\label{redefEom2}\\
\dot{B}'+\frac{3(1+4z^3S+z^4S')}{2(z+z^4S)}\dot{B}=&j_{\dot{B}}\,,\label{redefEom3}\\
A''+\frac{4}{z}A'+\frac{2}{z^2}A=&j_A\,,\label{redefEom4}
\end{align}
\end{subequations}
with the source functions given by
\begin{subequations}\label{source}
\begin{align}
j_S&=-\frac{9}{2}zB^2-3z^2BB'-\frac{1}{2}z^3B'^2\,,\label{source1}\\
j_{\dot{S}}&=-\frac{2(5S+2z^3S^2+zS')}{z^2+z^5S}\,,\label{source2}\\
j_{\dot{B}}&=\frac{3(1+z^4\dot{S})(3B+zB')}{8z^2(1+z^3S)}\,,\label{source3}\\
j_A&=-\frac{6(S(4+z^3S)+z^4\dot{B}(1+z^3S)^2(3B+zB')+zS'-z\dot{S}(1-2z^3S-z^4S'))}{(z+z^4S)^2}\,. \label{source4}
\end{align}
\end{subequations}
The relation between dot-derivative and time-derivative, originally given in (\ref{Eq:dot}), turns into
\begin{equation}\label{redefDot}
\dot{B}=-\frac{1}{2}\partial_v B+\frac{1}{4}B'+\frac{3}{4z}B+\frac{3}{4}z^2AB+\frac{1}{4}z^3B'A\,.
\end{equation}
The boundary conditions for the redefined fields read
\begin{subequations}\label{BC}
\begin{align}
S(z=0,v)&=0\,,\qquad S'(z=0,v)=0\,,\label{BC1}\\
\dot{S}(z=0,v)&=a_4\label{BC2}\,,\\
\dot{B}(z=0,v)&=B'(z=0,v)\,,\label{BC3}\\
A(z=0,v)&=0\,,\qquad A'(z=0,v)=a_4\,.\label{BC4}
\end{align}
\end{subequations}
For a given set of initial data \{$B(z,v_0)$, $a_4$\} and fixed boundary metric the system of equations (\ref{redefEom}) allows for the following solution strategy:
\begin{enumerate}
\item With the initial conditions \{$B(z,v_0)$, $a_4$\} and the boundary conditions (\ref{BC}) we first solve (\ref{redefEom1}) for $S$ on the initial time slice.
\item For given $S$ and the boundary condition (\ref{BC2}) we solve (\ref{redefEom2}) for $\dot{S}$. 
\item Having $B$, $S$ and $\dot{S}$ we next solve (\ref{redefEom3}) for $\dot{B}$ using the boundary condition (\ref{BC3}). 
\item With $B$, $S$, $\dot{S}$ and $\dot{B}$ and the boundary conditions in (\ref{BC4}) we solve (\ref{redefEom4}) for $A$.
\item Finally we integrate (\ref{redefDot}) to get the new $B$ on the next time slice and repeat the whole procedure all over again.
\end{enumerate}
For constant $v$ we solve each of the equations (\ref{redefEom}) with a pseudo-spectral method \cite{boyd2001chebyshev} where the $N+1$ grid points $u_i$ on an interval $[a,b]$ are located at
\begin{equation}
z_i=\frac{1}{2}\big((a+b)+(a-b)\,\mathrm{cos}(i\pi/N)\big) \qquad i=1,\ldots,N+1.
\end{equation}
These points can be used to construct the entries of the spectral differentiation matrix $D_{ij}$ \cite{Trefethen:2000:SMM:343368}:
\begin{subequations}
\begin{align}
D_{00}&=\frac{2N^2+1}{6}\,,\qquad D_{NN}=-\frac{2N^2+1}{6}\,,\\
D_{jj}&=\frac{-z_j}{2(1-z_j^2)}\qquad j=1,\ldots,N-1\,,\\
D_{ij}&=\frac{c_i}{c_j}\frac{(-1)^{i+j}}{(z_i-z_j)}\qquad j\ne j \quad i,j=0,\ldots,N\,,
\end{align}
\end{subequations}
where $c_0=c_N=2$ and otherwise $c_i=1$.
The derivative of a function $f(z_i)\equiv f_i$ on the spectral grid points is obtained by multiplication with this differentiation matrix
\begin{equation}
f'_i=D_{ij}f_j\,.
\end{equation}
This allows to turn each of the equations (\ref{redefEom}) into a system of linear equations. 
For example the second equation in (\ref{redefEom}) translates to
\begin{equation}
L_{ij}\dot{S}_j=(j_{\dot{S}})_i\,,
\end{equation}
where the matrix $L_{ij}$ is given by
\begin{equation}
L_{ij}=D_{ij}+\mathrm{diag}\Big[{\frac{2z^2(3S+zS')}{1+z^3S}}\Big]_{ij}\,,
\end{equation}
and the source vector $(j_{\dot{S}})_i$  by
\begin{equation}
(j_{\dot{S}})_i=\Big[-\frac{2(5S+2z^3S^2+zS')}{z^2+z^5S}\Big]_i\,.
\end{equation}
The boundary condition $\dot{S}_1=a_4$ is implemented by setting $(j_{\dot{S}})_1=a_4$ and $L_{1j}=\delta_{1j}$.
The solution vector $\dot{S}_i$ is then obtained by multiplying the inverse of $L_{ij}$ with the source vector
\begin{equation}
\dot{S}_i=[L_{ij}]^{-1}(j_{\dot{S}})_j\,.
\end{equation}
The equations for $S_i$, $A_i$ and $\dot{B}_i$ can be solved in the same way.

To advance the solution for $B$ to the next time slice it  is sufficient to  use a simple fourth order Runge-Kutta method \cite{Press:2007:NRE:1403886} 
\begin{equation}
B(z,v+\delta v)=B(z,v)+\delta v\, \Big(\tfrac{1}{6}k_1+\tfrac{1}{3}k_2+\tfrac{1}{3}k_3+\tfrac{1}{6}k_4\Big).
\end{equation}
where the coefficients $k_i$ are given by
\begin{subequations}
\begin{align}
k_1&= \partial_v B(z,v)\,,\\
k_2&= \partial_v\big(B(z,v)+\tfrac{1}{2}k_1\big)\,,\\
k_3&=\partial_v\big(B(z,v)+\tfrac{1}{2}k_2\big)\,,\\
k_4&=\partial_v\big(B(z,v)+k_3\big)\,,
\end{align}
\end{subequations}
and $\partial_v B$ is computed from \eqref{redefDot}.
The solution for a specific example of the case discussed in this section is presented in Chapter \ref{chap:Aniso}. Details about the numerical implementation of colliding shock waves can be found in \cite{vanderSchee:2014qwa,Attems:2016tby}. For the latter case we present simulation results in Chapter \ref{Chap:Shocks} and Chapter \ref{Chap:QNEC}.

\section{Finding Geodesics and Extremal Surfaces}\label{Sec:NumericSurface}
Mathematically, finding extremal surfaces with fixed boundary conditions in asymptotically AdS spacetimes boils down to solving a two-point boundary value problem.
The main difference to the initial value problem discussed in the previous section is that giving boundary conditions at the starting point is not sufficient to uniquely determine a solution. There are two standard approaches used to solve such problems namely the shooting method and the relaxation method \cite{Press:2007:NRE:1403886}.
In the following two subsections we discuss in some detail both of these methods and provide simple Mathematica implementations in Appendix \ref{App:Shoot} and Appendix \ref{App:Relax}.

\subsection{Shooting Method}\label{Sec:Shooting}
In this section we discuss a simple shooting method which is usually good enough to solve for minimal surfaces in time independent situations with boundary conditions coming from regularly shaped entangling regions.

In shooting we start with a guess for the initial conditions and integrate the differential equation up to some matching point which we want to hit.
The solution obtained from the initial guess will typically not hit the desired matching point, but there will be some mismatch $F_i$ between the point we hit and the aim.
The strategy is to iteratively improve the guess by a correction vector $\delta V_j$ for the initial conditions until the desired matching point is hit to a certain accuracy.
The correction vector can be obtained by inverting the following relation
\begin{equation}\label{Eq:shooting}
S_{ij}\delta V_j=-F_j\,,
\end{equation}
where the Jacobian is usually not known analytically and needs to be approximated
\begin{equation}\label{Eq:shootingJ}
S_{ij}=\frac{\partial F_i}{\partial V_j}\approx\frac{F_i(V_0,\ldots,V_j+\Delta V_j,\ldots)-F_i(V_0,\ldots,V_j,\ldots)}{\Delta V_j}\,.
\end{equation}
The correction vector is used to generate the next guess
\begin{equation}
V^{new}_i=V^{old}_i+\alpha\,\delta V_i\,,
\end{equation}
To stabilize the first iterations, where the discrepancy $\mathrm{max}|F_i|$ is typically large, it is sometimes necessary to use $\alpha<1$ which has the side effect of slowing down convergence. Usually one can set $\alpha=1$ as soon as the discrepancy is sufficiently small to speed up convergence again.

Let us now apply this procedure to the case of minimal surfaces in an AdS$_{d+1}$ Schwarzschild black brane geometry
\begin{equation}
d s^2=\frac{1}{z^2}\left(-(1-M z^{d})d v^2-2 d z d v + d \vec{x}^2\right)\,,
\end{equation}
where the temperature of the dual state is given by $T= \frac{d}{4\pi}\sqrt[d]{M}$.
Like in the pure AdS case we consider strip-shaped entangling regions \eqref{Eq:stripe} for which the area functional reduces to a geodesic equation in the relevant three-dimensional auxiliary spacetime
\begin{equation}\label{Eq:ds2aux}
d\tilde{s}^2=\frac{1}{z^{2(d-1)}}\left(-(1-M z^{d})d v^2-2 d z d v + d x^2 \right)\,.
\end{equation}
For this simple setting it is sufficient to stick to an affine parametrization [$X^{\alpha}(\tau)=(Z(\tau),V(\tau),X(\tau))$, $\dot{X}^\alpha\dot{X}^\beta\tilde{g}_{\alpha\beta}\equiv 1$] in which the minimal surface equation takes the form
\begin{equation}\label{Eq:geodesicShoot}
\ddot{X}^\alpha(\tau)+\Gamma^{\alpha}_{\beta\gamma}(X^{\delta}(\tau))\dot{X}^\beta(\tau)\dot{X}^\gamma(\tau)=0\,,
\end{equation}
where $\Gamma^{\alpha}_{\beta\gamma}$ denote the Christoffel symbols associated to the auxiliary spacetime \eqref{Eq:ds2aux}.

In this case we only need to optimize one parameter, which is the $z$-position $z_*$ of the turning point of the geodesic in the bulk.
To solve \eqref{Eq:geodesicShoot} we shoot from an initial guess $X^\mu_{ini}(0)$ and try to hit a point $(Z(\tau_\pm),T(\tau_\pm),X(\tau_\pm))=(z_{cut},t_0,\pm l/2)$ at the cutoff surface located at $z_{cut}$.
We initialize the procedure with a guess for the initial conditions at the turning point of the surface located at $x=0$
\begin{subequations}
\begin{align}
Z(0)&=z_*^{(0)}, &T(0)&=t_0, &X(0)&=0 \,,\\
Z'(0)&=0,      &T'(0)&=0,  &X'(0)&=1 \,.
\end{align}
\end{subequations}
Using these initial conditions we numerically integrate the geodesics equation up to some large value of the affine parameter $\tau_{end}\approx \mathcal{O}(10^6)$.
This can be done, for instance, using standard tools like Mathematicas NDSolve (for details see Appendix \ref{App:Shoot}). 
Next we have to extract the boundary separation at a fixed cutoff $z_{cut}$ from the solution of the first shot and compute the mismatch vector.
To do that we first solve numerically for the parameter values $\tau_{\pm}$ at the desired cutoff $z_{cut}$
\begin{equation}
Z(\tau_\pm)=z_{cut}\,.
\end{equation}
This can be done conveniently using, e.g Mathematicas NSolve.
Having $\tau_{\pm}$ we can determine the separation at the cutoff
\begin{equation}
l_{cut}^{(0)}=X(\tau_+)-X(\tau_-)\,.
\end{equation}
The mismatch in our simple case reduces to a single function that measures the discrepancy between the boundary separation of the guess and the desired separation of the true solution
\begin{equation}
F^{(0)}=|l-l^{(0)}_{cut}|\,.
\end{equation}
Typically, the mismatch after the initial shot will not be sufficiently small and we have to compute corrections to the initial guess
\begin{equation}
z_*^{(1)}=z_*^{(0)}+\alpha \delta z_*^{(0)}\,,
\end{equation}
where we also included the weight $\alpha$.
The corrections in subsequent iterations are obtained from inverting \eqref{Eq:shooting} and using \eqref{Eq:shootingJ}, which for the case at hand gives
\begin{equation}
\delta z_*^{(i)}=-\frac{F^{(i)} \Delta z_*}{F^{(i)}_\Delta-F^{(i)}}\,,
\end{equation}
where $F^{(i)}_{\Delta}$ is the mismatch obtained from the guess $z_*^{(i)}+\Delta z_*$. The constant shift $\Delta z_*$ in the finite difference approximation needs to be sufficiently small $\Delta z_*\approx 10^{-10}$ to make high accuracies accessible.
We typically iterate this procedure until we reach $F<10^{-15}$.
\begin{figure}[htb]
\center
 \includegraphics[width=0.45\linewidth]{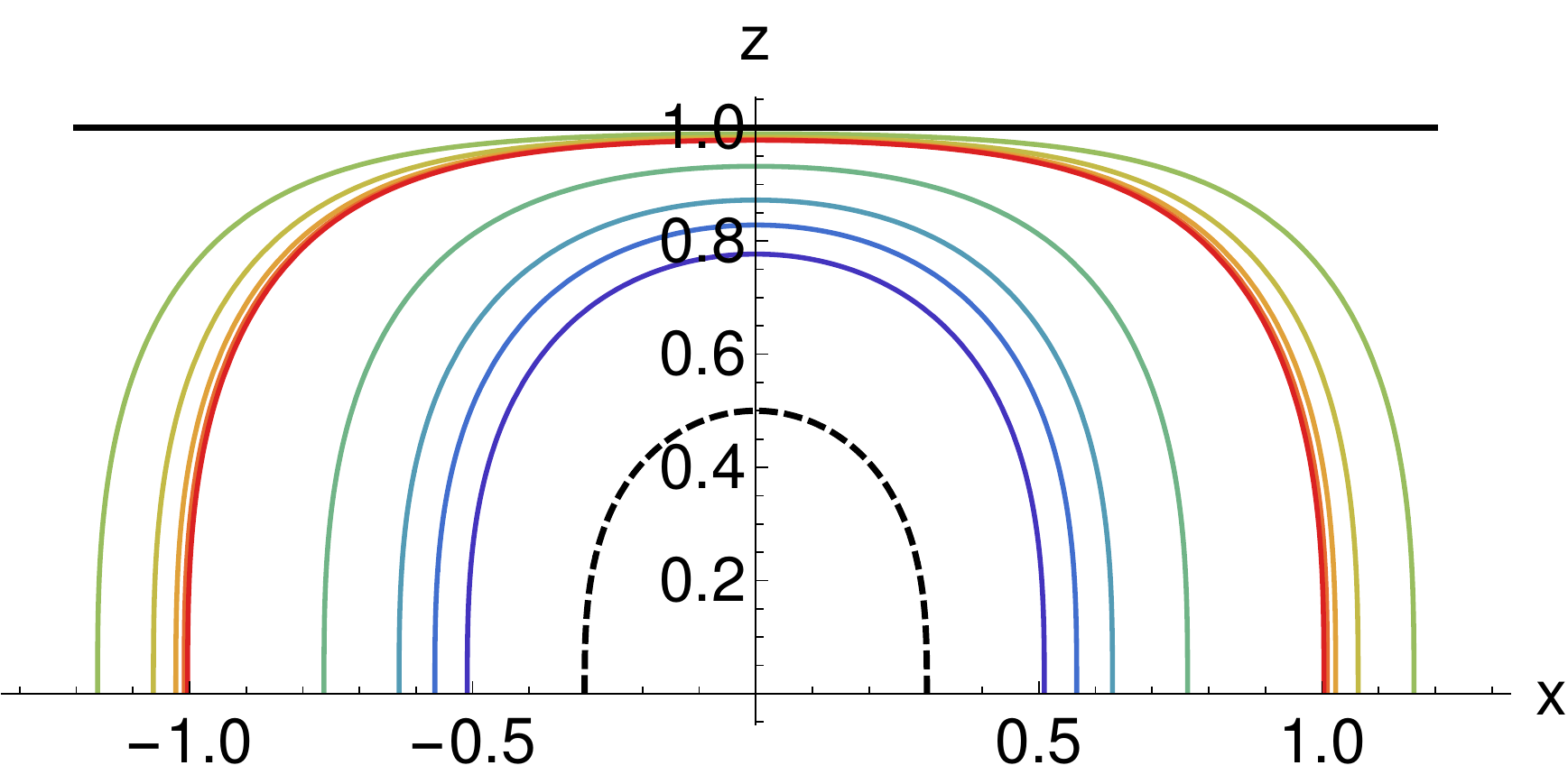} \quad\includegraphics[width=0.45\linewidth]{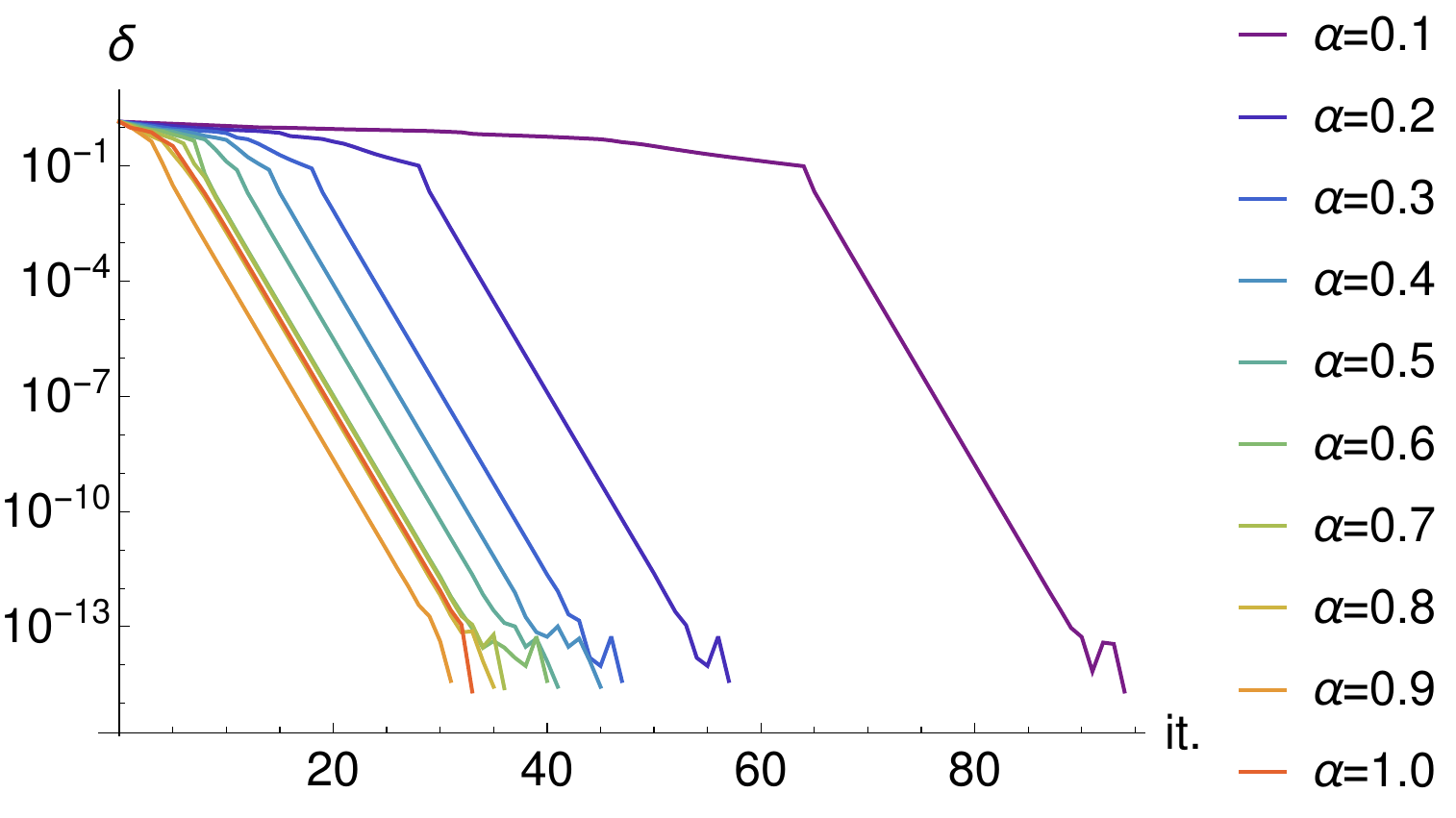}
\caption{Left: Iterations in the shooting method. The black dashed curve is the result obtained from the initial guess $z_*=0.5$ and the colored lines, from purple to red, correspond to subsequent iterations, where the red curve is the converged result. Right: Error $\delta$ over the number of iterations for different values of the weight $\alpha$. In this example the optimal weight that leads to the minimal number of iterations is $\alpha=0.9$.}
\label{Fig:iterationsShoot}
\end{figure}

In Figure \ref{Fig:iterationsShoot} we show the convergence behavior of the shooting method discussed above. In the plot on the left hand side we show the surfaces obtained in subsequent iterations, starting in violet and ending with the red curve which is the converged solution. The initial solution is given by the black dashed line and the horizon is given by the black solid line.
On the right hand side we show the mismatch over the number of iterations for a given value of the weight $\alpha$. In this example the optimal value for $\alpha$ which results in the minimal number of iterations is $\alpha=0.9$.

\subsection{Relaxation Method}\label{Sec:Relaxation}
In relaxation methods differential equations are replaced by approximate finite difference equations (FDEs) on a discrete set of points. The solution is determined by starting with an initial guess and improving it iteratively. 
In this iterative procedure the result is said to relax to the true solution. 

The equation relevant for our present purposes is the non-affine geodesic equation \eqref{Eq:GeoNonAffine} which is a set of three coupled second order differential equation in the three variables $X^{\alpha}(\sigma)=(Z(\sigma),V(\sigma),X(\sigma))$ representing the embedding of the surface in the ambient spacetime
\begin{equation}
\ddot{X}^\alpha(\sigma)+\Gamma^{\alpha}_{\beta\gamma}(X^{\delta}(\sigma))\dot{X}^\beta(\sigma)\dot{X}^\gamma(\sigma)=J(\sigma)\dot{X}^\alpha(\sigma)\,.
\end{equation}
First we reduce the order of the set of differential equations to first order by promoting $\dot X^\alpha(\sigma)\equiv P^\alpha(\sigma)$ to a separate variable
\begin{subequations}
\begin{eqnarray}\label{Eq:geodesicDiscrete}
0&=&\dot X^\alpha(\sigma)-P^{\alpha}(\sigma)\,,\\
0&=&\dot{P}^\alpha(\sigma)+\Gamma^{\alpha}_{\beta\gamma}(X^{\delta}(\sigma))P^\beta(\sigma)P^\gamma(\sigma)-J(\sigma)P^\alpha(\sigma)\,.
\end{eqnarray}
\end{subequations}
Next we define a grid $\sigma_i=h\,i$ with $N$ points of equidistant spacing $h=\tfrac{\sigma_N-\sigma_1}{N}$ with $i=1,\ldots,N$. The upper and lower bound of this grid are given by $\sigma_1\!=\!\sigma_-$ and $\sigma_N\!=\!\sigma_+$ respectively, with $\sigma_{\pm}$ given in \eqref{Eq:sigmaCut}.
The discredized version of the embedding functions and their first derivatives on this grid are written as
\begin{subequations}
\begin{eqnarray}
X^\alpha(\sigma_i)&\equiv& X^\alpha_i=(Z_i,V_i,X_i)\,,\\
P^\alpha(\sigma_i)&\equiv& P^\alpha_i=(P^z_i,P^v_i,P^x_i)\,.
\end{eqnarray}
\end{subequations}
The finite difference representation of the geodesic equation (\ref{Eq:geodesicDiscrete}) on the interior points $i=1,\ldots,N-1$ of the grid are given by
\begin{subequations}
\begin{eqnarray}\label{FDE}
0&=&E_i^{2\alpha-1}= X^\alpha_{i+1} - X^\alpha_{i} -h \bar{P}^\alpha_i\,,\label{FDE1}\\
0&=&E_i^{2\alpha}= P^\alpha_{i+1}-P^\alpha_{i}- h\bar{J}_i \bar{P}^\alpha_{i}+ h\sum\limits_{\beta\gamma}(\bar{\Gamma}^\alpha_{\beta\gamma})_i \bar{P}^\beta_{i}\bar{P}^\gamma_{i}\,,\label{FDE2}
\end{eqnarray}
\end{subequations}
where $E_i^k$ is the residual at point $i$ in equation $k$; quantities with bar are averaged via $\bar{X}^\alpha_i=\frac{X^\alpha_{i}+X^\alpha_{i+1}}{2}$ and $\bar{P}^\alpha_i=\frac{P^\alpha_{i}+P^\alpha_{i+1}}{2}$; the Christoffel symbols $(\bar{\Gamma}^\alpha_{\beta\gamma})_i\equiv \Gamma^{\alpha}_{\beta\gamma}(\bar{X}^{\delta}_i)$ are evaluated from the averaged metric functions;
the explicit form of the Jacobian $\bar{J}_i\equiv J(\tfrac{\sigma_{i+1}+\sigma_i}{2})$ is given in \eqref{Eq:Jacobian}.
The equations \eqref{FDE1} and \eqref{FDE2} represent a set $6\,(N-1)$ algebraic relations for $6N$ unknowns $X_i^\alpha$ and $P_i^\alpha$. The missing six equations that are necessary to make the system solvable are obtained from the boundary conditions. These are imposed at a fixed cutoff surface located at $z=z_{cut}$
\begin{subequations}
\begin{align}
 \qquad\qquad\qquad E_0^1&=Z_1-z_{cut}\,, &  E_N^1&=Z_N-z_{cut}\,,\qquad\qquad\qquad\\
E_0^3&=V_1-v_{0}\,,   &  E_N^3&=V_N-v_{0}\,,\\
E_0^5&=X_1+l/2\,,     &  E_N^5&=X_N-l/2\,,
\end{align}
\end{subequations}
where the remaining components of the vectors $E^{2\alpha}_{0,N}$ are zero.\footnote{Note, that we also could have instead imposed boundary conditions for $P^\alpha_{1,N}$, or mixed boundary conditions, including both, $P^\alpha_{1,N}$ and $X^\alpha_{1,N}$ to provide the missing six equations. However, for the kind of problems we are interested in this thesis, our choice in the main text is the most natural one.}
To initialize the procedure we need an initial guess for $X_i^\alpha$ and $P_i^\alpha$ which we generate from the discredized version of the AdS$_{d-1}$ solutions given in \eqref{nonAffine2}
\begin{subequations}
\begin{eqnarray}
Z_i&=&z_*(1-\sigma_i^2)\,,\\
V_i&=&v_0-Z_i\,,\\
X_i&=&\mathrm{sgn}(\sigma_i)\Big(-\frac{l}{2} + \frac{(Z_i)^{d}}{d z_*^{d-1}}\, {}_2F_1 \left[ \tfrac{1}{2},\tfrac{d}{2(d-1)},\tfrac{3d-8}{2d-6};\left(\tfrac{Z_i}{z_*}\right)^{2(d-1)}\right]\Big)\,,
\end{eqnarray}
\end{subequations}
The initial guess will in general not satisfy these FDEs very well, i.e.\! the residua $E^k_i$ will be rather large. To quantify the deviation of a given trial solution to the true solution we use the following measure
\begin{equation}
\delta=\frac{\sum_{i,k} |E_i^k|}{6N}\,.
\end{equation}
For notational reasons it is convenient to collect $X^\alpha_i$ and $P^\alpha_i$ into a single vector $Y^k_i\equiv(Z_i,P^z_i,V_i,P^v_i,X_i,P^x_i)$.
The strategy is to compute increments $\Delta Y^k$ for $Y^k$ such that $Y^k_{\rm new}=Y_{\rm old}^k+\Delta Y^k$ is an improved approximation to the previous solution $Y_{\rm old}^k$. This we do iteratively until we typically reach 
\begin{equation}
\delta<10^{-15}\,.
\label{Eq:error}
\end{equation}
Equations for the increments are obtained by demanding the first order Taylor expansion of the FDEs with respect to small changes in the coordinates to vanish 
\begin{equation}
E_i^k(Y_i^k+\Delta Y_i^k,Y_{i+1}^k+\Delta Y_{i+1}^k)\approx E_i^k(Y_i^k,Y_{i+1}^k)+\sum\limits_{n}\left(\frac{\partial E_i^k}{\partial Y_i^n}\Delta Y_i^n+\frac{\partial E_{i}^k}{\partial Y_{i+1}^n}\Delta Y_{i+1}^n\right)\,.
\end{equation} 
For a solution we want the updated value $E(Y+\Delta Y)$ to be zero, which gives a set of algebraic equations for the increments on the interior points
\begin{equation}
-E_i^k=\sum\limits_{n=1}^{6}\hat{S}^k_{i,n}\Delta Y_i^n+\sum\limits_{n=7}^{12}\tilde{S}^k_{i,n}\Delta Y_{i+1}^{n-6}\,,
\end{equation}
where
\begin{equation}
\hat{S}_{i,n}^k\equiv\frac{\partial E_i^k}{\partial Y_i^n},\qquad \tilde{S}_{i,n+6}^k\equiv\frac{\partial E_i^k}{\partial Y_{i+1}^n}\,\qquad n=1,...,6\,.
\end{equation}
Similarly one obtains algebraic relations at the boundary
\begin{equation}
-E_0^k=\sum\limits_{n=1}^{6}\hat{S}^k_{0,n}\Delta Y_1^n\,,\qquad-E_N^k=\sum\limits_{n=1}^{6}\tilde{S}^k_{N,n}\Delta Y_N^n\,,
\end{equation}
with
\begin{equation}
\hat{S}_{0,n}^k\equiv\frac{\partial E_0^k}{\partial Y_1^n},\qquad \tilde{S}_{N,n}^k\equiv\frac{\partial E_N^k}{\partial Y_{N}^n}\,\qquad n=1,...,6\,.
\end{equation}

Combining $\boldsymbol{\tilde{S}}$ and $\boldsymbol{\hat{S}}$ into a single matrix $\boldsymbol{S}$ gives a linear system which has to be solved for the correction vector $\Delta \boldsymbol{Y}$
\begin{equation}
\boldsymbol{S}.\Delta \boldsymbol{Y}=-\boldsymbol{E}\,.
\end{equation}
Here it is crucial to note that the matrix $\boldsymbol{S}$ typically has a huge number of vanishing entries, i.e. it can be brought into a sparse representation which allows for an efficient inversion. We do this following exactly the procedure described in the section on relaxation methods in reference \cite{Press:2007:NRE:1403886}. Mathematica provides convenient commands to set up such sparse matrices which we have used in our example code in Appendix \ref{App:Relax}.

The correction $\Delta Y^k$ generated from the first order Taylor expansion is in general only an improvement close to the true solution. We account for this by introducing a weight $\alpha$ that modifies the correction in each relaxation step 
\begin{equation}
Y^k_{\rm new}=Y^k_{\rm old}+\alpha\Delta Y^k .
\end{equation}
We choose the weight $\alpha$ such that the full correction is used only close to the true solution 
\begin{equation}
\alpha =
  \begin{cases}
    0.5 & \quad \text{if } \delta \ge 10^{-3}\,,\\
    1    & \quad \text{else }\,. \\
  \end{cases}
\end{equation}
We illustrate the convergence behavior of our code for the AdS-Vaidya spacetime
\begin{equation}\label{Eq:VaidyaMetric}
d\tilde{s}^2=\frac{1}{z^{2(d-1)}}\left(-\big(1-M(v) z^{d}\big)d v^2-2 d z d v + d x^2 \right)\,,
\end{equation}
where we parametrize the profile of the mass shell by
\begin{equation}\label{Eq:VaidyaMass}
M(v)=\frac{1}{2}\Big(1+\tanh(a v)\Big)\,.
\end{equation}
\begin{figure}[hbt]
\center
 \includegraphics[width=0.45\linewidth]{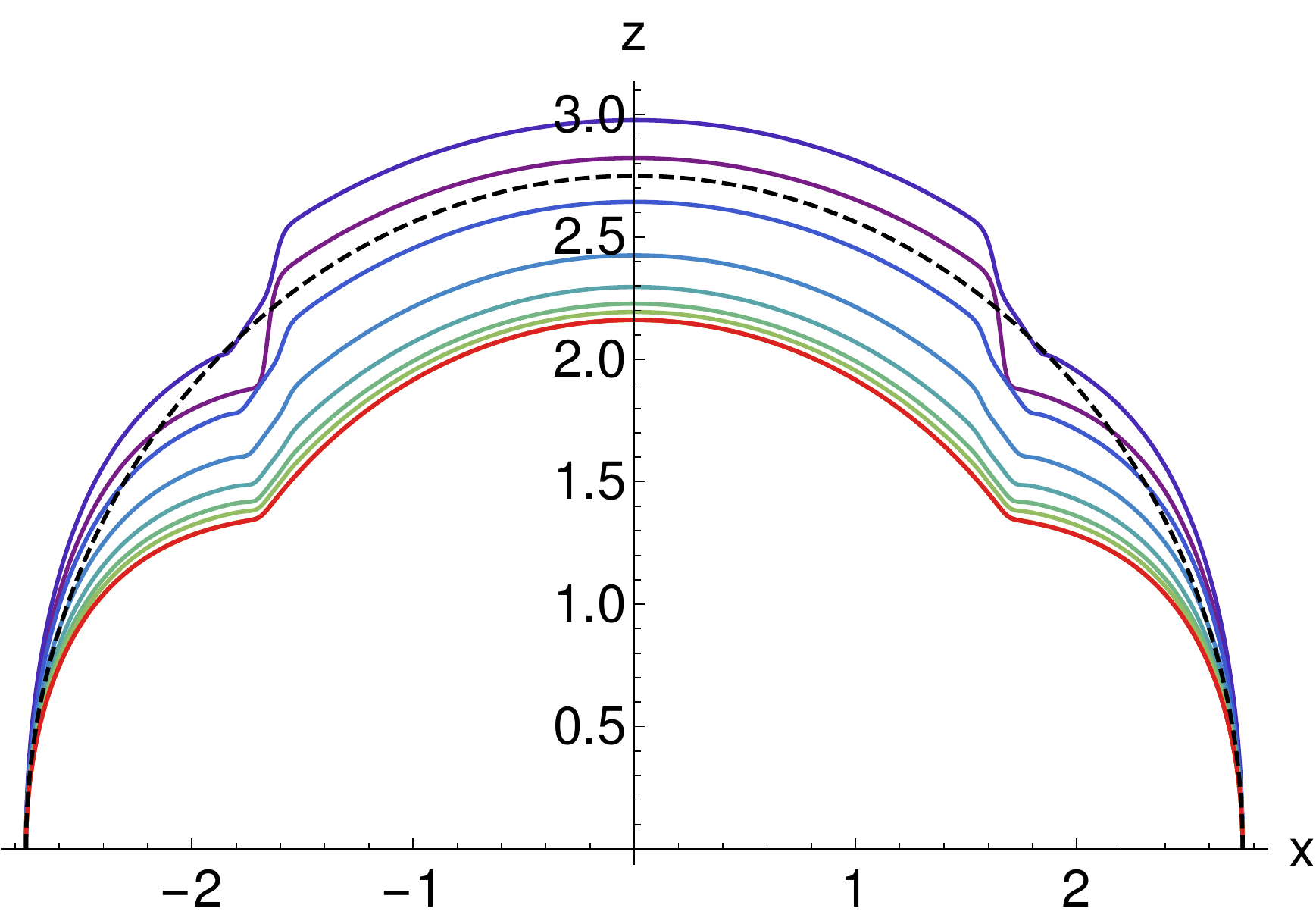} \quad\includegraphics[width=0.45\linewidth]{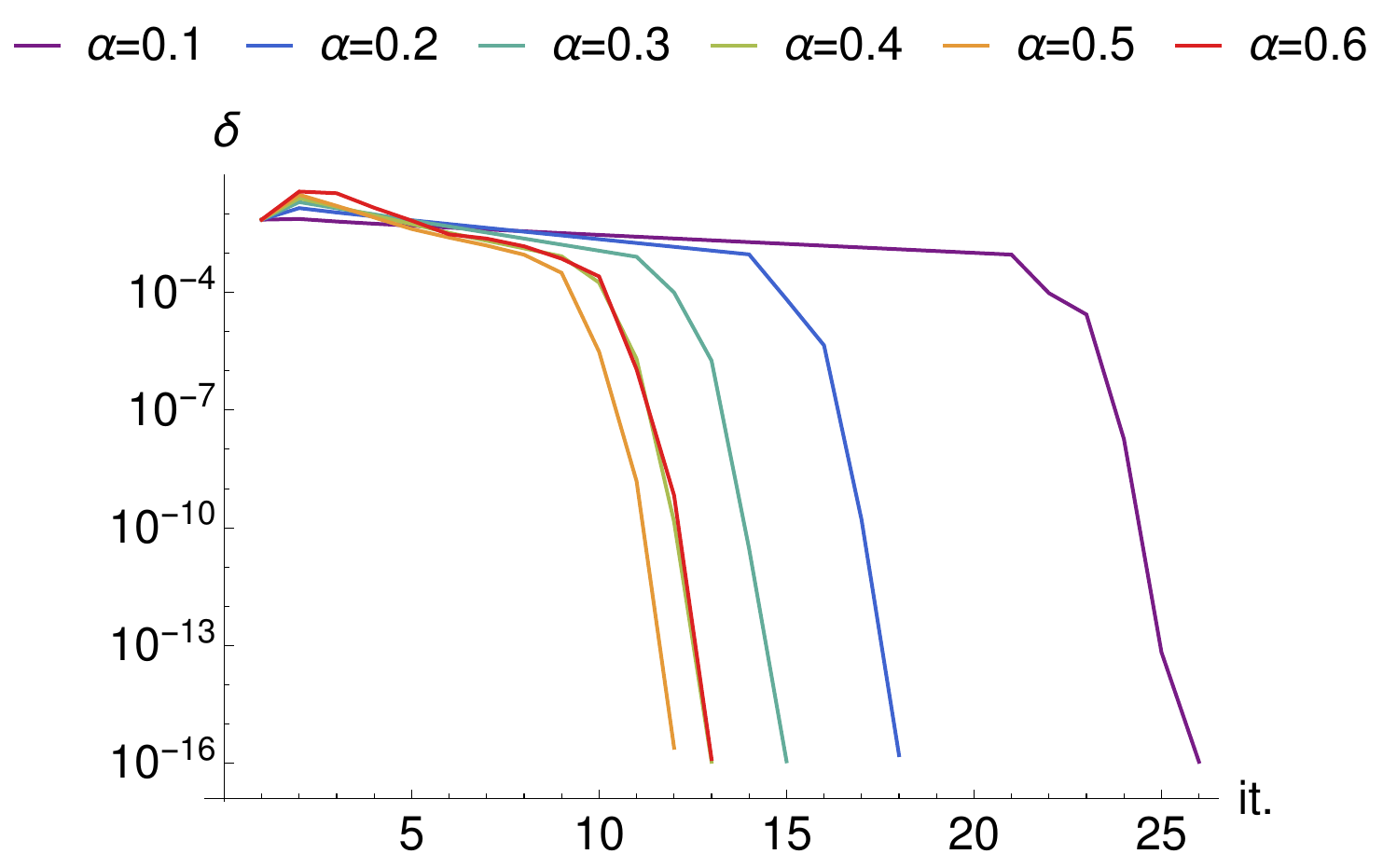}
\caption{Left: Surfaces obtained in subsequent relaxation steps with fixed boundary time $v_0=2$ and separation $l=5.5$ in a AdS$_3$ Vaidya spacetime with $a=30$. Right: Error $\delta$ over the number of iterations for different values of the weight $\alpha$. In this example the optimal weight, that leads to the minimal number of iterations, is $\alpha=0.5$.}
\label{Fig:iterations}
\end{figure}

On the left side of Figure \ref{Fig:iterations} we show the surfaces starting with the initial guess which is the black dashed line together with subsequent iterations from the purple to the final solution which is the red line. On the right hand side we show the error measure $\delta$ for different choices of the weight $\alpha$. For this example the optimal choice is $\alpha=0.5$ for which after 12 iterations the error criterion $\delta<10^{-15}$ is fulfilled.
Note that in general the optimal value for $\alpha$ strongly depends on the chosen ansatz and the detail of the spacetime in which the surface resides. For more complicated examples smaller values of $\alpha$ often make the relaxation more stable. 

\begin{figure}[htb]
\center
 \includegraphics[width=0.45\linewidth]{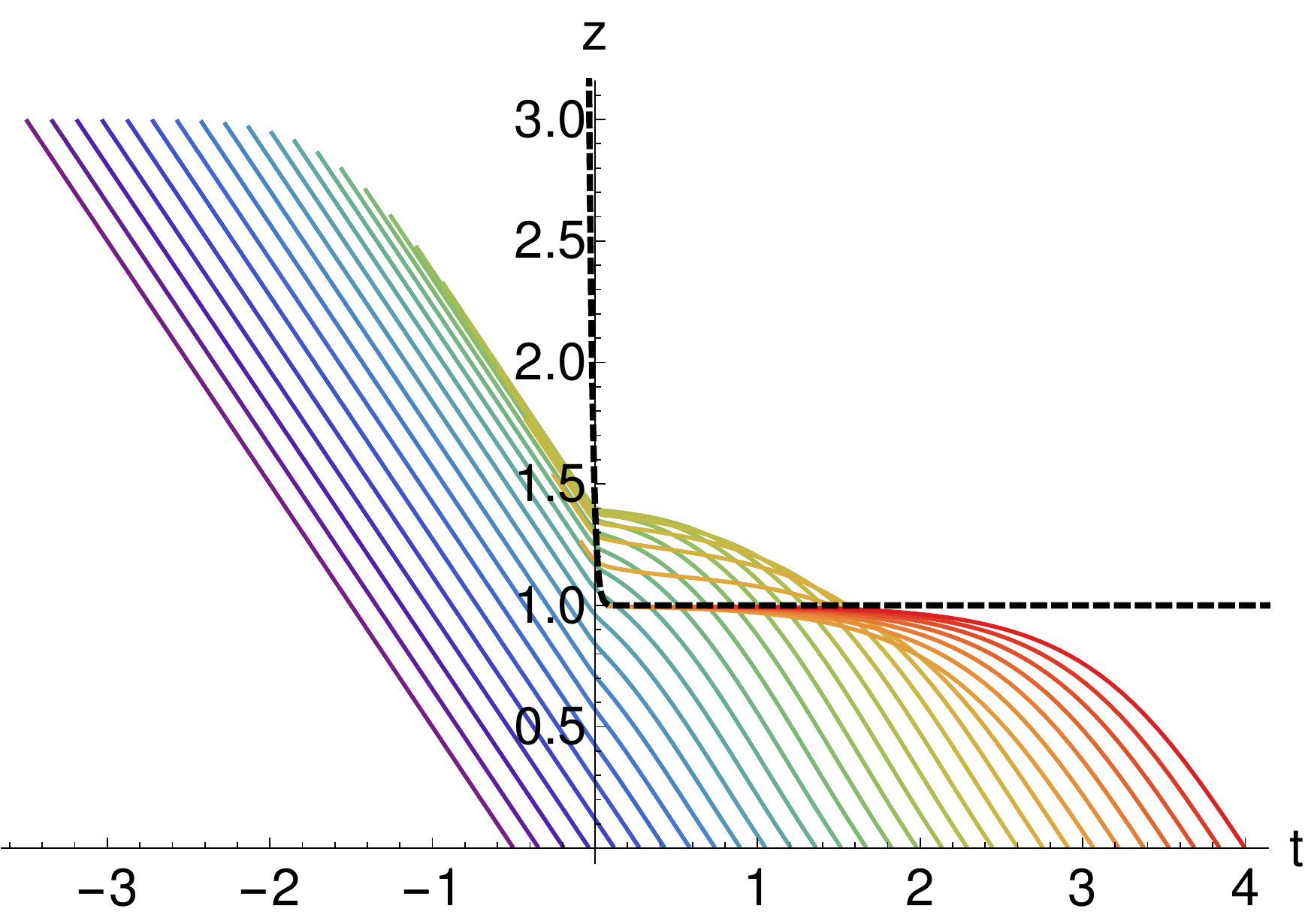} \quad\includegraphics[width=0.45\linewidth]{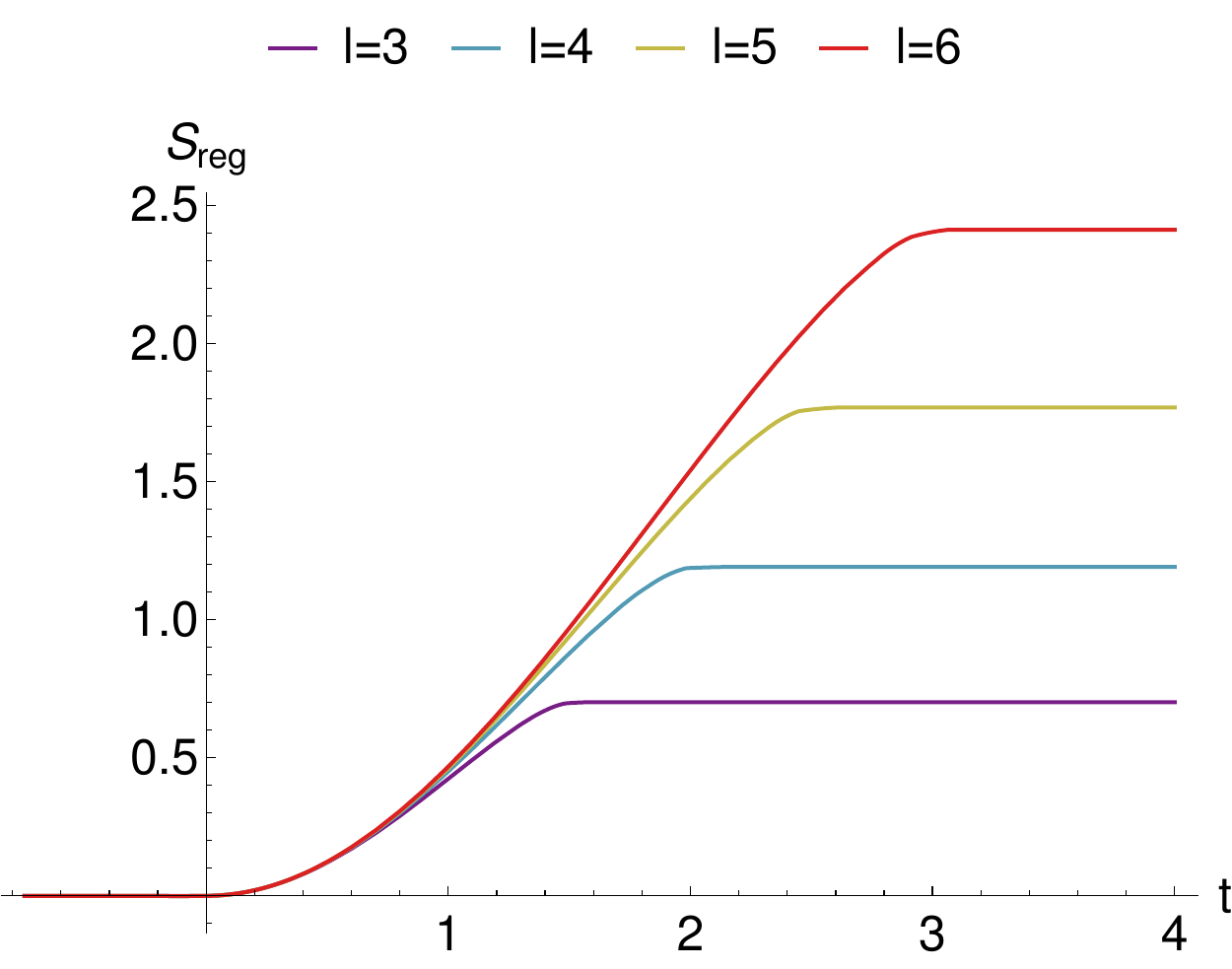}
\caption{Left: Minimal surfaces at different times and fixed boundary separation $l=6$ in the AdS$_3$ Vaidya geometry with $a=30$. The black dashed line indicates the radial position of the apparent horizon. Right: Vacuum subtracted entanglement entropy $S_{ren}=S-S_{vac}$ as function of time for different separations $l$. $S_{ren}$ nicely reproduces the scaling behavior of \eqref{Eq:CFTquench}. }
\label{Fig:Vaidya1}
\end{figure}

For the time evolution we use as ansatz at time $t_{i}=t_{i-1}+\delta t$ the geodesic from the previous time step $t_{i-1}$.
With a step size of $\delta t \approx 0.05$ usually less than five relaxation steps are sufficient to reach the accuracy of (\ref{Eq:error}). 
In Figure \ref{Fig:Vaidya1} we show the results for the time evolution of such surfaces and the corresponding entanglement entropy computed from their surface areas.
In strong contrast to the static case, where the surfaces always remain outside the horizon, in the time dependent Vaidya spacetime the surfaces can cross the horizon as shown in the plot on the left side. This happens when the surfaces probe a region of the spacetime which is highly dynamic like it is around $t=0$, where the black hole rapidly forms. At later times the surfaces remain entirely outside the horizon because they only probe the final black hole geometry which is almost static in this case. 
In the plot on the right hand side we show the vacuum subtracted entanglement entropy $S_{reg}$ as a function of time for different widths of the entangling region. It nicely shows the linear growth $S_{reg}\propto t$ for $t<l/2$ and the constant behavior for $t>l/2$ such as expected from \eqref{Eq:CFTquench} obtained from the CFT calculation \cite{Calabrese:2005in}. The linear growth has no obvious explanation in the geometric picture, but the constant behavior can easily be understood from the fact that at $t>l/2$ the surfaces remain entirely in the almost static black hole spacetime.

\begin{figure}[htb]
\center
 \includegraphics[width=0.45\linewidth]{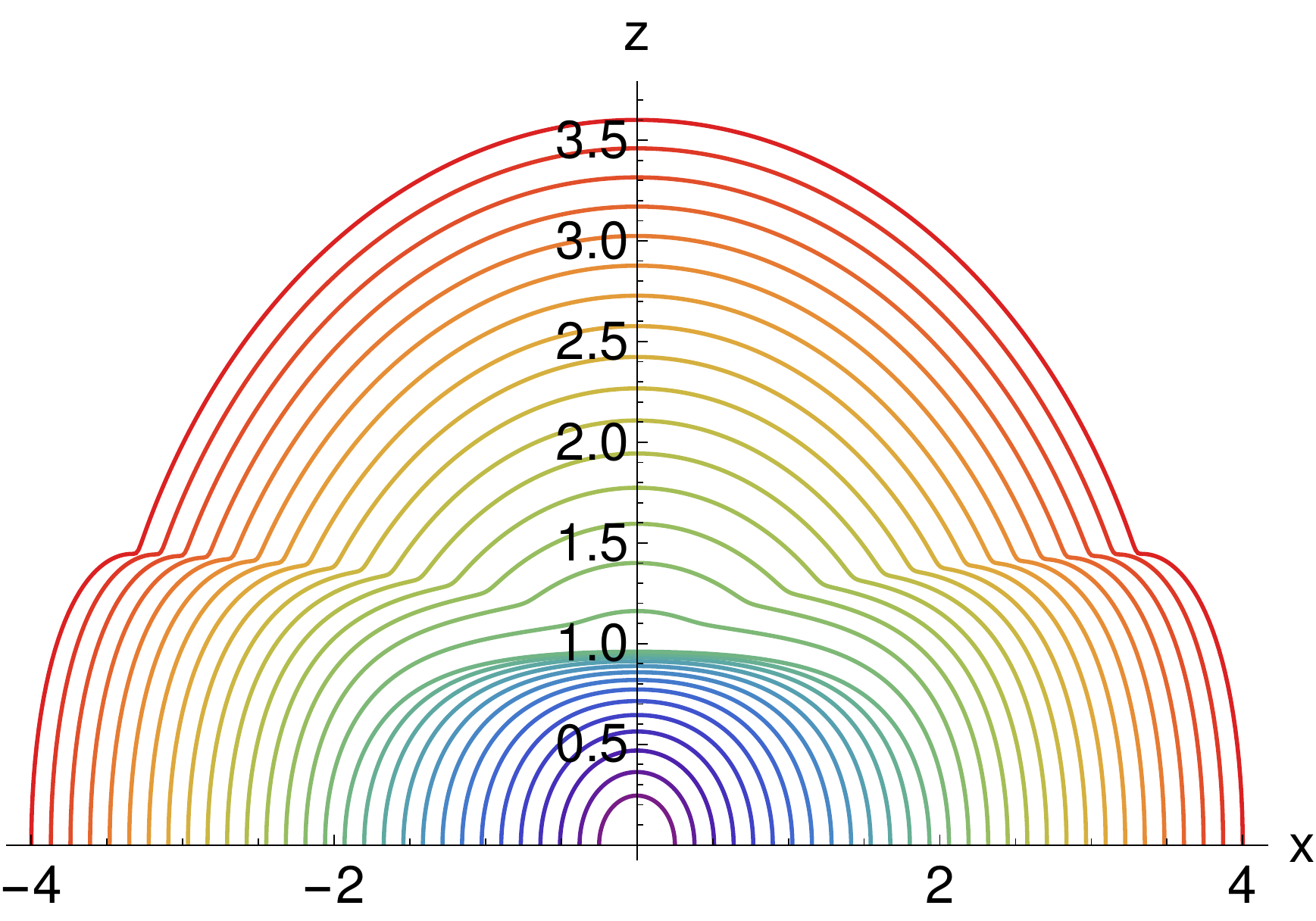} \quad\includegraphics[width=0.45\linewidth]{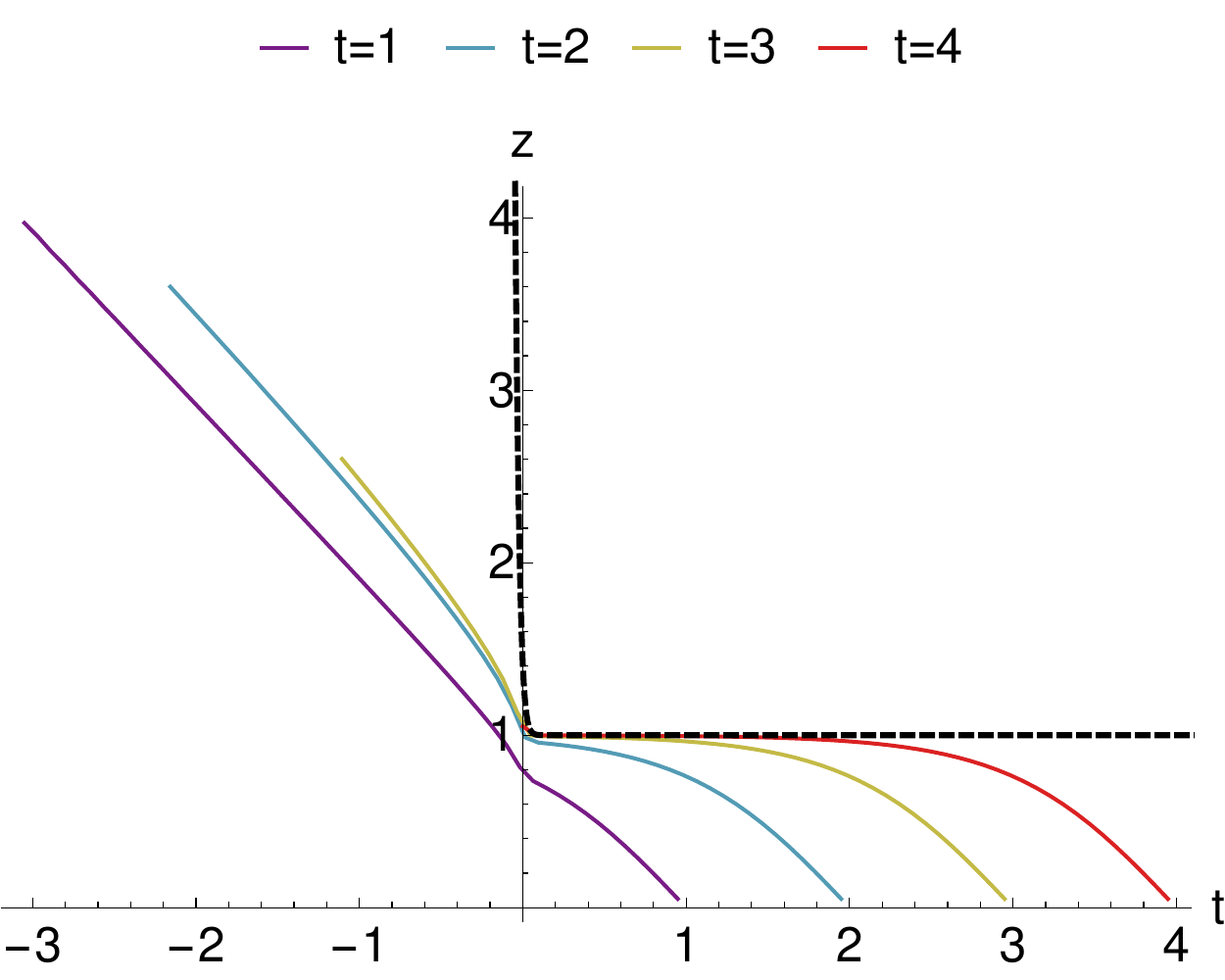}
\caption{Left: Minimal surfaces for different boundary separation $l$ and fixed boundary time $t=2$ in the AdS$_3$ Vaidya geometry with $a=30$. Right: location of the central point, located at $y=0$, of a family of surfaces with different separation but fixed boundary time.  The central point of all the surfaces remains outside the apparent horizon given by the black dashed line.}
\label{Fig:Vaidya2}
\end{figure}

In Figure \ref{Fig:Vaidya2} (left) we show surfaces at fixed time $t=2$ and different separations on the boundary. Surfaces with large separation are highly distorted, because they probe regions of the spacetime where the curvature strongly varies.
Interestingly, the central point, located at $y=0$, of all the surfaces remains always outside the horizon. This can be seen in the plot on the right hand side of Figure \ref{Fig:Vaidya2}, where each colored line represents the location of the central point of a family of surfaces with different separation but fixed boundary time.

The metric of the Vaidya spacetime is known in closed form and can be written explicitly into the relaxation code. In cases where no analytic solution is available, like for the anisotropic black brane or the shock wave geometries presented in later chapters, we feed our relaxation code an interpolated form of the metric which we generate by solving the Einstein equations numerically, using the spectral method explained in the first section of this chapter. 

\section{Computing QNEC Numerically}\label{Sec:QNECVaidya}
In this section we discuss our method to compute the variation of the entanglement entropy with respect to a lightlike deformation such as required in QNEC.
With the relaxation method, introduced in the previous section, this becomes rather simple. 
We use again the AdS$_3$ Vaidya spacetime of \eqref{Eq:VaidyaMetric} with \eqref{Eq:VaidyaMass} to illustrate our method.

\begin{figure}[htb]
\center
 \includegraphics[width=0.45\linewidth]{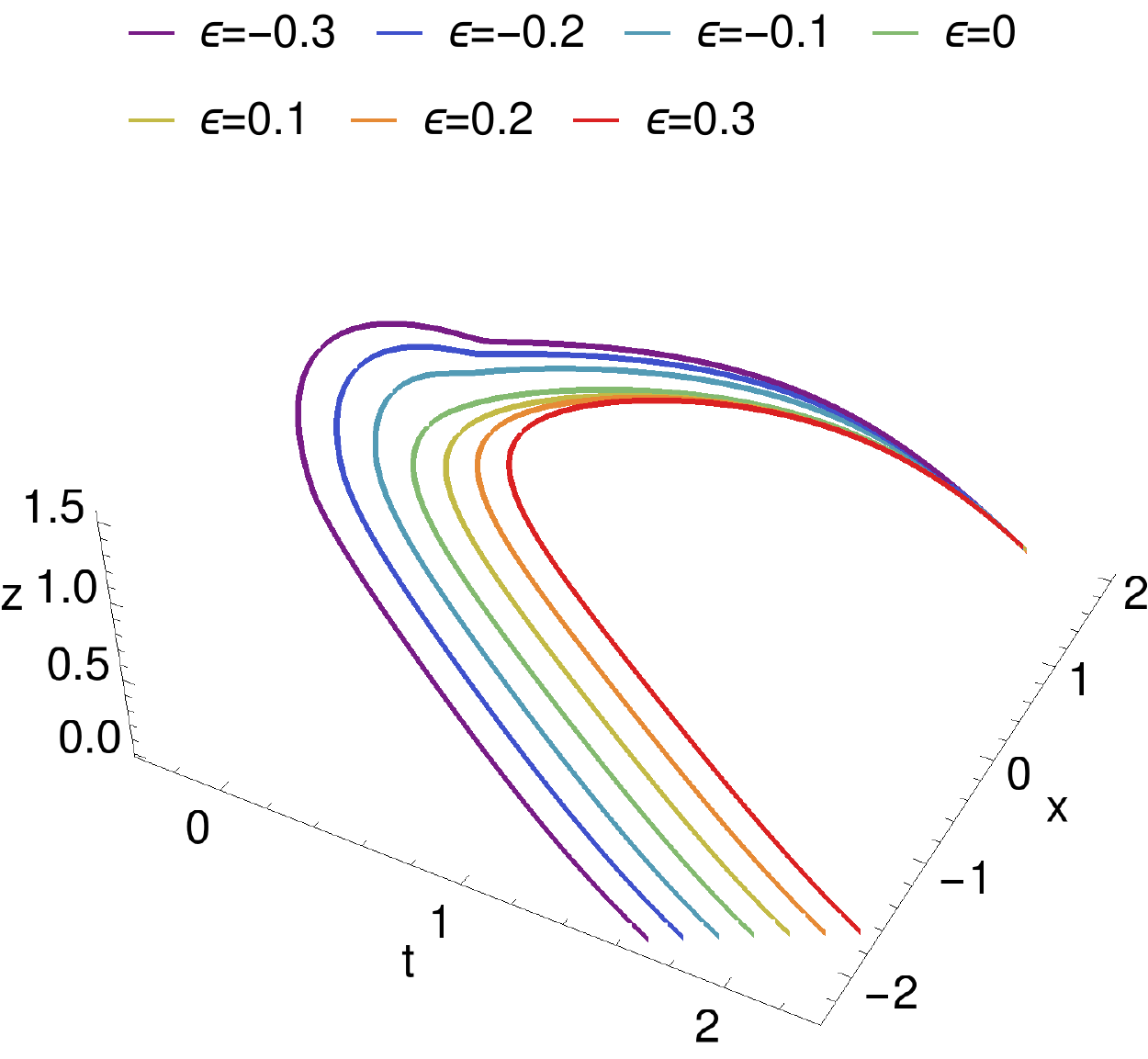} \quad\includegraphics[width=0.45\linewidth]{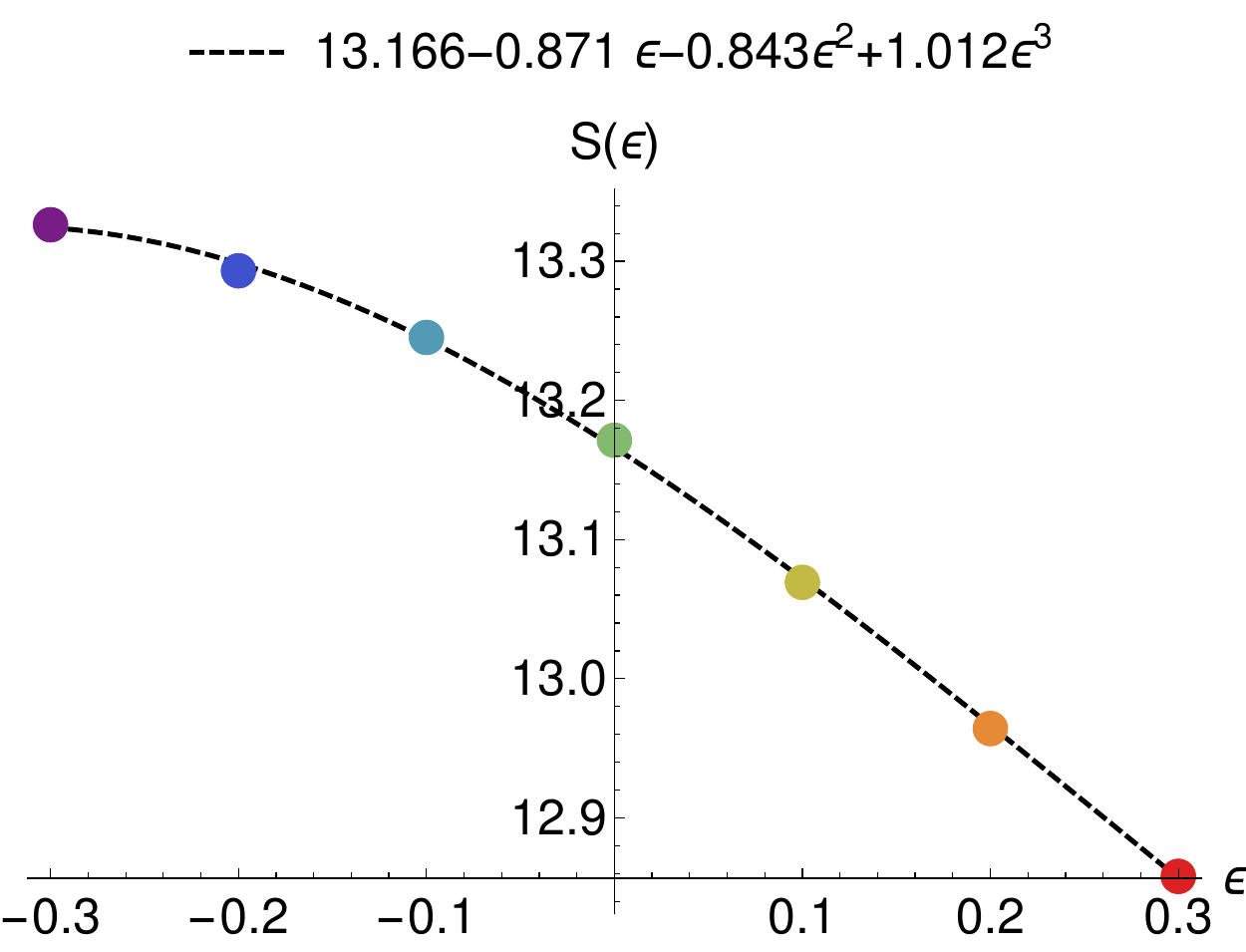}
\caption{Left: Family of extremal surfaces with one endpoint deformed along the lightlike vector $k^\mu_+=\epsilon*(1,1)$ in the AdS$_{3}$ Vaidya geometry with $a=30$. 
Right: Discrete values of the entanglement entropy such as computed from the areas of the surface family shown in the left plot. The black dashed line is a third order polynomial fit generated from the discrete values of the entanglement entropy. }
\label{Fig:VaidyaQNECgeo}
\end{figure}

First we generate a family of extremal surfaces with one endpoint shifted in equidistant steps of size $\epsilon$ along the desired lightlike vector $k^\mu_{\pm}=(1,\pm1)$, using our relaxation code.
In Figure \ref{Fig:VaidyaQNECgeo} (left) we show such a family of surfaces with shifted endpoints. Figure \ref{Fig:VaidyaQNECgeo} (right) shows the corresponding values of the entanglement entropy computed from the area of each surface, together with a third order polynomial fit $S\approx c_0+c_1 \epsilon+c_2 \epsilon^2+c_3 \epsilon^3$ from which we compute the desired derivatives at $\epsilon=0$, such as required in the QNEC formula.
Note that the values of $\epsilon$ in these plots are exaggerated for illustrative purposes. The typical values for $\epsilon$ we use in our simulations are $\mathcal{O}(10^{-3})$.

It is instructive to study QNEC as function of time and separation in our simple Vaidya example.
As argued in \cite{Leichenauer:2018obf}, QNEC in $d>2$ is saturated by all states. However, as noted in \cite{Khandker:2018xls}, QNEC in $d=2$ does not need to saturate in the presence of bulk matter.
More precisely, in $d=2$ QNEC is not saturated if the corresponding RT-surface probes through regions of the geometry where the bulk energy momentum tensor is non-vanishing. This behavior is nicely confirmed by in our example with the Vaidya spacetime, which has an in-falling matter shell defined by a non-vanishing energy momentum tensor located close to $t=0$.\footnote{Note that for our choice of Eddington-Finkelstein coordinates the in-falling shell resides at a fixed value of the advanced time coordinate.}
\begin{figure}[htb]
\center
 \includegraphics[width=0.45\linewidth]{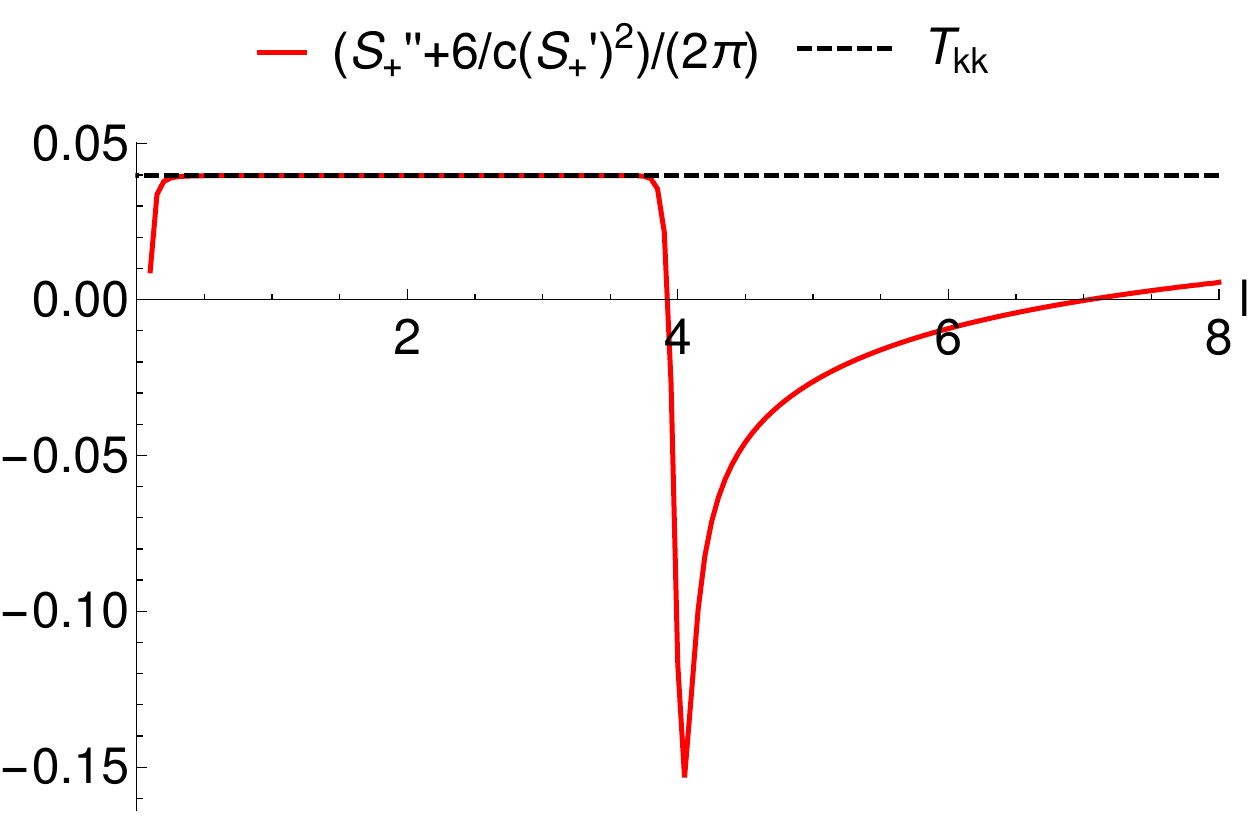} \quad\includegraphics[width=0.45\linewidth]{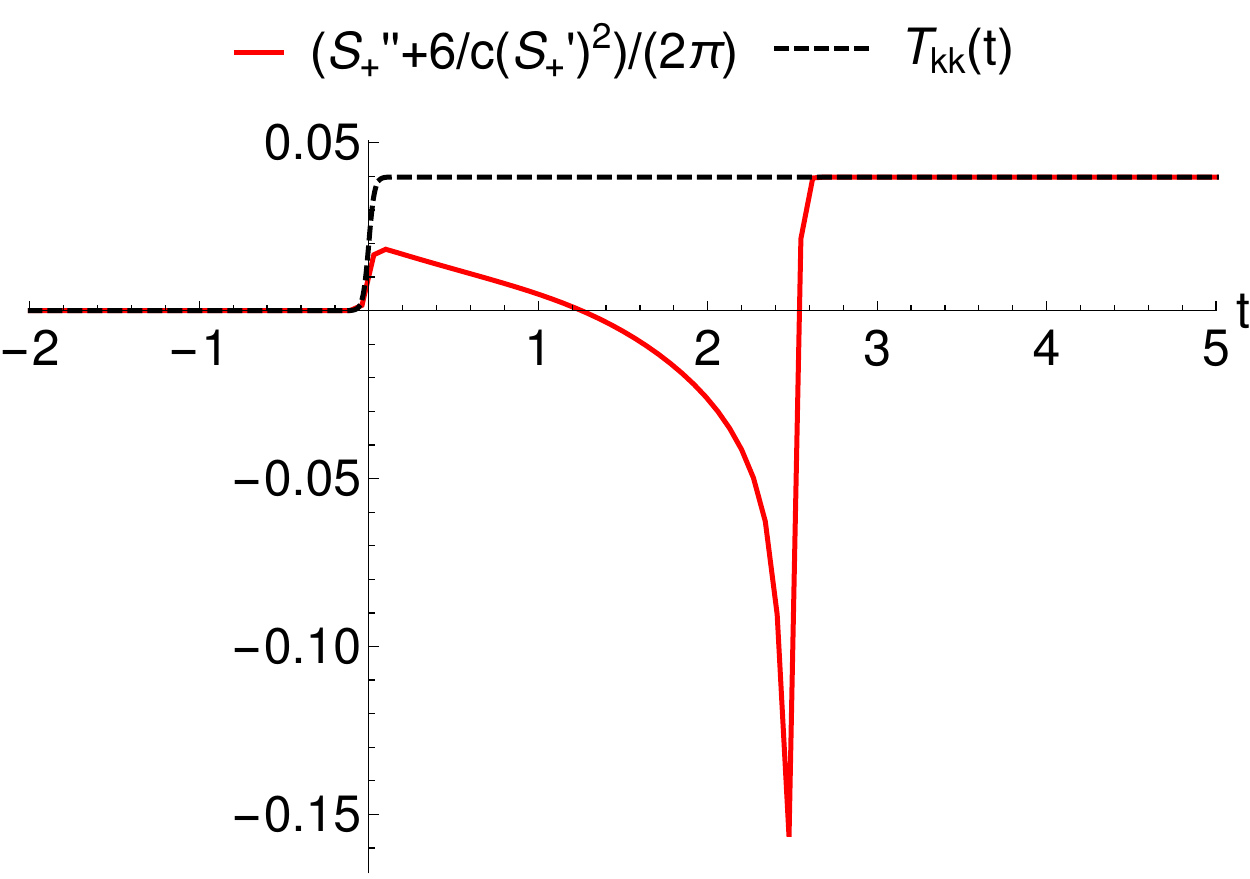}
\caption{Left: QNEC for positive deformation ($k^\mu_+\propto(1,1)$) as function of separation and fixed boundary time $t=2$ in the Vaidya geometry with $a=30$. The black dashed line is $T_{++}(t=2)$. Right: QNEC as function of time and fixed separation $l=5.0$.}
\label{Fig:VaidyaQNEC}
\end{figure}

In Figure \ref{Fig:VaidyaQNEC} (left) we plot QNEC at $t=2$ as function of the size of the entangling region. Note that surfaces anchored at $t=2$ with separations $l<t+2$ are too short to cross the in-falling shell at $t=0$ and QNEC is saturated. The apparent non-saturation in this plot at very small $l$ is a finite-cutoff effect in our numerics which is non-physical and QNEC actually saturates all the way down to $l=0$. At $l\approx 4$ the QNEC becomes suddenly non-saturated and one can show that it diverges in the limit of $\delta$-like shells obtained by sending $a\to \infty$. This is the point where the dip of the RT-surface touches the matter shell. For $t>2l$ QNEC increases again without saturating, because the surface then always crosses the in-falling matter shell.

In Figure \ref{Fig:VaidyaQNEC} (right) we plot QNEC as function of time for constant separation $l=5$. Here the situation is similar. For $t<0$ the surface resides entirely in the vacuum region of the spacetime and QNEC is saturated trivially. At $t=0$ the surface starts to cross the matter shell and QNEC is not saturated. At $t=l/2=2.5$ the dip of the surface crosses the matter shell and QNEC becomes sharply peaked. For $t>l/2$ the surface is not long enough to touch the matter shell any more and QNEC saturates again. In Appendix \ref{App:MathematicaQNEC} we give a simple Mathematica implementation of the 
numerical method presented in this section.
This closes our discussion on numerical methods.
In the following chapters we will apply these methods to more complicated examples.

\chapter{Entanglement Entropy in an Anisotropic System}\label{chap:Aniso}
In this chapter we determine holographically two-point correlators of gauge invariant operators with large conformal weights and entanglement entropy of strips for a time dependent anisotropic five-dimensional asymptotically Anti-de~Sitter spacetime. At the early stage of evolution where geodesics and extremal surfaces can extend beyond the apparent horizon all observables vary substantially from their thermal value, but thermalize rapidly. At late times we recover quasinormal ringing of correlators and holographic entanglement entropy around their thermal values, as expected on general grounds. We check the behavior of holographic entanglement entropy and correlators as function of the separation length of the strip and find agreement with the exact expressions derived in the small and large temperature limits.
Results displayed in this chapter are published in \cite{Ecker:2015kna}.

\section{Anisotropic Asymptotically AdS$_5$ Spacetimes}
In this section we review the most important details of the model first introduced in \cite{Chesler:2008hg} and studied further in 
\cite{Chesler:2009cy, Heller:2012km, Heller:2013oxa, Chesler:2013lia}. 

The five-dimensional bulk metric that introduces anisotropy between the longitudinal ($y$) and transverse ($\vec{x}=(x_1,x_2)$) directions with an $\mathcal{O}(2)$ rotational invariance in the transverse plane is the one we have introduced in \eqref{Eq:metricAniso} which we repeat here for convenience,

\begin{equation}\label{Eq:Metric}
d s^2=-A(r,v)d v^2+2d rd v +S ^2(r,v)\Big( e^{-2B(r,v)} d y^2 +e^{B(r,v)} d\vec{x}^2 \Big)\,.
\end{equation}

The Einstein equations given in \eqref{Eq:Einstein} have to be solved for special initial conditions and appropriate boundary conditions. 
There are, at least, two ways to create a far-from-equilibrium state. 
On the one hand one can turn on a time dependent anisotropy function at the boundary $B(r=\infty,v)=B_0(t)$ as in the original works of \cite{Chesler:2008hg, Chesler:2009cy} and let the system evolve. In this case the boundary metric is curved and the conformal anomaly is present \cite{Henningson:1998gx}. 
On the other hand one can specify the initial state in the absence of external sources by specifying the metric in the bulk on the initial time slice \cite{Heller:2012km} with a flat boundary geometry. 
For simplicity, in the following  we will study the setup where the boundary metric is flat and time independent. 

For our initial data  we follow \cite{Heller:2013oxa, Chesler:2013lia} and choose for the anisotropy function on the initial time slice 
\begin{equation}
B(r,v_0)=\frac{\beta}{r^4}\exp \left[-\Big(\frac{1}{r}-\frac{1}{r_0}\Big)^2/\omega^2\right]\,,
\label{eq:initial}
\end{equation}
with $\beta=6.6$, $r_0=4$ and $w=1$.
In addition  the initial conditions have to be supplemented  with a value for the coefficient $a_4$ which sets the energy density of the initial state, for which we take  $a_4=-1$,  corresponding to an equilibrium temperature $T=1/\pi$, as we now recall.

At late times we expect isotropization, $B=0$. In that case we recover the usual static AdS black brane solution as follows. Solving \eqref{eq:E5} and using residual gauge transformations yields $S=r$. This implies $S^\prime=1$ and $\dot S=\tfrac12\, A$. Solving \eqref{eq:E1} then yields $A=r^2\,(1-1/r^4)$, where we fixed the integration constant such that $a_4=-1$. The other equations are either trivial, \eqref{eq:E2} and \eqref{eq:E4}, or redundant, \eqref{eq:E3}. The result for $A$ is the usual Killing norm for the static AdS black brane. Surface gravity is given by $\kappa = \tfrac12\,A^\prime\big|_{r=1}=2$ so that the Hawking temperature is $T=\kappa/(2\pi)=1/\pi$.

In the generic anisotropic case, $B\neq 0$, we solve the Einstein equations \eqref{Eq:Einstein} numerically for the initial conditions \eqref{eq:initial}.
In this background we then study the evolution of two-point correlation functions for operators of large conformal weights and the entanglement entropy. This in turn requires us to determine the background sufficiently far beyond the apparent horizon.

We solve the Einstein equations \eqref{Eq:Einstein} using pseudo spectral methods such as described in Chapter \ref{Chap:Numerics}. We do not fix the location of the apparent horizon. This facilitates the study of geodesics and extremal surfaces that reach behind the apparent horizon, which is of relevance for two-point functions and entanglement entropy. For that reason we want a large computational domain in the holographic coordinate $z=1/r$. 
In all the computations we took $z\in [0,~1.6]$ with the final position of the horizon located at $z=1$.
For the time evolution it is sufficient to use a fourth order Runge-Kutta method with time steps $\delta t=10^{-3}$. 
All the computations in this chapter were done with the open source software GNU Octave \cite{octave:2014}. 
\subsection{Two-Point Correlators}\label{se:2.2}

The equal time two-point function for an operator of large conformal weight $\Delta$ can be computed via a path integral as \cite{Balasubramanian:1999zv, Festuccia:2005pi}
\begin{equation}
\langle \mathcal{O}(t, \vec{x})\mathcal{O}(t, \vec{x}')\rangle=\int  \mathcal{D P}\, e^{i \Delta \mathcal{L(\mathcal{P})} }\approx \!\!\!\sum_{\textrm{\tiny geodesics}}\!\!\! e^{-\Delta L_g}\approx e^{-\Delta L}\,,
\end{equation}
where the  integral is a sum over all possible paths with endpoints at $(t, \vec{x}')$ and $(t, \vec{x})$ and $\mathcal{L(P)}$ is the proper length of the path.  The first approximation neglects perturbative corrections and is the so-called geodesic approximation, which holds in the limit when the conformal weight of the operator is large. The conformal weight effectively plays the role of $1/\hbar$ in usual perturbative expansions of path integrals. Then it can be shown that the sum over all paths reduces to a sum over all geodesics where $L_g$ denotes the length of the corresponding geodesic. To leading order only the geodesic with the smallest value of $L_g$ contributes, whose length we denote by $L$, which explains the second approximation\footnote{
For a comparison of the  two-point correlation function obtained by  using the ``extrapolate" dictionary and the geodesic approximation  in AdS$_3$ Vaidya spacetime see \cite{Keranen:2014lna}.}. It neglects instanton corrections.

However, the length of the geodesic has a divergence originating from the asymptotically AdS boundary and therefore needs to be  renormalized. We choose to subtract the length of a geodesic in the static  black brane background, which we denote by $L_{\textrm{\tiny therm}}$. In terms of the renormalized length $\delta L=L-L_{\textrm{\tiny therm}}$ the two-point function becomes
\begin{equation}
\langle \mathcal{O}(t, \vec{x})\mathcal{O}(t, \vec{x}')\rangle\sim e^{-\Delta \delta L}\,.
\end{equation} 
This means that we can obtain  the time evolution of two-point functions by looking at spacelike  geodesics that are anchored at the boundary at fixed separation $l$ and calculating their length at different times.
Due to the anisotropy in the system  we only solve for the subset of correlation functions that are either separated in the longitudinal direction or in the transverse directions.

To this end we let all the coordinates depend on one parameter $\sigma$, which lies in the interval $\sigma\in [-\sigma_m,~ \sigma_m]$.
To obtain the lengths of the geodesics we have to solve the geodesic equation for the two subspaces given by the line elements 

\begin{subequations}
\begin{eqnarray}
d s^2_{x_{1,2}}&=& -A d v^2-\frac{2}{z^2}  d z d v +S^2 e^{B}d x_{1,2}^2\,, \\
d s^2_y&=&  -A d v^2-\frac{2}{z^2}  d z d v +S^2 e^{-2B} d y^2\,.
\end{eqnarray}
\end{subequations}
For the separation in the transverse direction the geodesics end at $(v(\pm \sigma_m)=t, \allowbreak \,x_1(\pm \sigma_m)=\pm x_0/2, \, x_2(\pm \sigma_m)=0, \,y (\pm \sigma_m)=0)$, where $t$ is the boundary time.
Similarly, for the longitudinal separation we take  $(v(\pm \sigma_m)=t,~y(\pm \sigma_m)=\pm x_0/2,~\vec{x}(\pm \sigma_m)=0)$. 
With this choice of boundary conditions the lengths of the geodesics in the background (\ref{Eq:Metric}) are given by
\begin{subequations}
\begin{eqnarray}
L_{x_{1,2}}&=&\int_{-\sigma_m}^{\sigma_m} d\sigma \sqrt{-A (v')^2-\frac{2}{z^2} z'v'+S^2 e^{B}(x'_{1,2}) ^2}\,,\\
L_y&=&\int_{-\sigma_m}^{\sigma_m} d\sigma \sqrt{-A (v')^2-\frac{2}{z^2} z'v'+S^2 e^{-2B}(y') ^2}\,,
\end{eqnarray}
\end{subequations}
where prime denotes the derivative with respect to $\sigma$.

We can only study geodesics after some advanced time $v>v_{\textrm{\tiny min}}$ with  boundary separations  below a maximal separation $l<l_{\textrm{\tiny max}}$. 
This comes from the fact that by solving Einstein's equations numerically we have to choose a finite computational domain. 
Also, by  specifying  the initial state in the entire bulk on the initial time slice  the advanced time interval at our disposal is $v\in [v_0, \infty]$. 
As the geodesics reach into the bulk they bend back in advanced time leaving the computational domain for advanced times $v<v_{\textrm{\tiny min}}$ as well as extending too far into the bulk for separations $l>l_{\textrm{\tiny max}}$. 

To compute two-point functions we need to find curves of extremal length in a curved spacetime whose endpoints reside on fixed positions on the boundary of that spacetime.
These curves are solutions to the geodesic equation subject to boundary conditions at the endpoints.
For numerical reasons it turns out to be convenient to use a non-affine parameter $\sigma$ , where $\tau=\tau(\sigma)$ is the usual affine parametrization, $\frac{d X^\mu}{d\tau}\frac{d X^\nu}{d\tau}g_{\mu\nu}=1$.
In terms of $\sigma$ the geodesic equation reads
\begin{equation}
\ddot X^\mu + \Gamma^\mu{}_{\alpha\beta} \dot X^\alpha \dot X^\beta = J \dot X^\mu\,,
\label{Eq:GeodescEquation}
\end{equation}
where $\dot X^\mu=\frac{d X^\mu}{d\sigma}$ and $J=\frac{d^2\tau}{d\sigma^2}/\frac{d\tau}{d\sigma}$ denotes the Jacobian which originates from the change in parametrization (see Appendix \ref{App:AreaFunct}). 
This form of the geodesic equation gives us the freedom to choose parametrizations resulting in better convergence behavior of the relaxation algorithm than the affine parametrization does. In physical terms the right hand side in \eqref{Eq:GeodescEquation} introduces a fictitious viscous force that enhances numerical convergence.
In our case the geodesic equation \eqref{Eq:GeodescEquation} is given by a set of three coupled non-linear ODEs of second order for the geodesic coordinates $V,Z,X$. 

This set of equations is a two-point boundary value problem, which is usually either solved with shooting methods or relaxation methods \cite{Press:2007:NRE:1403886}. We do not shoot but relax, following the procedure described in Chapter \ref{Chap:Numerics}, where we state explicitly the geodesics used as initial guess for the relaxation algorithm and the corresponding Jacobian used in the numerical simulation. Since we are interested in one-parameter families of geodesics (evaluated at different constant time slices) we can take the solution for the $n^{\rm th}$ family member as initial guess for the $(n+1)^{\rm st}$ family member. 

\subsection{Holographic Entanglement Entropy}

In time dependent systems the covariant entanglement entropy \cite{Hubeny:2007xt} for some boundary region $A$  is obtained  by extremizing the 3-surface functional  
\begin{equation}\label{Eq:area1}
{\cal A}=\int d^3\sigma\sqrt{\det\Big(\frac{\partial X^{\mu}}{\partial\sigma^a}\frac{\partial X^{\nu}}{\partial\sigma^b}g_{\mu\nu}\Big)}\,,
\end{equation}
that ends on the boundary surface $A$.
In the dual field theory the entanglement entropy is then conjectured to be given by \cite{Ryu:2006bv,Ryu:2006ef,Hubeny:2007xt}
\begin{equation}
S_{\textrm{\tiny EE}}=\frac{\cal A}{4 G_N}\,.
\end{equation}

Usually the boundary regions of interest are either a sphere or a strip that has finite extent in one direction and infinite extent in the other two directions. 
In spacetimes with spherical symmetry  in the three spatial dimensions the problem of finding the extremal area functional \eqref{Eq:area1} effectively reduces to finding geodesics. 
In our case where spherical symmetry is broken this is not the case anymore. For example, finding the extremal area for a  spherical  boundary region would require to solve non-linear coupled PDEs. 
However, in the  case of a strip with finite extent either in the transverse or longitudinal direction it is  possible to reduce the problem to finding  geodesics in a suitable auxiliary spacetime, as we now demonstrate.

We introduce two scalar fields $\phi_i(x^\alpha)$ and write the line element as
\begin{equation}
d s^2=g_{\mu\nu}d x^\mu d x^\nu =h_{\alpha \beta} d x^\alpha d x^\beta +\phi_1^2d x_2^2+\phi_2^2 d x_3^2\,,
\end{equation}
where $h_{\alpha\beta}$ is a three-dimensional metric with coordinates $(v, r, x_1)$ where $x_1$ represents the coordinate we choose to have finite spatial extent, i.e.~either $x_\parallel$ or one of $x_\perp$. The remaining (non-compact) coordinates are then denoted by $x_2$, $x_3$, which we choose to be two of our three world-volume coordinates; the third one is denoted by $\sigma$. Parametrizing the three-dimensional coordinates as $x^{\alpha}=(v(\sigma), r(\sigma), x_1(\sigma))$, the area functional \eqref{Eq:area1} can be written as 
\begin{equation}
{\cal A} = \int d x_3 \int d x_2\int d\sigma \sqrt{\phi_1^2 \phi_2^2 h_{\alpha\beta}\frac{\partial x^{\alpha}}{\partial\sigma}\frac{\partial x^{\beta}}{\partial\sigma} }\,.
\label{eq:cg}
\end{equation}
Performing the integration over the Killing coordinates $x_2$ and $x_3$ yields a (possibly infinite) constant volume factor through which we are going to divide. Thus, instead of calculating entanglement entropy we calculate a entanglement entropy density per Killing volume. The problem of extremizing the three-surface corresponding to a boundary region $A$ of strip-topology is then reduced to a one-dimensional problem.

In fact, from the expression \eqref{eq:cg} one can see that the problem of finding the extremal three-surfaces reduces to finding geodesics of the conformal metric
\begin{equation}
d\tilde{s}=\tilde{h}_{\alpha\beta}d x^\alpha d x^\beta=\phi_1^2\phi_2^2 h_{\alpha\beta}d x^\alpha d x^\beta\;.
\end{equation}
The three-dimensional conformal metrics for separation in the transverse and longitudinal directions for which we have to solve the geodesic equation in our case are given by
\begin{subequations}
 \label{Eq:ds2TansLong}
\begin{eqnarray}
d\tilde{s}^2_{x_{1,2}}&=&S^4 e^{-B}   \left( -A d v^2+2 d r d v +S^2 e^{B}d x_{1,2}^2 \right)\,, \\
d\tilde{s}^2_y&=&S^4 e^{2B}   \left( -A d v^2+2 d r d v +S^2 e^{-2B} d x_y ^2 \right)\,.
\end{eqnarray}
\end{subequations}

The numerical procedure discussed for the two-point functions works also for entanglement entropy. Namely, for our problem at hand the evaluation of extremal surfaces is reduced to the evaluation of geodesics in some auxiliary spacetime. 

\section{Results}

In this section we display and discuss our main results. In all figures where a time axis is plotted we measure the boundary time $t$ or the bulk advanced time $v$ in units of the temperature of the final black brane, $T=1/\pi$.
To make the approach to thermal equilibrium most transparent we use normalized  quantities for the geodesic length  $L_\mathrm{ren}$ and entanglement entropy $S_\mathrm{ren}$ defined by 
\begin{subequations}\label{ren}
\begin{eqnarray}
L_\mathrm{ren}=\frac{L-L_\mathrm{th}}{L_\mathrm{th}}\,,\\
S_\mathrm{ren}=\frac{S-S_\mathrm{th}}{S_\mathrm{th}}\,,
\end{eqnarray}
\end{subequations}
where $L$ ($S$) is the unrenormalized length (entanglement entropy) and $L_\mathrm{th}$ ($S_\mathrm{th}$) is the corresponding thermal value.

\subsection{Background Geometry and Holographic Stress Tensor}\label{se:4.1}
\begin{figure}
\begin{center}
\hspace*{-0.4truecm}\includegraphics[scale=.45]{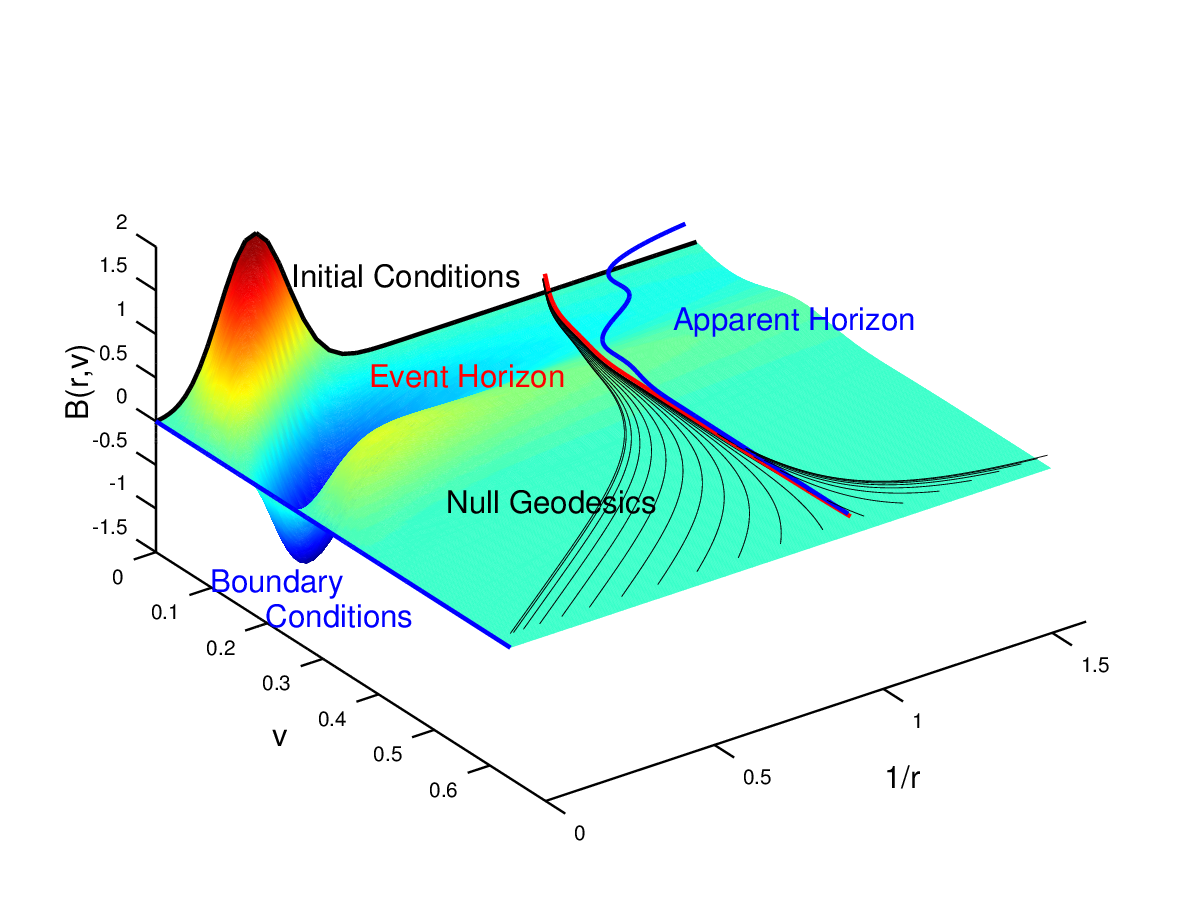}\hspace*{-0.8truecm}\includegraphics[scale=.35]{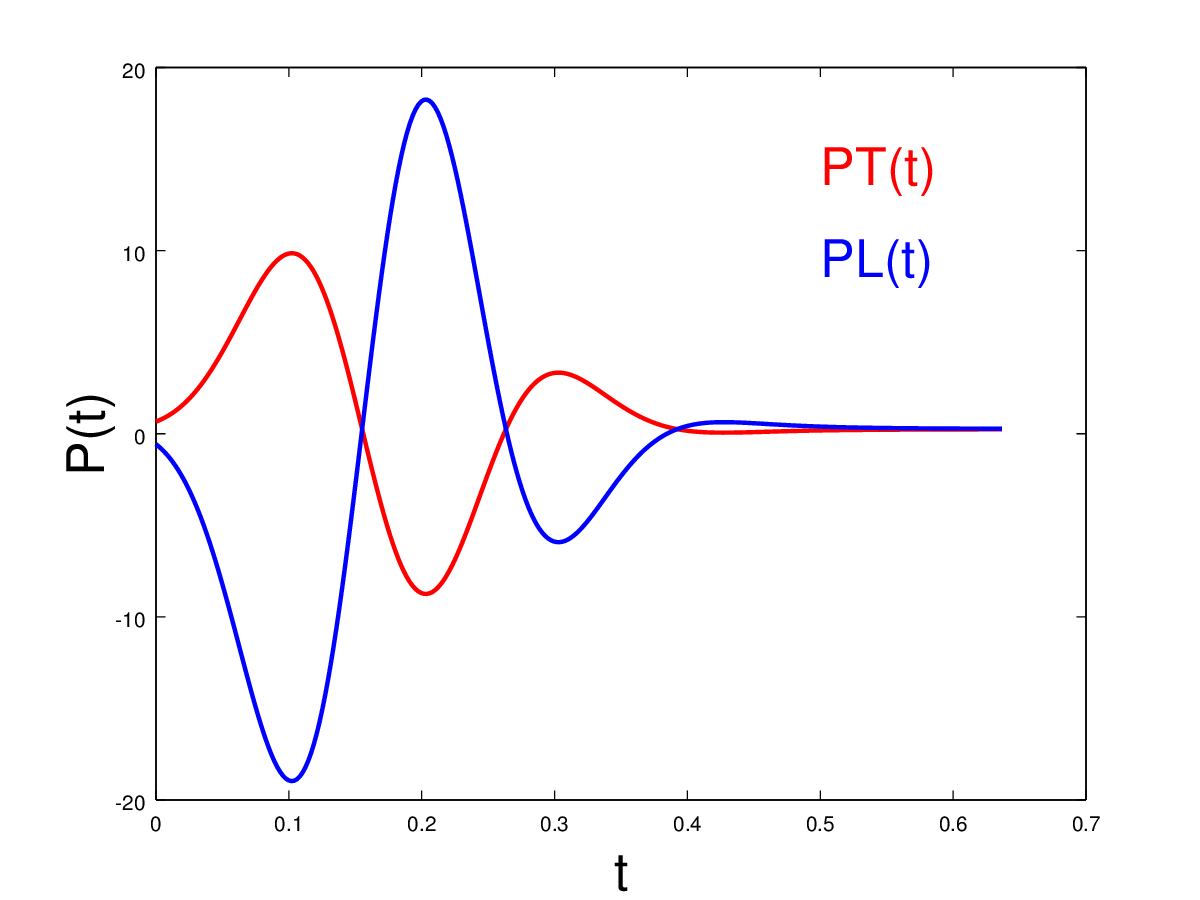}
\caption[Anisotropy function and pressures.]{\label{EMT} 
Left: anisotropy function $B(r,v)$. Right: transverse and longitudinal pressure. 
 }
\end{center}
\end{figure}

Figure \ref{EMT} displays the most salient features of the background geometry. The left figure plots the anisotropy function $B(r,v)$ and displays the regions outside and inside the apparent horizon, as well as the event horizon. The black lines depict a null congruence of geodesics close to the event horizon to exhibit their ingoing/outgoing nature. The right figure plots transversal and longitudinal pressures as function of boundary time $t$. Note the quick thermalization of the pressure components.

\subsection{Two-Point Correlators}\label{se:4.2}

\begin{figure}
\begin{center}
\hspace*{-0truecm}\includegraphics[scale=.4]{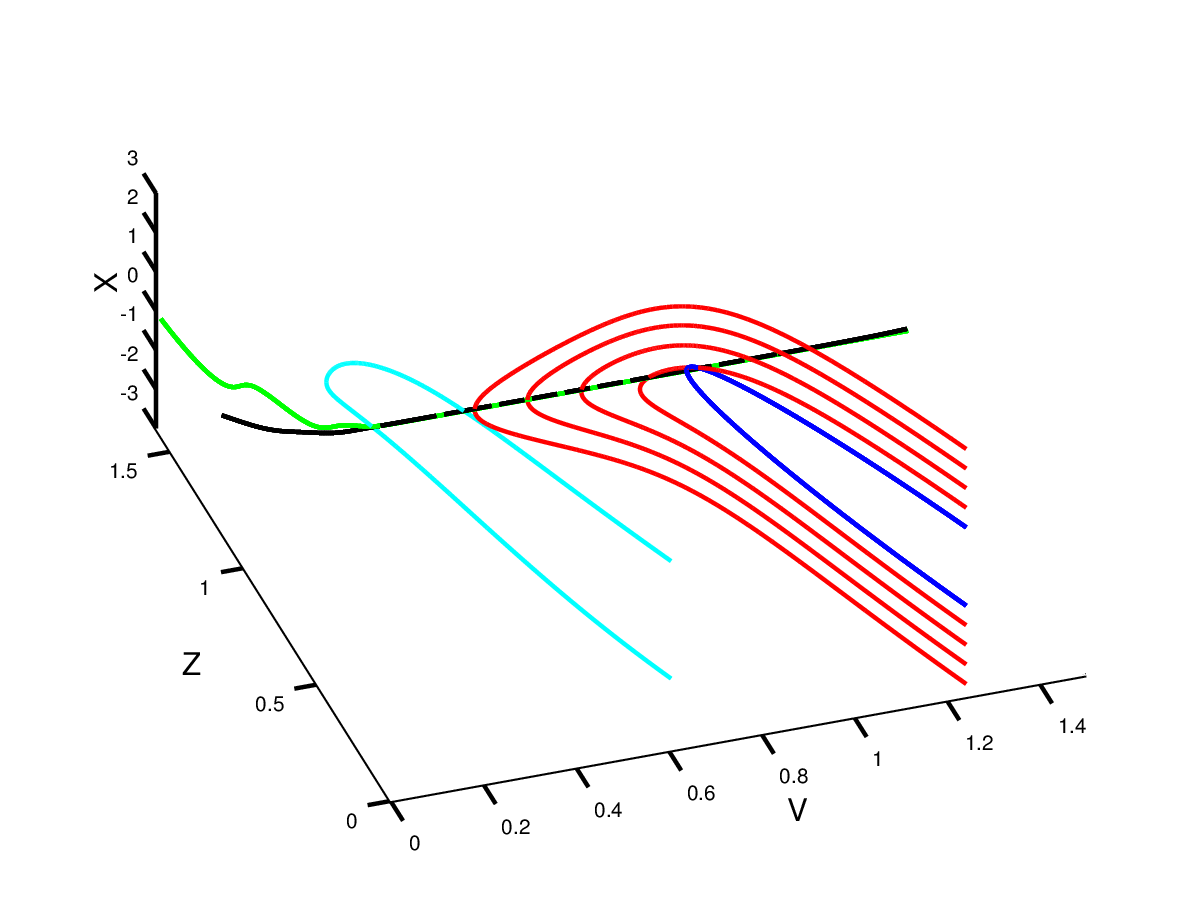}\hspace*{-1truecm}\includegraphics[scale=.40]{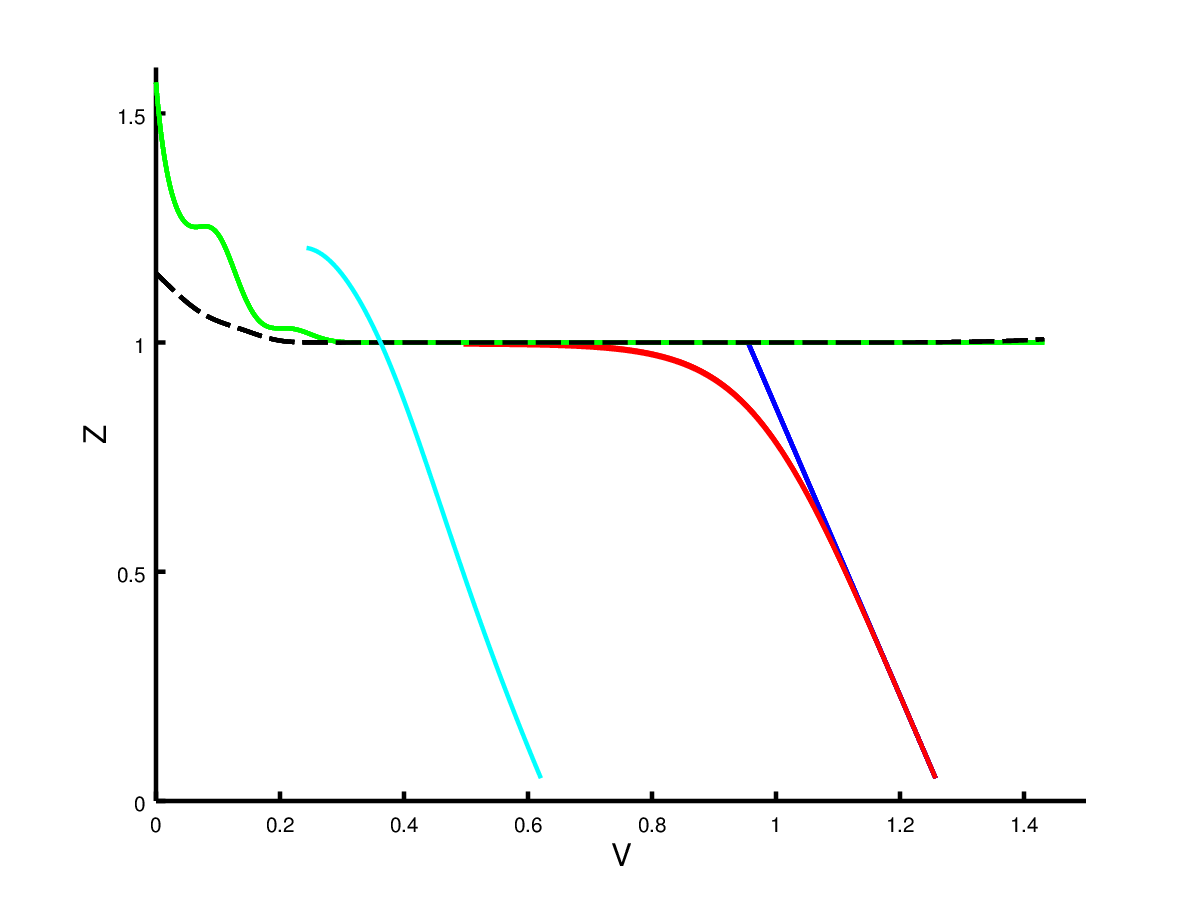}
\caption[Geodesics and their relative position to event and apparent horizon.]{\label{2PFHorizon} 
The green (black-dashed) line indicates the $z$-position of the apparent (event) horizon; 
the dark blue curve is the Poincar\'e patch AdS geodesic we use to initialize the simulation; red curves are geodesics with different boundary separation probing the thermal regime (none of them crosses the apparent horizon); the cyan curve in the left part of each plot is a geodesic which probes the non-thermal regime and reaches beyond event and apparent horizon. three-dimensional plot (left) and view in $x$-direction (right).
 }
\end{center}
\end{figure}

As we have noted before geodesics can extend beyond the apparent horizon. This is made explicit in Figure~\ref{2PFHorizon} where the blue curve serves as our initial guess for the relaxation code. The red geodesics at late times approach the apparent horizon without crossing it. At sufficiently early times (and sufficiently large separation) the geodesics cross the apparent horizon, an example of which is depicted by the cyan curve.

\begin{figure}
\begin{center}
\includegraphics[scale=.57]{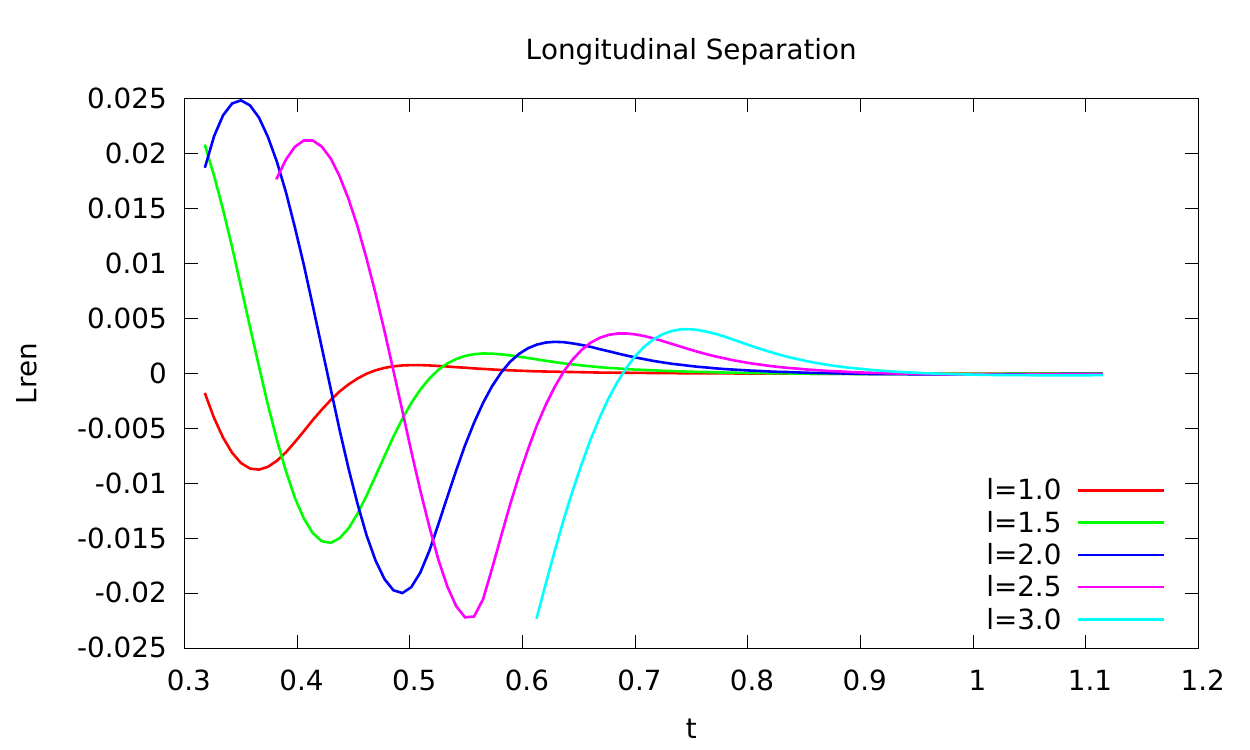}\includegraphics[scale=.57]{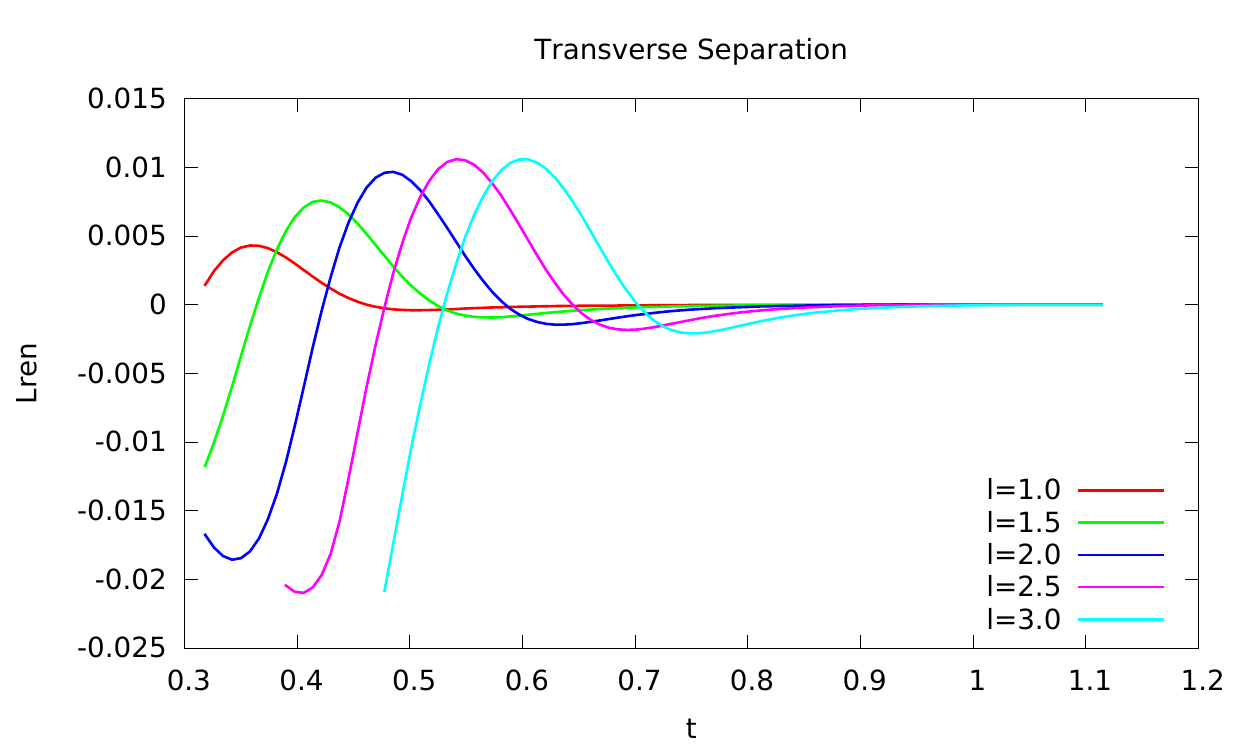}
\caption[Renormalized length of geodesics.]{\label{2PntFun} 
Renormalized length of geodesics for different separations in longitudinal and transverse directions.
 }
\end{center}
\end{figure}

The evolution of the renormalized lengths in the transverse and longitudinal directions  for different separations are  depicted in Figure~\ref{2PntFun}. Depending on the separation the two-point functions start at $t=t_{\rm min}$ which is the time when the geodesics extend beyond the computational domain.

\begin{figure}
\begin{center}
\includegraphics[scale=.6]{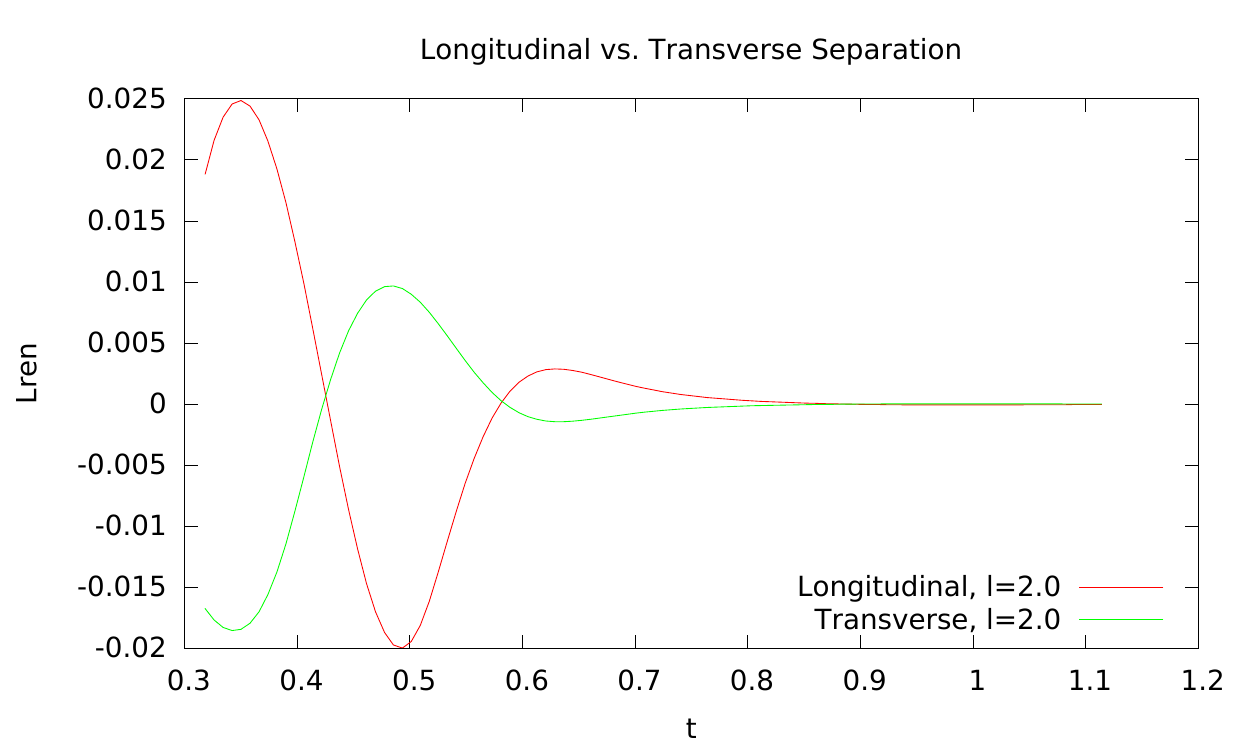}
\caption[Comparison of longitudinal and transverse geodesic lengths.]{\label{Comp2PntFun} 
Comparison of longitudinal and transverse geodesic lengths for the same boundary separation.
 }
\end{center}
\end{figure}

The first observation is that the transverse and longitudinal directions oscillate out of phase as shown in Figure~\ref{Comp2PntFun}. The same feature is seen in the transverse and longitudinal pressure. 
By comparing the thermalization times of the one-point functions, i.e.~the expectation value of the stress energy tensor with the two-point functions we see that the two-point functions thermalize as expected  later. 
Also, the thermalization time increases if the boundary separation is increased.

\subsection{Holographic Entanglement Entropy}\label{se:4.3} 

The extremal surface equations --- which we mapped to geodesic equations in an auxiliary spacetime --- are solved again by a relaxation method. We observe the same qualitative features as for geodesics in Figure~\ref{2PFHorizon} above: at early times extremal surfaces can extend beyond the apparent horizon, while at sufficiently late times they approach it from the outside without crossing. However, there are also notable differences to geodesics, which we discuss now.

As discussed in Section \ref{Sec:BasicExamples} conformal geodesics reach much farther  into the bulk  compared to the pure AdS case. 
Therefore the boundary separations we can study for  the entanglement entropy are smaller compared to the two-point functions. 
This is also the reason why for the same boundary separation the entanglement entropy reaches equilibrium later as the two-point functions. 
For the same boundary separation and at the same boundary time conformal geodesics reach deeper into the bulk and further back in time and therefore are more sensitive to out of equilibrium effects which are most noticeable  at early times.
In addition the shape of the curves differs from the two-point functions and we exhibit these features now in some plots.

\begin{figure}
\begin{center}
\includegraphics[scale=.55]{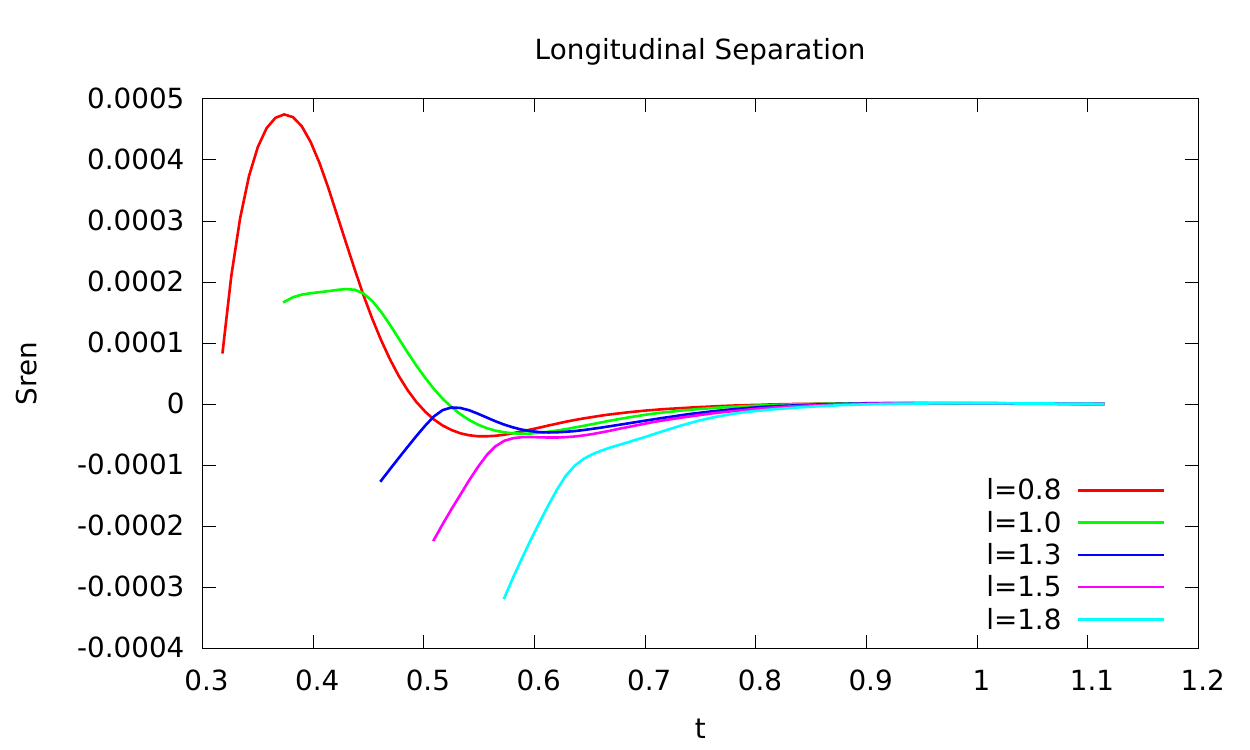}$\;\;\;\;$\includegraphics[scale=.55]{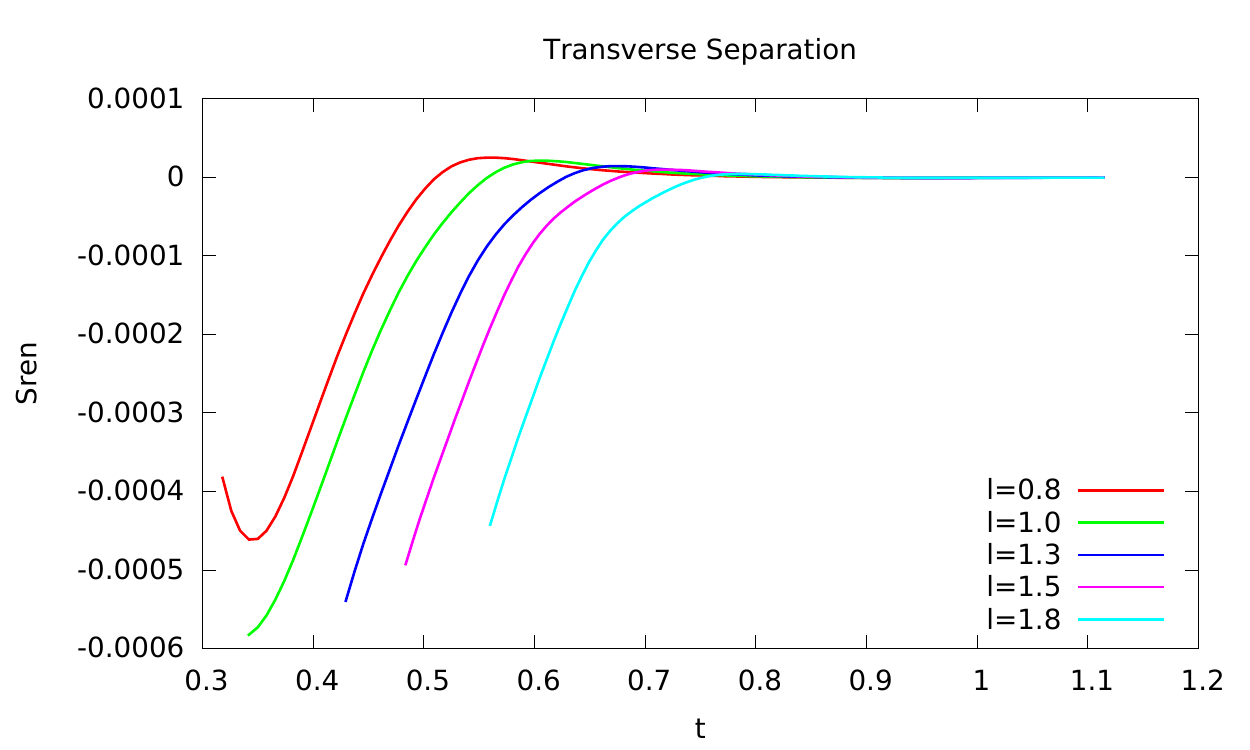}
\caption{\label{2PntFun2} 
Longitudinal and transverse entanglement entropy for different separations. 
 }
\end{center}
\end{figure}

Figure \ref{2PntFun2} plots entanglement entropy for different separations in longitudinal and transverse directions. Comparison with Figure~\ref{2PntFun} shows that the oscillations are less pronounced for entanglement entropy.

\begin{figure}
\begin{center}
\includegraphics[scale=.6]{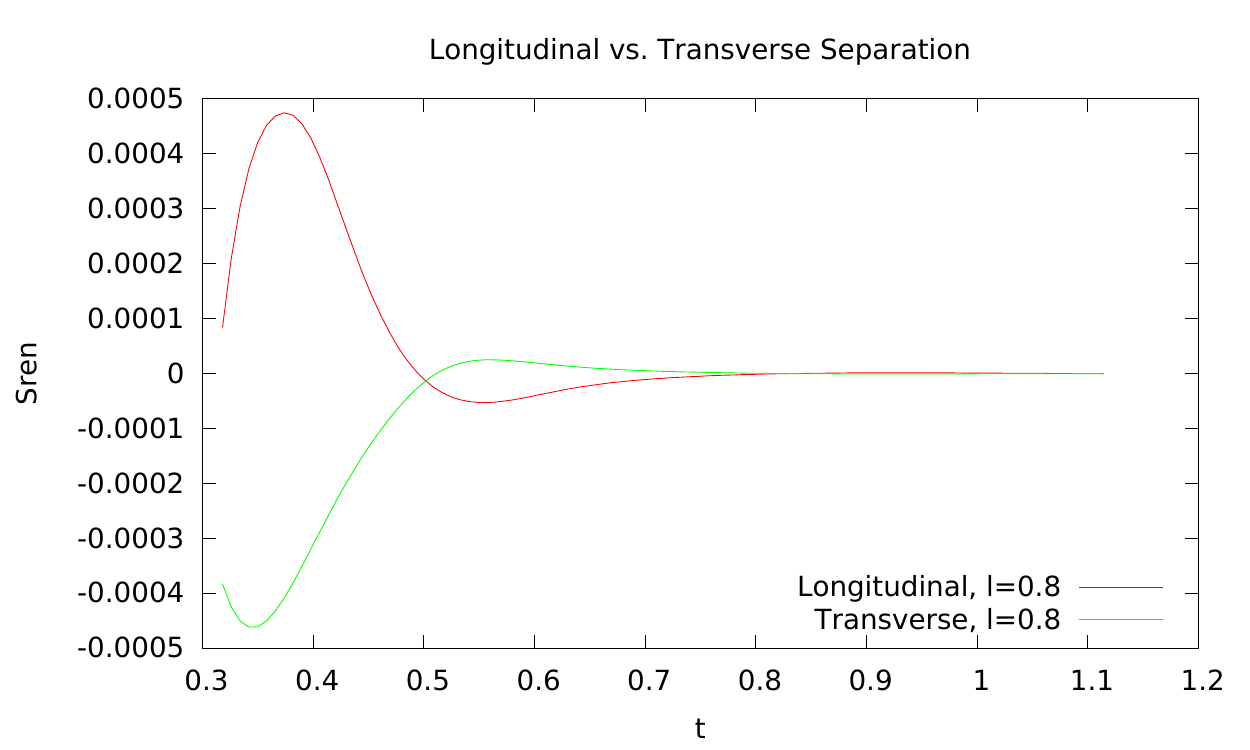}
\caption{\label{Comp2EE} 
Longitudinal and transverse entanglement entropy for same separation. 
 }
\end{center}
\end{figure}

Figure \ref{Comp2EE} plots entanglement entropy for a fixed separation in longitudinal and transverse directions. Again the behavior of the curves is out of phase, in the sense that maxima of one curve correspond to minima of the other. Comparison with Figure~\ref{Comp2PntFun} shows again that the oscillations are less pronounced for entanglement entropy.

\section{Late Time Behavior and Quasinormal Modes}\label{se:4}

After the early far-from-equilibrium phase the geometry   relaxes to  the static Schwarz\-schild black brane solution. As noted in \cite{Heller:2013oxa,Chesler:2013lia} the anisotropy of the system is exponentially damped and at sufficiently late times 
one enters the  linearized regime.
In this regime the approach to equilibrium is accurately described by the lowest lying quasinormal mode (QNM) which characterizes the response of  the system  to infinitesimal metric perturbations.  
 In the case at hand the relevant channel for the gravitational fluctuations is the spin two symmetry channel which coincides with the fluctuations of a massless scalar field in the static black brane geometry. The asymptotic response of the pressure anisotropy then takes the form 
\begin{equation}
b_4(t)\sim \mathrm{Re}\left[ c_1 e^{-i\,\omega_1 t}\right]\,,
\end{equation}
with some constant $c_1$ and  the  complex frequency $\omega_1$ of the  lowest QNM  given by \cite{Starinets:2002br, Kovtun:2005ev} 
\begin{equation}
\frac{\omega_1}{\pi T}=\pm 3.119452-2.746676 \,i\,.
\end{equation}
On the field theory side QNMs appear as poles in the retarded Green function \cite{Birmingham:2001pj,Son:2002sd,Kovtun:2005ev,Berti:2009kk}. It is therefore expected that also the late time behavior of the correlation functions  obtained in the previous section is described by the lowest QNM. 
We now show that this is indeed the case. 

\begin{figure}
\begin{center}
\includegraphics[scale=.55]{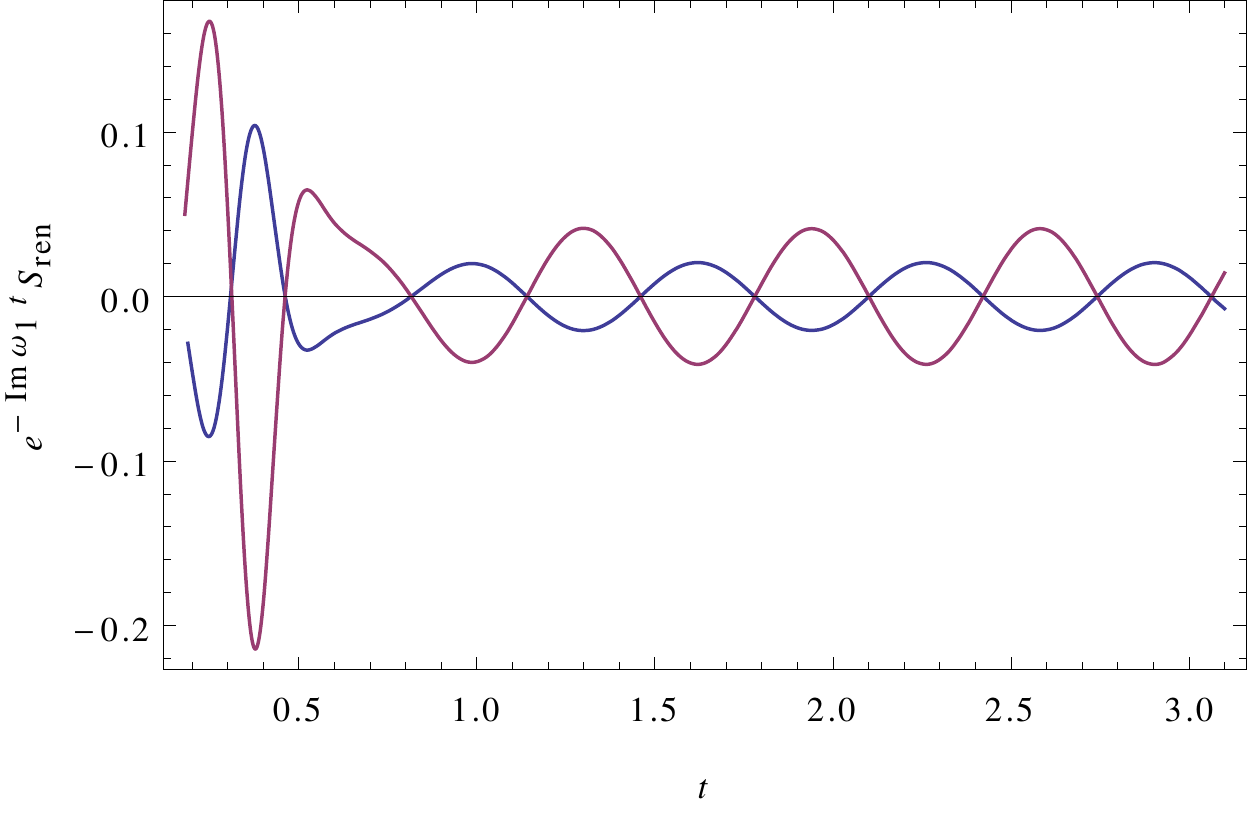}$\;\;\;\;\;$\includegraphics[scale=.55]{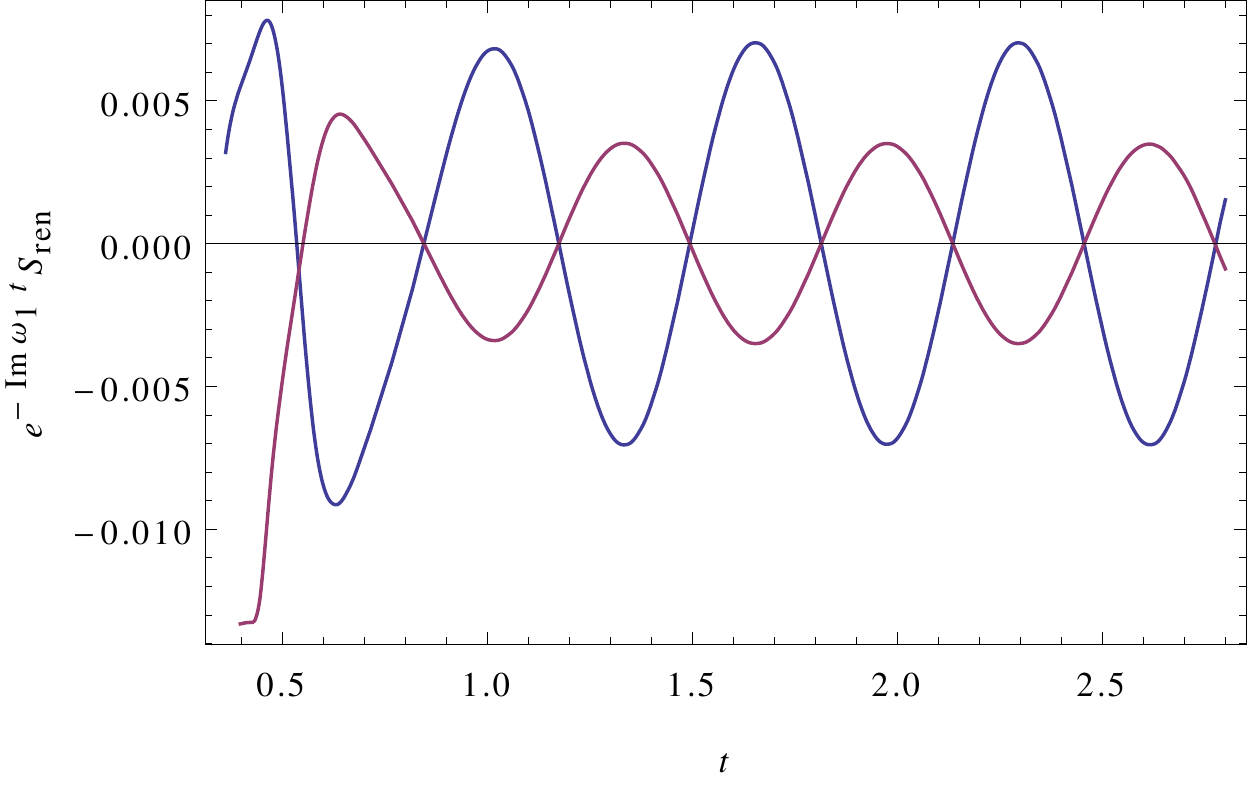}
\caption[Correlators and entanglement entropy multiplied by lowest QNM.]{\label{QNM} 
Left: Renormalized geodesic length for longitudinal (red) and transverse (blue) separation   $l=1$  multiplied by the imaginary part of the lowest QNM. Right: Renormalized entanglement entropy for the same parameters as on the left.
 }
\end{center}
\end{figure}

In Figure \ref{QNM} (left) we plot the renormalized geodesic length multiplied with the imaginary part of the   lowest QNM
$e^{-\mathrm{Im}\left[\omega_1 t \right]}L_{\textrm{ren}}$ for transverse and longitudinal separations.  
One clearly sees that after a short period of time the evolution of the correlator is accurately described by the ring-down of the black brane with constant amplitude and frequency. 
The connection between the late time behavior of correlation functions  and QNMs was previously also  observed in \cite{Balasubramanian:2012tu,Ishii:2015gia,David:2015xqa}.

It turns out that  entanglement entropy also follows this pattern. 
In \cite{Bhattacharya:2013bna} a connection between QNMs and the behavior of the entanglement entropy was found. 
From linearized Einstein equations one can derive a differential equation for the first order correction $\Delta S_A$ of the entanglement entropy describing its change when a given ground state is excited. By imposing in-falling boundary conditions  at the horizon one obtains a QNM dispersion relation putting a constraint on entanglement entropy. 
With our numerical solution we can demonstrate that the late time behavior of entanglement entropy indeed follows the QNM ring-down even without imposing in-falling boundary conditions. 
In Figure~\ref{QNM} (right) we show the entanglement entropy multiplied with $e^{-\mathrm{Im}\left[\omega_1 t \right]} S_{\textrm{ren}}$  for the infinite strip with finite separation in longitudinal and transverse direction. As for the correlation function, at late times, the entanglement entropy shows quasinormal ringing with constant amplitude and frequency. 
These oscillations show that  entanglement entropy need not approach its thermal value from below but rather shows oscillatory behavior around its thermal value.

\begin{figure}
\begin{center}
\includegraphics[scale=.55]{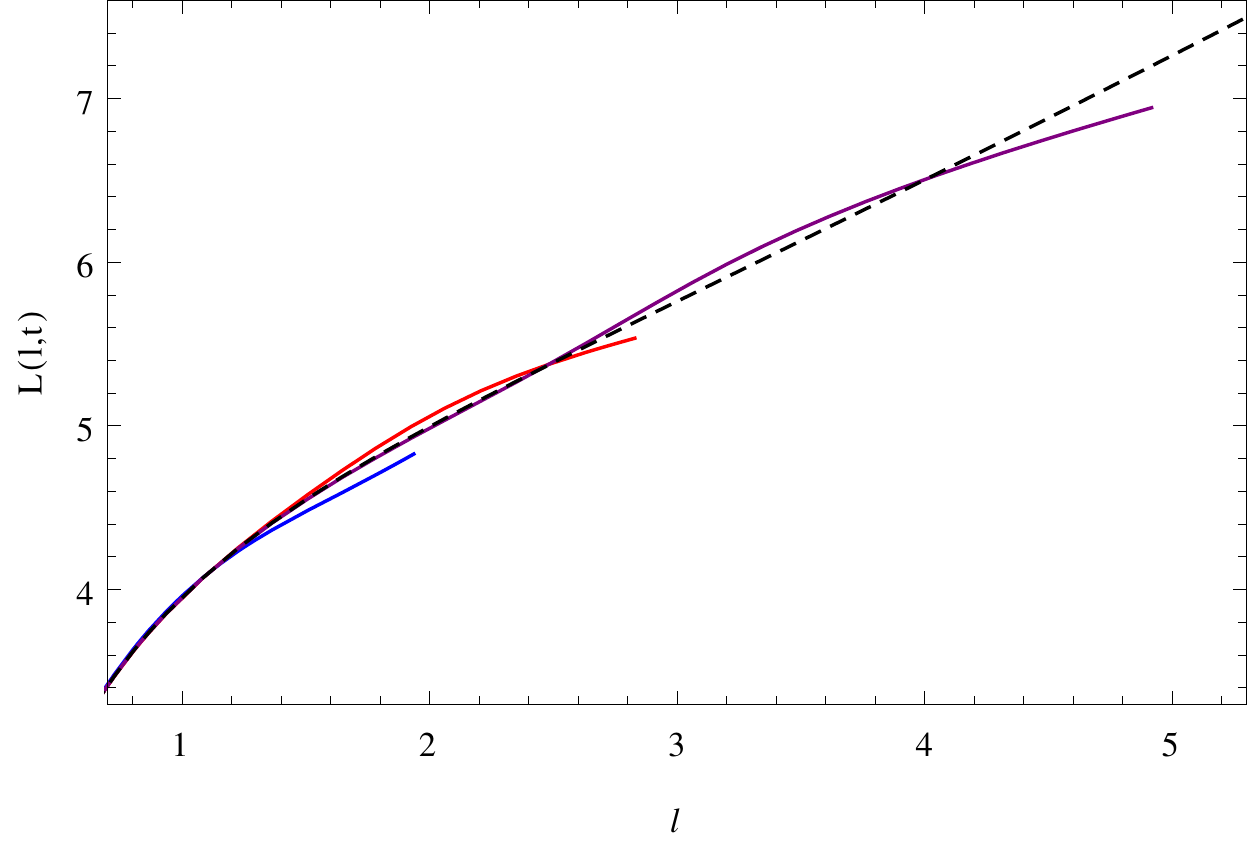}$\;\;\;\;\;$\includegraphics[scale=.55]{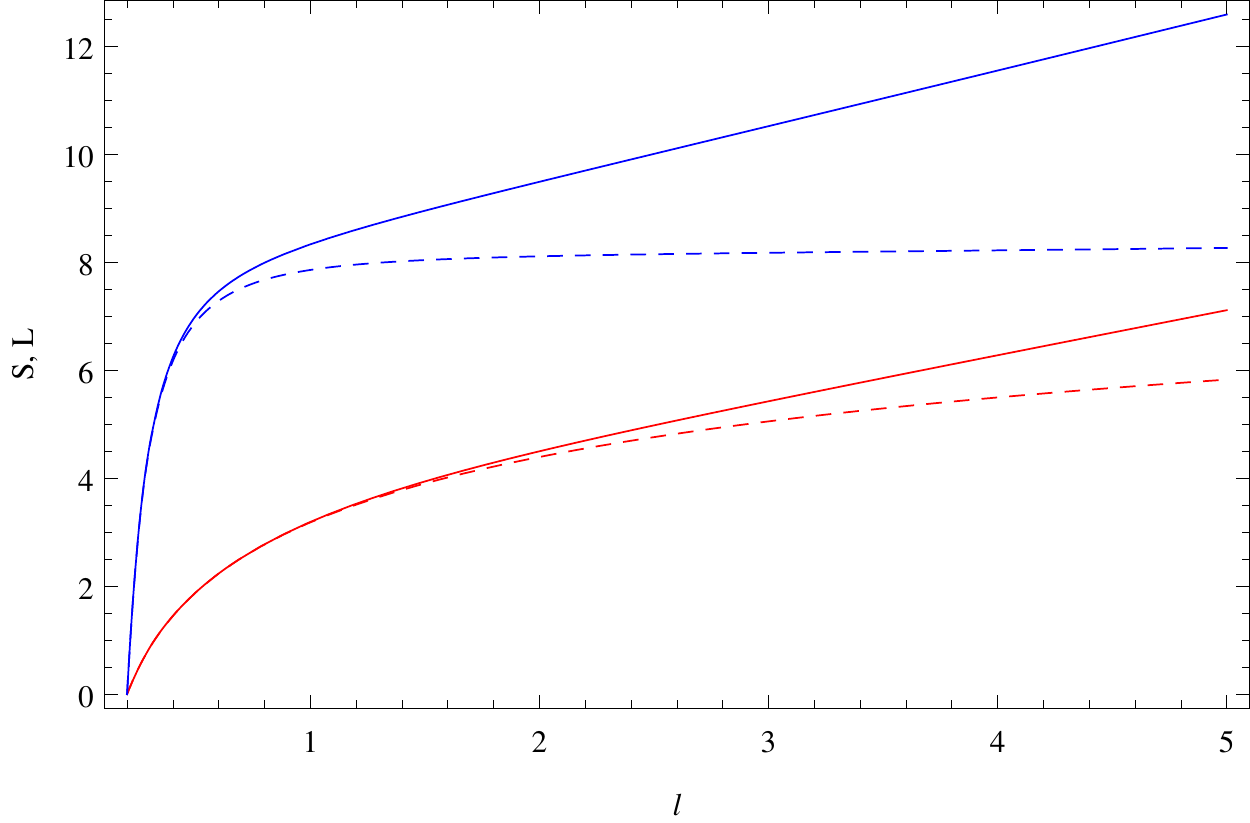}
\caption[Late time correlators and entanglement entropy as function of separation.]{\label{fixedT} 
Left: Renormalized length of geodesics as a function of their boundary  separation in  transverse directions for different fixed boundary times $t=0.5,\;1,\;1.5$ (endpoints from left to right). The curves terminate  when the geodesics leave the computational domain. 
The black dashed line shows  the thermal limit. 
Right: Entanglement entropy (blue) and geodesic  length  (red) in the  thermal  (solid) and  zero temperature (dashed) limit.  
 }
\end{center}
\end{figure}

To conclude this section we finally study the departure of the length of the geodesics and entanglement entropy from equilibrium for different times as a function of the boundary separation. This time we normalized the length of geodesics by subtracting a cutoff dependent piece
\begin{equation}
L(l,t)=L_\mathrm{bare}(l,t,z_c)-L_\mathrm{bare}(l=0.1,t,z_c)\,,
\end{equation}
where $z_c$ is the value of the cutoff.
The case for the two-point function with separation in the transverse direction is  displayed in Figure~\ref{fixedT}. 
Out of equilibrium effects manifest themselves as oscillations around the thermal value.  
The curves terminate when the geodesics leave the computational domain, so for early times we only have access to rather small boundary separations. The same effect is seen for entanglement entropy.

In the thermal limit the scaling for the  geodesic length   at small and  large  boundary separation is dictated by conformal symmetry and is proportional to
$2\log(l/2)$ and $2 l$ respectively.   
At large separation the entanglement entropy also scales linearly with the separation length, whereas  at small separation it is proportional to $1/l^2$. All these results agree precisely with the perturbative expressions derived in the limits of small and large temperatures \cite{Ryu:2006ef, Fischler:2012ca}. 
Our numerical results are shown in Figure~\ref{fixedT} where we also plot the corresponding zero temperature results, which coincide with the thermal curves for small separations.

\enlargethispage{0.7truecm}

\chapter{Entanglement Entropy in Shock Wave Collisions}\label{Chap:Shocks}
In this chapter we study the time evolution of two-point functions and entanglement entropy in strongly anisotropic, inhomogeneous and time dependent ${\cal N}=4$ super Yang--Mills theory in the large $N$ and large 't~Hooft coupling limit using AdS/CFT. On the gravity side this amounts to calculating the length of geodesics and area of extremal surfaces in the dynamical background of two colliding gravitational shock waves, which we do numerically. We discriminate between three classes of initial conditions corresponding to wide, intermediate and narrow shocks, and show that they exhibit different phenomenology with respect to the nonlocal observables that we determine. 
Our results permit to use (holographic) entanglement entropy as an order parameter to distinguish between the two phases of the cross-over from the transparency to the full-stopping scenario in dynamical Yang--Mills plasma formation, which is frequently used as a toy model for heavy ion collisions.
The time evolution of entanglement entropy allows to discern four regimes: highly efficient initial growth of entanglement, linear growth, (post) collisional drama and late time (polynomial) fall off. Surprisingly, we found that two-point functions can be sensitive to the geometry inside the black hole apparent horizon, while we did not find such cases for the entanglement entropy.
Results displayed in this chapter are published in \cite{Ecker:2016thn}.
\section{Gravitational Shock Waves in Asymptotically AdS$_5$}\label{se:2}

The holographic setup we consider describes the collision of two sheets of energy having Gaussian shape in their direction of motion and which are homogeneous in the remaining two spatial directions.
These shocks serve as caricatures of two highly Lorentz contracted nuclei  which approach each other at the speed of light and induce non-Abelian plasma formation as they collide. 

On the gravity side the corresponding five-dimensional bulk metric is rotationally invariant and homogeneous in the transverse plane ($x_1,x_2$) but inhomogeneous in the longitudinal direction $y$, which is the direction of motion of the shocks. 
The metric ansatz  in Edding\-ton--Finkelstein coordinates reads
\begin{equation}\label{metric}
d s^2=-Ad v^2 +S ^2\Big( e^{-2B} d y^2 +e^{B} d\vec{x}^2 \Big)+2d v(d r + Fd y)\;,
\end{equation}
where the functions $A,~S,~B$ and $F$ depend on the holographic coordinate $r$, (advanced) time $v$ and longitudinal coordinate $y$, but are independent from the transversal coordinates $\vec{x}$. 
The equations of motion can be found e.g. in \cite{Chesler:2010bi} and are solved near the boundary by
\begin{subequations}\label{Eq:asymptShock}
\begin{eqnarray}
A&=&r^2+2\xi r+ \xi^2-2\partial_v\xi+\frac{a_4}{r^2}+\frac{\partial_v a_4-4\xi a_4}{2r^3}+\mathcal{O}(r^{-4})\,,\\
B&=&\frac{b_4}{r^4}+\frac{15\partial_v b_4+2\partial_y f_4 -60\xi b_4}{15r^5}+\mathcal{O}(r^{-6})\,,\\
S&=&r+\xi-\frac{4\partial_y f_4 + 3\partial_v a_4}{60r^4}+\mathcal{O}(r^{-5})\,,\\ 
F&=&\partial_y\xi +\frac{f_4}{r^2}+\frac{4\partial_v f_4+\partial_y a_4 -10\xi f_4}{5r^3}+\mathcal{O}(r^{-4})\,,
\end{eqnarray}
\end{subequations}
where $\xi(v,y)$ encodes  the residual diffeomorphism freedom $r\to r+\xi(v,y)$. 
It is possible, though not necessarily numerically convenient, to choose $\xi=0$. 

As usual the normalizable modes $a_4(v,y)$, $b_4(v,y)$ and $f_4(v,y)$ are undetermined by the near-boundary expansion and require a solution of the full bulk dynamics. These coefficients in the asymptotic expansion determine the expectation value of the conserved and traceless stress energy tensor in the dual field theory \cite{deHaro:2000xn}
\begin{equation}\label{Eq:EMTshock}
\langle T^{\mu\nu}\rangle =\frac{N_c^2}{2\pi^2}
\begin{pmatrix}
  \mathcal{E}                & \mathcal{\mathcal{S}}                & 0                                & 0                               \\
  \mathcal{\mathcal{S}}      & \mathcal{\mathcal{P}}_\parallel      & 0                                & 0                               \\
  0                          & 0                                    & \mathcal{\mathcal{P}}_\perp      & 0                               \\
  0                          & 0                                    & 0                                & \mathcal{\mathcal{P}}_\perp     \\
 \end{pmatrix}\,,
\end{equation}
where 
\begin{equation}
\mathcal{E}=-\frac{3}{4}a_4\,,\qquad \mathcal{P}_\parallel = -\frac{1}{4}a_4-2b_4\,,\qquad \mathcal{P}_\perp = -\frac{1}{4}a_4 +b_4\,,\qquad \mathcal{S} = -f_4\,.
\end{equation}

\subsection{Initial Conditions}

The pre-collision  geometry  describing  two shocks moving in $\pm \tilde{y}$-direction can be written down in Fefferman-Graham coordinates ($\tilde r$, $\tilde t$, $\tilde y$, $\vec{\tilde x}$)  as follows \cite{Janik:2005zt}
\begin{equation}
\label{ansatzFG}
d s^2=\tilde{r}^2\eta_{\mu\nu}d \tilde{x}^\mu d \tilde{x}^\nu +\frac{1}{\tilde r^2}\Big(d \tilde{r}^2+h(\tilde{t}+\tilde{y})(d \tilde{t}+d \tilde{y})^2+h(\tilde{t}-\tilde{y})(d \tilde{t}-d \tilde{y})^2\Big),
\end{equation}
where $\eta_{\mu\nu}$ denotes the usual four-dimensional Minkowski boundary metric 
and $h(\tilde{t}\pm\tilde{y})$ is  an arbitrary function for which we choose a Gaussian of width $\omega$ and amplitude $\mu^3$
\begin{equation}
\label{IC}
h(\tilde{t}\pm\tilde{y})=\frac{\mu^3}{\sqrt{2\pi\omega^2}}e^{-\frac{(\tilde{t}\pm\tilde{y})^2}{2\omega^2}}\,.
\end{equation}
In this gauge the non-vanishing components of the energy momentum tensor read
\begin{equation}
\tilde T^{\tilde t \tilde t}=\tilde T^{\tilde y \tilde y}=h(\tilde{t}-\tilde{y}) + h(\tilde{t}+\tilde{y})\,,\qquad \tilde T^{\tilde t \tilde y}=h(\tilde{t}-\tilde{y}) - h(\tilde{t}+\tilde{y})\,,
\end{equation}
and describe two lumps of energy with maximum overlap at $\tilde t=0$.
At early times $\tilde t \ll -w$, when the shocks $h(\tilde{t}\pm\tilde{y})$ have negligible overlap, the line-element (\ref{ansatzFG}) is close to an exact solution to  Einstein's equations, but around $\tilde t=0$ their dynamics can only be computed numerically.

We do this for three  different initial conditions $h_{n,i,w}(\tilde y)$ describing qualitatively different situations that we shall refer to as narrow, intermediate and  wide shocks,  where in all cases the initial position of the shocks is at $\tilde{y}_0=\pm7/4$. 
For the width of the shocks we take $\omega_{n,i,w}=0.1,~0.25,~0.5$ and we will display all our results in units of $\mu$.

For the numerical  evolution scheme the initial data needs to be transformed to Edding\-ton--Fin\-kel\-stein coordinates $(r,v,y,x_1,x_2)$ by solving for radially in-falling  null geodesics in the background  (\ref{ansatzFG}), leading to ordinary differential equations, which  are solved   for appropriate boundary conditions at the boundaries of the radial domain.
We omit a discussion of the numerical details concerning this coordinate transformation and the subsequent evolution and refer the reader to \cite{vanderSchee:2014qwa,Chesler:2013lia}, where the full procedure is explained.

\subsection{Evolution of the Energy Momentum Tensor}\label{se:2.2}

\begin{figure}
\hspace{-0.cm}
\includegraphics[scale=.19]{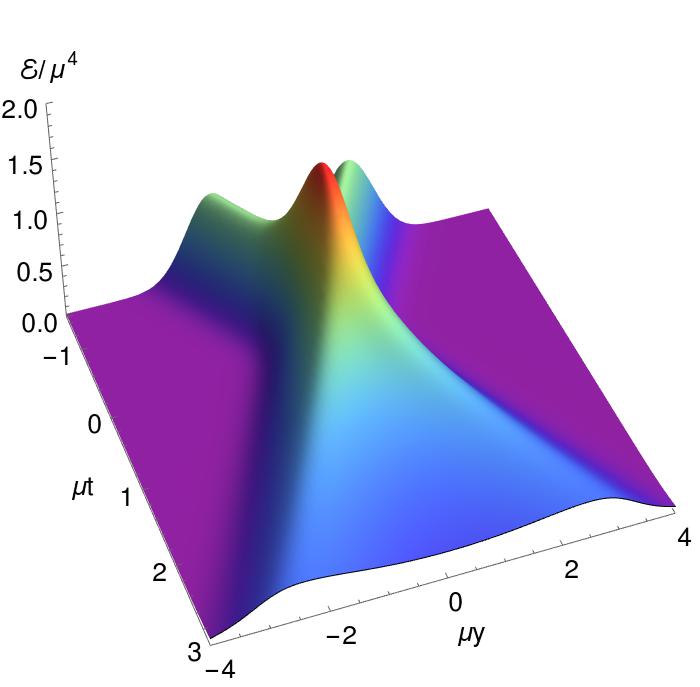}\includegraphics[scale=.19]{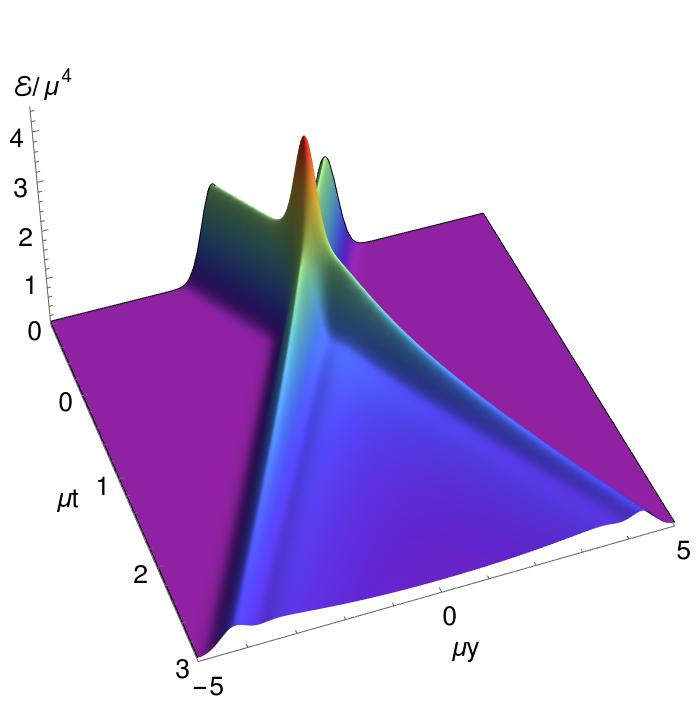}\includegraphics[scale=.19]{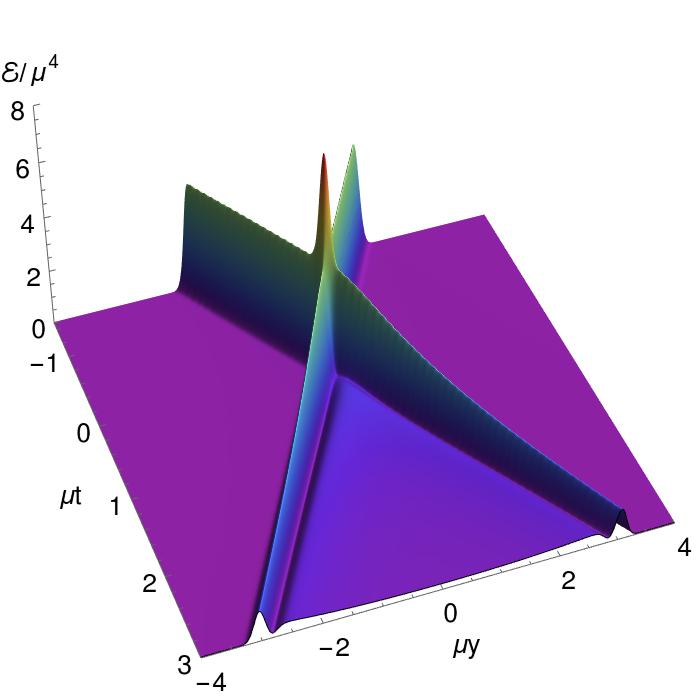}
\caption[Anisotropy function and pressures.]{\label{EMTevolution} 
Evolution of the energy density $\epsilon/\mu^4$ as a function of time $t$ and longitudinal coordinate $y$ for wide, intermediate and narrow shocks (from left to right).}
\end{figure}

The time evolution of the energy momentum tensor for colliding shocks has been studied extensively in \cite{Chesler:2010bi,Casalderrey-Solana:2013aba,Casalderrey-Solana:2013sxa,Chesler:2013lia}.
In Figure~\ref{EMTevolution} we show the evolution of the energy density $\mathcal{E}(t,y)$ extracted from the numerical evolution for the different initial conditions stated above.
As discussed in \cite{Casalderrey-Solana:2013aba} the energy density behaves qualitatively different in collisions of narrow shocks and in those of wide shocks. This cross-over is not only of academic interest, but also for applications, since it was argued that the narrow shocks describe more adequately the situation at LHC, while the wide shocks are more suitable for RHIC  \cite{Casalderrey-Solana:2013aba} (see also \cite{vanderSchee:2015rta}). We list below some relevant properties that differ between wide and narrow shocks:
\begin{itemize}
\item Narrow shocks exhibit transparency, in the sense that they pass through each other and, even though their shape gets altered and they decay, they continue to move at the speed of light after the collision. By contrast, wide shocks realize a full-stopping scenario, in the sense that the energy density is localized mostly in the central region after the collision, and the shocks themselves not only change their shape but also get slowed down. Wide shocks then lead to initial conditions for hydrodynamics where all velocities are close to zero, i.e. there is a hydrodynamical explosion in close similarity to the Landau model of heavy ion collisions \cite{Landau:1953gs}.
\item Narrow shocks can yield regions of negative energy density after the collision right behind the original shocks on the lightcone. Curiously, this region does not admit a local restframe \cite{Arnold:2014jva}, but also does not violate general principles of quantum field theory, such as the averaged null energy condition \cite{Ford:1999qv}. At $y=0$ after the shocks pass through each other, the energy density grows at early times as $\mathcal{E}=2 \mu^6 t^2+\mathcal{O}(t^5)$, which implies pressures equal to $\mathcal{\mathcal{P}}_\parallel/\mathcal{E}=-3$ and $\mathcal{\mathcal{P}}_\perp/\mathcal{E}=2$.
This feature was first observed for $\delta$-like shock waves analytically \cite{Grumiller:2008va} and then numerically for sufficiently narrow Gaussian profiles \cite{Casalderrey-Solana:2013aba}. 
By contrast, for the wide shocks the energy density and pressures remain positive everywhere.

\end{itemize}
Given the substantial differences in local observables one may expect that the characteristic features for narrow and wide shocks also show up in nonlocal observables, like two-point functions and entanglement entropy. In the remainder of this chapter we verify this expectation by explicit computations, starting with the two-point functions in the next section.

\section{Two-Point Functions}

Within AdS/CFT the equal time two-point function of operators $\mathcal{O}$ with large conformal weight $\Delta$ can be computed from the length $L$ of spacelike geodesics in the bulk geometry \cite{Balasubramanian:1999zv, Festuccia:2005pi} via
\begin{equation}
\langle \mathcal{O}(t, \vec{x})\mathcal{O}(t, \vec{x}')\rangle=\int  \mathcal{D P}\, e^{i \Delta \mathcal{L(\mathcal{P})} }\approx \!\!\!\sum_{\textrm{\tiny geodesics}}\!\!\! e^{-\Delta L_g}\approx e^{-\Delta L}\;.
\end{equation}
In asymptotically AdS the length of a geodesic which is attached to the boundary is infinite and a regularization scheme must be adopted.
A natural way to regularize is to subtract the length $L_0$ of a geodesic in AdS corresponding to the vacuum value of the correlator 
\begin{equation}
L_{\textrm{\tiny reg}}=L-L_0 \;.
\end{equation}

For illustrative purposes we set $\Delta=1$ when we display our results which is the same as interpreting $L_{\textrm{\tiny reg}}$ to be given in units of $\Delta$.
Thus, the two-point functions we compute are defined as follows
\begin{equation}\label{2pfreg}
\langle \mathcal{O}(t, \vec{x})\mathcal{O}(t, \vec{x}')\rangle_{\textrm{\tiny reg}}= e^{-L_{\textrm{\tiny reg}}}\;.
\end{equation}

In order to obtain the geodesic length we solve the geodesic equation numerically with a relaxation algorithm such as described in Chapter \ref{Chap:Numerics}.
\subsection{Geodesics in the Shock Wave Geometry}\label{se:3.1}

For simplicity we restrict our attention to geodesics that only  extend  along the $y$-direction  and not along the transverse directions ($x_1,x_2$), i.e.\ we consider geodesics in the three-dimensional bulk-subspace 
\begin{eqnarray}\label{submetric}
d s^2_y&=&  -A d v^2-\frac{2}{z^2}  d z d v +2 Fd yd v +S^2 e^{-2B} d y^2,
\end{eqnarray}
where   $z=1/r$.
To find these geodesics we solve the (non-affine) geodesic equation
\begin{equation}
\ddot X^\mu + \Gamma^\mu{}_{\alpha\beta} \dot X^\alpha \dot X^\beta = J \dot X^\mu,
\label{eq:geo}
\end{equation}
subject to the following boundary conditions at $z=0$
\begin{equation}
X^\mu(\pm1)\equiv(V(\pm1),Z(\pm1),Y(\pm1))=(t,0,\pm l/2),
\label{eq:para}
\end{equation}
where $X^\mu(\sigma)$ are the embedding functions of the geodesic and dots denote derivatives with respect to the non-affine parameter $\sigma\in [-1,1]$.
%
As initial guess we use again the pure AdS geodesics of \eqref{Eq:Ansatz2PF1}--\eqref{Eq:Ansatz2PF3} and the corresponding expression for the Jacobian given in \eqref{Eq:Jacobian}.
We assume the boundary separation to be centered around $y=0$. Describing off-central geodesics requires some straightforward modifications of our formulas. 

The bulk part of the geodesic length, which is the contribution from $z>z_{\textrm{\tiny cut}}$, follows from integrating the line element (\ref{submetric}) and the corresponding line element for pure AdS 
\begin{subequations}\label{Lbulk}
\begin{eqnarray}
L^{\textrm{\tiny bulk}}&=&\int_{\sigma_-}^{\sigma_+} d\sigma \sqrt{-A \dot{V}^2-\frac{2}{Z^2}\dot{Z}\dot{V}+2F\dot{V}\dot{Y} +S^2 e^{-2B}\dot{Y}^2},\\
L^{\textrm{\tiny bulk}}_0&=&\int_{\sigma_-}^{\sigma_+} d\sigma \frac{1}{Z_0}\sqrt{- \dot{V_0}^2-2 \dot{Z}_0 \dot{V}_0 + \dot{Y}_0^2},
\end{eqnarray}
\end{subequations}
where the metric functions $(A,B,S,F)$ have  to be evaluated along the geodesic $X^\mu(\sigma)$. 
In order to realize an IR-cutoff at a given value $z_{\textrm{\tiny cut}}$ the range of the non-affine parameter $\sigma\in [\sigma_-,\sigma_+]$ is bounded as usual
\begin{equation}
\sigma_\pm=\pm\sqrt{1-\frac{2 z_{\textrm{\tiny cut}}}{l}}.
\end{equation}
The near boundary part of the geodesic length, which is the contribution from $0\le z \le z_{\textrm{\tiny cut}}$, can be extracted from the near boundary solution of the geodesic equation.
Near $z=0$ the embedding functions and the Jacobian can be expressed in terms of a power series in $z$
\begin{equation}\label{asympAnsatz}
 Z(z)=z,\qquad V(z)=\sum_{n=1}^{n_{\textrm{\tiny max}}}v_n z^n,\qquad Y(z)=\sum_{n=1}^{n_{\textrm{\tiny max}}}y_n z^n,\qquad J(z)=\sum_{n=1}^{n_{\textrm{\tiny max}}}j_n z^{n-2}\,,
\end{equation}
The coefficients $(t_n,y_n,j_n)$ in \eqref{asympAnsatz} can be computed by solving the geodesic equation order by order in $z$, which leads to the following expressions
\begin{subequations}\label{asympSolution2PF}
\begin{eqnarray}
 Z(z) & = & z\,, \\
 V(z) & = & v_0-z+v_2 z^2+\left(v_2 y_2^2-v_2^3\right) z^4+O\left(z^5\right)\,, \\
 Y(z) & = & \frac{l}{2}+y_2 z^2+\left(y_2^3-v_2^2 y_2\right) z^4+O\left(z^5\right)\,, \\
 J(z) & = & \frac{1}{z}+\left(4 v_2^2-4 y_2^2\right) z+O\left(z^5\right)\,.
\end{eqnarray}
\end{subequations}
Here we fixed the leading coefficients by the boundary conditions (\ref{eq:para}), but the coefficients  $v_2$ and $y_2$ cannot be determined by
a near boundary expansion. This is analogous to the normalizable modes of the metric, which are also sensitive to the full bulk geometry.
The pure AdS solution is given by
\begin{subequations}\label{asympGeo0}
\begin{eqnarray}
 Z_0(z)&=&z\,,\\
 V_0(z)&=&t_0-z\,,\\
 Y_0(z)&=&\pm \sqrt{(l/2)^2-z^2}\nonumber\\
       &=&\pm \Big( \frac{l}{2}-\frac{z}{l}-\frac{z^4}{l^3}\Big)+\mathcal{O}(z^{6})\,,\\
 J_0(z)&=&\frac{1}{z}-\frac{4}{l^2}z-\frac{16}{l^4}z^3+\mathcal{O}(z^{5})\,.
\end{eqnarray}
\end{subequations}
which hence has $v_2=0$ and $y_2=\mp 1/l$. We can now compute the near boundary expansion of the geodesic length, which for one branch is given by
\begin{eqnarray}\label{bdryL}
L^{\textrm{\tiny bdry}}-L_0^{\textrm{\tiny bdry}}&=& \int_{0}^{z_{\textrm{\tiny cut}}} d z  \left(-\frac{2}{l^2}-2 v_2^2+2 y_2^2\right)z\nonumber\\
&+&\left(-\frac{a_4}{2}-\frac{6}{l^4}-12 v_2^2 y_2^2+6
   v_2^4+6 y_2^4\right)z^3 +O\left(z^5\right)\,,
\end{eqnarray}
where the leading AdS divergent $\tfrac{1}{z}$ term nicely cancels out.
The regularized geodesic length $L_{\textrm{\tiny reg}}$, which we need to evaluate \eqref{2pfreg}, is the sum of the bulk contribution and the near boundary contribution 
\footnote{In practise we do not compute the near boundary term, as the extraction of $v_2$ and $y_2$ would be numerically as hard as taking a small enough $z_{cut}$
such that this term is small. We have included this formula for completeness, and will later see that a similar procedure does work for entanglement entropy.}  
\begin{equation}\label{Lreg}
L_{\textrm{\tiny reg}}=(L^{\textrm{\tiny bulk}}-L_0^{\textrm{\tiny bulk}})+(L^{\textrm{\tiny bdry}}-L_0^{\textrm{\tiny bdry}})\;.
\end{equation}
When using \eqref{Lreg} to evaluate \eqref{2pfreg} numerically one has to ensure that the results are, to some required accuracy, independent of the discretization and the cutoff.
 We require this accuracy to be of the same order as the maximal residual ($=10^{-5}$) we allow in the geodesic equation and below which we stop to iterate the relaxation procedure.
We checked the convergence of the two-point function with the gridsize in the range from 50 up to 400 gridpoints and find that for more than $200$ gridpoints the change is smaller than $\mathcal{O}(10^{-5})$ which is the same order as the allowed residual. 
Based on this analysis we choose $200$ gridpoints to discretize our geodesics and set $z_{\textrm{\tiny cut}}=0.075$ in all the calculations in this chapter. 
\subsection{Evolution of Two-Point Functions}\label{se:3.2}
In this section we present our numerical results for two-point functions in holographic shock wave collisions.
Before we discuss the actual results let us start with some remarks regarding the computational domain used in these simulations.
As input for the relaxation algorithm we provide numerical results of the shock wave metric in a finite domain in $(t,y,z)$.  This computational domain, in which we can solve the geodesic problem, is bounded by $\mu t\in[-1.5,6]$, $\mu y\in[-5,5]$, where in the radial coordinate we have chosen the apparent horizon as a natural bound $z\in[0,1.08 z_{\textrm{\tiny AH}}]$.
That means whenever we display geodesics which reach beyond this radial domain, which can happen as we discuss below, an extrapolated version of the metric is used\footnote{
For the narrow shocks the computational domain does not reach behind the horizon, so there extrapolation is always used (note that the fact that the geodesic crosses the horizon or not is not affected by this extrapolation).}.
For a given choice of boundary conditions $(\mu t,\mu l)$ the final shape of the geodesic in the bulk is a priori unknown, i.e.\ initially we do not know if the geodesic resides entirely within or extends beyond the computational domain in which the metric is known. Therefore  finding a feasible set of parameters $(\mu t,\mu l)$ for a given computational domain requires some trial and error. 
The geodesics bend back in advanced time as they reach into the bulk, leaving the computational domain for too early boundary times. Therefore we can display our results only in a finite time near the collision time $t=0$ where all geodesics with different boundary separation lie in the computational domain.
All these points apply accordingly to the entanglement entropy simulation.

\begin{figure}
\begin{minipage}[t]{0.4\textwidth}
\vspace{-0.2cm}
\hspace{-0.2cm}
\includegraphics[width=\textwidth]{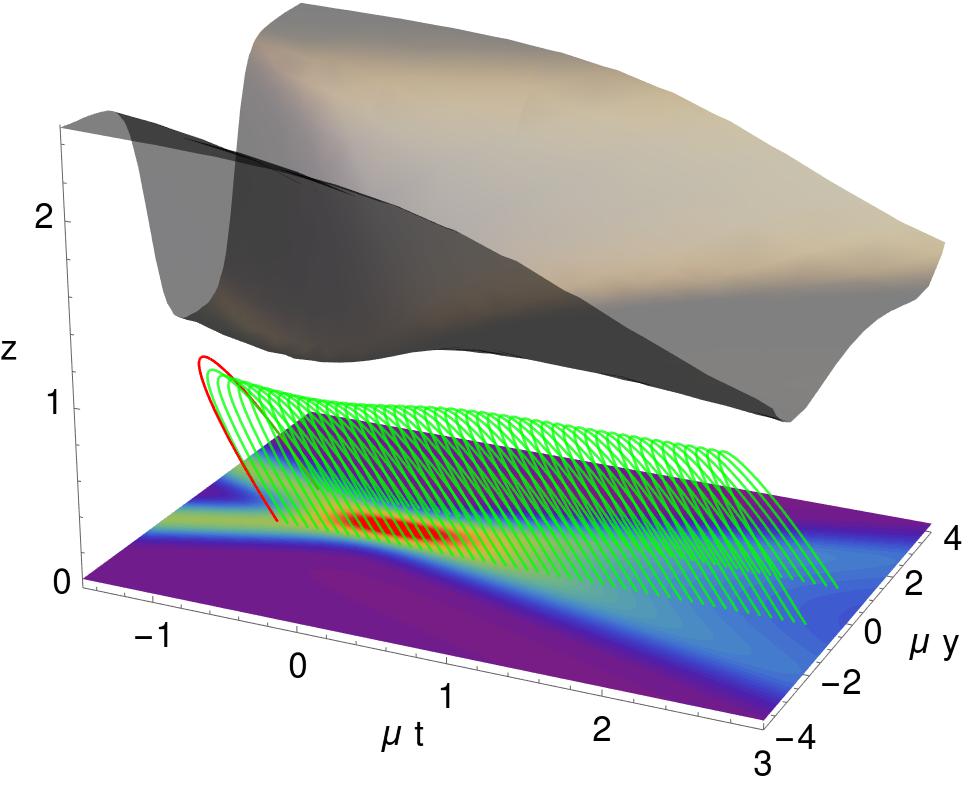}
\end{minipage}
\begin{minipage}[t]{0.4\textwidth}
\vspace{0.8cm}\includegraphics[width=\textwidth]{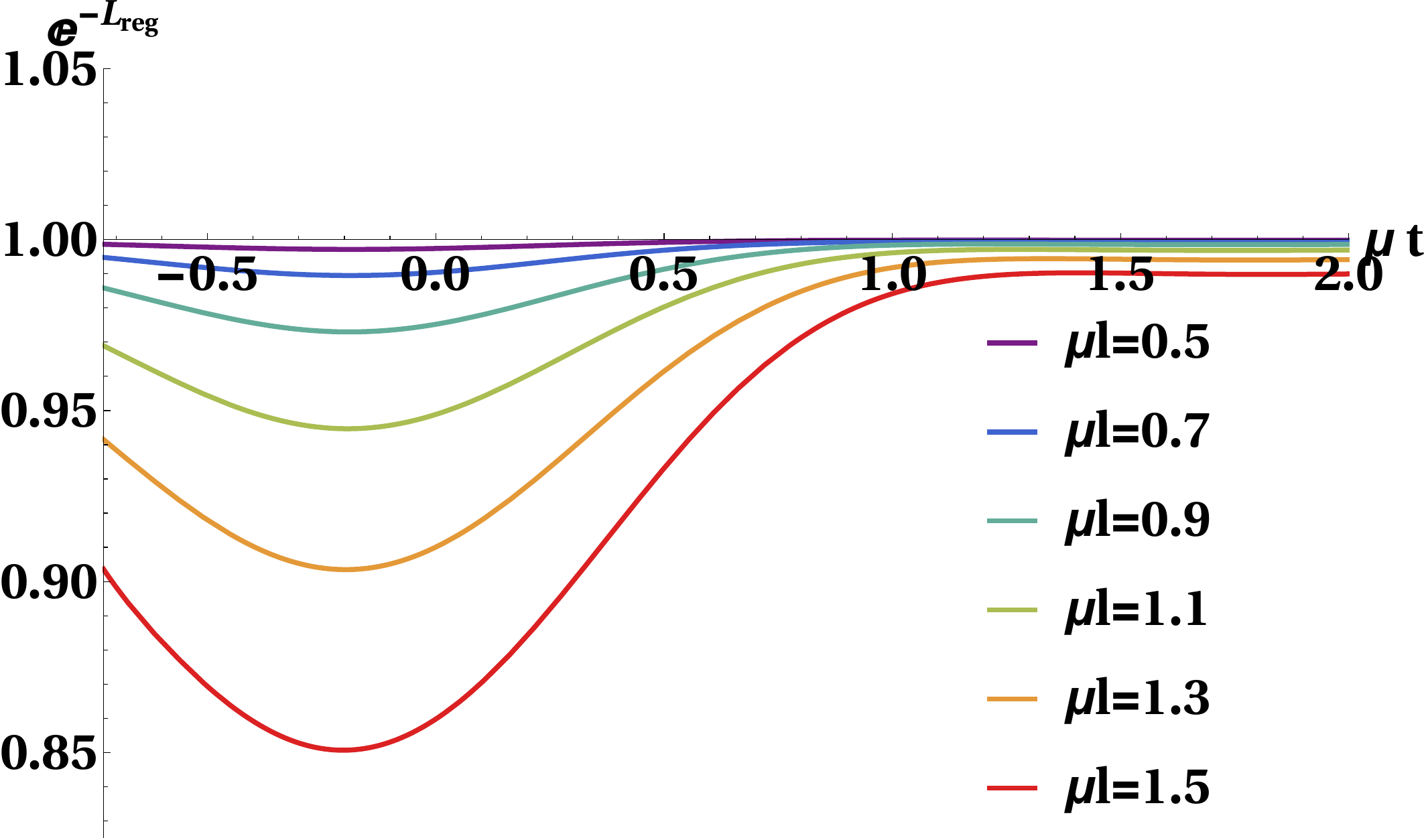}
\end{minipage}
\begin{minipage}[t]{0.4\textwidth}
\vspace{-0.0cm}
\hspace{-0.2cm}
\includegraphics[width=\textwidth]{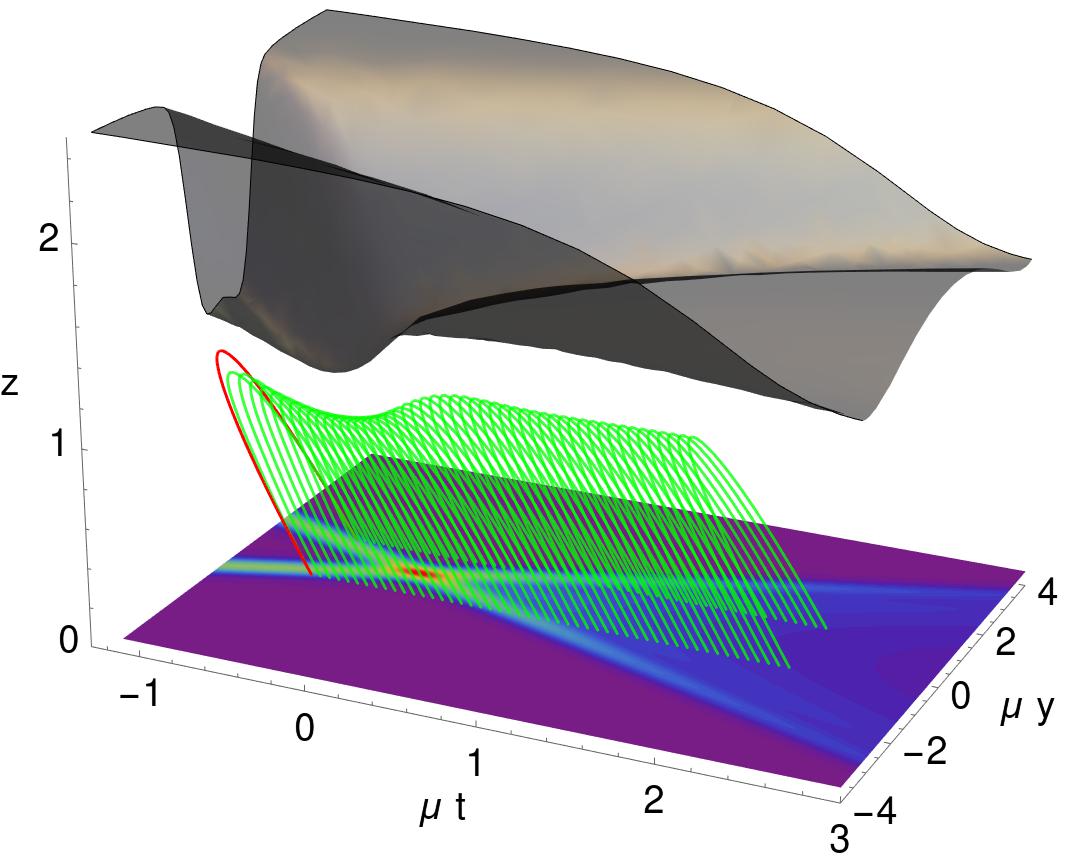}
\end{minipage}
\begin{minipage}[t]{0.4\textwidth}
\vspace{0.8cm}\includegraphics[width=\textwidth]{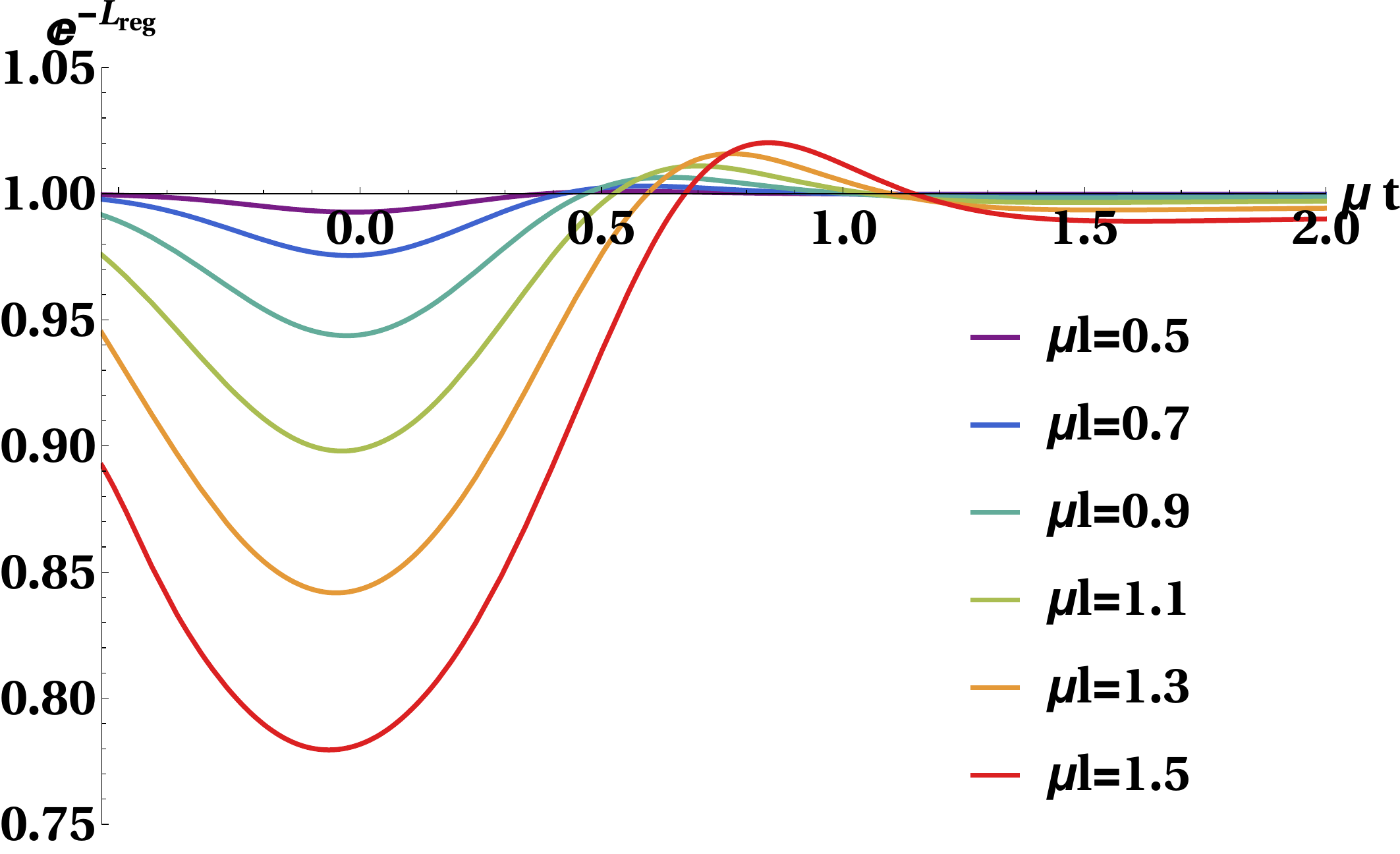}
\end{minipage}
\begin{minipage}[t]{0.4\textwidth}
\vspace{-0.0cm}
\hspace{-0.2cm}
\includegraphics[width=\textwidth]{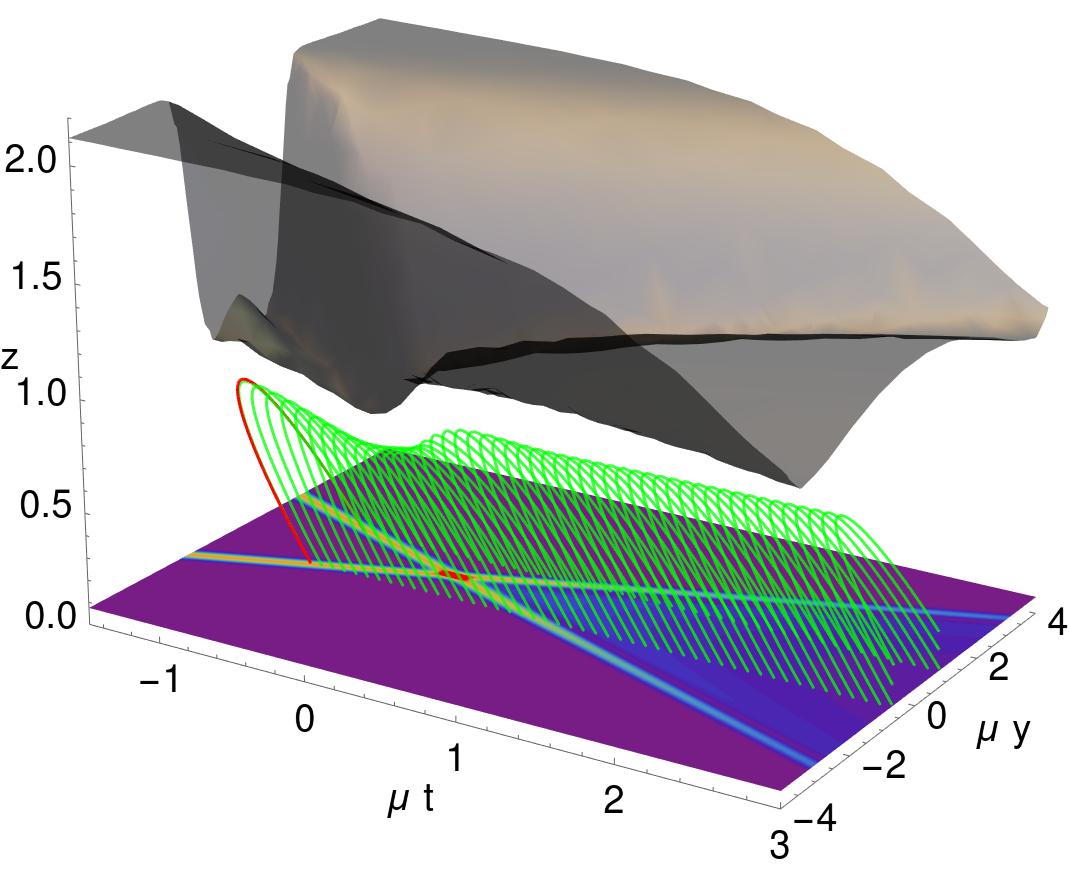}
\end{minipage}
\begin{minipage}[t]{0.4\textwidth}
\vspace{0.8cm}
\hspace{2.8cm}
\includegraphics[width=\textwidth]{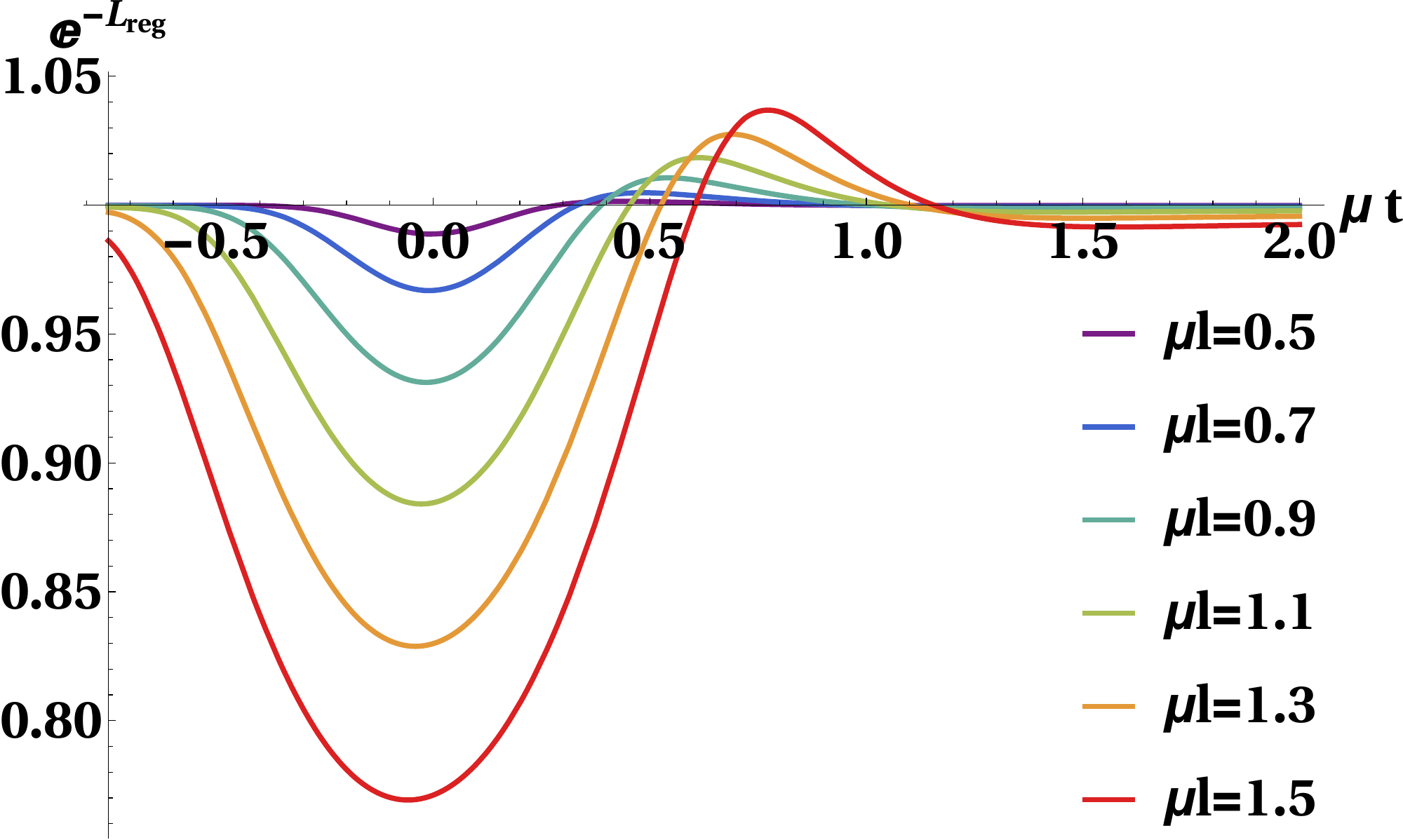}
\end{minipage}
\caption[Geometrical setup and two-point function (time-evolution).]{Left: Summary of the geometrical setup.
The black surfaces represent the radial position $z_{\textrm{\tiny AH}}(t,y)$ of the apparent horizon; red curves are AdS geodesics used for the initialization, the  green lines are geodesics ($\mu l=1.5$) for various time steps and at $z=0$ we show a density plot of the energy density for wide, intermediate and narrow shocks (top to bottom).
Right:  Corresponding evolution of the two-point function for different boundary separation $\mu l$.}\label{2PFfixed}
\end{figure}

For the time evolution it is of advantage, after using the pure AdS geodesic at the initial time, to use the previous solution to initialize the next time step. This approach turns out to be numerically extremely efficient and the relaxation algorithm reveals its full power, since in most cases the result at a given time is an excellent estimate at the next time step. A time step of $\Delta t=0.1$ allows to resolve nicely the shape of the two-point function and reduces the required number of iterations almost to a minimum. Usually two iterations are sufficient to achieve relative residuals in the geodesic equation which are $<10^{-5}$ and in many cases even one or two orders smaller. 

We follow the same logic when we compute the evolution in the boundary separation, where this approach is not only numerically efficient but also crucial to reach large separations. Undeformed ansatz geodesics of large separation typically reach far beyond the radial domain and finding the true solution using such geodesics to initialize the relaxation inevitably fails. We circumvent this problem by initializing with an ansatz of small separation ($\mu l=0.2$), which comfortably resides within the computational domain. Then we increase step by step the boundary separation and use the result for a given separation as ansatz for the next separation step. By using a step size of $\Delta l=0.1$ we can nicely resolve the shape of the two-point function and the relaxation usually converges again after two iterations. Since the relaxed geodesics are typically strongly deformed in direction away from the apparent horizon, i.e.\ the upper bound of the radial domain, we can reach separations which were inaccessible by simply relaxing the corresponding ansatz geodesic.

We like to discuss first the results from the time evolution before we go to the evolution in the separation.
In Figure~\ref{2PFfixed} (left) the whole setup for wide, intermediate and narrow  shocks is displayed. 
The dark surface represents the radial position of the apparent horizon $z_{\textrm{\tiny AH}}(t,y)$.
The evolution of the energy density of the boundary conformal field theory is shown by a contour  plot located  at the boundary $z=0$. The green lines are geodesics at  various time steps  for a given separation. For narrower shocks the minimum of the apparent horizon is  closer to the boundary and  the influence on  the shape of the geodesics is bigger. 
One can see that the tips of the geodesics tend to avoid the apparent horizon and the evolution of the tips show a similar shape as the apparent horizon.
Once the profile of the geodesics is found the evolution of the two-point functions can be extracted by computing their length.   
\begin{figure}
\begin{minipage}[t]{0.4\textwidth}
\vspace{-0.6cm}
\hspace{-0.5cm}
\includegraphics[width=\textwidth]{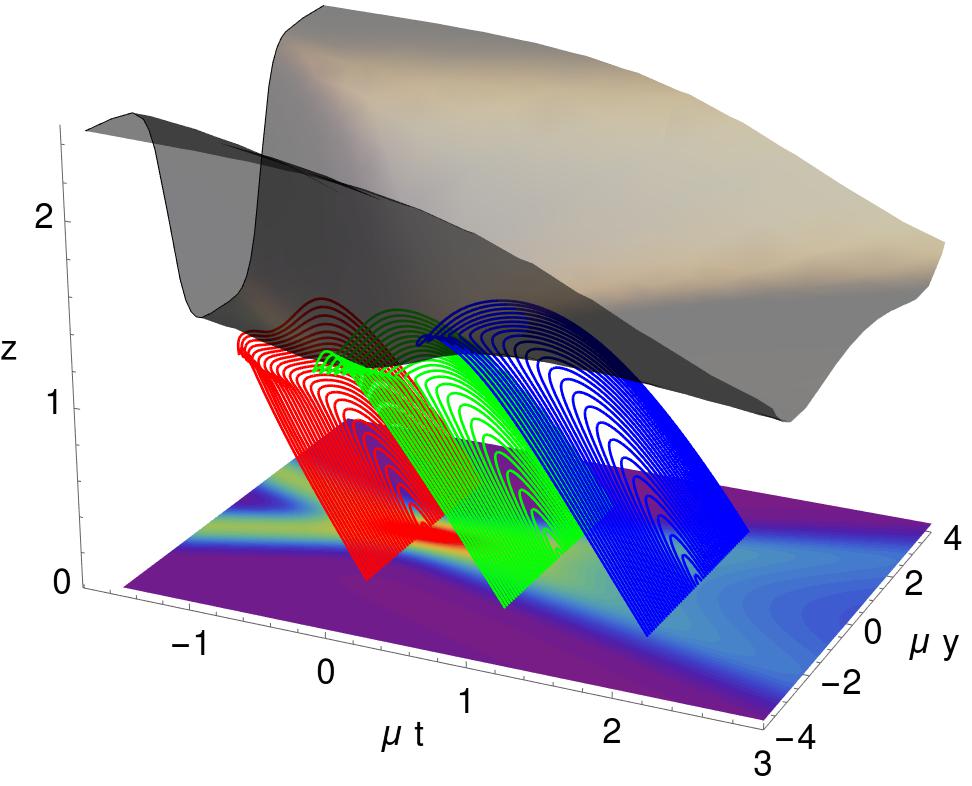}
\end{minipage}
\begin{minipage}[t]{0.5\textwidth}
\vspace{0.1cm}
\includegraphics[width=0.85\textwidth]{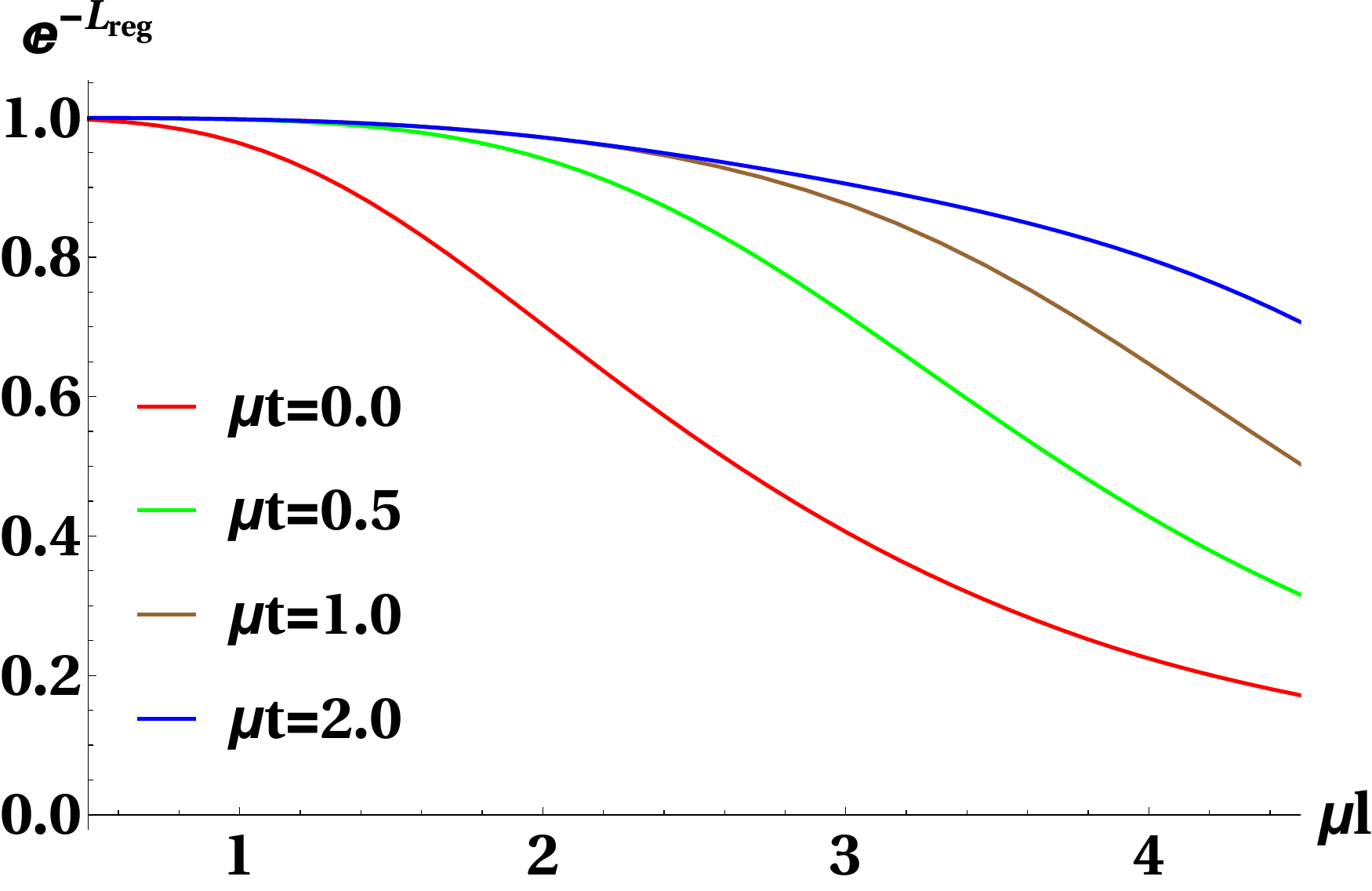}
\end{minipage}
\begin{minipage}[t]{0.4\textwidth}
\hspace{-0.5cm}
\includegraphics[width=\textwidth]{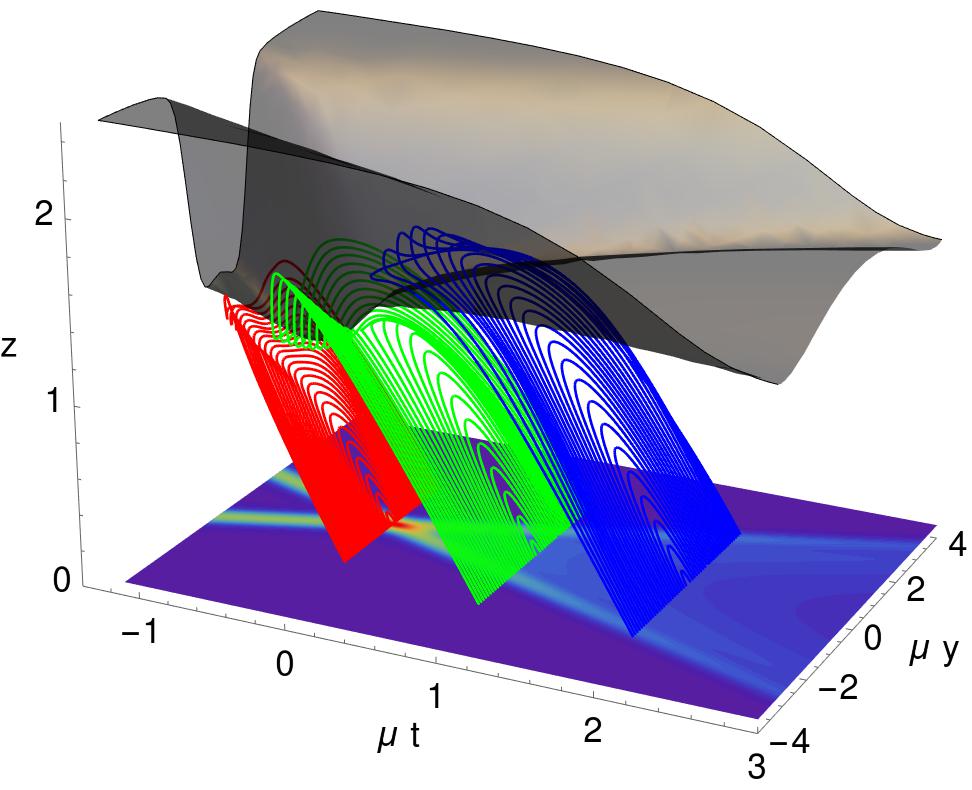}
\end{minipage}
\begin{minipage}[t]{0.5\textwidth}
\includegraphics[width=0.85\textwidth]{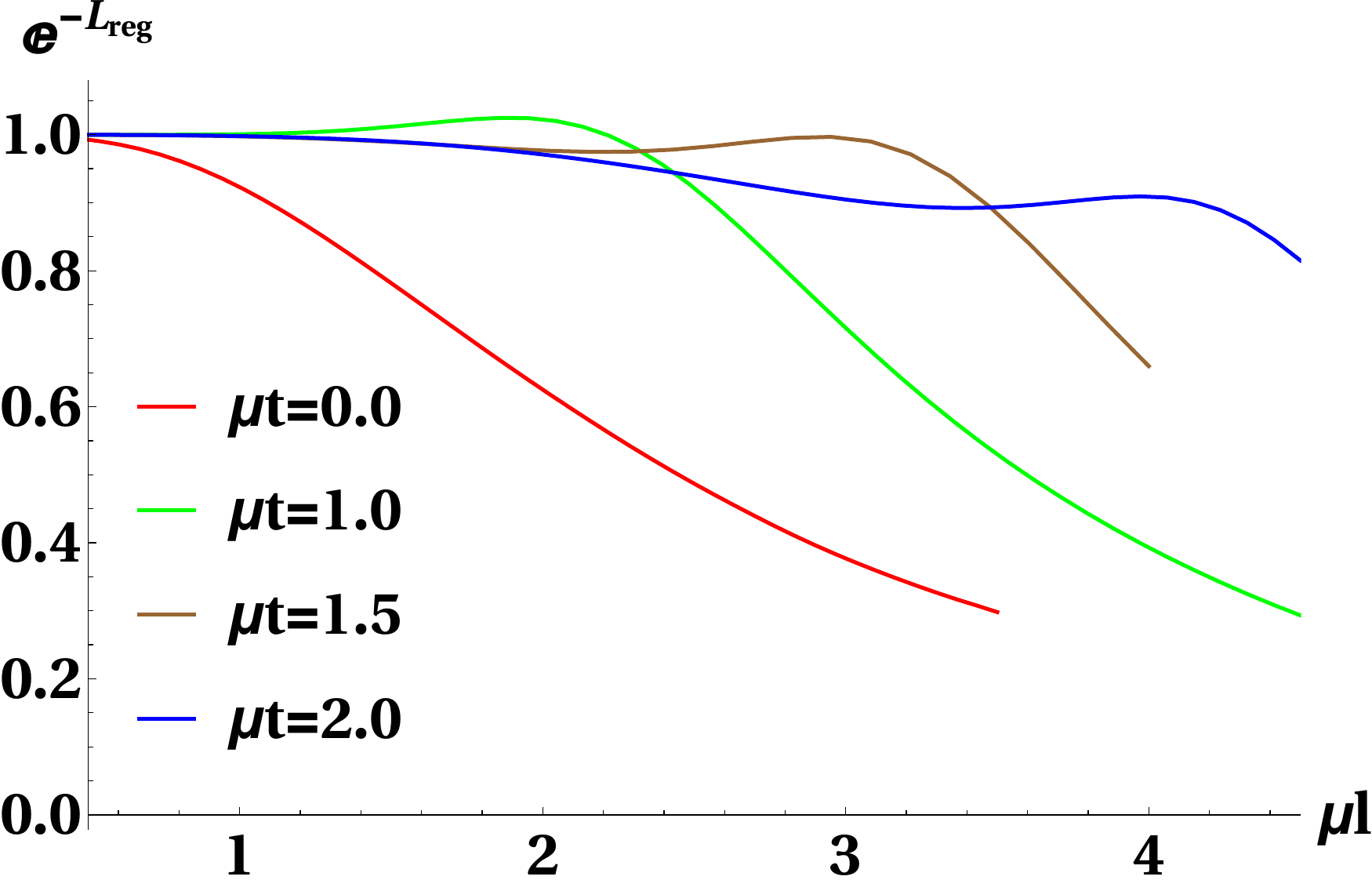}
\end{minipage}
\begin{minipage}[t]{0.4\textwidth}
\hspace{-0.5cm}
\includegraphics[width=\textwidth]{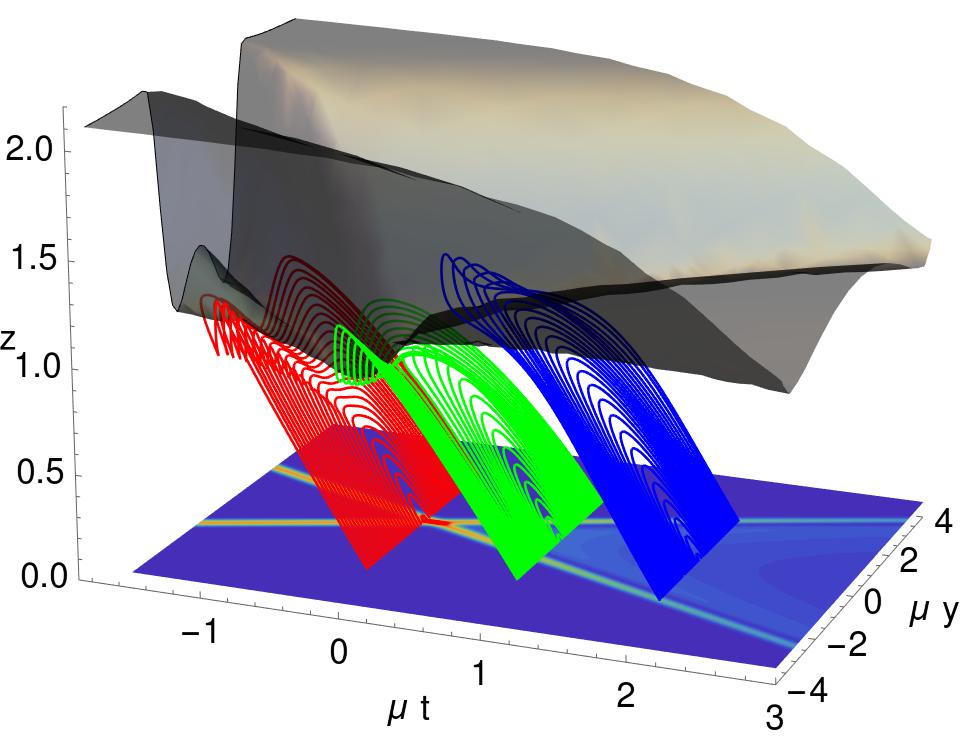}
\end{minipage}
\begin{minipage}[t]{0.5\textwidth}
\hspace{1.3cm}
\includegraphics[width=0.85\textwidth]{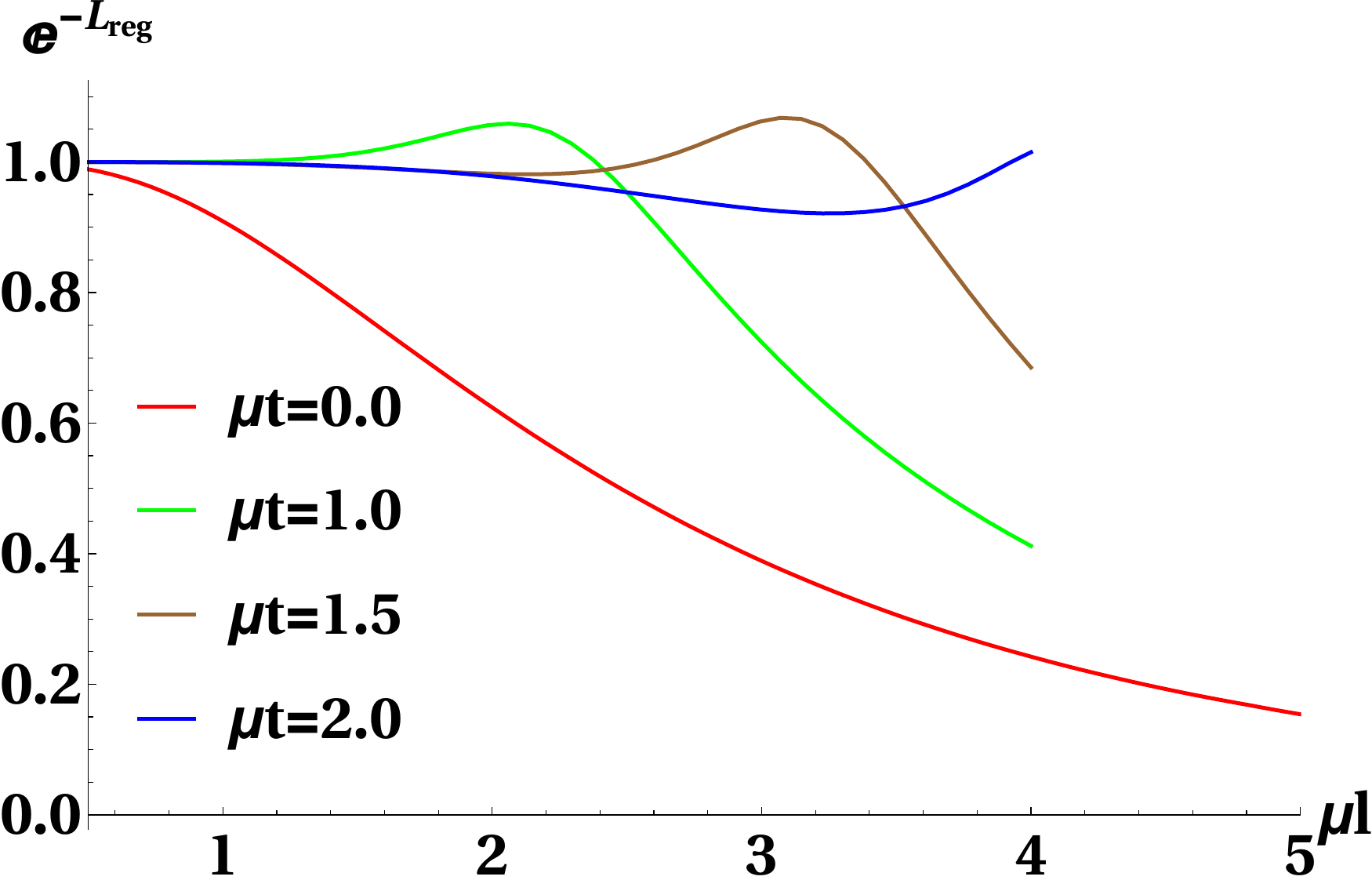}
\end{minipage}
\caption[Geometrical setup and two-point function (boundary separations).]{Left: Summary of the geometrical setup.
The black surfaces represent the radial position $z_{\textrm{\tiny AH}}(t,y)$ of the apparent horizon; red, green and blue curves are geodesics of various separations at $\mu t=0$, $\mu t=1$ and $\mu t=2$ respectively and at $z=0$ we show a contour plot of the energy density for wide, intermediate and narrow shocks (top to bottom).
Right:  Corresponding evolution of the two-point function with the boundary separation $\mu l$ at different times.}\label{2PFLevo}
\end{figure}
On the right hand side of Figure~\ref{2PFfixed} the evolution of the two-point functions for various boundary separations  for the different  geometries are  displayed.

Let us now summarize the most salient features in the time evolution of the two-point function during a holographic shock wave collision.
\begin{itemize}
 \item \textbf{rapid onset of linear de-correlation:} 
The system starts in some correlated state. As the shocks are getting closer more and more short range correlations are destroyed and the system rapidly starts to de-correlate in a linear fashion until a local minimum is reached.
The rapid onset of the linear regime is clearly visible for the narrow shocks in Figure~\ref{2PFfixed}, where for intermediate and wide shocks the onset lies outside our computational domain for larger separations, but the linear regime is still visible. 
 For intermediate and narrow shocks the minimum is located close to $t= 0$ where the energy density is maximal. For wide shocks this minimum is reached significantly before $t=0$. 
\item \textbf{premature de-correlation:} A careful tracking of the position of the minimum as a function of the boundary separation reveals that it is shifted to earlier times as the separation increases. This effect, which is very small and therefore hardly visible in Figure~\ref{2PFfixed}, is a robust feature of all three kinds of shocks that we have studied.
 \item \textbf{linear correlation restoration:} During the collision, when the shocks interact, new correlations are formed in the system. As the shocks move outwards ($t>0$), the correlations are linearly restored for all three kinds of shocks. 
 \item \textbf{correlation overshooting of narrow shocks:} After the linear restoration regime, the correlations in wide and narrow shocks approach their final values in very different ways.
For intermediate and narrow shocks the correlations significantly overshoot their final values before they finally approach them from above.
In the case of wide shocks this effect is strongly damped and the correlations approach their initial value almost monotonically from below.
\end{itemize}

We switch now to the scaling of the two-point function with the separation. The holographic setup and the results for the evolution of the two-point function are displayed in Figure~\ref{2PFLevo}.
At the collision time ($\mu t=0$) the two-point function falls off monotonically with the separation in all three cases, although the corresponding geodesics are strongly deformed. For the wide shocks this behavior persists also at later times, where due to the weaker influence of the shocks the correlations fall off more slowly.
For intermediate and narrow shocks an additional maximum appears at $\mu t>0$ which is more pronounced for narrow shocks. The position of this additional maximum is centered around the position of the outgoing shocks.
It is suggestive that narrow shocks which pass through each other almost transparently remain correlated for some time after the collision while wide shocks stop each other before they explode hydrodynamically and the correlations are completely lost.  
This motivated us to study the correlations between the shocks themselves, which we do systematically in Section \ref{app:3}. There we find that the correlations between intermediate and narrow shocks significantly grow after the collision before they start to decay, where the correlations between wide shocks decay immediately. 

\begin{figure}
\hspace{-0.cm}
\includegraphics[width=4.7cm]{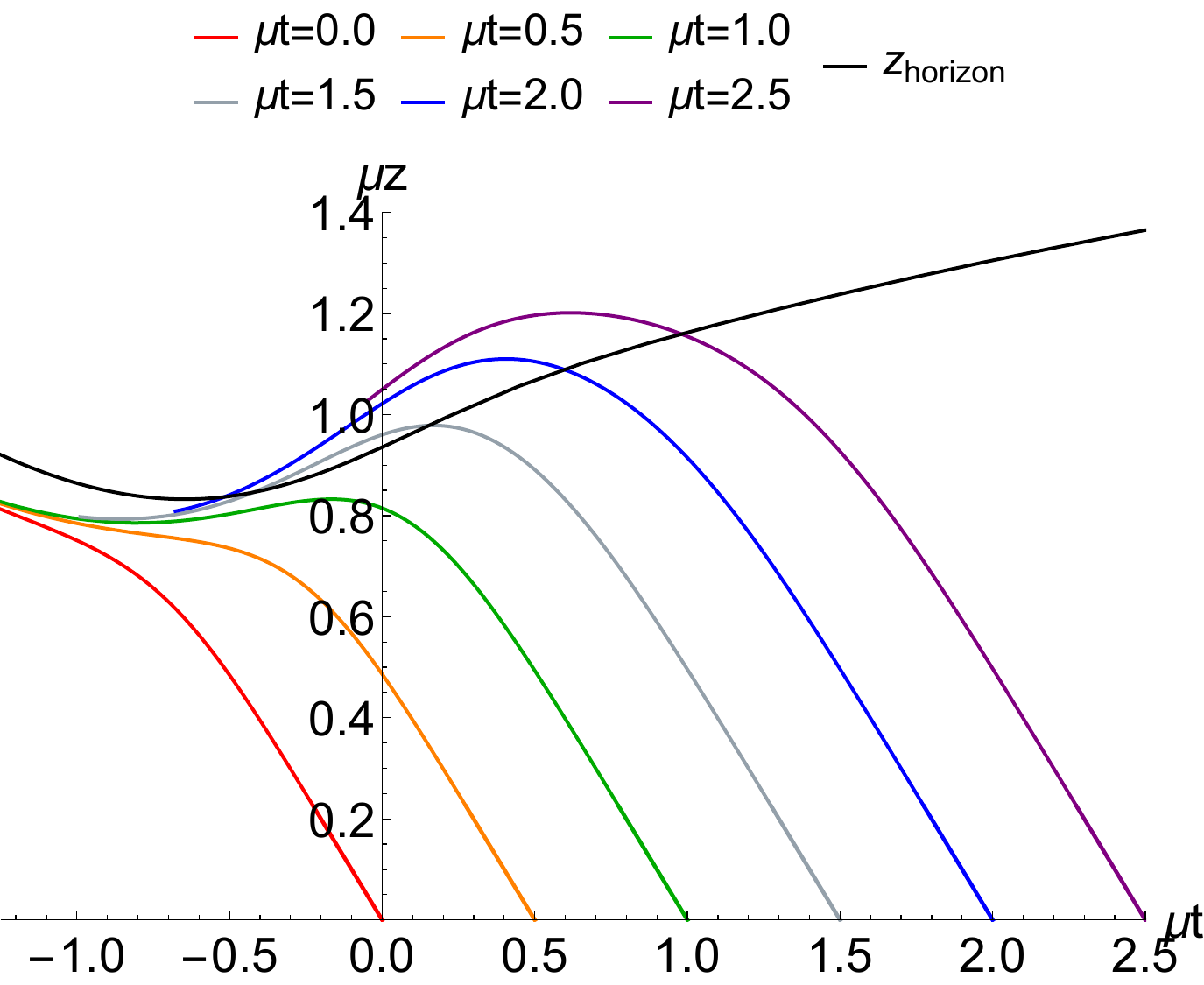}$\;$\includegraphics[width=4.7cm]{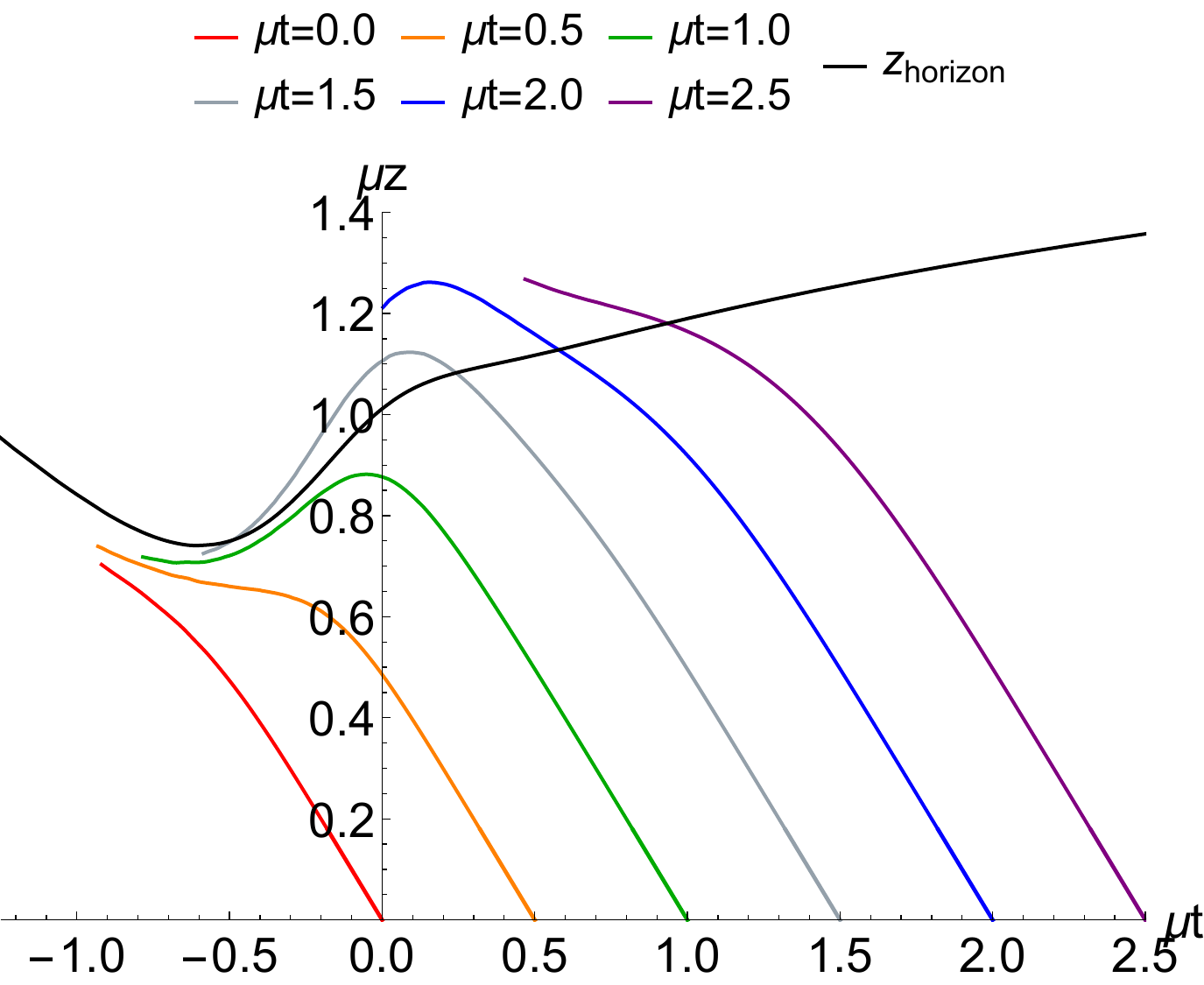}$\;$\includegraphics[width=4.7cm]{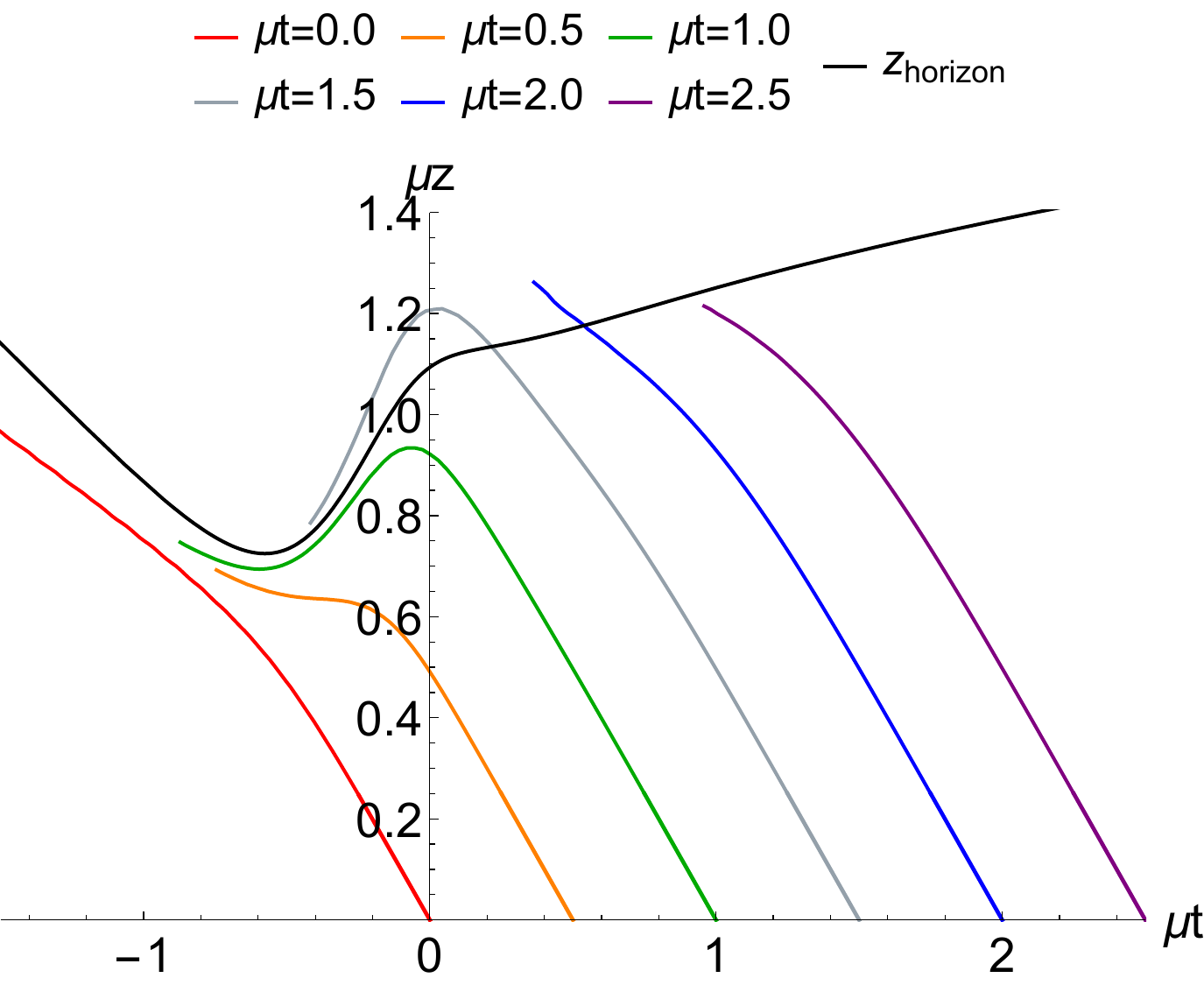}
\caption[Geodesics can extend beyond the apparent horizon.]{\label{GeodesicDip} 
The $z$-position of the geodesics at $y=0$ for several times and separations, starting with $l=0$ near the boundary, and increasing going towards the end of the curve. We show wide, intermediate and narrow shocks (from left to right). The $z$-position of the apparent horizon at $y=0$ is shown in black. At late times and sufficiently large boundary separation in all three cases (wide, intermediate and narrow shocks) geodesics can reach behind the apparent horizon, whereas for early times they reside outside the horizon entirely.}
\end{figure}

Interestingly, for larger separations the geodesics remain outside the horizon for early times, but they cross the horizon after a time of around $\mu t=1.5$. 
 This can be seen from the blue curves in Figure~\ref{2PFLevo} and is displayed more clearly in Figure~\ref{GeodesicDip} where we plot the tip of the geodesic located at $y=0$, for different separations and the position of the apparent horizon at $y=0$.
 This happens for all the initial conditions (wide, intermediate, narrow) we have studied and is in strong contrast to the entanglement entropy case where we do not find extremal surfaces which cross the horizon.
The crossing after a time of $\mu t=1.5$ is perhaps counterintuitive since geodesics are expected to remain outside the horizon when the system is close to equilibrium. Indeed, hydrodynamics applies after a time $\mu t=0.89$ \cite{Casalderrey-Solana:2013aba}, which is well before the crossing of the geodesics.
At later times presumably the geodesics indeed remain outside again, though our numerics did not allow to determine the precise time at which this is the case. 
 
\subsection{Correlations of Colliding Shocks}\label{app:3}

 Instead of studying the time evolution of the two-point function  between  two fixed points  in space, in the context of heavy ion collisions it might be more interesting to actually study the correlation between the two shocks itself. In order to do so, the endpoints of the geodesics follow the maxima of the energy density.
 
\begin{figure}
\begin{minipage}[t]{0.5\textwidth}
\vspace{-0.8cm}
\hspace{-0.0cm}
\includegraphics[width=\textwidth]{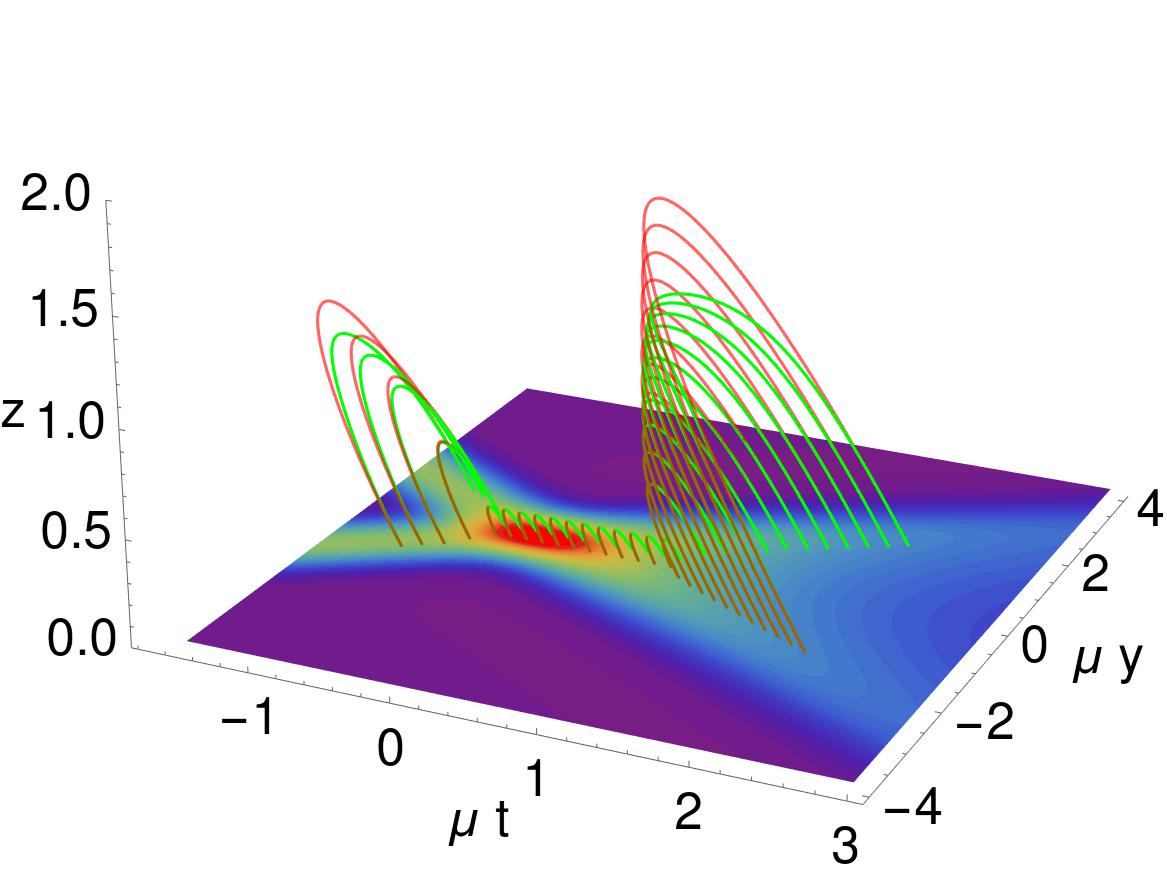}
\end{minipage}
\begin{minipage}[t]{0.5\textwidth}
\vspace{0.1cm}
\hspace{0.5cm}
\includegraphics[width=0.85\textwidth]{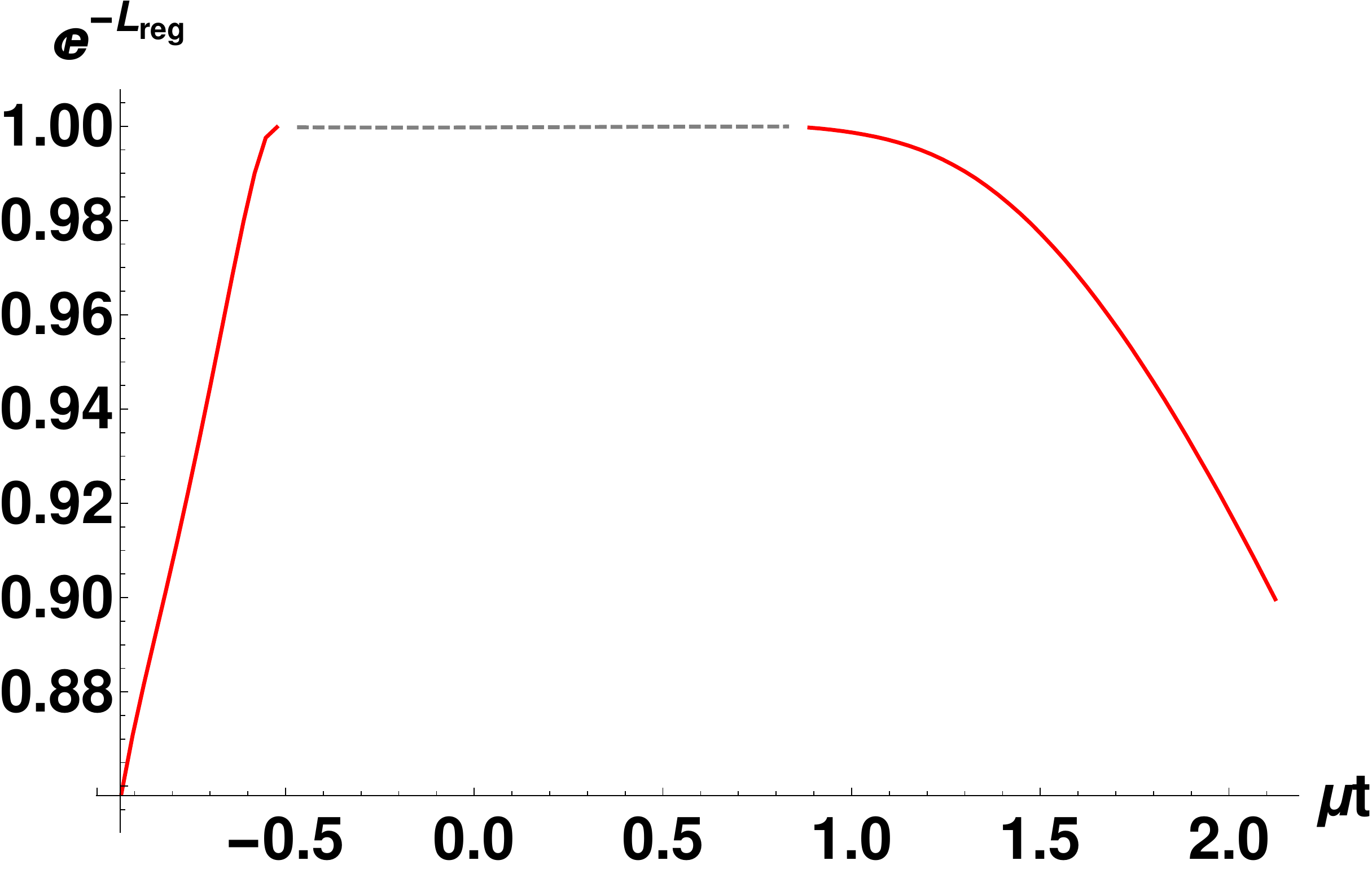}
\end{minipage}
\begin{minipage}[t]{0.5\textwidth}
\hspace{-0.0cm}
\includegraphics[width=\textwidth]{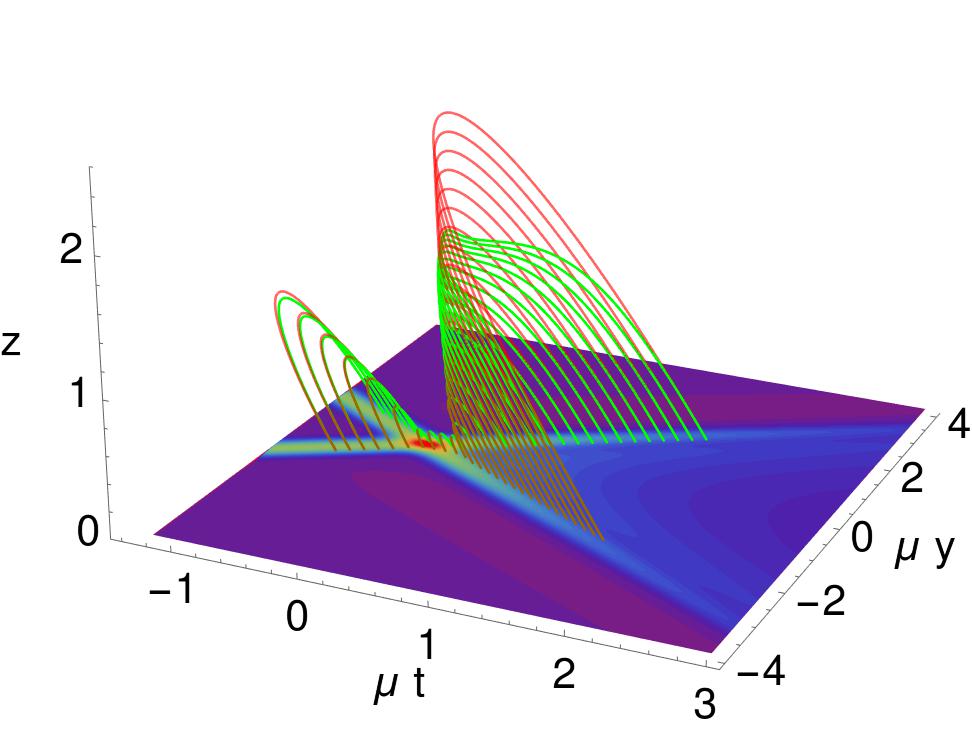}
\end{minipage}
\begin{minipage}[t]{0.5\textwidth}
\hspace{0.5cm}
\includegraphics[width=0.85\textwidth]{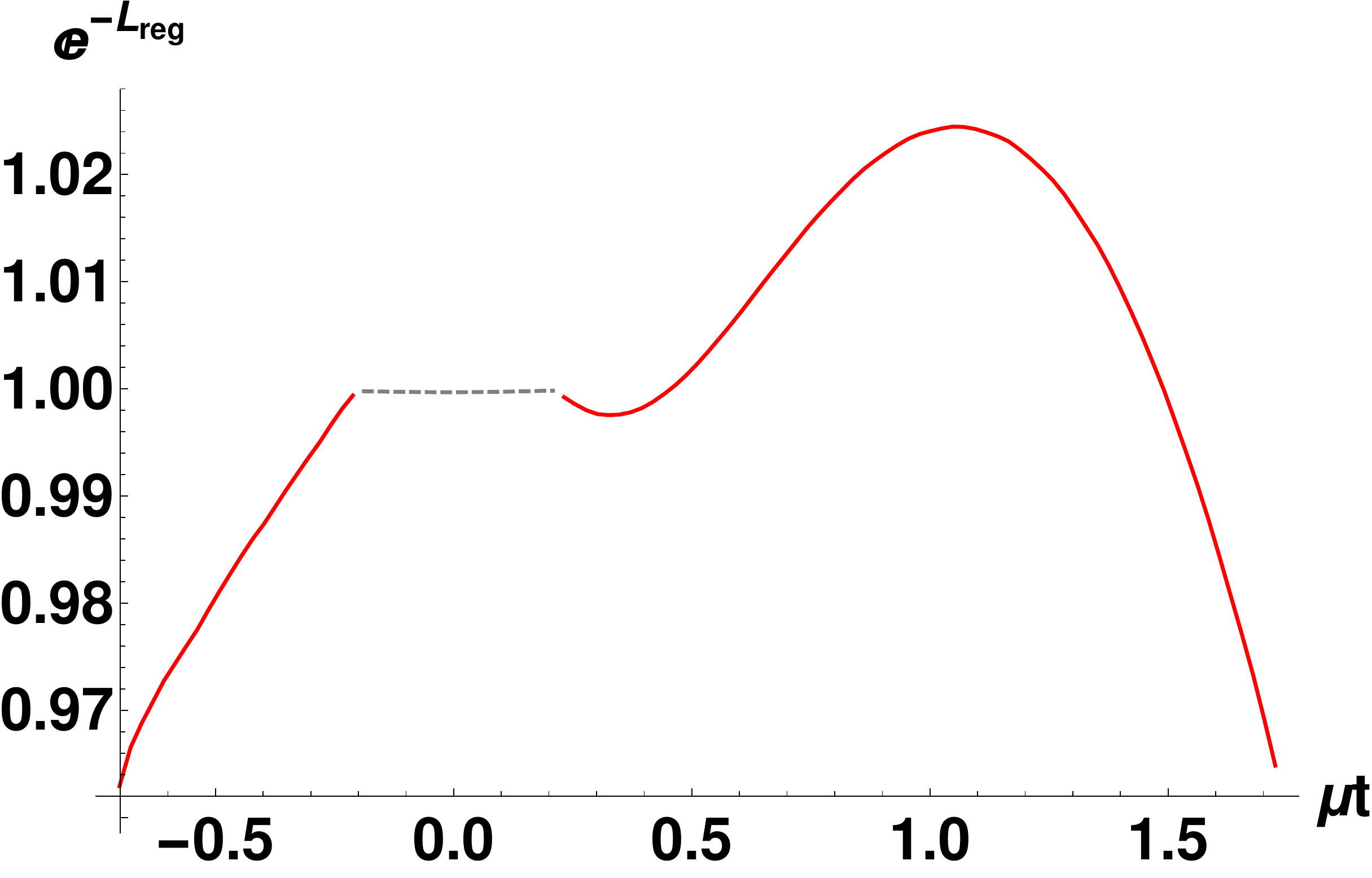}
\end{minipage}
\begin{minipage}[t]{0.5\textwidth}
\hspace{-0.0cm}
\includegraphics[width=\textwidth]{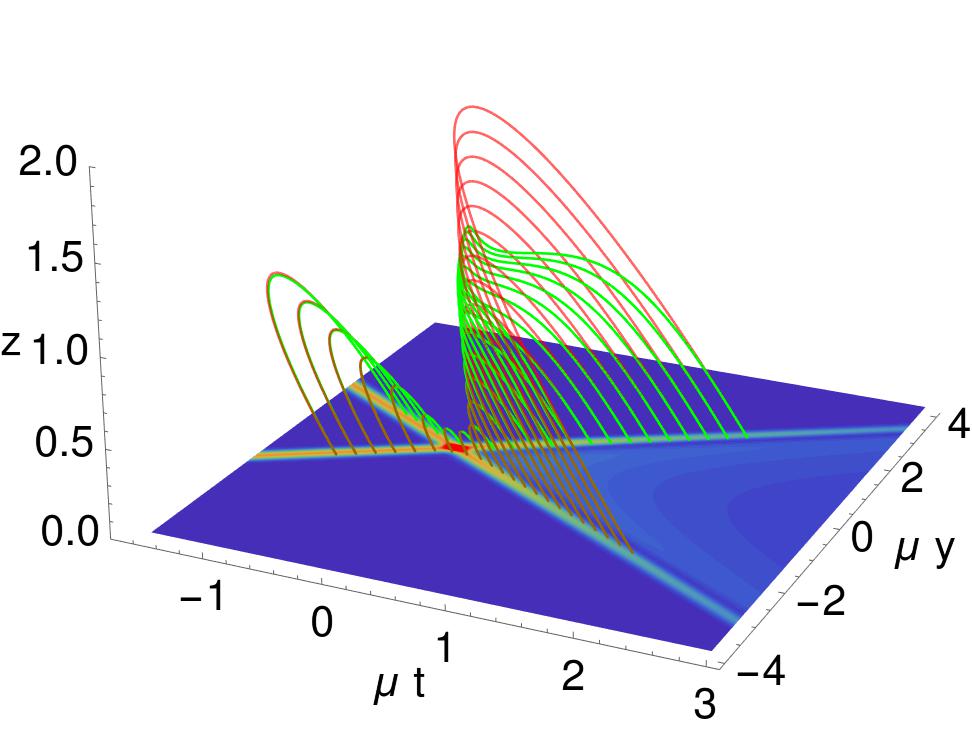}
\end{minipage}
\begin{minipage}[t]{0.5\textwidth}
\hspace{0.5cm}
\includegraphics[width=0.85\textwidth]{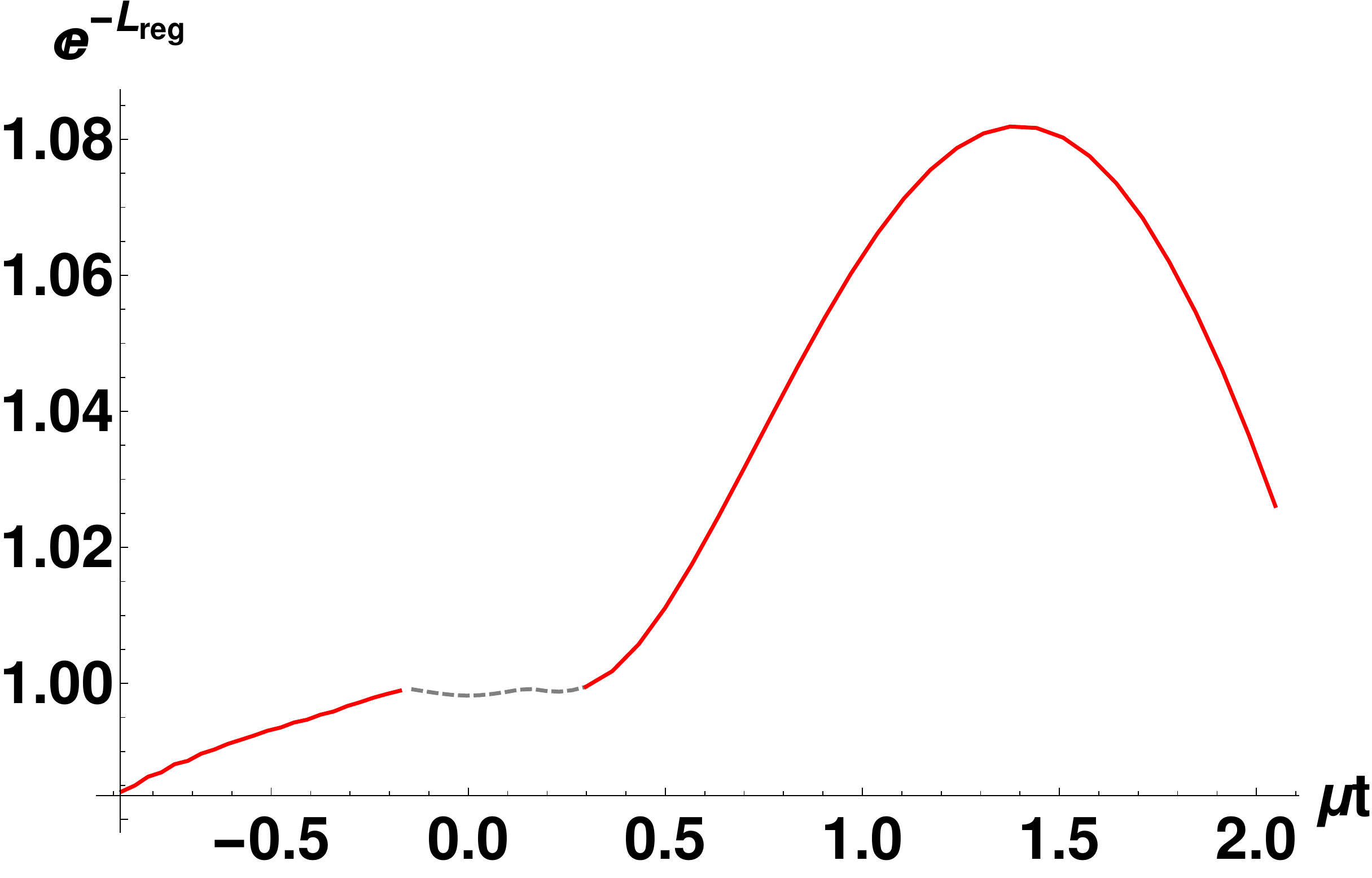}
\end{minipage}
\caption[Correlations of colliding shocks.]{Left: Time evolution of geodesics in the shock wave geometry (green) for wide, intermediate and narrow shocks (top to bottom) pure AdS geodesics (red) with endpoints attached to the position of the maxima in the energy density. Right: Time evolution of the correlation between the shocks; dashed lines indicate the region where only a central maximum in the energy density exist and the separation is fixed to $3*z_{\textrm{\tiny cut}}$.
 }\label{2PFfollow}
\end{figure}

When the separation of the endpoints becomes smaller than three times the cutoff we fix the endpoints to this value until the distance between the two maxima after the collision exceed this value.
  The results are displayed in Figure~\ref{2PFfollow}, where the geometrical situation is displayed on the left hand side and the time evolution of the two-point-functions on the right hand side.
 
As already discussed in Section \ref{se:3.2}, for wide shocks the behavior is qualitatively different than for intermediate and narrow shocks. 

As the two wide shocks approach each other their correlation increases almost linearly until it reaches a plateau, which is the point when  the separation of the endpoints is smaller than three times  the cutoff. 
Once the shocks separate again from each other their correlation decreases. 
  
As the shocks get narrower the initial growth slows down  because the shocks start to overlap later. 
After the fixed separation period  a local minimum appears after which the correlations continue to grow to reach another maximum which appears later for narrow shocks. 
  In addition, the maximum correlation is highest for narrow shocks. 

This behavior is reminiscent of the full stopping and transparency scenario for wide and narrow shocks considered in \cite{Casalderrey-Solana:2013aba}. 
 As the wide  shocks start to interact the energy density starts to pile up and all the energy density is contained in a small region after which hydrodynamical explosion occurs. This behavior is also encoded in the two-point function which reaches a maximum and can only decrease when hydrodynamic explosion occurs. 

For narrower shocks the situation is different. The shocks almost move through each other. Their shape gets altered but no hydrodynamic explosion occurs. The shocks separate from each other and plasma between them forms resulting in a growth of the correlation also after the collision. At sufficiently late times, when the shocks are separated far enough and a hydrodynamical description is applicable the two-point function decreases rapidly. 

To summarize, there is a general pattern appearing. As the shocks become narrower the initial growth slows down, the maximum  correlation increases and  occurs later. 

\section{Entanglement Entropy}
In this section we monitor the evolution of entanglement entropy. 
In time dependent systems the covariant entanglement entropy \cite{Hubeny:2007xt} for some boundary region $A$  is obtained  by extremizing the three-surface functional  
\begin{equation}\label{Eq:area}
{\cal A}=\int d^3\sigma\sqrt{\det\Big(\frac{\partial X^{\mu}}{\partial\sigma^a}\frac{\partial X^{\nu}}{\partial\sigma^b}g_{\mu\nu}\Big)}\,,
\end{equation}
that ends on the boundary surface $A$.
In the dual field theory the entanglement entropy is then conjectured to be given by \cite{Ryu:2006bv,Ryu:2006ef,Hubeny:2007xt}
\begin{equation}
S_{\textrm{\tiny EE}}=\frac{\cal A}{4 G_N}\,.
\end{equation}
Under certain circumstances the problem of finding extremal surfaces can be reduced to finding geodesics in an auxiliary space-time and the problem of solving non-linear partial differential equation can be circumvented \cite{Ecker:2015kna}. 
In the case at hand this can be achieved by considering a stripe entangling region with finite extent in the longitudinal direction $y$ and infinite extent in the homogeneous transverse directions $(x_1,x_2)$ for which (\ref{Eq:area}) simplifies to
\begin{equation}\label{areastripe}
{\cal A}=\int d x_1 \int d x_2 \int d\sigma\sqrt{\Omega^2h_{\mu\nu}\frac{\partial X^{\mu}}{\partial\sigma}\frac{\partial X^{\nu}}{\partial\sigma}}=V \tilde{L}\,.
\end{equation}
The surface functional (\ref{areastripe}) suffers from two kinds of infinities, one from the integral $V=\int d x_1 \int d x_2$ over the homogeneous directions and another one from the infinite geodesic length $\tilde{L}$ in the auxiliary spacetime $\Omega^2h_{\mu\nu}$.
Since the infinite volume factor $V$ contains no dynamical information these singularities are avoided by considering entanglement entropy densities $\frac{S_{\textrm{\tiny EE}}}{V}$.
Analogous to the two-point function we regularize the geodesic length $\tilde{L}$ by subtracting the corresponding auxiliary vacuum contribution $\tilde{L}_0$.
The observable we compute is the regularized entanglement entropy density per Killing volume in units of 4$G_N$ 
\begin{equation}\label{eereg}
S_{\textrm{\tiny reg}}=4G_N\Big(\frac{S_{\textrm{\tiny EE}}}{V}-\frac{S_{\textrm{\tiny EE}}^0}{V_0}\Big)=\tilde{L}-\tilde{L}_0\,.
\end{equation}

\subsection{Geodesics in the Auxiliary Spacetime}\label{se:3.1a}
Our aim is to compute the entanglement entropy for a stripe region with finite extent in $y$-direction and infinite extent in $(x_1,x_2)$ using formula (\ref{eereg}).
Therefore we have to find geodesic lengths $\tilde{L}$ and $\tilde{L}_0$ in the corresponding auxiliary spacetimes.
The auxiliary spacetime, which is related to the metric (\ref{metric}) by a conformal factor $\Omega^2=S^4 e^{2B}$, reads

\begin{equation}\label{confsubmetric}
d\tilde{s}_y^2=S^4 e^{2B}   \big(-A d v^2-\frac{2}{z^2}  d z d v +2 Fd yd v +S^2 e^{-2B} d y^2 \big)\,.
\end{equation}
This time we initialize the relaxation algorithm with a geodesic in Poincar\'e patch AdS times a conformal factor $\Omega_0^2=\tfrac{1}{z^4}$
\begin{equation}\label{confadsmetric}
d \tilde{s}_0^2=\frac{1}{z^6}\left(-d v^2-2 d z d v + d y^2\right)\,.
\end{equation}
Using the ansatz (\ref{nonAffine2}) and the corresponding Jacobian (\ref{Eq:Jacobian}) in the relaxation algorithm allows us to compute geodesics in the auxiliary spacetime (\ref{confsubmetric}).

The bulk parts of the geodesic lengths in \eqref{eereg}, which are the contributions from $z>z_{\textrm{\tiny cut}}$, follow from integrating the line elements (\ref{confsubmetric}) and (\ref{confadsmetric}) 
\begin{subequations}
\begin{eqnarray}
\tilde{L}^{\textrm{\tiny bulk}}&=&\int_{\sigma_-}^{\sigma_+} d\sigma S^2 e^{B}\sqrt{-A \dot{V}^2-\frac{2}{Z^2}\dot{Z}\dot{V}+2F\dot{V}\dot{Y} +S^2 e^{-2B}\dot{Y}^2},\\
\tilde{L}_0^{\textrm{\tiny bulk}}&=&\int_{\sigma_-}^{\sigma_+} d\sigma \frac{1}{Z_0^3}\sqrt{- \dot{V_0}^2-2 \dot{Z}_0 \dot{V}_0 + \dot{Y}_0^2},
\end{eqnarray}
\end{subequations}
where in this case the bounds of the integral $\sigma_\pm$, implementing the IR-cutoff at $z\!=\!z_{\textrm{\tiny cut}}$, are given by
\begin{equation}
\sigma_\pm=\pm\sqrt{1-\frac{z_{\textrm{\tiny cut}}}{z_*}}\,,
\end{equation}
with $z_*$ such as defined above \eqref{Eq:sigmaCut}.

We build the near boundary part ($0\le z \le z_{\textrm{\tiny cut}}$), like for the two-point function, from the asymptotic solution of the geodesic equation in the conformal spacetime, which
leads to the following near-boundary expansion
\begin{subequations}\label{asympSolutionEE}
\begin{eqnarray}
 Z(z) & = & z\,, \\
 V(z) & = & t_0-z+v_4 z^4+\frac{a_4 z^5}{5}+\mathcal{O}\left(z^6\right)\,, \\
 Y(z) & = & \frac{l}{2}+y_4 z^4+\frac{f_4 z^5}{5}+\mathcal{O}\left(z^6\right)\,, \\
 J(z) & = & \frac{3}{z}+\left(2 a_4-4 b_4\right) z^3+\mathcal{O}\left(z^6\right)\,,
\end{eqnarray}
\end{subequations}
where the normalizable modes $a_4(v,y)$, $b_4(v,y)$ and $f_4(v,y)$ are evaluated at $v=t_0$ and $y=\pm\tfrac{l}{2}$.  We again have two undetermined constants $v_4$ and $y_4$, which now appear two orders higher than for the case of the two-point function. Again we also have the analytic solution in the auxiliary pure AdS spacetime
\begin{subequations}
\begin{eqnarray}\label{asympAnsatzEE}
Z_0(z)&=&z\,,\\
V_0(z)&=&t-Z_0(z)\,,\\
Y_0(z)&=&\pm\Big(-\frac{l}{2} + \frac{W Z_0(z)^4}{4}\, {}_2F_1 \left[ \tfrac{1}{2},\tfrac{2}{3},\tfrac{5}{3}; W^2 Z_0(z)^6\right]\Big)\nonumber\\
      &=&\pm\big(-\frac{l}{2}+\frac{W}{4}z^4\big)+\mathcal{O}(z^{10})\,,\\
J_0(z)&=&\frac{3-6W^2z^6}{z-W^2z^7}=\frac{3}{z}-3W^2z^5+\mathcal{O}(z^{11}).
\end{eqnarray}
\end{subequations}
The near boundary contribution to the geodesic length  for both endpoints evaluates to
\begin{eqnarray}
\tilde{L}^{\textrm{\tiny bdry}}-\tilde{L}_0^{\textrm{\tiny bdry}}&=& \big(b_4-\frac{a_4}{2}\big)z+\big(\partial _tb_4-\frac{7 \partial _ta_4}{20}\big)z^2\nonumber\\
                   &+& \frac{1}{120} (20 \partial _y\partial _tf_4-13 \partial _t^2a_4+70 \partial _t^2b_4+7 \partial _y^2a_4+2 \partial _y^2b_4+960 y_4^2-960 t_4^2)z^3\nonumber\\
                   &+& \mathcal{O}(z^{4}),
\end{eqnarray}
where the divergent term cancels out again. Now this formula is clearly more useful, as the two leading contributions do not depend on the unknown coefficients $v_4$ and $y_4$, which hence allows to reduce the cutoff dependence significantly.
The regularized entanglement entropy of \eqref{eereg} is the sum of the bulk contribution and the near boundary contribution   
\begin{equation}\label{Sreg}
S_{\textrm{\tiny reg}}=(\tilde{L}^{\textrm{\tiny bulk}}-\tilde{L}_0^{\textrm{\tiny bulk}})+(\tilde{L}^{\textrm{\tiny bdry}}-\tilde{L}_0^{\textrm{\tiny bdry}})\;.
\end{equation}
As for the two-point function we checked the convergence of $S_{\textrm{\tiny reg}}$ with the gridsize in the range from 50 up to 400 gridpoints and find again that for more than $200$ gridpoints the change in $S_{\textrm{\tiny reg}}$ is smaller than $\mathcal{O}(10^{-5})$ which is the same order as the allowed residual we choose in the relaxation algorithm.

To achieve cutoff independence of $S_{\textrm{\tiny reg}}$ turns out to be more delicate than for the two-point function. Now for a range $z_{\textrm{\tiny cut}}=[0.05,0.1]$ we obtain a slightly worse cutoff dependence of $\mathcal{O}(10^{-3})$ which is however sufficient for our qualitative studies where $S_{\textrm{\tiny reg}}=\mathcal{O}(10^{-1})$ and the influence of the cutoff can be estimated to be $\approx 1 \%$.
Again we choose $200$ gridpoints to discretize our geodesics and set $z_{\textrm{\tiny cut}}=0.075$ in all the calculations we present in this chapter.

\subsection{Evolution of Entanglement Entropy}

\begin{figure}
\begin{center}
\hspace*{-0.0cm}\includegraphics[scale=.23]{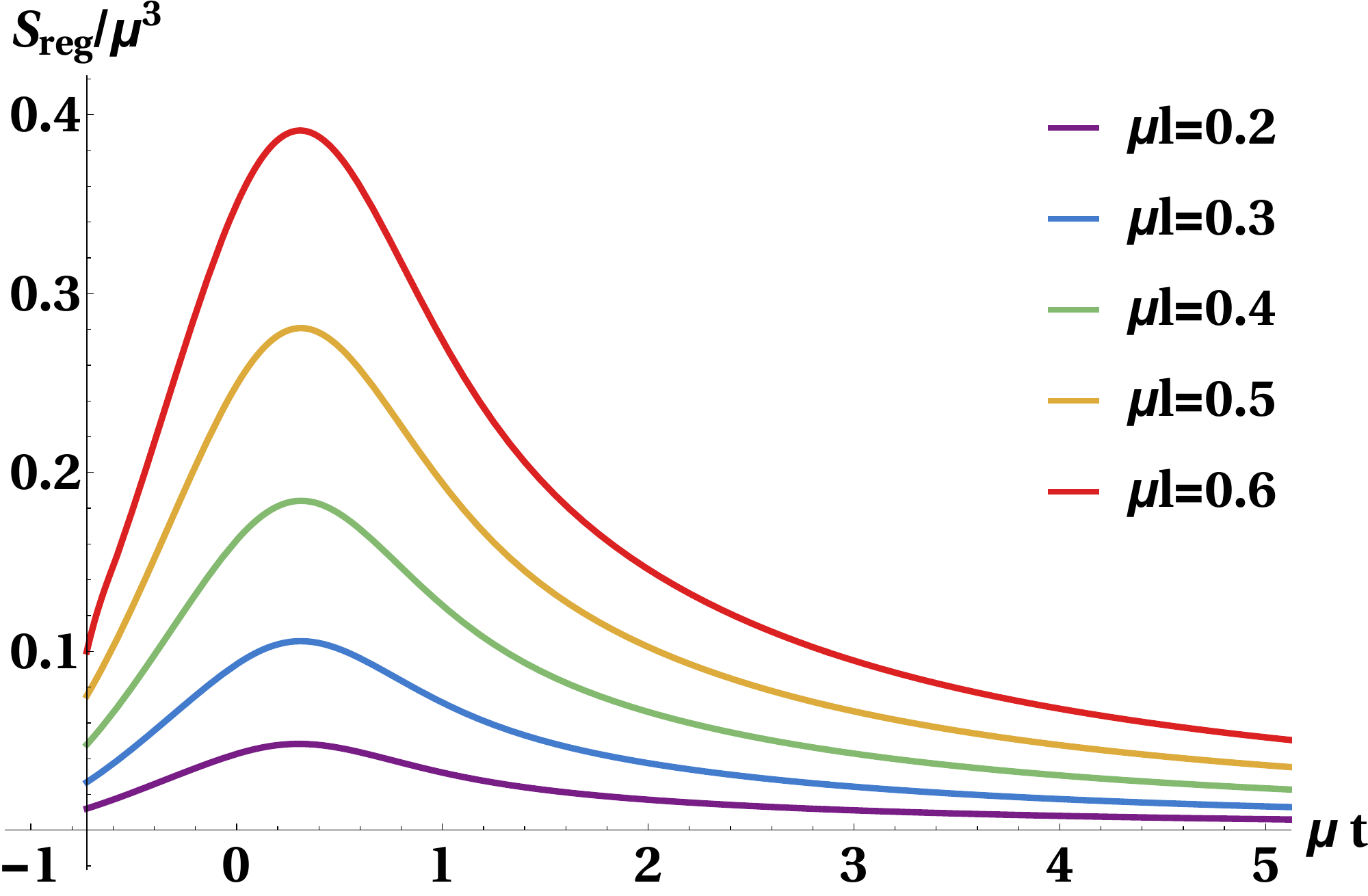}$\;\;$\includegraphics[scale=.23]{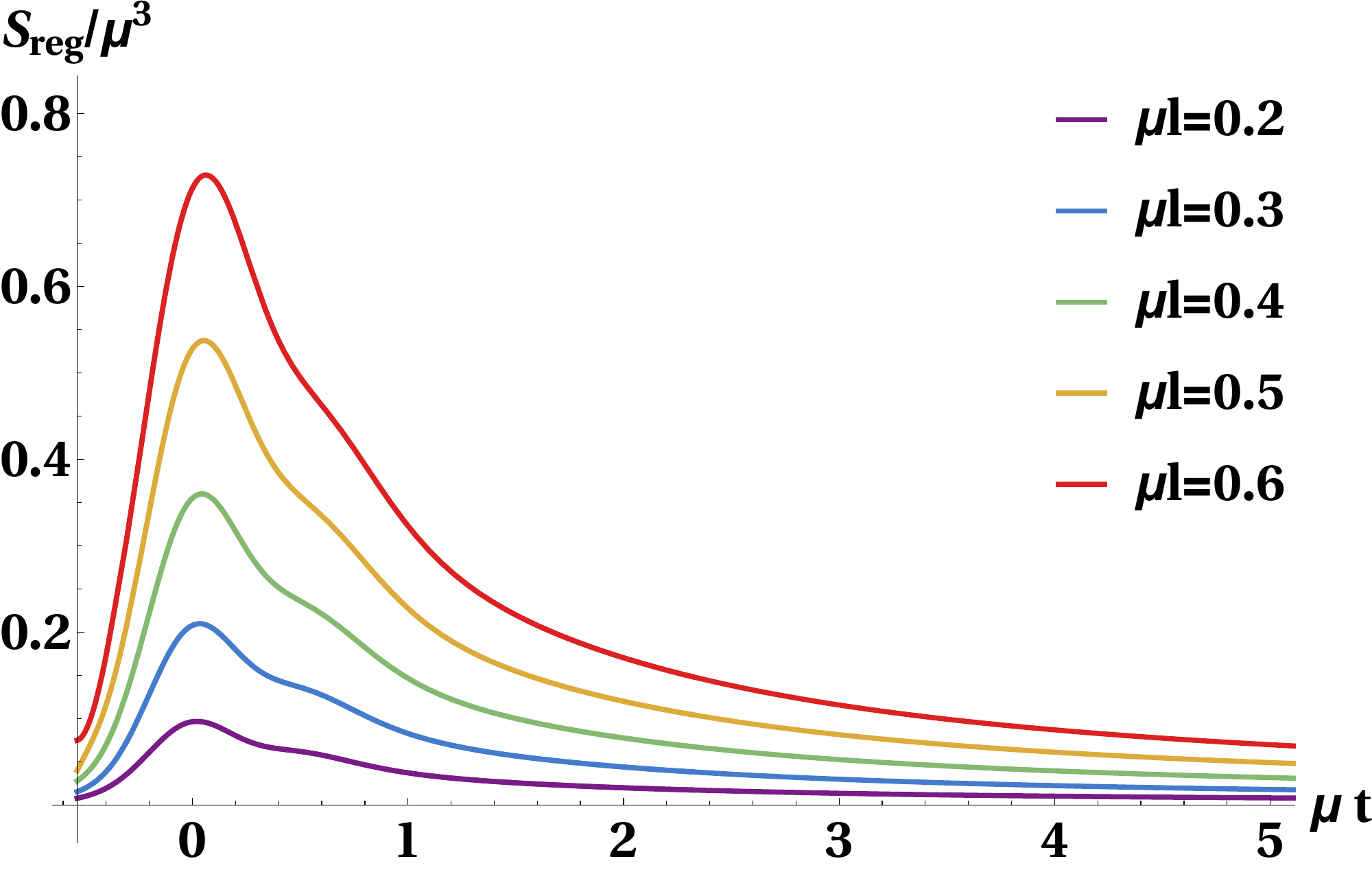}$\;\;$\includegraphics[scale=.23]{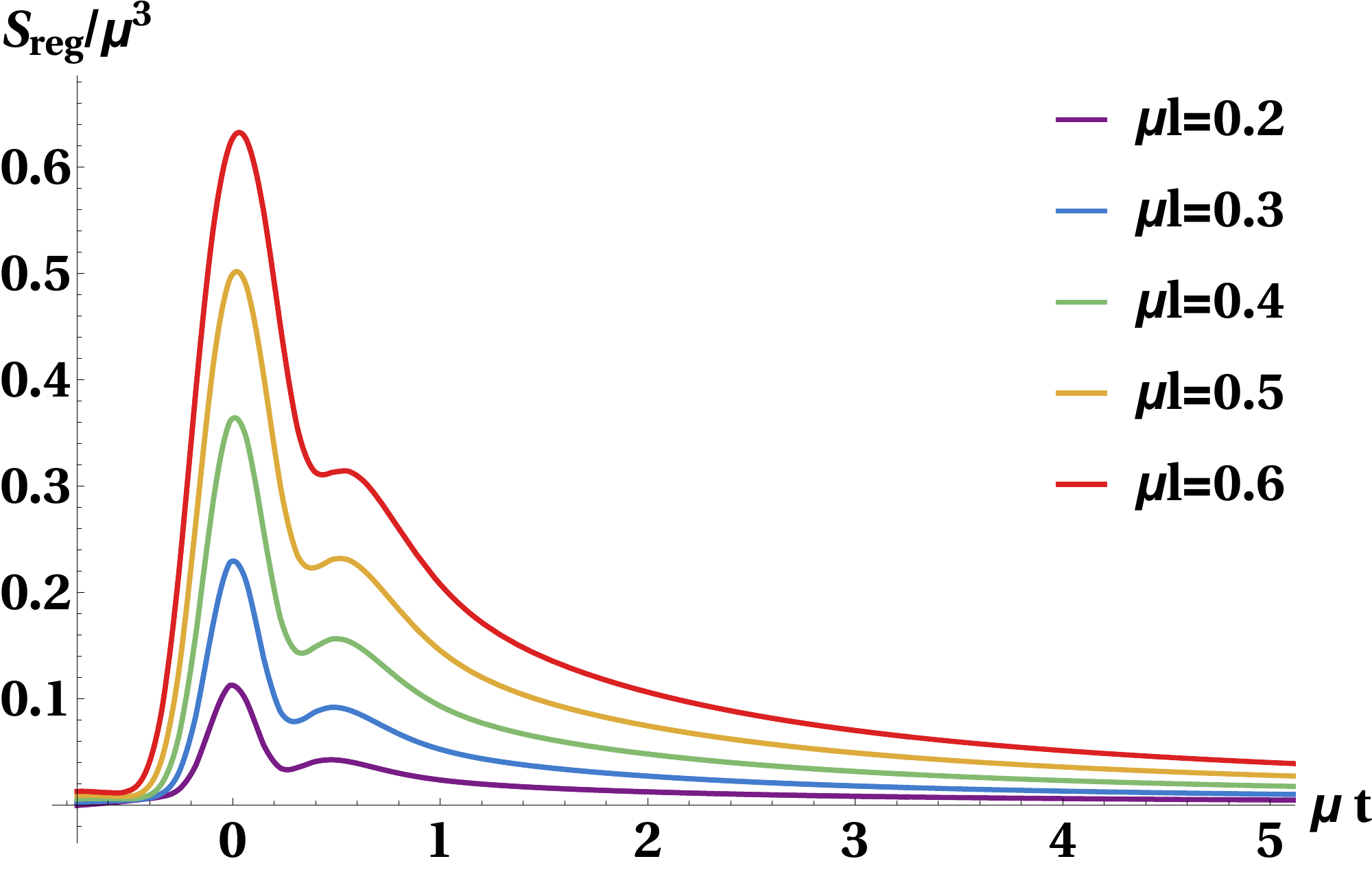}
\caption[Time evolution of entanglement entropy.]{\label{EEevolution} 
Evolution of the entanglement entropy for different separations of  width $\mu l$ of the stripe region for wide  (left), intermediate  (middle) and narrow  (right) shocks. 
 }
\end{center}
\end{figure}

In this section we present our numerical results for the entanglement entropy.
The shape of entanglement entropy as a function of time originates from a complicated interplay between the different metric functions appearing in the energy momentum tensor. 
However, most features can be understood in terms of  energy density and pressures. 
In Figure~\ref{EEevolution} we display the time evolution of entanglement entropy for various separations in the two different scenarios. 
It can be characterized by four distinct regions:
\begin{enumerate}
\item \textbf{rapid initial growth}: 
Once some energy density enters the entangling region the rapid initial growth starts.
The narrower the shocks the more rapidly the initial growth happens, because the rate at which the energy density enters the entangling region is bigger than for wider shocks. 
\item \textbf{linear growth}: The linear growth starts when the two shocks start to overlap and the energy piles up, with a steeper slope for larger separations. This is the same behavior as the post-local equilibration growth after a global quench \cite{Liu:2013iza}.
The maximum occurs with a short delay compared to the maximum energy deposited in the entangling region, with a more pronounced  delay for wider shocks.
\item \textbf{post collisional regime}: The post collisional regime is quite different for the three cases considered. 
For wide shocks the entanglement entropy falls off without any additional features. 
In the case of intermediate shocks a small shoulder appears. In the case of narrow shocks this shoulder turns into a new feature, where an additional minimum appears and the entanglement entropy starts growing again until a second maximum is reached. 
The minimum happens approximately at a time when the longitudinal pressure becomes negative. The existence or absence of a minimum of entanglement entropy in this regime thus serves as an order parameter to discriminate between narrow and wide shocks. 
\item \textbf{late time regime}: At late times we find a polynomial fall off behavior
\begin{equation}\label{lateEE}
S_{\textrm{\tiny reg}}\approx a_{w,i,n} (\mu t)^{-b_{w,i,n}}\;,
\end{equation}
where the coefficient  $a_{w,i,n}$ depends on the initial conditions and the separation.
In Table 5.1 we give the late time behavior extracted from the time interval $\mu t=[2,6]$ for different separations.
The late time behavior can be compared to the late time behavior of an effective entropy density 
\begin{equation}
s_{\textrm{\tiny eff}}(t)=\int\limits_{-l/2}^{l/2} \mathrm{d}y\, S^3(r_h,t,y)\;,
\end{equation}
where the function $S$ is evaluated at the position of the apparent horizon and integrated over the same intervals as for the entanglement entropy. 
The late time behavior is displayed in Table 5.2 and barely depends on the separation. 
It is expected on general grounds that at very late times and large separations, far beyond our computational domain,  the effective entropy density and entanglement entropy show the same fall off behavior. 
\end{enumerate}

\begin{figure}
\begin{minipage}[t]{0.45\textwidth}
\vspace{-0.8cm}
\hspace{-0.0cm}
\includegraphics[width=0.9\textwidth]{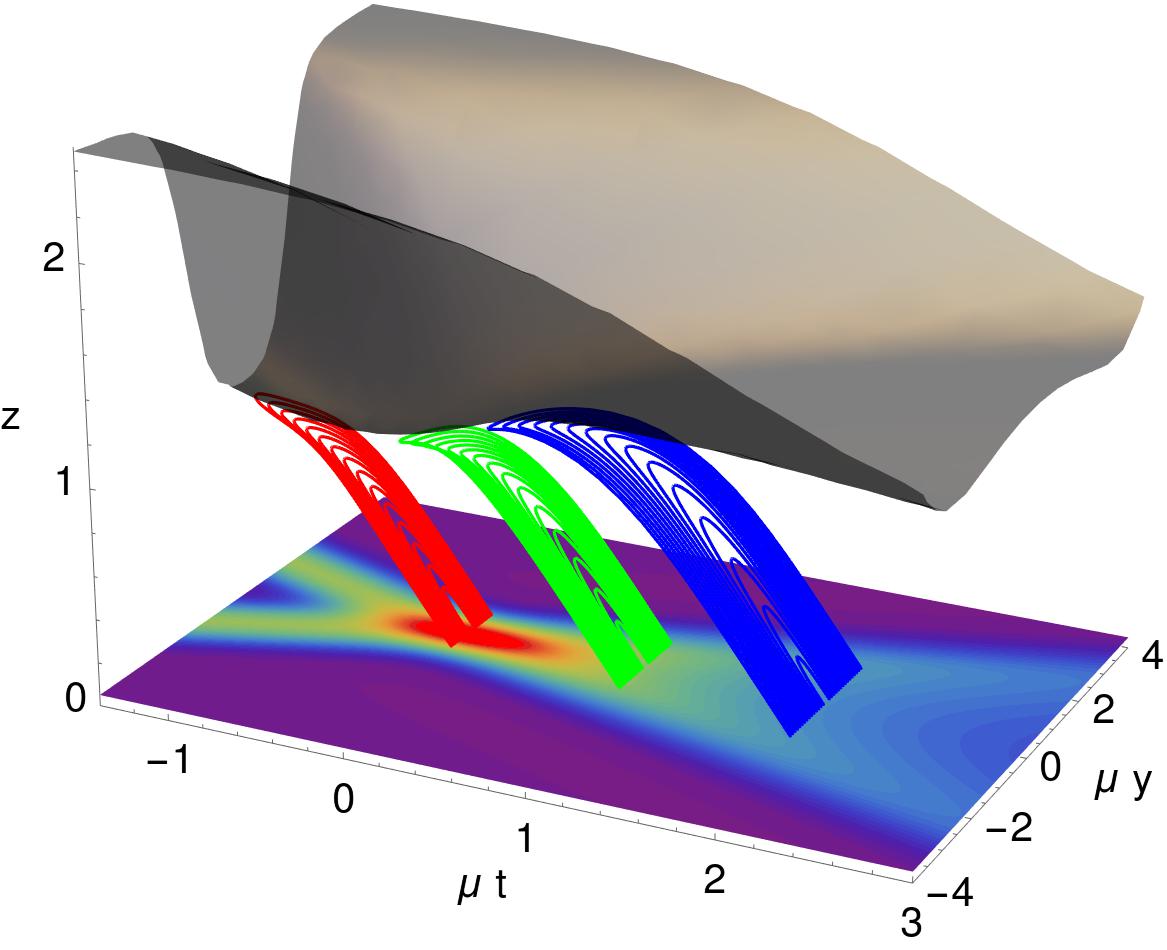}
\end{minipage}
\begin{minipage}[t]{0.45\textwidth}
\vspace{0.1cm}
\includegraphics[width=0.85\textwidth]{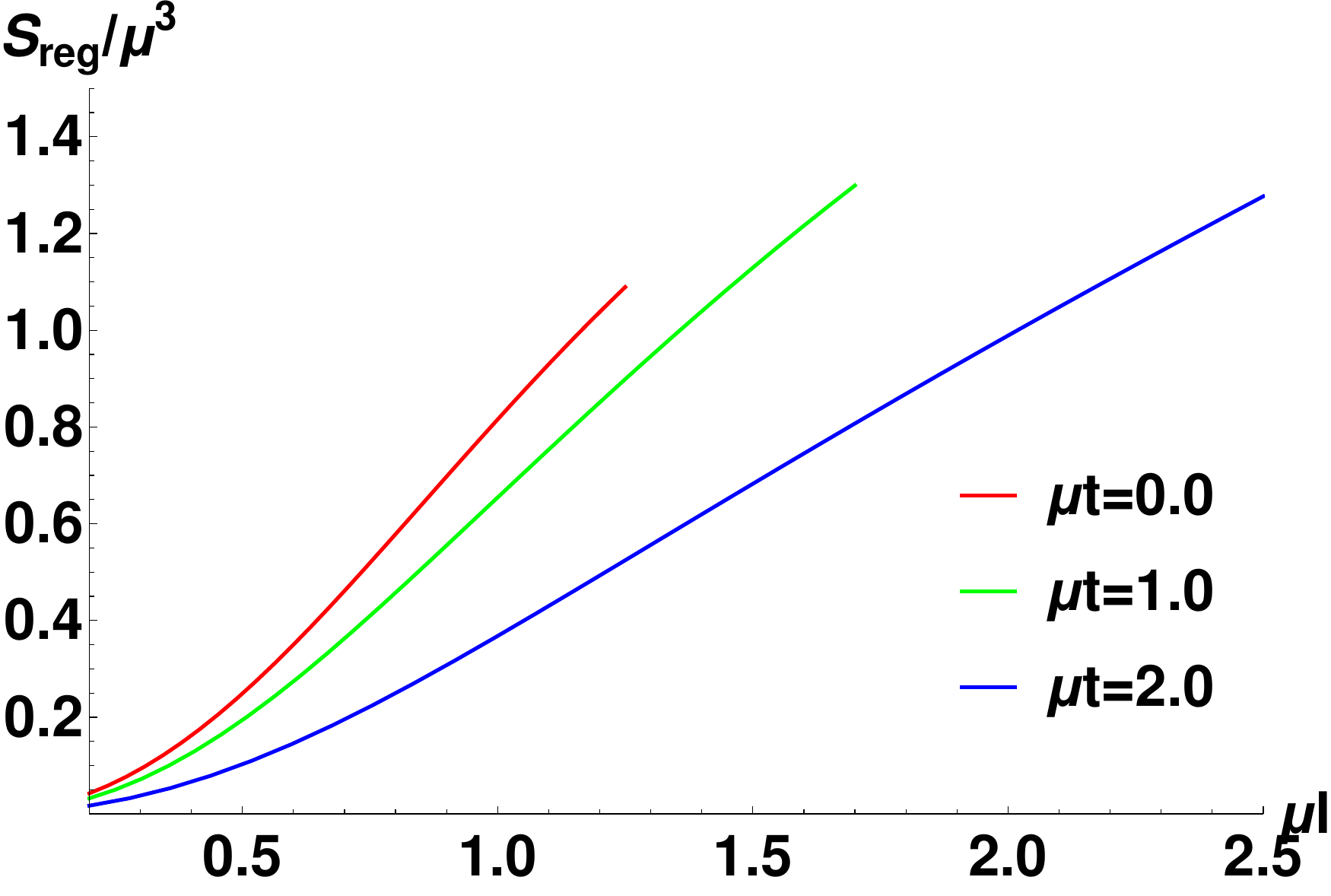}
\end{minipage}
\begin{minipage}[t]{0.45\textwidth}
\hspace{-0.0cm}
\includegraphics[width=0.9\textwidth]{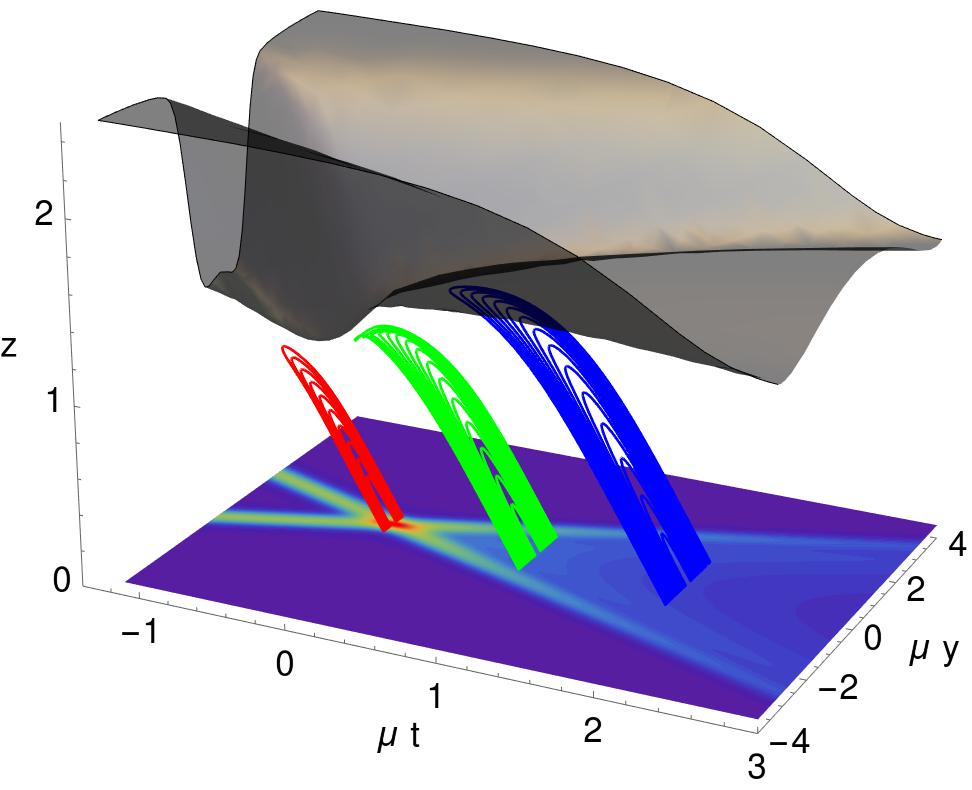}
\end{minipage}
\begin{minipage}[t]{0.45\textwidth}
\includegraphics[width=0.85\textwidth]{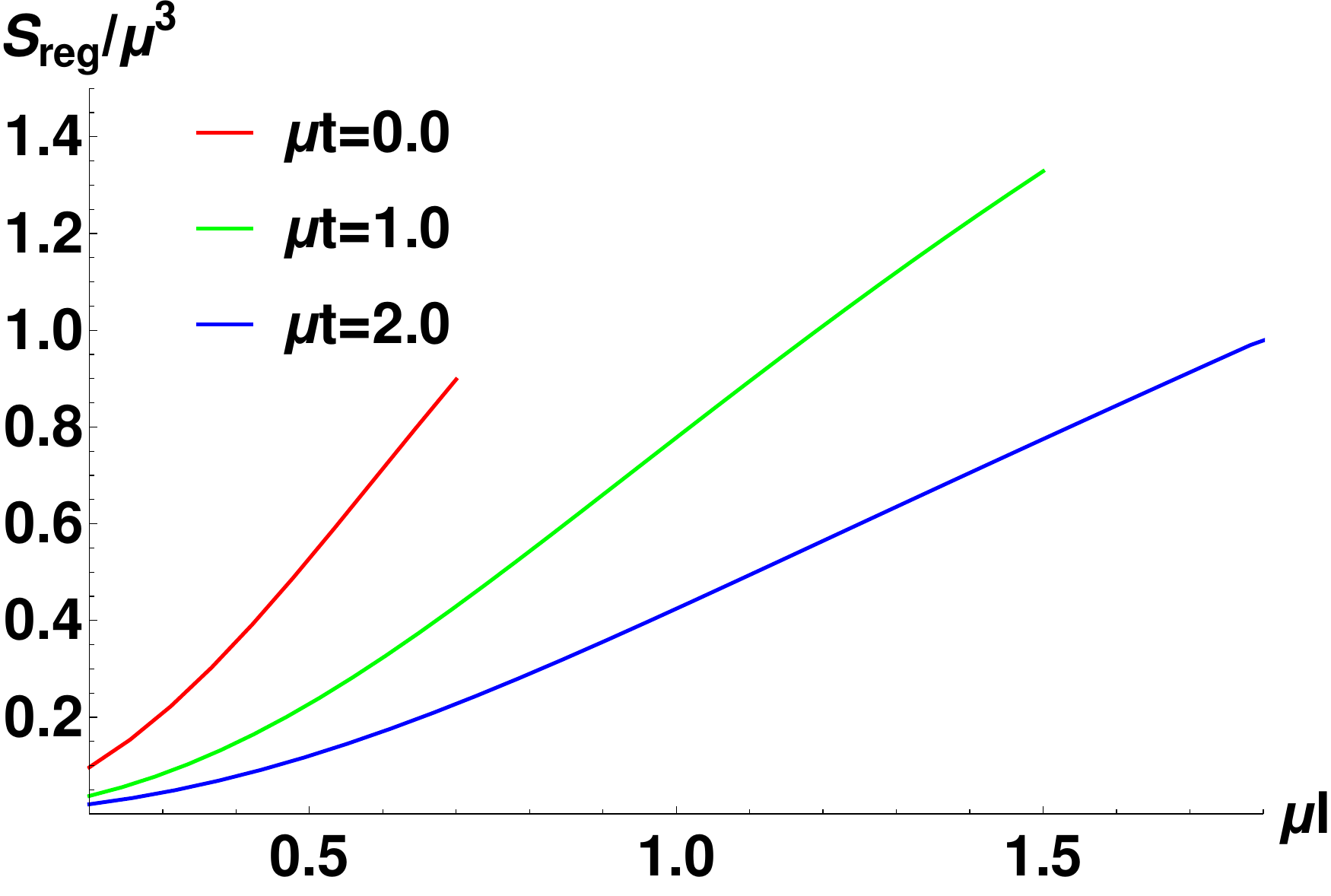}
\end{minipage}
\begin{minipage}[t]{0.45\textwidth}
\hspace{-0.0cm}
\includegraphics[width=0.9\textwidth]{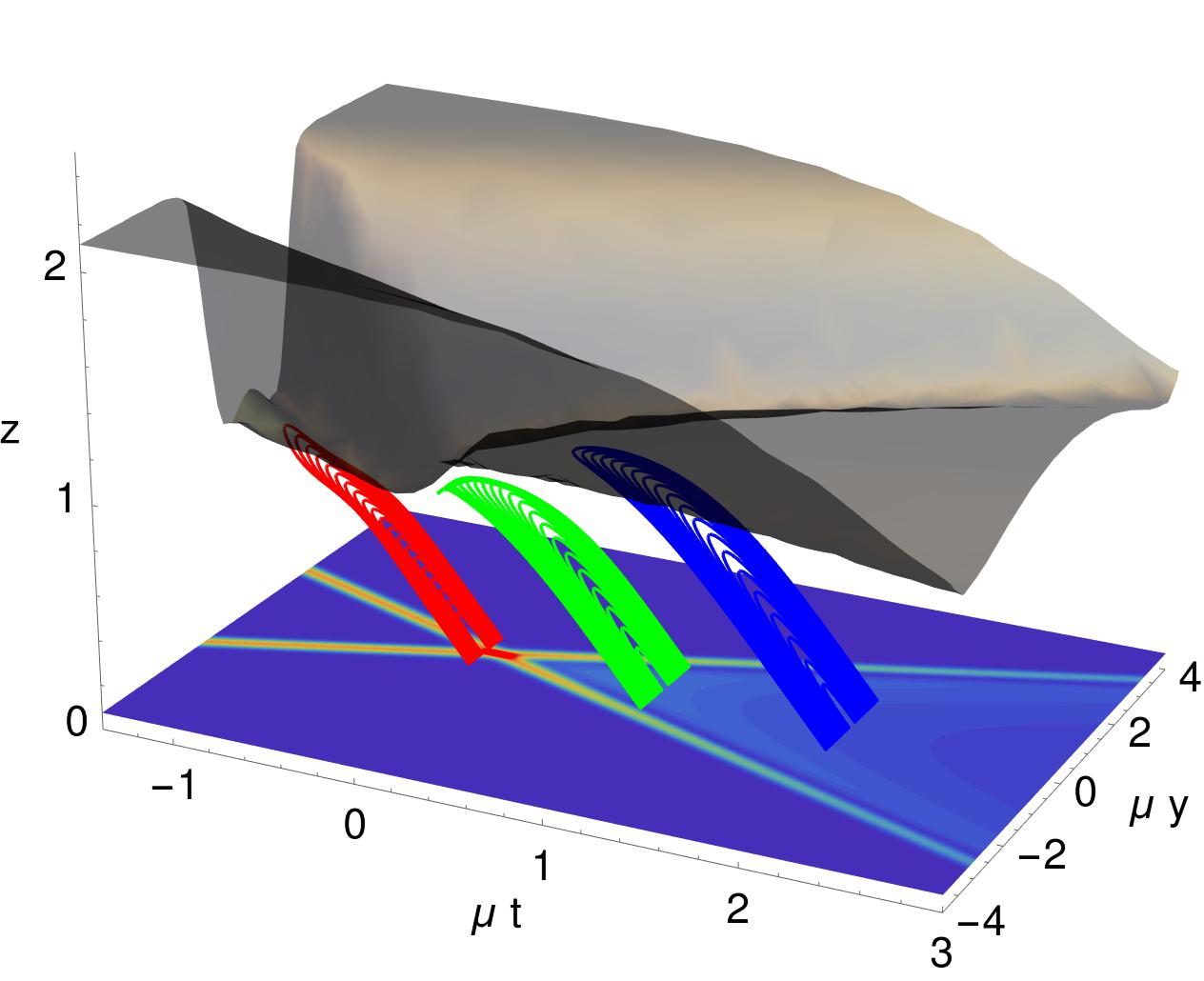}
\end{minipage}
\begin{minipage}[t]{0.45\textwidth}
\hspace{1.3cm}
\includegraphics[width=0.85\textwidth]{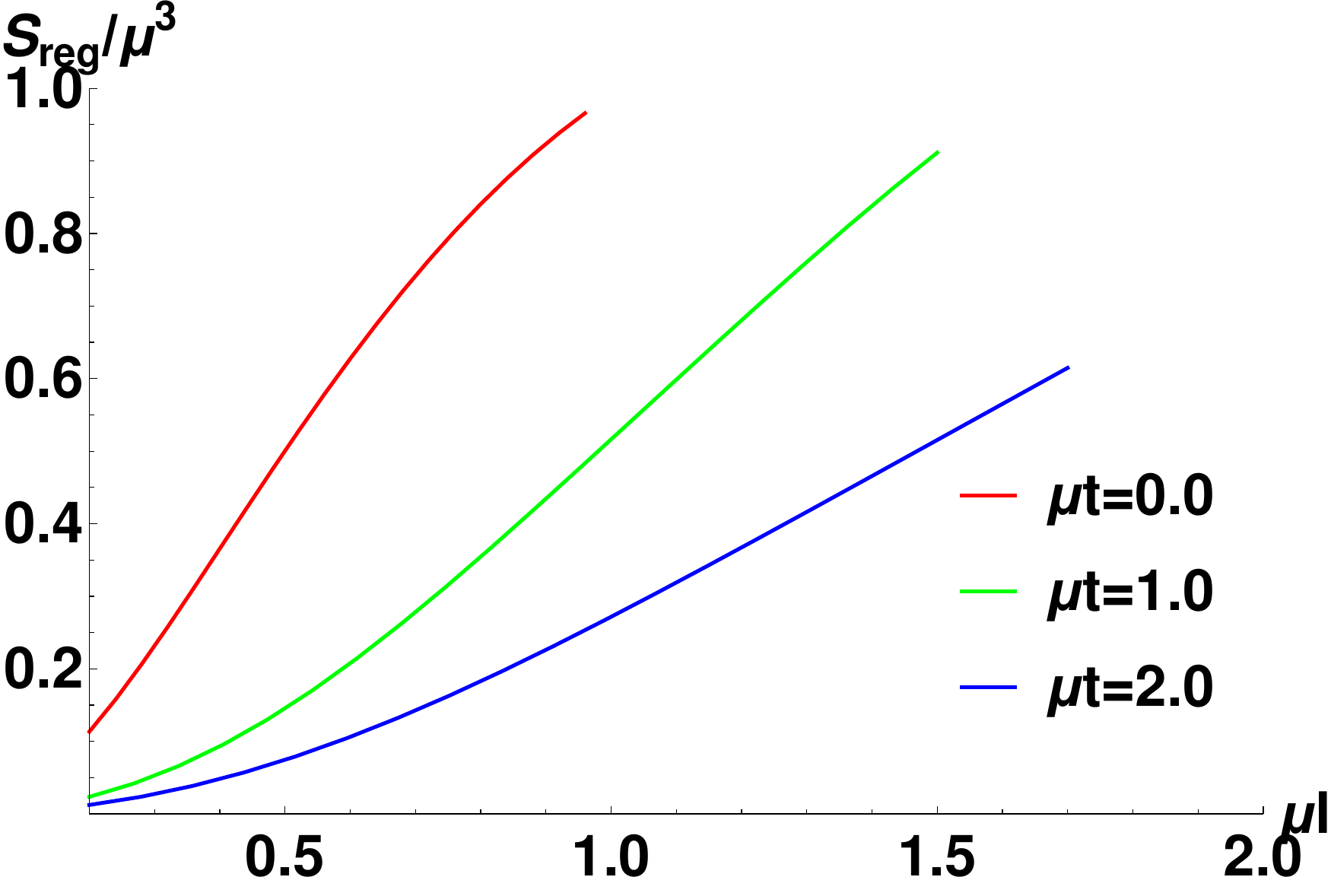}
\end{minipage}
\caption[Geometrical setup and entanglement entropy (boundary separations).]{Left: Summary of the geometrical setup.
The black surfaces represent the radial position $z_{\textrm{\tiny AH}}(t,y)$ of the apparent horizon; red, green and blue curves are geodesics of various separations at $\mu t=0$, $\mu t=1$ and $\mu t=2$ respectively and at $z=0$ we show a contour plot of the energy density for wide, intermediate and narrow shocks (top to bottom).
Right:  Corresponding evolution of the entanglement entropy with the boundary separation $\mu l$ at different times.}\label{EELevo}
\end{figure}

\begin{figure}
\hspace{-0.cm}
\includegraphics[width=4.7cm]{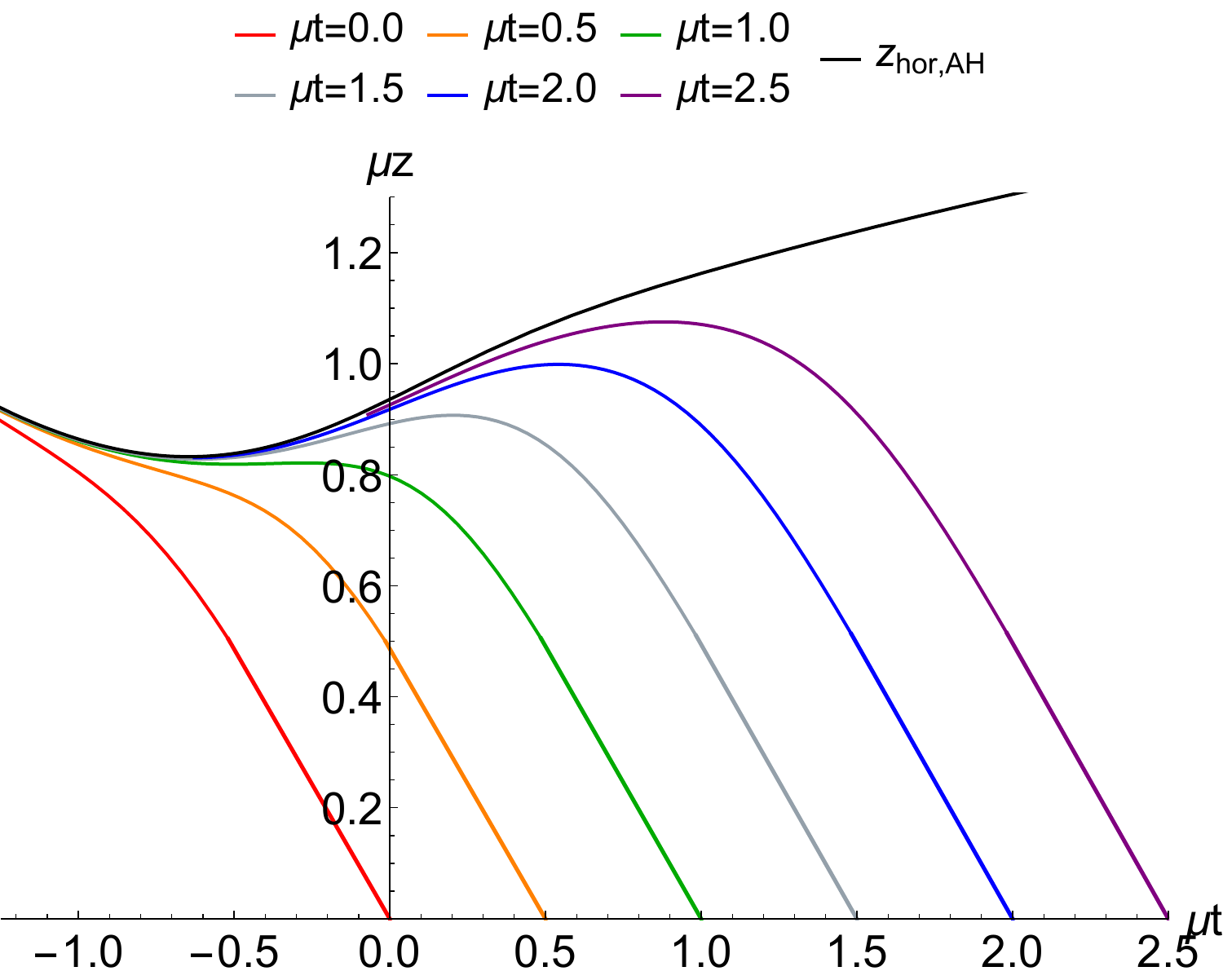}$\;\;$\includegraphics[width=4.7cm]{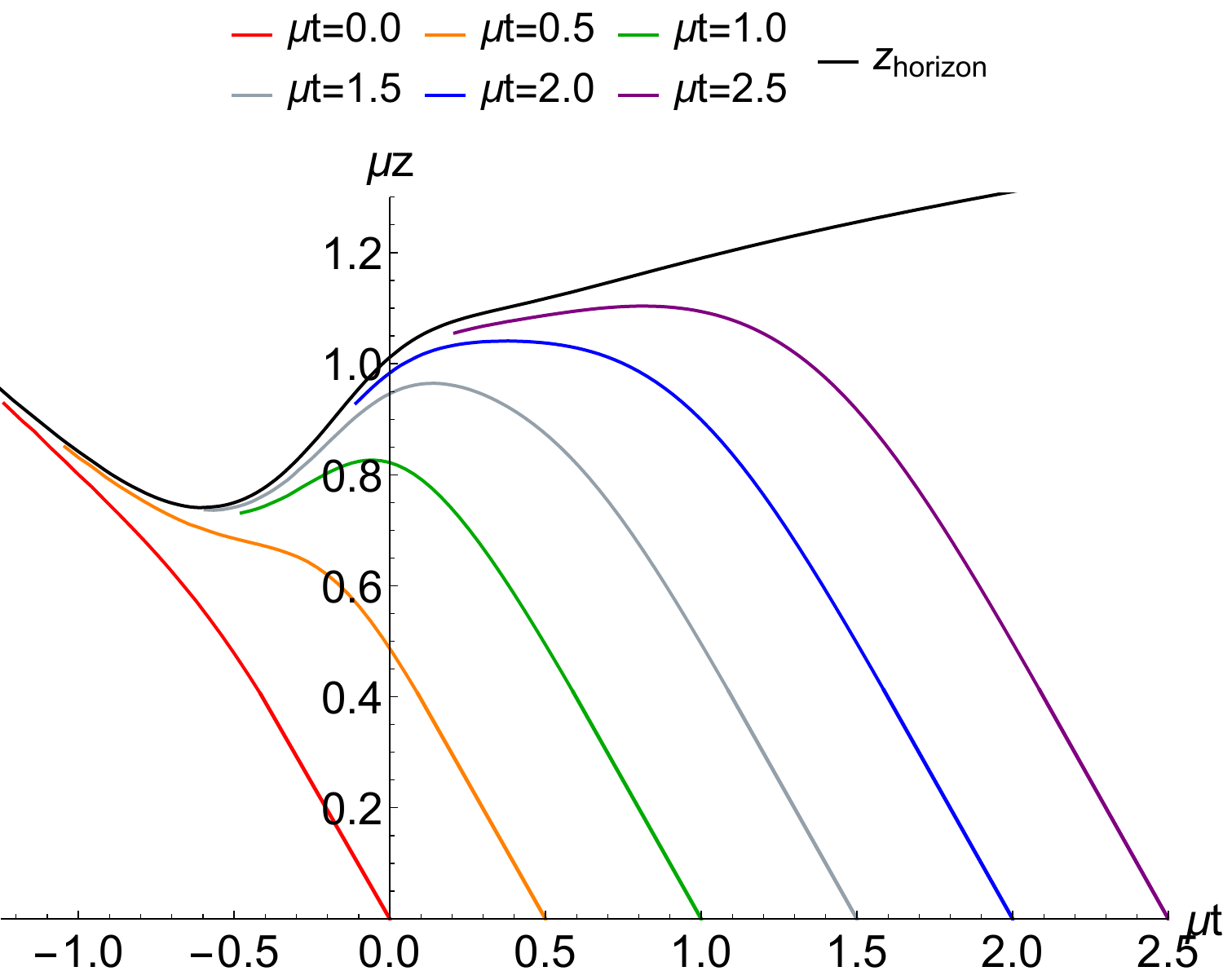}$\;\;$\includegraphics[width=4.7cm]{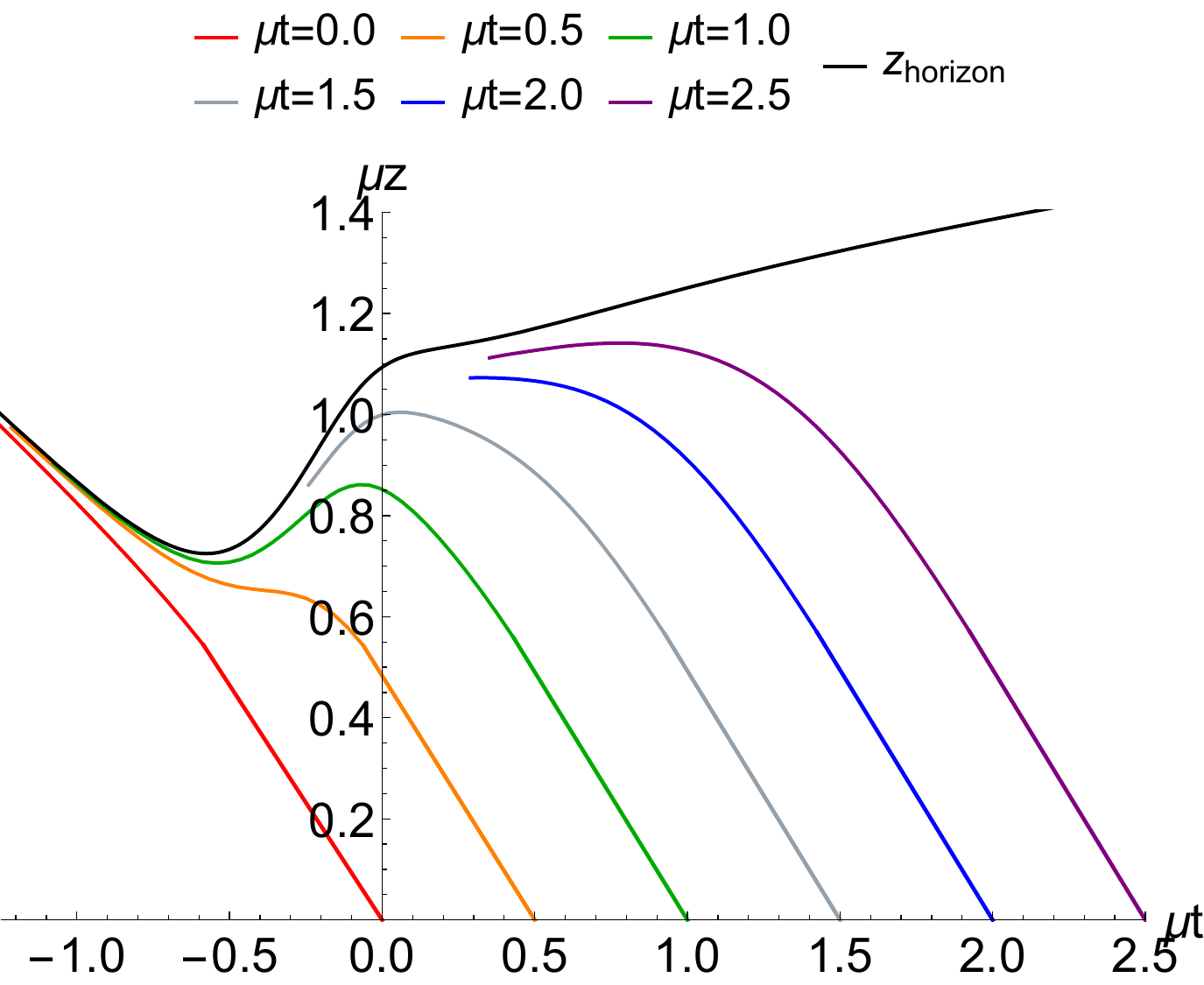}
\caption[Conformal geodesics used for holographic entanglement entropy.]{\label{GeodesicEEDip} 
The $z$-position of the geodesics at $y=0$ for several times and separations, starting with $l=0$ near the boundary, and increasing going towards the end of the curve. We show wide, intermediate and narrow shocks (from left to right). The $z$-position of the apparent horizon at $y=0$ is shown in black. In all the cases we studied the geodesics do not cross the horizon.}
\end{figure}

\begin{table}[t]\label{Tab:fit}
 \caption[Late time fit of entanglement entropy.]{Late time fit of the entanglement entropy in the time range $\mu t\in[2.0,6.0]$.}
 \centering
 \begin{tabular}{lcccccc}  \hline\hline
  $\mu l$ & $a_w$ & $a_i$ & $a_n$ & $b_w$ & $b_i$ & $b_n$ \\
  \hline
  0.5     & 0.202 & 0.171 & 0.158 & -1.136 & -0.978 & -1.074 \\
  1.0     & 0.709 & 0.602 & 0.564 & -1.092 & -0.961 & -1.035 \\
  1.5     & 1.276 & 1.099 & 1.031 & -1.036 & -0.952 & -0.982 \\
 \hline
\end{tabular}
\end{table}

\begin{table}[t]\label{Tab:fit1}
 \caption[Late time fit of the effective entropy density.]{Late time fit of the effective entropy density in the time range $\mu t\in[2.0,6.0]$.}
 \centering
 \begin{tabular}{lcccccc} 
 \hline\hline
  $\mu l$ & $a_w$ & $a_i$ & $a_n$ & $b_w$ & $b_i$ & $b_n$ \\
  \hline
  0.5     & 1.042 & 0.696 & 0.665 & -1.107 & -0.749 & -0.952 \\
  1.0     & 2.035 & 1.430 & 1.372 & -1.088 & -0.766 & -0.971 \\
  1.5     & 2.924 & 2.244 & 2.241 & -1.054 & -0.795 & -1.027 \\
 \hline
\end{tabular}
\end{table}

Let us now discuss the results from the evolution in the separation. The geometrical setup and the evolution in the separation at different times are shown in Figure~\ref{EELevo}.
Analogous to Figure \ref{GeodesicDip} we show in Figure \ref{GeodesicEEDip} again the position of the tip of the extremal surface, this time for the entanglement entropy.
Surprisingly, contrary to case of the two-point function we never see the tip crossing the horizon, and in fact it always closely follows the horizon for larger separations. This is again perhaps counter-intuitive, since one would usually think about the entanglement entropy as a more `nonlocal' quantity than the two-point functions, and hence probing deeper in the bulk. Indeed, this is the case for pure AdS and also for thermal AdS, but in this case for large enough separations the two-point function at the same time and length probes deeper in the bulk than the entanglement entropy.

Of course our simulations only probed a limited set of times and lengths for our extremal surfaces and hence we cannot make a general statement if the entanglement entropy never probes beyond the apparent horizon in geometries produced by shock wave collisions. Nevertheless, we think we have strong evidence that this is so, mainly since increasing the lengths at our chosen times clearly moves the tip of the surface along the horizon. We furthermore checked that extremal surfaces centered around $y \neq 0$ behave similarly, so that the property is not due to our symmetric set-up.

\chapter{QNEC in Shock Wave Collisions}\label{Chap:QNEC}
In this chapter we  present a series of explicit computations of QNEC in a strongly coupled QFT, including vacuum, thermal equilibrium, a homogeneous far-from-equilibrium quench as well as a colliding system that violates NEC. For vacuum and the thermal phase QNEC is always weaker than NEC. While for the homogeneous quench QNEC is satisfied with a finite gap, 
we find the interesting result that the colliding system can saturate QNEC, depending on the null direction.
Results displayed in this chapter are published in \cite{Ecker:2017jdw}.

\section{Computing QNEC} 
We determine QNEC holographically by studying the gravitational dual, where entanglement entropy of a region in the CFT can be computed using the RT-formula \cite{Ryu:2006bv,Hubeny:2007xt,Dong:2016hjy}
\begin{equation}\label{eq:RT}
S_\text{EE}=\frac{\mathcal{A}}{4G_N}=\frac{N_c^2}{2\pi} \mathcal{A}\equiv N_c^2\, \mathcal{S}_\text{EE}
\end{equation}
Here $\mathcal{A}$ is the area of an extremal co-dimension 2 surface in the bulk which is homologous to the entangling region in the boundary and $G_N$ is Newton's constant. 
The prescription was proven in the static case \cite{Lewkowycz:2013nqa} and has survived many tests in dynamical situations \cite{Hubeny:2007xt,Callan:2012ip,Wall:2012uf,Dong:2016hjy}.

All our examples use five-dimensional metrics of the form
\begin{equation}
d s^2=2d t\,(F d y-d z/z^2)-Ad t^2+R^2\big(e^Bd\mathbf{x}_{\perp}^{2}+e^{-2B}d y^2\big)
\label{eq:lalapetz}
\end{equation}
where $A$, $B$, $F$ and $R$ can depend on boundary coordinates $t$, $y$ and the AdS radial coordinate $z$. Near the AdS$_5$ boundary at $z=0$ these functions can be expanded as
\begin{subequations}
\begin{align}
    A &= z^{-2} + a_4(t,\,y)z^2 + {\cal O}(z^3)\,, \\
    B &= b_4(t,\,y)z^4 + {\cal O}(z^5)\,, \\
    F &= f_4(t,\,y)z^2 + {\cal O}(z^3)\,, \\
    R &= z^{-1} + {\cal O}(z^4) \,.
\end{align}
\end{subequations}
They have normalizable modes $a_4$, $b_4$ and $f_4$, from which the projection of the stress tensor can be determined \cite{deHaro:2000vlm} as
\begin{equation}
\label{eq:SEtensor}
\frac{1}{N_c^2} \langle T_{\mu\nu}k^\mu_\pm k^\nu_\pm\rangle \equiv \mathcal{T}_{\pm\pm} = \frac{1}{2\pi^2}(-a_4-2 b_4 \pm 2 f_4),
\end{equation}
with null vectors $k_\pm^\mu = \delta_t^\mu\pm \delta_y^\mu$ at the boundary $z=0$.

In this chapter all our entangling regions are infinite strips along the perpendicular directions $x_\perp$ and hence are specified fully by their endpoints $S_\text{EE}(t_L, y_L; t_R, y_R)$ with a corresponding separation $L=y_R-y_L$. For these regions the extremal surface equation reduces to a geodesic equation in an auxiliary spacetime, which simplifies the computation considerably (see \cite{Ecker:2015kna,Ecker:2016thn} for a detailed description of the numerical procedure to find the relevant geodesics~\footnote{In this work we mostly work in a gauge where the horizon is at $z=1$ and the geodesics are parametrized by $t$ and the angle in the $y$-$z$-plane. To solve the equations we used a 5$^{\textrm{th}}$ order finite difference scheme with order 100 grid points. We verified our results with other gauges and numerical settings.}). The lengths of the geodesics then give the entropy density per transverse area. An important subtlety in computing \eqref{eq:RT} is its UV divergence. We regulate it by putting a cutoff at $z_\text{cut}=0.01$ and verifying that none of the physics presented in this chapter depends on the cutoff~\footnote{There is a subtlety in taking the functional derivative in \eqref{Eq:QNEC}, which can be UV-divergent if not taken in the right direction \cite{Bousso:2015mna}. 
We only deform the entangling region in the longitudinal direction, which then avoids this divergence.}.

After computing entanglement entropy it is straightforward to evaluate QNEC 
\begin{equation}\label{eq:QNEC2}
\langle T_{kk}\rangle\ge \frac{1}{2\pi \sqrt{h}}\,S''\,,\qquad\forall\quad k^\mu k_\mu =0\,,
\end{equation}
 at some point $(t,\,y)$ for the null vectors $k_\pm^\mu$. This is done by computing $\partial_\lambda^2S_\text{EE}(t+\lambda,y\pm\lambda;\,t,y+L)$ at $\lambda=0$, which yields $S''/\sqrt{h}$ in \eqref{eq:QNEC2}~\footnote{In this chapter all set-ups are invariant under $y\to-y$ so without loss of generality we only vary the left point of the entangling region.
}.

\begin{figure*}[ht!]
\center
\includegraphics[width=1.0\textwidth]{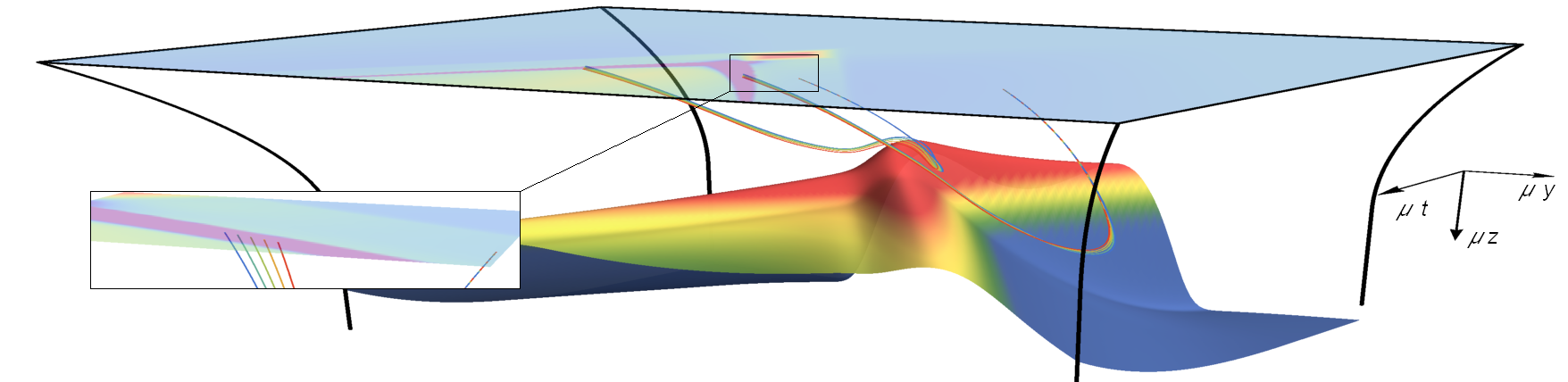}\\
\vspace{0.5cm}
\includegraphics[width=0.6\linewidth]{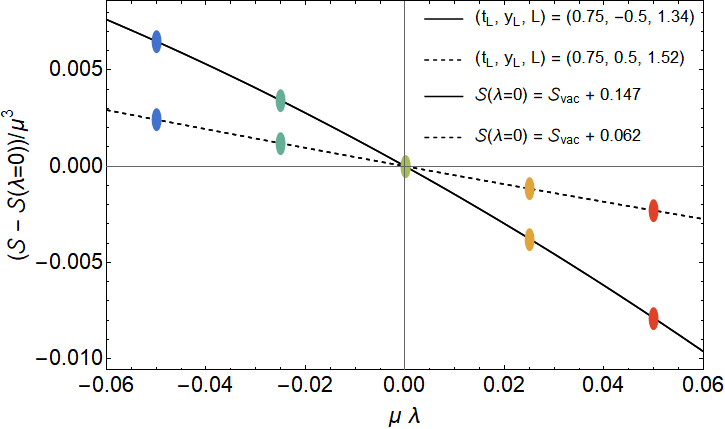}
\caption{
Top: Two families of extremal surfaces at representative locations $(t_L,\,y_L)=(0.75,\,\pm 0.5)$ for the shock wave geometry. The families correspond to a null variation at a point where the classical NEC is violated (purple region at $z=0$ or black region in Figure~\ref{fig:ShockWaves}). The family starting at $y=-0.5$ hovers just above the apparent horizon (colored surface) and hence has larger entropy, as well as more negative $\mathcal{S}''_\pm$ (bottom Figure, see also Figure~\ref{fig:QNECshocks2}).
\label{fig:QNECshocks}}
\end{figure*}

It is instructive to examine QNEC from a near-boundary perspective, where it is possible to prove QNEC \cite{Koeller:2015qmn}. Close to the boundary point $(t_L,\,y_L)$ an extremal surface is given by
$t(z)=t_L+\lambda-z+t_4(\lambda)z^4+a_4z^5/5+O(z^6)$,
$y(z)=y_L\pm\lambda-z+y_4(\lambda)z^4+f_4z^5/5+O(z^6)$,
where $t_4$ and $y_4$ also depend on $(t_R,\,y_R)$ and are undetermined in a near-boundary analysis. Extremal surfaces are stationary under perturbations, so variations of extremal surfaces only yield boundary terms. A simple geometric argument then gives
$\partial_\lambda\mathcal{A}=-4t_4(\lambda)\pm4y_4(\lambda)$,
which leads to the second variation 
\begin{equation}\label{eq:displayed}
S''_\pm = (\pm 4 \partial_\lambda y_4 - 4 \partial_\lambda t_4) / (4 G_N)\,.
\end{equation}
Comparing the results \eqref{eq:SEtensor} and \eqref{eq:displayed} with each other shows that inclusive QNEC does not hold or saturate automatically, but may do so for suitable functions $a_4$, $b_4$, $f_4$, $y_4$ and $t_4$.

Since we perturb in a null direction the leading contribution to the distance between the two extremal surfaces separated by $\lambda$ vanishes. We have two subleading contributions, coming from the subleading terms in the extremal surface and metric expansions respectively:
\begin{align}
\Delta s^2 &= |x^\mu(t_L, y_L, z) - x^\mu(t_L + \lambda, y_L + \lambda, z)|  \nonumber \\
&=z^2 \lambda^2 (-2 b_4 \pm 2 f_4 - a_4 \mp 2 \partial_\lambda y_4 + 2 \partial_\lambda t_4) 
\label{eq:ds2proof}
\end{align}
Assuming the classical NEC in the bulk spacetime and using that the deformation along $\lambda$ is null, it can be shown \cite{Wall:2012uf} that the distance between the surfaces has to be spacelike, i.e., $\Delta s^2 \geq 0$, also called `entanglement nesting property'. This condition reduces precisely to QNEC in \eqref{eq:QNEC2}, see \cite{Koeller:2015qmn}.
Equation \eqref{eq:ds2proof} is useful for us, not only to illustrate why in holography we expect QNEC to be valid, but also to independently verify QNEC from a bulk perspective.
This is done by explicitly computing the distance between two nearby extremal surfaces and comparing this with QNEC determined as described next.

To evaluate QNEC in practise we evaluate the second derivative by computing $\mathcal{S}_{\textrm{\tiny{EE}}}$ for five equidistant values of $\lambda$ between $-0.05$ and $0.05$. We then obtain four estimates of $\mathcal{S}''_\pm$ by generating a quadratic fit through all five points, the first three points, the middle three points and the last three points, thereby both obtaining a mean estimate as well as a numerical error. 

Figure~\ref{fig:QNECshocks} shows an example of a family of surfaces for $k_+^\mu$ at $t_L=t_R=0.75$, $y=0.5$ and $L=1.0$, including the apparent horizon of the shock wave collisions and the (violation of) NEC in the boundary theory. On the right we display entanglement entropy of the five surfaces, having their vacuum contribution subtracted.

To obtain the full QNEC result it is necessary to add the vacuum contribution again. This is straightforward, since for a strip the vacuum entanglement entropy per transverse area is known analytically \cite{Ryu:2006bv},
\begin{equation}\label{eq:Svac}
\mathcal{S}_\text{EE} =\frac{1}{2\pi} \bigg( \frac{1}{z_\text{cut}^2} - \frac{1}{2c_0^3l^2} \bigg)\,,\qquad c_0=\frac{3\Gamma[1/3]^3}{2^{1/3}(2\pi)^2}\,,
\end{equation}
where $l=\sqrt{(L\pm\lambda)^2-\lambda^2}$ is the proper length of the (boosted) strip. Taking the second derivative with respect to $\lambda$ at $\lambda = 0$ gives
\begin{equation}\label{eq:Sppvac}
\frac{1}{2\pi}\,\mathcal{S}''_\pm = -\frac{1}{\pi^2c_0^3L^4}\approx -\frac{0.06498}{L^4} \,.
\end{equation}
From \eqref{eq:Sppvac} it is clear that the CFT vacuum satisfies QNEC in a trivial way, especially for small $L$, while it saturates QNEC in the limit $L\to\infty$.

\section{Results}

\subsection{Thermal Plasma} 

We first consider a homogeneous thermal equilibrium state with dual description in terms of the AdS$_5$ Schwarzschild black brane that has $A=1/z^2-({\pi}T)^4z^2$, $R=1/z$ and $B=F=0$, where the energy density is related to the temperature by $T^0_0=3N_c^2\pi^2T^4/8$. The null projections of the energy momentum tensor, $\mathcal{T}_{\pm\pm}$, are the same for both lightlike directions due to parity symmetry,
\begin{equation}
\frac{1}{N_c^2} \langle T_{\mu\nu}k^\mu_\pm k^\nu_\pm\rangle \equiv \mathcal{T}_{\pm\pm} = \frac{\pi^2}{2}\,T^4\approx 0.0507 \,\pi^4 T^4\,.
\end{equation}
In this case $\mathcal{S}''_+\!=\!\mathcal{S}''_-$, which can be understood by realizing that the plasma is time-reversal invariant. That means we can invert the $k_t$ component and invariance of the second derivative under $k_\mu\to-k_\mu$ yields the identity.

\begin{figure}[ht]
\center
\includegraphics[width=0.6\linewidth]{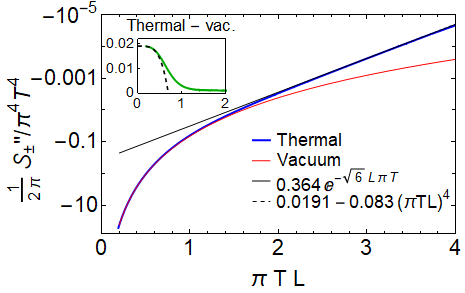}
\caption{
$\mathcal{S}''_\pm$ for the thermal state as a function of strip length (blue). For small $L$ the curve follows the vacuum result (\eqref{eq:Sppvac}, red) whereas for large length $\mathcal{S}''_\pm$ approaches zero exponentially (black). 
Since $\mathcal{S}''_\pm<0$ and $\mathcal{T}_{\pm\pm}>0$ QNEC is obviously satisfied.
}
\label{fig:QNECblackbrane}
\end{figure}

In Figure~\ref{fig:QNECblackbrane} we show that at small length $\mathcal{S}''_\pm$ approaches the vacuum result, while for large $L$ it approaches zero from below exponentially fast. Since $\mathcal{T}_{\pm\pm}$ is positive we see that QNEC is easily satisfied for all lengths and  never saturates. Analytic calculations in the appendix confirm our numerical results at small and large $L$.

\subsection{Far-From-Equilibrium Quench}

Now we consider a quenched far-from-equilibrium system where a homogeneous shell of null dust is injected in the gravitational dual \cite{AbajoArrastia:2010yt}, leading to the AdS$_5$ Vaidya spacetime
\begin{align}
A=z^{-2} - M(t) z^2\,,\qquad M(t)\equiv\tfrac{1}{2}\left(1+\tanh(2 t)\right)\,.
\label{eq:MVaidya}
\end{align}
Equation~\eqref{eq:MVaidya} realizes a homogeneous quench of the vacuum at $t\!=\!-\infty$ to a thermal state with $T\!=\!\frac{1}{\pi}$ at $t\!=\!\infty$. The corresponding projection of the energy momentum tensor is time dependent, with $\mathcal{T}_{\pm\pm}=\frac{1}{2\pi^2}M(t)$. The Vaidya geometry is not invariant under time inversion, so $\mathcal{S}''_\pm$ are distinct from each other.

\begin{figure}[ht]
\center
\includegraphics[width=0.48\linewidth]{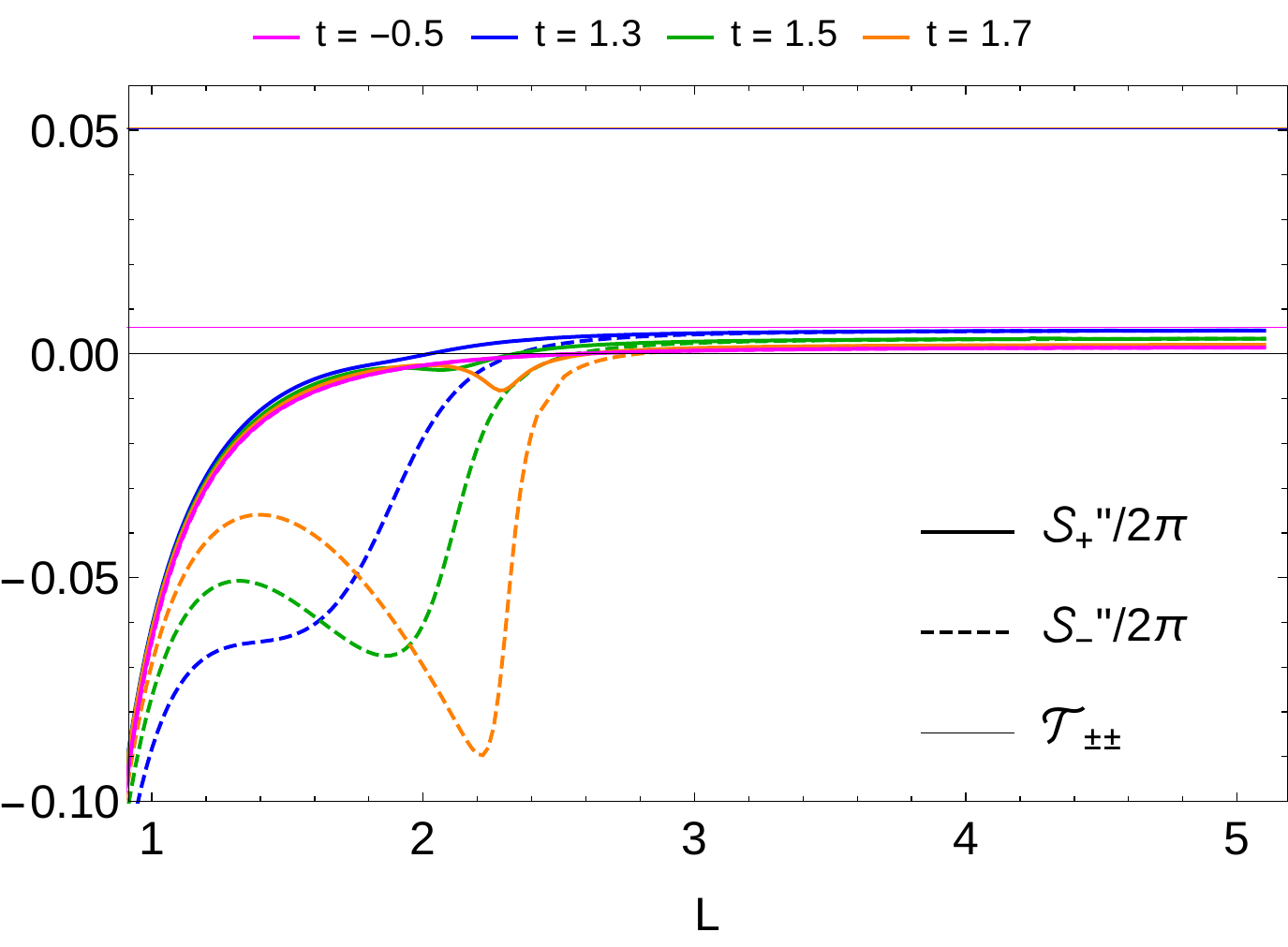}$\;\;$
\includegraphics[width=0.48\linewidth]{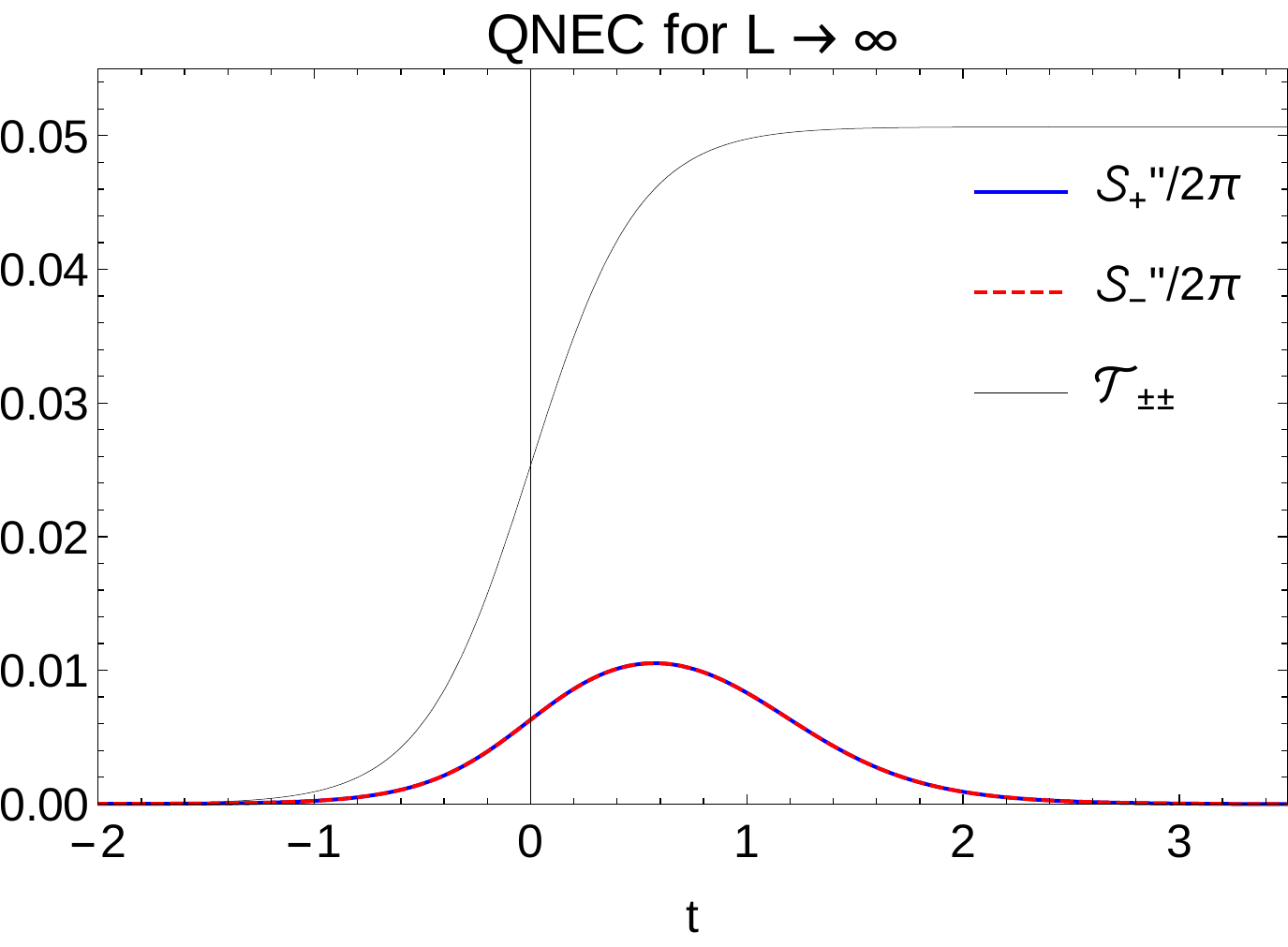}
\caption{ 
Left: $\mathcal{S}''_\pm$ at four different times as a function of separation $L$ together with the corresponding $\mathcal{T}_{\pm\pm}$ (for the quenched geometry \eqref{eq:lalapetz} with \eqref{eq:MVaidya} and $B\!=\!F\!=\!0$).
Right: Time evolution of $\mathcal{T}_{\pm\pm}$ and the long length limit of $\mathcal{S}''_\pm/(2\pi)$. Growth and settling down of $\mathcal{S}''_\pm/(2\pi)$ happens later than for $\mathcal{T}_{\pm\pm}$.
}
\label{fig:Vaidya}
\end{figure}

In Figure~\ref{fig:Vaidya} (left) we show $\mathcal{S}''_\pm$ versus the length of the strip at four different times.
For small lengths these curves again approach the vacuum result, but at intermediate lengths there is a clear difference between $\mathcal{S}''_+$ and $\mathcal{S}''_-$, whereby in particular $\mathcal{S}''_-$ can develop a pronounced local minimum.
For large lengths we find that $\mathcal{S}''_+$ and $\mathcal{S}''_-$ asymptote to equal values.
In Figure~\ref{fig:Vaidya} (right) we plot  these asymptotic values  as a function of time, where we see that QNEC is always satisfied, and $\mathcal{S}''_\pm$ reaches a maximum slightly after the time of the quench.
We also see that QNEC settles down to its thermal value later than the stress-tensor itself. 

\begin{figure}[ht]
\center
\includegraphics[width=0.6\linewidth]{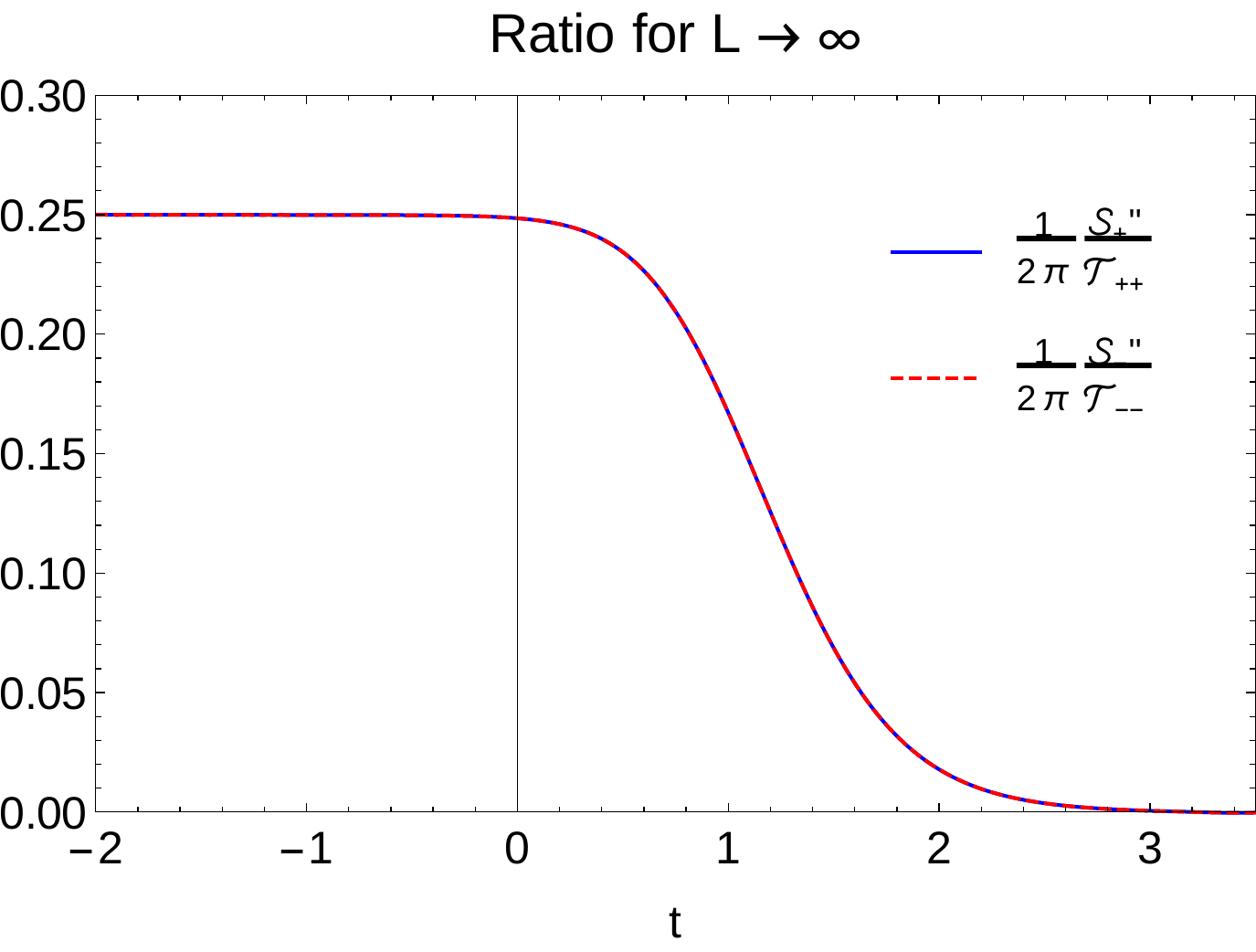}
\caption{Ratio of the two sides of the QNEC inequality \eqref{eq:QNEC2}. Curiously the ratio asymptotes to $0.25$ at early times and never grows above that value. QNEC is still non-trivial for a time of order $1/({\pi}T)$ after the geometry has already settled down.
}
\label{fig:QNECvaidya2}
\end{figure}

Even though the geometry is only slightly perturbed at early times, we curiously see that the ratio of $\mathcal{S}''_\pm/(2\pi)$ versus $\mathcal{T}_{\pm\pm}$ reaches a constant value of about 0.25, see Figure~\ref{fig:QNECvaidya2}.
This setting is the first case where QNEC is stronger than NEC, i.e.~we find $\mathcal{S}''_\pm > 0$.
Nevertheless, QNEC never saturates, even at early times where both sides approach 0.

\newcommand{\lcc}{x}

\begin{figure*}[ht]
\center
\includegraphics[width=0.47\textwidth]{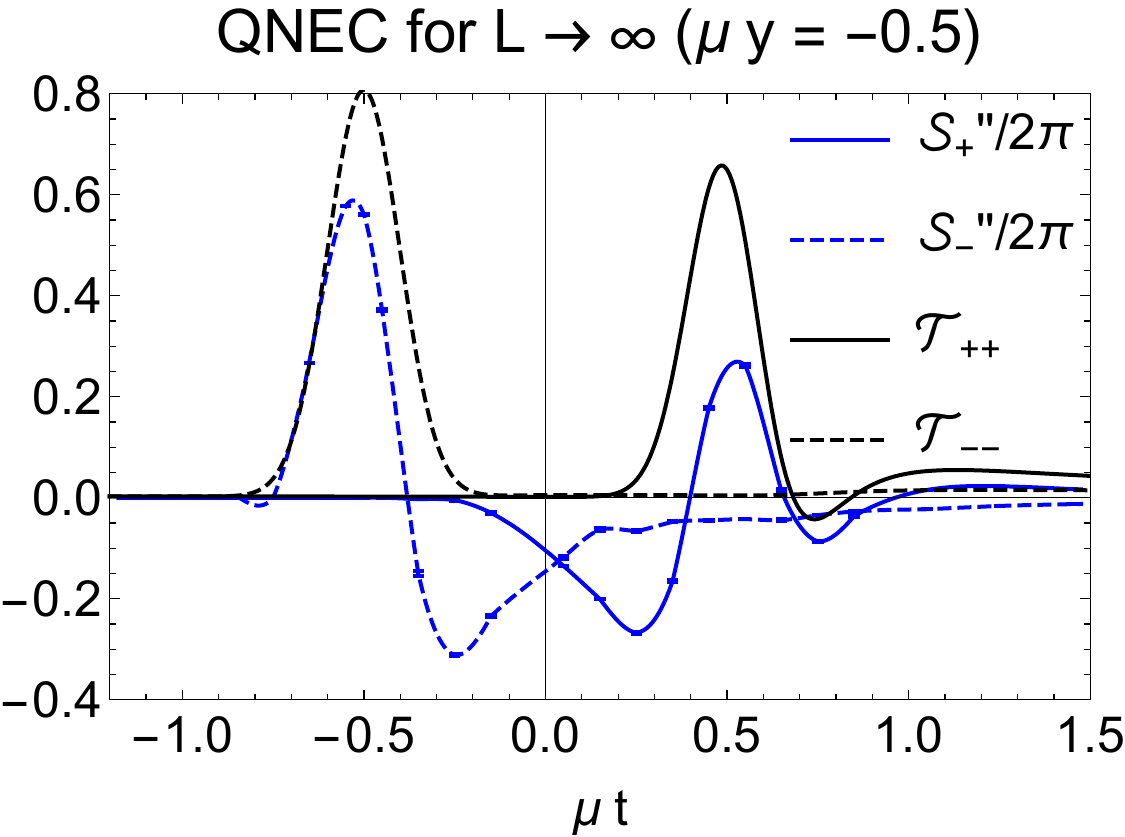}$\;\;$
\includegraphics[width=0.47\textwidth]{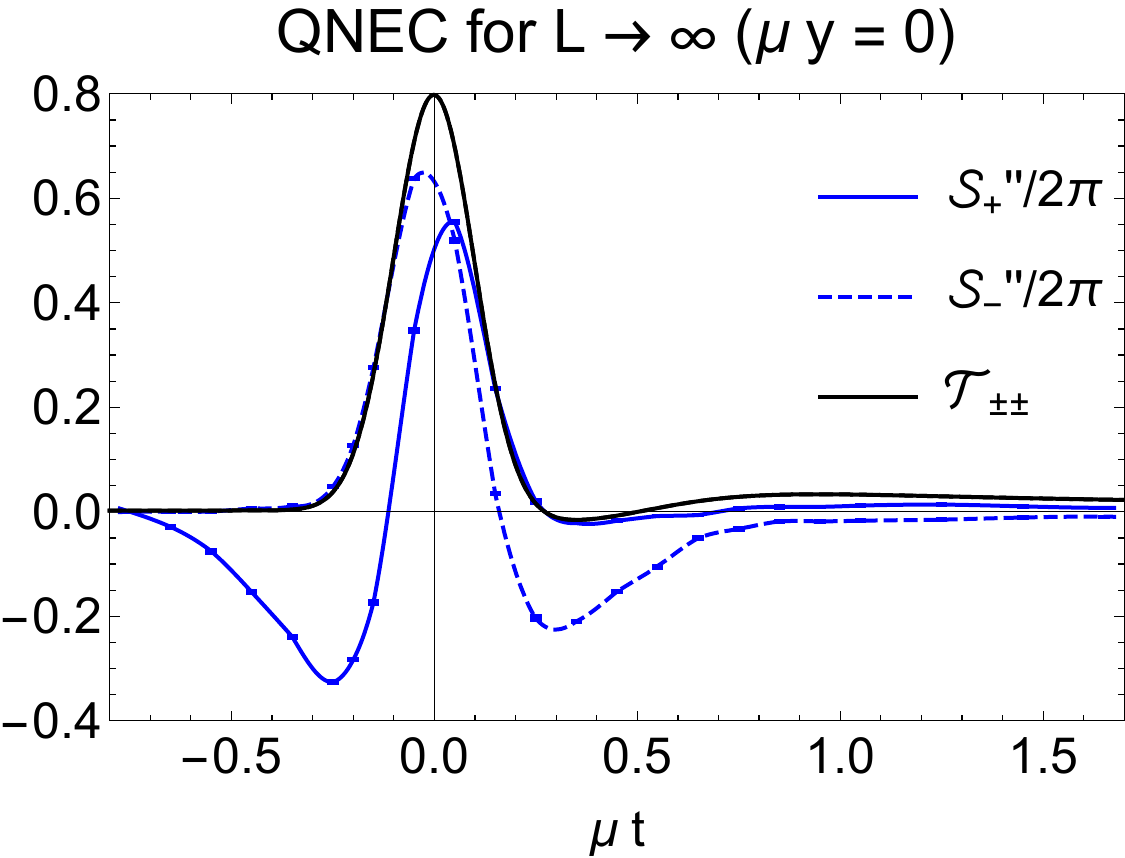}
\includegraphics[width=0.47\textwidth]{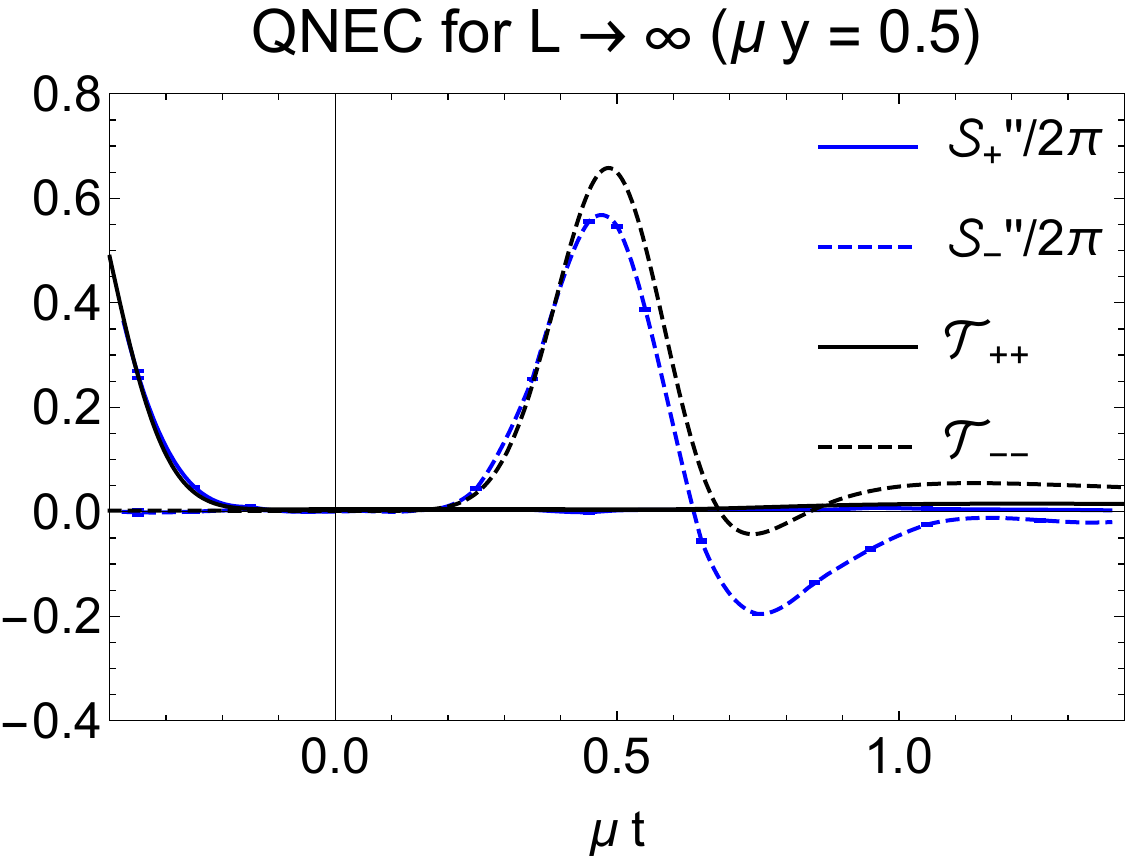}
\caption{Large $L$ limit of QNEC as a function of time for $y=-0.5$ (left), $y=0$ (middle) and $y=0.5$ (right). Strikingly, depending on the direction of $k_\mu$ all cases show a saturation of QNEC in the far-from-equilibrium regime, where in the center case first the $k_-$ direction saturates, after which it transitions to the $k_+$ direction, which saturates when NEC is violated ($\mathcal{T}_{\pm\pm}<0$).
}
\label{fig:QNECshocks2}
\end{figure*}

\subsection{Shock Wave Collision} 

\begin{figure}[htb]
\center
\includegraphics[width=0.5\linewidth]{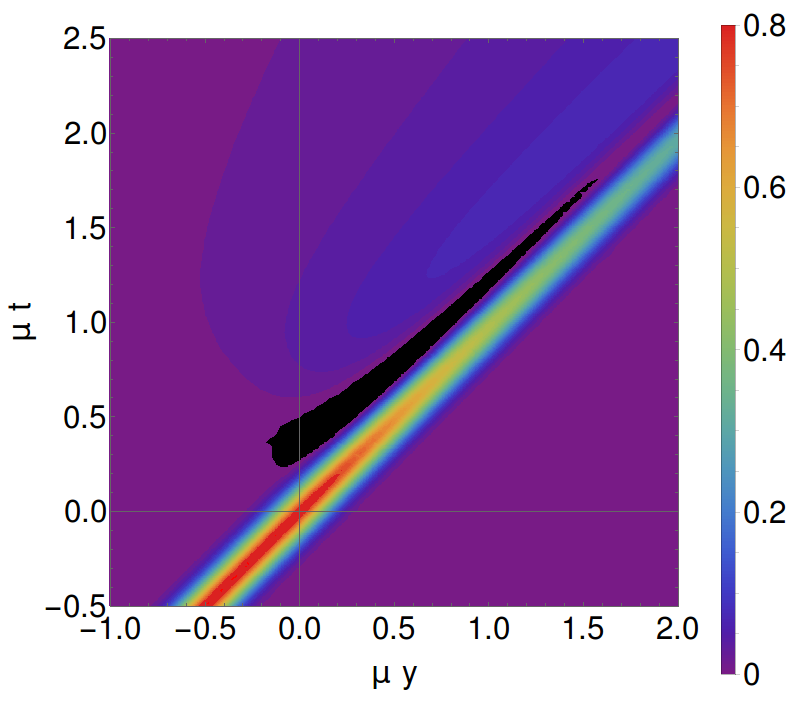}$\;\;$
\caption{%
Contour plot of ${\cal{T}}_{--}$ with NEC violation in the black region. }
\label{fig:ShockWaves}
\end{figure}

\begin{figure}[htb]
\center
\includegraphics[width=0.7\linewidth]{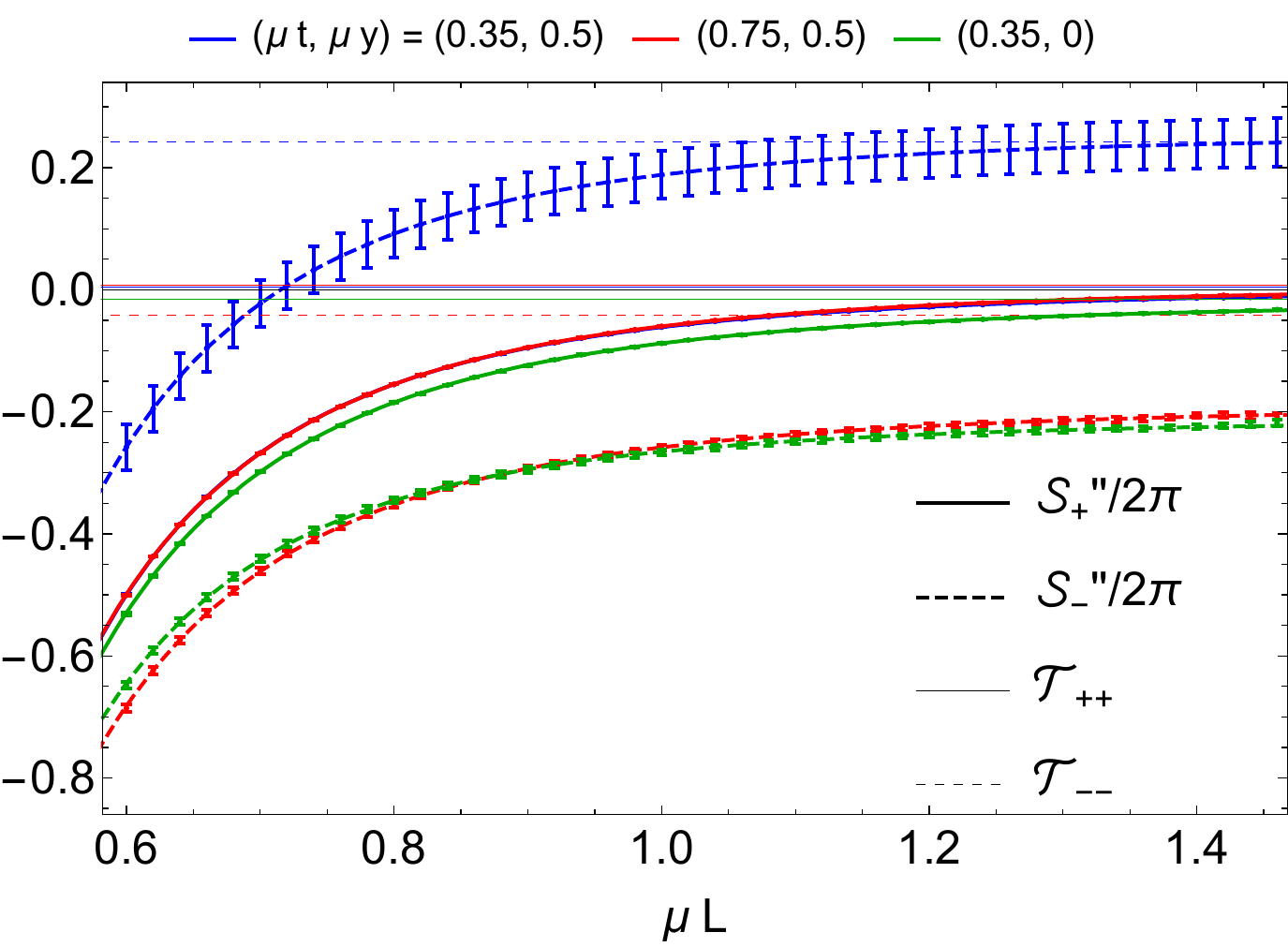}
\caption{%
QNEC terms as a function of $L$ for three representative points in the shock wave geometry (see Figure~\ref{fig:QNECshocks}). Dashed blue saturates QNEC, even though NEC is positive. Dashed red violates NEC, but $\mathcal{S}''_{-}$ is even smaller and no saturation occurs in $-$ direction, while it occurs in $+$ direction.}
\label{fig:ShockWaves2}
\end{figure}

The richest example presented here analyzes QNEC for the CFT state dual to colliding gravitational shock waves. This in particular leads to regions where the ordinary NEC is violated \cite{Arnold:2014jva} and hence gives a perfect setting to examine QNEC. Colliding shock waves are dual to planar sheets of energy moving at the speed of light and fully characterized by their only non-zero component of the boundary stress-energy tensor $\mathcal{T}_{\pm\pm}=1/2\pi^2h_\pm(\lcc_\pm)$, with $\lcc_\pm=t{\pm}y$, where $h_\pm(\lcc_\pm)=\mu^3\exp[-\lcc_\pm^2/2w^2]/\sqrt{2{\pi}w^2}$ and ${\mu}w=0.1$. 
We determined the functions $A$, $B$, $F$ and $R$ in the metric \eqref{eq:lalapetz} numerically in previous work \cite{Casalderrey-Solana:2013aba} and use these results here as input for our evaluation of entanglement entropy and QNEC.

Figure~\ref{fig:QNECshocks} shows the bulk shock wave evolution, whereby the colors at $z=0$ represent (the violation of) NEC (see also \cite{Arnold:2014jva} and Figure~\ref{fig:ShockWaves}). Figure~\ref{fig:ShockWaves2} shows analogous $\mathcal{S}''_\pm$ versus $L$ plots at three representative points, noting that $\mathcal{T}_{++}$ differs from $\mathcal{T}_{--}$ at $y \neq 0$. The red curve is at the location where NEC is significantly violated, with $\mathcal{T}_{--}\!=\!-0.04\mu^4$, while QNEC is satisfied, with $\mathcal{S}''_{-}/(2\pi)$ asymptoting to $-0.19\mu^4$. For $k_\mu\!=\!k_+$ NEC is satisfied, but QNEC is saturated, with $\mathcal{T}_{++}\!=\!\mathcal{S}''_{+}/(2\pi)\!=\!0.01\mu^4$ for $L \to \infty$.

Figure~\ref{fig:QNECshocks2} shows the asymptotic behavior of QNEC for $\mu y=-0.5,\,0.0\text{\;and\;}0.5$ (recall that $\pm0.5$ are distinct from each other due to our choice of varying the left point of the strip in \eqref{eq:QNEC2}).

Strikingly, at $y=0$ we find QNEC saturation in the far-from-equilibrium regime for $k_-$ at negative times, which transitions to saturation for $k_+$ at positive times. During the hydrodynamic phase at ${\mu}t>0.8$ there is no saturation. For $y=0.5$ we have the non-trivial result that QNEC is saturated for both $k_-$ and $k_+$ as the outgoing shock passes around ${\mu}t=0.3-0.5$. Lastly, for $y=-0.5$ the entangling region encompasses most of the collision region and we do not find saturation for $t>0$.

\section{Discussion} 

Our main result of this chapter is the saturation of (inclusive) QNEC in far-from-equilibrium regions created during shock wave collisions. This saturation is non-trivial and not seen in other systems we studied. For vacuum and thermal states QNEC is weaker than NEC, since $\mathcal{S}''$ is always negative. For a homogeneous quench QNEC is stronger than NEC, but the ratio of both sides of the inequality never exceeds $0.3$. In shock wave collisions QNEC is never saturated in the hydrodynamic regime, but it is saturated in the far-from-equilibrium region, regardless of whether NEC is valid. Reference \cite{Koeller:2017njr} (see also \cite{Casini:2017roe}) conjectures that saturation of QNEC can lead to a simplified expression for (part of) the modular Hamiltonian of a half-space in vacuum.

Even in vacuum QNEC is non-trivial, as for our strip the entanglement entropy term scales as $\mathcal{S}_\pm''\propto-1/L^4$, which has a UV divergence as $L\to0$. This makes the inequality trivially satisfied in the small length limit, and it is hence an interesting question whether QNEC also holds if one looks at a more physical quantity, such as the vacuum subtracted entanglement entropy. None of the proofs of QNEC apply for that case, but for all points where we checked QNEC we found that this stronger condition also holds. 

QNEC is a remarkable quantum inequality, and examples such as the ones studied in this Chapter will help to further explore its more general implications as well as applications such as holographic descriptions of strongly coupled quantum matter.

\chapter{Closing Remarks}\label{Chap:ClosingRemarks}

\section{Summary}
In the thesis at hand I studied the time evolution of equal time two-point functions, the holographic entanglement entropy and the quantum null energy condition of various strongly coupled far-from-equilibrium states in $\mathcal{N}=4$ Super-Yang-Mills theory via their higher-dimensional gravity duals.
After reviewing in Chapter \ref{Chap:Theoretical Background} the necessary basics of entanglement entropy, string theory, the AdS/CFT correspondence, the holographic entanglement entropy and the quantum null energy condition, I gave in Chapter \ref{Chap:Numerics} a detailed description of the numerical techniques used in the thesis.
The holographic prescription of two-point functions and entanglement entropy amounts to finding geodesics and extremal surfaces in the dual gravitational background which usually requires numerical techniques.
My main tool to determine geodesics and extremal surfaces is the so-called relaxation method which I explained in detail in Section \ref{Sec:NumericSurface}. I provided a ready-to-use and easily adaptable Mathematica implementation of the relaxation method in Appendix \ref{App:Relax} and also a simple version of the shooting method in Appendix \ref{App:Shoot}, where the latter turned out to be sufficient in time independent situations.
These computer codes can also be downloaded from my homepage \href{http://christianecker.com/}{{christianecker.com}}.

In addition to the numerical studies, I presented a number of cases where the entanglement entropy and the quantum null energy condition can be determined in closed form.
This includes several basic examples for entanglement entropy in vacuum presented in Section \ref{Sec:BasicExamples} and two specific examples of the quantum null energy condition for thermal states in CFT$_2$ (Section \ref{Sec:QNECCFT2} and Appendix \ref{App:QNECsat}) and CFT$_4$ (Appendix \ref{App:QNECCFT4}), where I studied the later perturbatively in the small and large size limits of the entangling region. 

Furthermore, I provided in Section \ref{Sec:Relaxation} and Section \ref{Sec:QNECVaidya} basic numerical studies of entanglement entropy and the quantum null energy condition of globally quenched states in CFT${_2}$ which are holographically dual to AdS$_3$ Vaidya geometries. 
In this case the entanglement entropy follows nicely the well-known scaling behavior obtained from a direct CFT$_2$ calculation. My simulations also confirmed the effect of non-saturation of the quantum null energy condition in cases where the associated extremal surfaces cross the infalling matter shell in the Vaidya geometry.

The first advanced examples presented in Chapter~\ref{chap:Aniso} are homogeneous but anisotropic finite temperature far-from-equilibrium states in strongly coupled $\mathcal{N}=4$ Super-Yang-Mills theory. I studied the time evolution of two-point functions and entanglement entropy of these states. 
I found that the entanglement entropy approaches its equilibrium value in an oscillatory manner and not monotonically from below, as it is for thermodynamical reasons the case for the thermal entropy.
Furthermore, I have shown that the entanglement entropy close to equilibrium is well described by the lowest quasinormal mode in the spin-two channel.

In the second advanced example in Chapter~\ref{Chap:Shocks} I applied the methods developed for the anisotropic system to the more complicated inhomogeneous system of colliding gravitational shock waves.
In this case I made a careful comparison between different initial conditions, describing narrow, intermediate and wide shocks and discovered characteristic features of two-point functions and entanglement entropy that allow to discriminate between the kind of initial conditions used.

In the final advanced example in Chapter~\ref{Chap:QNEC} I studied the quantum null energy condition in the shock wave system, which itself can violate the classical null energy condition for sufficiently narrow shocks, and showed that the quantum null energy condition holds. Furthermore, I made the unexpected observation that the integrated version of the quantum null energy condition can saturate in situations where the energy momentum tensor is still far from equilibrium.

In the next section I draw some conclusions based on the lessons learned from the results summarized above.
\section{Conclusion}


Today holography is the only tool available to study the full non-linear dynamics of strongly coupled quantum systems from first principles. The holographic mapping of complicated quantum dynamics to classical gravitational dynamics allows to study the time evolution of local and non-local observables like the energy momentum tensor, two-point functions and entanglement entropy, which is not accessible via a pure field theoretic approach. 

Many of the analytic and numeric calculations in this thesis are pioneering achievements in the sense that they represent first numerical studies of entanglement entropy and the quantum null energy condition in various highly non-trivial far-from equilibrium situations, thereby also providing an important proof of principle for the methods used.

Especially the studies of holographic shock wave collisions, which can be seen as holographic toy models for heavy ion collisions, can provide valuable insights into the physics of the quark gluon plasma.
In particular collisions of narrow shocks, which model collisions at LHC energies, show interesting features like regions of negative energy density in the forward light cone resulting in a violation of the classical null energy condition.
My studies successfully show that in this context the quantum null energy condition is a correct replacement for the null energy condition that is not only satisfied but can also saturate which might have detectable consequences in heavy ion experiments.

The specific applications mentioned here are motivated by the holographic modelling of the quark gluon plasma. However, the methods developed are applicable in many contexts where entanglement entropy plays an important role, like in quantum gravity, information theory and condensed matter physics.
I made my computer codes available (\href{http://christianecker.com/}{{http://christianecker.com/}}) to the scientific community, hoping that other researches profit from my implementation efforts and find novel applications that go beyond those presented in this thesis.
In the next section I will provide some ideas of possible generalizations and extensions of my work.

\section{Outlook}
There are plenty of generalizations and extensions possible for the studies presented in this thesis.
I list some of them and provide suggestions how they might be approached.

\paragraph{Two-Dimensional CFTs} 
In two-dimensional CFTs the quantum null energy condition has the special form \eqref{Eq:QNEC} which, as shown in Section \ref{Sec:QNECCFT2}, saturates in vacuum and thermal states dual to BTZ geometries.
In an upcoming work \cite{Ecker:2018xx} we are able to prove saturation for an even larger class of excited states dual to the so-called Ba\~nados family of bulk geometries \cite{Banados:1998gg} which includes far-from-equilibrium systems studied by other authors \cite{Bhaseen:2013ypa,Erdmenger:2017gdk}.
However, as pointed out in \cite{Khandker:2018xls}, in two dimensions the quantum null energy condition does not need to saturate if the bulk theory contains matter. A concrete example is the Vaidya setup presented in Section \ref{Sec:QNECVaidya}.
Finding an exact relation, if there is one, between the bulk energy momentum tensor and the amount of non-saturation remains an interesting outstanding challenge.

\paragraph{Finite Coupling Corrections}
Another possibility is to to follow \cite{Dong:2013qoa,Bhattacharyya:2014yga} and study the influence of finite coupling corrections on entanglement entropy and the quantum null energy condition which requires to include higher derivative terms in the bulk theory.
The authors of \cite{Grozdanov:2016zjj} provide an interesting example, namely a finite coupling corrected version of the shock wave system discussed in this thesis. It would be highly interesting to see, if finite coupling correction also lead to a softening of the top-down thermalization behavior \cite{Baier:2012ax,Waeber:2018bea} in the entanglement entropy.

\paragraph{Higher-Dimensional Surfaces}
A numerical challenge would be to generalize my relaxation code to two or three-dimensional surfaces. In this case one has to solve partial differential equations which is in principle possible with relaxation.
However, the numerical robustness, efficiency, and feasibility of such a higher-dimensional relaxation procedure remains to be shown.
Having a higher-dimensional surface relaxation code at hand would open many possibilities, including non-trivial checks of saturation of the quantum null energy condition in $d>2$ dimensional holographic CFTs argued in \cite{Leichenauer:2018obf}.
Based on my current experience, saturation of the quantum null energy condition in higher dimensions seems highly non-trivial and it would be extremely interesting to study the geometric mechanisms responsible for saturation, and how they are encoded in the shape of extremal surfaces.

\paragraph{Beyond the Semiclassical Limit}
It would be interesting to construct explicit examples which include quantum corrections to the holographic entanglement entropy such as suggested in \cite{Barrella:2013wja,Faulkner:2013ana}. 
In certain three-dimensional bulk theories the leading order quantum correction to the geometry is known explicitly \cite{Martinez:1996uv} and it is therefore straightforward to compute the shift $\delta A$ in \eqref{Eq:QantumCorrEE} with my methods. The open challenge is to construct also the remaining terms in \eqref{Eq:QantumCorrEE}.
The quantum corrected entanglement entropy, together with the corrected boundary stress tensor, can then directly be used to study the quantum null energy condition in two dimensions beyond the limit of infinite central charge.

\paragraph{Beyond Relativistic CFTs}
In this thesis I focus exclusively on relativistic CFTs and their gravity duals in the supergravity approximation.
Beyond those exist many other quantum field theories of interest which are not relativistic CFTs or relevant deformations thereof.
One example are Galilean 2$d$ CFTs which are conjectured field theory duals of flat space Einstein gravity in three dimensions \cite{Bagchi:2016bcd}. For these theories explicit expressions for the holographic stress tensor \cite{Bagchi:2015wna} and the entanglement entropy \cite{Bagchi:2014iea,Basu:2015evh,Jiang:2017ecm} are known.
Another example are warped CFTs in two dimensions with holographic duals in terms of topological massive gravity \cite{Detournay:2012pc}. Also for these theories the entanglement entropy formula is available \cite{Castro:2015csg,Song:2016gtd}.
Like their relativistic CFT$_2$ cousins, Galilean and warped CFTs in 2$d$ have infinite dimensional symmetry algebras which allow for a high level of analytic control.
It would be highly interesting to construct a non-relativistic counterpart to the relativistic quantum null energy condition for such theories.
Some ingredients essential to the relativistic version, like entanglement entropy and the stress tensor, are available for these non-relativistic theories. However, more conceptional questions, like what the pendant of a lightlike deformation or projection in a non-relativistic theory might be, still need to be addressed.

\paragraph{} 
As mentioned, the numerical techniques developed and applied in this thesis could have further applications and lead to additional generalizations not envisaged here. It seems likely that in the forthcoming years numerical holography will remain an invaluable tool to unravel holographic mysteries and to provide novel applications of the AdS/CFT correspondence.

\begin{appendices}
\chapter{Variation of the Area Functional}\label{App:AreaFunct}
We write the line element of the asymptotic $\mathrm{AdS}_{d+1}$ geometry in the following way
\begin{equation}
ds^2=G_{\mu\nu}dx^\mu dx^\nu\,.
\end{equation}
A $(d-1)$-dimensional surface in the bulk can be written in terms of embedding functions $X^\mu=X^\mu(\sigma^a,z)$ which are parametrized with $d-2$ intrinsic coordinates $\sigma^a$ and the bulk coordinate $z$.
The induced metric on the surface is given by
\begin{equation}\label{inducedMetric2}
H_{\alpha\beta}=\partial_\alpha X^\mu\partial_\beta X^\nu G_{\mu\nu}\,.
\end{equation}
The area functional can be written in terms of the induced metric
\begin{equation}
\mathcal{A}=\int dz d^{d-2}\sigma \sqrt{H[X]}.
\end{equation}
We are interested in stationary solutions $\delta\mathcal{A}=0$ which means we have to determine the variation of the surface functional with respect to the embedding functions
\begin{equation}\label{area2}
\delta \mathcal{A}=\int dz d^{d-2}\sigma \delta\left(\sqrt{H[X]}\right).
\end{equation}
Using $\delta \mathrm{det}(M) = \mathrm{det}(M) \mathrm{tr}(M^{-1}\delta M)$ the variation of the square root in \eqref{area} is given by
\begin{equation}
\delta \sqrt{H}=\frac{1}{2\sqrt{H}}\delta H=\frac{H}{2\sqrt{H}} H^{\alpha\beta} \delta H_{\alpha\beta}=\frac{1}{2} \sqrt{H}H^{\alpha\beta} \delta H_{\alpha\beta}\,.
\end{equation}
Plugging the explicit form of the induced metric into the variation of the area functional yields 
\begin{align}
\delta \mathcal{A}&=\int dz d^{d-2}\sigma \frac{1}{2} \sqrt{H}H^{\alpha\beta} \delta H_{\alpha\beta}\nonumber\\
                  &=\int dz d^{d-2}\sigma \frac{1}{2} \sqrt{H}H^{\alpha\beta} \delta \left(\partial_\alpha X^\mu\partial_\beta X^\nu G_{\mu\nu}\right)\nonumber\\
                  &=\int dz d^{d-2}\sigma \frac{1}{2} \sqrt{H}H^{\alpha\beta} \left(2 \partial_\alpha X^\mu G_{\mu\nu}\partial_\beta(\delta X^\nu)+\partial_\alpha X^\mu\partial_\beta X^\nu \delta G_{\mu\nu}\right)\,.
\end{align}
As a next step we perform a partial integration on the first term and the variation of the metric $\delta G_{\mu\nu}=(\partial_{X^\gamma}G_{\mu\nu})\delta X^\gamma$ which gives
\begin{align}
\delta \mathcal{A}&=\int dz d^{d-2}\sigma \Big\{\partial_\beta (\sqrt{H}H^{\alpha\beta}\partial_\alpha X^\mu G_{\mu\nu}\delta X^\nu)-\partial_\beta (\sqrt{H}H^{\alpha\beta}\partial_\alpha X^\mu G_{\mu\nu})\delta X^\nu\nonumber \\
                  &+\frac{1}{2} \sqrt{H}H^{\alpha\beta} \partial_\alpha X^\mu \partial_\beta X^\nu (\partial_{X^\gamma}G_{\mu\nu})\delta X^\gamma\Big\}\,.
\end{align}
Restricting to variations for which $\delta X^z=0$ the total derivative term can be dropped. Furthermore we perform the partial derivative of the metric $\partial_\beta G_{\mu\nu}=\partial_\beta X^\gamma\partial_{X^\gamma}G_{\mu\nu}$ and symmetrize the corresponding term in $\alpha$ and $\beta$ 
\begin{align}
\delta \mathcal{A}&=\int dz d^{d-2}\sigma \Big\{-\partial_\beta (\sqrt{H}H^{\alpha\beta}\partial_\alpha X^\mu) G_{\mu\nu}\delta X^\nu\nonumber \\
                  &-\frac{1}{2} \sqrt{H}H^{\alpha\beta} \partial_\alpha X^\mu \partial_\beta X^\gamma (\partial_{X^\gamma}G_{\mu\nu})\delta X^\nu\nonumber\\
                  &-\frac{1}{2} \sqrt{H}H^{\alpha\beta} \partial_\alpha X^\gamma \partial_\beta X^\nu (\partial_{X^\gamma}G_{\mu\nu})\delta X^\nu\nonumber\\
                  &+\frac{1}{2} \sqrt{H}H^{\alpha\beta} \partial_\alpha X^\mu \partial_\beta X^\nu (\partial_{X^\gamma}G_{\mu\nu})\delta X^\gamma\Big\}\,.
\end{align}
Relabeling indices and identifying the Christoffel symbol for the bulk metric $\Gamma^\rho_{\mu\nu}=\frac{1}{2}G^{\rho\gamma}(\partial_{X^\mu}G_{\nu\gamma}+\partial_{X^\nu}G_{\mu\gamma}-\partial_{X^\gamma}G_{\mu\nu})$ the previous expression takes the following form  
\begin{align}
\delta \mathcal{A}&=\int dz d^{d-2}\sigma \big\{-\partial_\alpha (\sqrt{H}H^{\alpha\beta}\partial_\beta X^\mu)-\sqrt{H}H^{\alpha\beta} \partial_\alpha X^\sigma \partial_\beta X^\nu \Gamma^{\mu}_{\sigma\nu}\big\}\delta X_\mu\,.
\end{align}
The surface which extremizes the surface functional has then to satisfy the equation
\begin{align}\label{Eq:ExtrSurf}
\frac{1}{\sqrt{H}}\partial_\alpha (\sqrt{H}H^{\alpha\beta}\partial_\beta X^\mu)+H^{\alpha\beta} \partial_\alpha X^\sigma \partial_\beta X^\nu \Gamma^{\mu}_{\sigma\nu}=0\,.
\end{align}

It is instructive to specialize \eqref{Eq:ExtrSurf} to the one-dimensional case $X^\mu=X^\mu(\sigma)$ which gives the non-affine geodesic equation
\begin{align}
0&=\frac{1}{\dot{\tau}}\left(\dot{\tau}\frac{1}{(\dot{\tau})^2}\dot{X}^\mu\right)+\frac{1}{(\dot{\tau})^2}\dot{X}^\sigma\dot{X}^\nu\Gamma^\mu_{\sigma\nu}\nonumber\\
 &=-J\dot{X}^\mu+\ddot{X}^\mu+\dot{X}^\sigma\dot{X}^\nu\Gamma^\mu_{\sigma\nu}\,,
\end{align}
where dot means derivative with respect to the non-affine parameter $\sigma$; in the first line we replaced the induced metric by $\dot{X}^\mu \dot{X}^\nu G_{\mu\nu}=:\dot{\tau}^2$; in the second line we made the identification $J\equiv\frac{\ddot{\tau}}{\dot{\tau}}$.

\chapter{QNEC Saturation in Thermal CFT$_2$}\label{App:QNECsat}
In this appendix we derive QNEC saturation in a thermal CFT$_2$ with $J=0$ using the minimal surface equations in BTZ coordinates directly.
We start with the line element of the non-rotating BTZ geometry
\begin{equation}\label{Eq:ds2BTZ}
ds^2=\frac{1}{z}\left(-f(z)dt^2+f(z)^{-1}dz^2+dy^2\right)\,, \quad f(z)=1-\hat Mz^2\,.
\end{equation}
Next we introduce the embedding functions $X^\mu(y)=(z(y),t(y),y)$ of a one-dimensional surface with endpoints located at $X^\mu(0)=(0,0,0)$ and $X^\mu(l+\lambda)=(0,\lambda,l+\lambda)$, where $l$ denotes the length of the entangling region and $\lambda$ parametrizes the lightlike deformation of one of the boundary points of the HRT-surface.
The surface has a turning point located at  $X^\mu(y_\ast)=(z_\ast,t_\ast,y_\ast)$, where $\dot{z}|_{y=y_\ast}=0$.
The corresponding area functional can be written as an integral over $y$
\begin{equation}\label{Eq:Area}
\mathcal{A}=\int_0^{l+\lambda} dy \,\mathcal{L}=\int_0^{l+\lambda} dy \sqrt{g_{\mu\nu}\dot{X}^\mu\dot{X}^\nu}\,,
\end{equation}
where $\dot{X}^\mu=\frac{d}{dy}X^\mu=(\dot{z},\dot{t},1)$ and the Lagrangian evaluates to
\begin{equation}
\mathcal{L}=\mathcal{L}(z,\dot{z},\dot{t})=\frac{1}{z}\sqrt{-f(z)\dot{t}^2+f(z)^{-1} \dot{z}^2+1}\,.
\end{equation}
Due to invariance of $\mathcal{L}$ under $y$- and $t$-translations there are two Noether charges
\begin{equation}\label{Noether1}
 Q_1=\frac{\partial{\mathcal{L}}}{\partial{\dot{t}}}\dot{t}+\frac{\partial{\mathcal{L}}}{\partial{\dot{z}}}\dot{z}-\mathcal{L}=\frac{2}{z\sqrt{1-\dot{t}^2f(z)+\dot{z}^2/f(z)}} \,,
\end{equation}
\begin{equation}\label{Noether2}
 Q_2=\frac{\partial{\mathcal{L}}}{\partial{\dot{t}}}=-\frac{2\dot{t}f(z)}{z\sqrt{1-\dot{t}^2f(z)+\dot{z}^2/f(z)}}\,.
\end{equation}
We can now define a new constant $\Lambda=-\frac{Q_2}{Q_1}$ and express $\dot{t}$ as follows
\begin{equation}
\dot{t}=\frac{\Lambda}{f(z)}\,.
\end{equation}
Next we rewrite the left hand side of (\ref{Noether1}) as $Q_1=\frac{2}{z_\ast N_\ast}$ and use the expression for $\dot{t}$ on the right hand side which then gives 
\begin{equation}
\dot{z}=\pm \frac{\sqrt{N_\ast^2 z_\ast^2 f(z)+z^2(\Lambda^2-f(z))}}{z}\,.
\end{equation}
There are two branches, $z_+(y)$ for $0\le y\le y_\ast$ and $z_-(y)$ for $y_\ast\le y \le l+\lambda$, because $z(y)$ is not single valued on the interval $y\in[0,l+\lambda]$.
We can now express the spatial separation on the boundary in terms of the following integral
\begin{eqnarray}
l+\lambda&=&\int_0^{l+\lambda}dy=\int_0^{y_\ast}dy+\int_{y_\ast}^{l+\lambda}dy\nonumber\\
         &=&\int_0^{z_\ast}\frac{dy}{dz_+}dz_++\int_{z_\ast}^0\frac{dy}{dz_-}dz_-=2\int_0^{z_\ast}\frac{dy}{dz_+}dz_+\,,
\end{eqnarray}
where we have used $z_-(y)=-z_+(y)$ in the last line.
It is convenient to transform to the new variable $x\equiv z_+/z_\ast$ such that the turning point of the surface is located at $x=1$ and the previous integral can be written as
\begin{equation}\label{int1}
\frac{l+\lambda}{2}=z_\ast\int_0^1 \frac{1}{\dot{x}}dx\,,
\end{equation}
with $\dot{x}$ given by
\begin{equation}
\dot{x}=\frac{\sqrt{N_\ast^2 f(z_\ast x)+x^2(\Lambda^2 -f(z_\ast x))}}{x}\,.
\end{equation}
This integral gives the result
\begin{equation}\label{Eq:I1}
\frac{l+\lambda}{2}=\frac{1}{\sqrt{\hat M}}\log\frac{\Lambda^2-1-\hat M(N_\ast^2-2)z_\ast^2+2z_\ast\sqrt{\hat M(\Lambda^2-(N_\ast^2-1)f(z_\ast))}}{\Lambda^2-(1-\sqrt{\hat M}N_\ast z_\ast)^2}\,.
\end{equation}
Similar to the spatial separation we can express the temporal separation on the boundary in terms of an integral
\begin{eqnarray}
\lambda&=&\int_0^\lambda dt=\int_0^{l+\lambda}\frac{dt}{dy}dy=\int_0^{y_\ast}\frac{dt}{dy}dy+\int_{y_\ast}^{l+\lambda}\frac{dt}{dy}dy\nonumber\\
&=&\int_0^{z_\ast}\dot{t}\frac{dy}{dz_+}dz_++\int_{z_\ast}^0\dot{t}\frac{dy}{dz_-}dz_-=2\int_0^{z_\ast}\dot{t}\frac{dy}{dz_+}dz_+\nonumber\\
&=&2z_{\ast}\int_0^1\frac{\dot{t}}{\dot{x}}dx\,.
\end{eqnarray}
Also this integral can be solved analytically and we find 
\begin{equation}\label{Eq:I2}
\lambda=\frac{1}{\sqrt{\hat M}}\log\frac{\Lambda^2-f(z_\ast)+\hat M z_\ast^2(\Lambda^2+N_\ast^2 f(z_\ast))+2\Lambda\sqrt{z_\ast^2 \hat M(\Lambda^2+N_\ast^2 f(z_\ast)-f(z_\ast))}}{f(z_\ast)(\Lambda^2+1+\sqrt{\hat M}N_\ast z_\ast)(\Lambda^2-1+\sqrt{\hat M}N_\ast z_\ast)}\,.
\end{equation}
Next we expand the integral (\ref{Eq:I2}) for small $\Lambda$
\begin{equation}
\lambda=\frac{2\Lambda z_\ast}{f(z_\ast)}+\mathcal{O}(\Lambda^3)\,.
\end{equation}
We can now solve the previous equation for $\Lambda$ and obtain an expression for $\Lambda(\lambda)$ which is valid when $\lambda$ is small
\begin{equation}
\Lambda\approx\lambda\frac{f(z_\ast)}{2z_\ast}\,.
\end{equation}
Now we can exponentiate the integral (\ref{Eq:I1}) and substitute for $\Lambda$
\begin{equation}
e^{\sqrt{\hat M}\frac{l+\lambda}{2}}=\frac{(\lambda^2-4z_\ast^2)f(z_\ast)}{\lambda^2 f(z_\ast)-4z_\ast^2(1-\sqrt{4\hat M z_\ast^2-\hat M\lambda^2 f(z_\ast))}+\hat M z_\ast^2)}\,.
\end{equation}
Expand this equation to second order in $\lambda$ gives an equation for $z_\ast$ which holds as long as $\lambda$ is small
\begin{equation}
0=4+4e^{l\sqrt{\hat M}}-\frac{8}{1-\sqrt{\hat M}z_\ast}+4e^{l\sqrt{\hat M}}\sqrt{\hat M} \lambda+\left(2\hat M\Big(e^{l\sqrt{\hat M}}-\frac{1}{1-\sqrt{\hat M}z_\ast}\Big)-\frac{\sqrt{\hat M}}{z_\ast}\right)\lambda^2\,.
\end{equation}
Solving for $z_\ast$ gives two solutions
\begin{subequations}
\begin{eqnarray}
z_{\ast,1}&=&-\frac{1-\eta}{\sqrt{\hat M}(1+\eta)}+\frac{2 \eta}{(1+\eta)^2}\lambda+\frac{\eta(3-2\eta+3\eta^2)\sqrt{\hat M}}{2(1-\eta)(1+\eta)^3}\lambda^2\,,\\
z_{\ast,2}&=&-\frac{\sqrt{\hat M}}{4(1-\eta)}\lambda^2\,,
\end{eqnarray}
\end{subequations}
where we have to pick the first solution $z_\ast=z_{\ast,1}$ because it has a finite $\lambda\to 0$ limit and we have introduced the notation $\eta=e^{\sqrt{\hat M}l}$.
For small values of $l$ the surface resides in the asymptotic AdS region where the embedding function describes a semi-circle of radius $l/2$. This is exactly what we find at leading order
\begin{equation}
z_\ast=\frac{l}{2}-\frac{\hat M l^3}{24}+\frac{\hat M^2 l^5}{240}-\frac{17\hat M^3 l^7}{40320}+\mathcal{O}(l^9)\,.
\end{equation}
Having explicit expressions for $\Lambda$ and $z_\ast$ we are equipped to solve the integral (\ref{Eq:Area}) and obtain a small $\lambda$ expansion of the area.
For QNEC we only need the first and second derivative at $\lambda=0$ of the integral, so it is sufficient to compute the $\mathcal{O}(\lambda^2)$ term.
Expressing the (regularized) area as integral over $x$ we obtain 
\begin{equation}
\mathcal{A}=\int_0^1dx\left(\frac{2N_\ast}{x\sqrt{(N_\ast^2-x^2)f(x z_\ast)+\Lambda^2 x^2}}-\frac{2}{x}\right)\,.
\end{equation}
Substituting $\Lambda$ and $z_\ast$ and expanding to second order in $\lambda$ results in a huge expression which we do not write down here.
We only give the relevant $\mathcal{O}(\lambda^2)$ solution of the integral
\begin{subequations}
\begin{eqnarray}
\mathcal{A}' &\equiv&\frac{d\mathcal{A}}{d\lambda}\Big|_{\lambda=0}=\sqrt{\hat M} \mathrm{coth}\left(\frac{l \sqrt{\hat M}}{2}\right)\,,\\
\mathcal{A}''&\equiv&\frac{d^2\mathcal{A}}{d\lambda^2}\Big|_{\lambda=0}=-\hat M \mathrm{csch}\left(\frac{l \sqrt{\hat M}}{2}\right)^2\,.
\end{eqnarray}
\end{subequations}
The relevant expression for QNEC gives 
\begin{subequations}
\begin{eqnarray}\label{Eq:SddEOM}
\frac{1}{2\pi} \Big( S''+\frac{6}{c}(S')^2\Big)&=&\frac{1}{8\pi G_N^{(3)}}\left(\mathcal{A}''+\frac{6}{c}\frac{1}{4 G_N^{(3)}}(\mathcal{A}')^2\right)\,,\nonumber\\
&=&\frac{c \hat M}{12\pi}\Big(\mathrm{coth}^2\left(l\sqrt{\hat M}/2\right)-\mathrm{csch}^2\left(l\sqrt{\hat M}/2\right)\Big)\,,\nonumber\\
&=&\frac{c}{12\pi} r_+^2\,,
\end{eqnarray}
\end{subequations}
where we have restored the units $\hat M=8\pi G_N^{(3)}M=r_+^2$ in the last step. We find exact agreement with $T_{\pm\pm}=\frac{c}{12\pi}r_+^2$ given in \eqref{Eq:Tkk}, which proofs QNEC saturation.

\chapter{QNEC for AdS\texorpdfstring{$_5$}{5} Schwarzschild Black Brane}\label{App:QNECCFT4}

Entanglement entropy for the AdS$_d$ Schwarzschild black brane was considered by Fischler and Kundu who gave infinite series representations in terms of ratios of $\Gamma$-functions \cite{Fischler:2012ca} and more recently by Erdmenger and Miekley \cite{Erdmenger:2017pfh} who expressed their results in closed form in terms of Meijer G-functions. For QNEC it is necessary to compute non-equal time entanglement entropy, which is not straightforward using these methods. We use a more pedestrian approach that allows straightforward generalization from entanglement entropy to QNEC as well as fast and precise numerical evaluation of QNEC at small and large separations. For sake of specificity we focus on $d=5$, but our methods and results can be generalized easily to arbitrary dimensions. In this way we shall recover the vacuum result for entanglement entropy \eqref{eq:Svac} and QNEC \eqref{eq:Sppvac} as well as the corresponding thermal results in the main text, see Figure~\ref{fig:QNECblackbrane}.

\newcommand{\hor}{(\pi T)}

The AdS$_5$ Schwarzschild black brane metric is given by
\begin{equation}\label{eq:s1}
d s^2 = \frac{1}{z^2}\,\Big(-f(z)\,d t^2 + \frac{d z^2}{f(z)} + d y^2 + d x_1^2+d x_2^2\Big)\,,
\end{equation}
with
\begin{equation}\label{eq:s2}
f(z)= 1 - (\pi T)^4 z^4 \,, 
\end{equation}
where $T$ is the Hawking temperature in the same units as in the main text.
For a strip the minimal area per transverse density functional reads
\begin{equation}\label{eq:s3}
{\cal A} =\int\limits_0^{\frac{\ell+\lambda}{2}-\omega}\!\!\!\!d y\, L(z,\,\dot z,\,\dot t)\,,
\end{equation}
with Lagrangian
\begin{equation}\label{eq:s4}
L(z,\,\dot z,\,\dot t) = \frac{2}{z^3}\,\sqrt{1+\frac{\dot z^2}{f(z)}-\dot t^2f(z)}\,,
\end{equation}
where the dimensionful quantity $\ell$ is the width of the strip in $y$-direction before deformation and $\lambda$ parametrizes the null deformation of the boundary interval with boundary points $(t_\pm,\,y_\pm)=(\pm\lambda/2,\,\pm(\ell+\lambda)/2)$. This means that for $\lambda=0$ we shall recover the entanglement entropy results for a strip of width $\ell$ centered around $y=0$ at the constant time-slice $t=0$. Moreover, $\omega$ denotes the cutoff on the holographic coordinate, such that $z(\ell/2-\omega)=z_\text{cut}\ll 1$, dots denote derivatives with respect to $y$ and the overall factor $2$ in \eqref{eq:s4} comes from the fact that we have two equally big contributions to the area by integrating $y$ from the midpoint $y=0$ to either of the endpoints $y_\pm=\pm(\ell+\lambda)/2$. 

Since the functional \eqref{eq:s3} respects translation invariance, $y\to y+y_0$, there is an associated Noether charge yielding a first integral,
\begin{equation}
Q_1 = L - \dot z \,\frac{\partial L}{\partial\dot z} - \dot t \, \frac{\partial L}{\partial\dot t}= \frac{2}{z^3\sqrt{1+\dot z^2/f(z)-\dot t^2 f(z)}} =: \frac{2}{z_\ast^3 N_\ast}\,,
\label{eq:s5}
\end{equation}
with 
$N_\ast=\sqrt{1-(\dot t^2 f)|_{z=z_\ast}} = \sqrt{1-\Lambda^2/f(z_\ast)}$
chosen such that at $z(y\to 0)=z_\ast$ we are at the tip of the extremal surface, $\dot z=0$. The constant $\Lambda=(\dot t f)|_{z=z_\ast}$ was introduced in anticipation of \eqref{eq:s25} below.

There is a second Noether charge following from $\partial_y (\partial L/\partial\dot t)=0$, yielding a constant of motion $\Lambda$
\begin{equation}\label{eq:s25}
Q_2 = \dot t f(z) =: \Lambda \,.
\end{equation}
Combining the two Noether charges $Q_{1,2}$ establishes an expression for $\dot z$
\begin{equation}\label{eq:s34}
\dot z = - \sqrt{\big(N_\ast^2 z_\ast^6/z^6-1\big)\,f(z) + \Lambda^2}\,.
\end{equation}

The values of the two Noether charges are fixed by the interval parameters $\ell$ and $\lambda$. Integrating \eqref{eq:s34} from the tip of the surface $z=z_\ast$ to the boundary $z=0$ and introducing the dimensionless variable $x=z/z_\ast$ yields
\begin{equation}\label{eq:s38}
\frac{\ell+\lambda}{2} = z_\ast\,\int\limits_0^1d x\,\frac{x^3}{R(x)}\,,
\end{equation}
with
$R(x):=\sqrt{(N_\ast^2-x^6)(1-(\pi Tz_\ast x)^4)+ \Lambda^2x^6}$.
Similarly, integrating $\dot t$ from $t=0$ to $t=\lambda/2$ (which again can be converted into a $z$-integration from the tip of the surface $z=z_\ast$ to the boundary $z=0$) yields
\begin{equation}\label{eq:s100}
\frac{\lambda}{2}=\Lambda\, z_\ast\int\limits_0^1d x\,\frac{x^3}{f(xz_\ast)\,R(x)} \,.
\end{equation}
For small $\ell$ it is useful to determine $\Lambda$ instead from
\begin{equation}\label{eq:s101}
\Lambda = \frac{\lambda}{\ell+\lambda+2z_\ast I_{\Delta}}\,,
\end{equation}
with 
\begin{equation}\label{eq:s102}
I_{\Delta} = \int\limits_0^1d x\,\frac{x^3}{R(x)}\bigg(\frac{1}{f(xz_\ast)}-1\bigg)\,.
\end{equation}
For QNEC we need to expand to order ${\cal O}(\lambda^2)$ but not higher, which means that in \eqref{eq:s101} we need to take into account only terms in $I_{\Delta}$ of order unity or linear in $\Lambda$, but no higher powers of $\Lambda$.

Inserting the first integrals \eqref{eq:s34}, \eqref{eq:s25} into the area functional \eqref{eq:s3} with \eqref{eq:s4} and \eqref{eq:s2} and expanding in powers of the cutoff $z_{\textrm{cut}}$ yields
\begin{equation}\label{eq:s36}
{\cal A} = \frac{1}{z_\text{cut}^2} + \frac{2}{z_\ast^2}\,\big(I_{\cal A}^{\lambda}-\tfrac12\big) + {\cal O}(z_{\textrm{cut}}^2)\,,
\end{equation}
with the finite contribution
\begin{equation}\label{eq:s37}
I_{\cal A}^{\lambda} = \int\limits_0^1d x\,\frac{1}{x^3}\,\bigg(\frac{N_\ast}{R(x)}-1\bigg)\,.
\end{equation}

The remaining task in order to get the area as function of the dimensionless product of temperature and strip width, $T\ell$, is to evaluate the integrals \eqref{eq:s37}, \eqref{eq:s102} and \eqref{eq:s38}. We consider first the limit of small widths, $T\ell \ll 1$, and then of large widths, $T\ell \gg 1$. These results will allow comparison with the numerical fits in the main text and in Figure~\ref{fig:QNECblackbrane}.

We start with the small width expansion $T\ell\ll 1$. Note that we have the chain of inequalities $0<\lambda/\ell\ll T\ell\ll 1$. As we shall see, all our results are expressed succinctly in powers of a single transcendental number, 
\begin{equation}\label{eq:c0}
 c_0 = \frac{3\Gamma[1/3]^3}{2^{1/3}(2\pi)^2} \approx 1.159595\,,
\end{equation}
which was already introduced in the main text \eqref{eq:Svac}. Perturbative evaluation of the integral \eqref{eq:s102} together with \eqref{eq:s101} yields
\begin{eqnarray}
\Lambda &=& \tfrac{\lambda}{\ell+\lambda}-(\pi Tz_\ast)^4\,\tfrac{4\pi c_0\lambda z_\ast}{15 \sqrt{3}(\ell+\lambda)^2}\nonumber\\
&+& (\pi Tz_\ast)^8\,\Big(\tfrac{16\pi^2 c_0^2\lambda z_\ast^2}{675(\ell+\lambda)^3}-\tfrac{2\lambda z_\ast}{3(\ell+\lambda)^2}\Big) \nonumber\\ 
&+& {\cal O}((T z_\ast)^{12}) + {\cal O}(\lambda^3/\ell^3)\,.
\label{eq:s105}
\end{eqnarray}
Similarly, evaluation of the integral \eqref{eq:s38} establishes a series expansion for $z_\ast$,
\begin{eqnarray}
\frac{z_\ast}{c_0\ell} &=& 1 + (\pi T\ell)^4\,\tfrac{2\pi c_0^6}{15\sqrt{3}} + (\pi T\ell)^8\, \big(\tfrac{4\pi^2 c_0^{12}}{135}-\tfrac{c_0^9}{6}\big)\nonumber\\
&+&\tfrac{\lambda}{\ell}\, \Big(1 - (\pi T\ell)^4\,\tfrac{2\pi c_0^6}{3\sqrt{3}} + (\pi T\ell)^8\, \big(\tfrac{4\pi^2c_0^{12}}{15}-\tfrac{3c_0^9}{2}\big)\Big)\nonumber\\
&+&\tfrac{\lambda^2}{\ell^2}\, \Big(-\tfrac12 + (\pi T\ell)^4\,\big(\tfrac{c_0^4}{6} - \tfrac{49\pi c_0^6}{45\sqrt{3}}\big) + (\pi T\ell)^8\, \big(\tfrac{c_0^8}{6}- \tfrac{71 c_0^9}{12}- \tfrac{c_0^{10} \pi}{5 \sqrt{3}} + \tfrac{2074 c_0^{12}\pi^2}{2025} \big) \Big)\nonumber\\
&+&{\cal O}((T\ell)^{12})+{\cal O}(\lambda^3/\ell^3)\,,
\label{eq:s106}
\end{eqnarray}
where we additionally expanded in powers of the dimensionless small parameter $\lambda/\ell$, keeping only the powers needed to determine QNEC. Finally, the area integral \eqref{eq:s37}, together with the other results above, leads to an expression for the area \eqref{eq:s36}
\begin{eqnarray}
{\cal A} &=& \tfrac{1}{z_{\textrm{cut}}^2} - \tfrac{1}{2c_0^3\ell^2} + \hor^4\ell^2\,\tfrac{\pi c_0^3}{5\sqrt{3}} + \hor^8\ell^6\,\big(\tfrac{c_0^6}{12} - \tfrac{2 c_0^9 \pi^2}{225}\big)\nonumber \\
&+&\tfrac{\lambda}{\ell}\,\Big(\tfrac{1}{c_0^3\ell^2} + \hor^4\ell^2\,\tfrac{2\pi c_0^3}{5\sqrt{3}} + \hor^8\ell^6 \big(\tfrac{c_0^6}{2} - \tfrac{4 c_0^9 \pi^2}{75}\big)\Big)\nonumber \\
&+&\tfrac{\lambda^2}{\ell^2}\,\Big(-\tfrac{2}{c_0^3\ell^2} + \hor^4\ell^2\,\tfrac{2\pi c_0^3}{15\sqrt{3}} + \hor^8\ell^6 \big(\tfrac{4 c_0^6}{3} - \tfrac{88 c_0^9 \pi^2}{675}\big)\Big)\nonumber\\
&+& {\cal O}(z_{\textrm{cut}}^2) + {\cal O}(T^{12}\ell^{10}) + {\cal O}(\lambda^3/\ell^3)\,.\label{eq:s107}
\end{eqnarray}
The first line recovers the entanglement entropy results of \cite{Fischler:2012ca, Erdmenger:2017pfh}. 
The second derivative of the area \eqref{eq:s107} with respect to $\pm\lambda$ evaluated at $\lambda=0$ yields the QNEC quantity $\mathcal{S}''_\pm$ used in the main text 
\begin{equation}
\frac{1}{2\pi}\,\mathcal{S}''_\pm = -\frac{1}{\pi^2c_0^3\ell^4} + \frac{\hor^4\,c_0^3}{15\sqrt{3}\pi} - \hor^8\ell^4\bigg(\frac{44c_0^9}{675}-\frac{2c_0^6}{3\pi^2}\bigg)+{\cal O}(T^{12}\ell^8)\,.
\label{eq:s108}
\end{equation}
This is our main result in the limit of small separations. For comparison with our numerical results in the main text we evaluate \eqref{eq:s108} using \eqref{eq:c0}
\begin{equation}\label{eq:s109}
\frac{1}{2\pi}\,\mathcal{S}''_\pm \approx -\frac{0.06498}{\ell^4} + 0.01910 - 0.08289\, \ell^4\,,
\end{equation}
where we set $\pi T=1$.
The number $0.06498$ reproduces the correct vacuum result \eqref{eq:Sppvac}, while the numbers $0.01910$ and $0.08289$ appear in the fit in the inset of Figure~\ref{fig:QNECblackbrane}. 

If $T\ell\gg 1$ then the holographic depth $z_\ast$ approaches the horizon, 
\begin{equation}\label{eq:s14}
z_\ast = \hor^{-1} (1-\epsilon)\qquad 0<\epsilon\ll 1\,.
\end{equation}
This means that we have again a small parameter that we can use for perturbative purposes, namely $\epsilon$. However, a technical difficulty is that integrals like \eqref{eq:s37} now acquire terms that diverge like $\ln\epsilon$ or $1/\epsilon$ due to the behavior of the integrands near the upper integration boundary $x=1$. Thus, we need to isolate these divergences as we expand around $\epsilon=0$. 
We encounter two types of delicate integrals. The first one is of the form
\begin{equation}\label{eq:int1}
I_1[h(x)]=\int\limits_0^1\frac{h(x)\,d x}{\sqrt{1-x}\,(1 - x + \epsilon x)^{3/2}} = \frac{2h(1)}{\epsilon} + {\cal O}(\ln\epsilon)\,,
\end{equation}
and the second one reads
\begin{eqnarray}
I_2[h(x)]&=&\int\limits_0^1\frac{h(x)\,d x}{\sqrt{(1-x)(1 - x + \epsilon x)}} \nonumber\\
         &=& -h(1)\,\ln\frac\epsilon4+ \int\limits_0^1d x \,\frac{h(x)-h(1)}{1-x} - \big(h(1)+h^\prime(1)\big)\,\frac\epsilon2\,\ln\frac\epsilon4 +{\cal O}(\epsilon)\,,\label{eq:int2}
\end{eqnarray}
where in both cases the function $h(x)$ must be (and in all our cases will be) Taylor-expandable around $x=1$. We have also simple explicit expressions for the subleading terms, but do not display them since we are not going to use them (with one exception). By virtue of the formulas above we now evaluate the three relevant integrals.

Let us start with the integral \eqref{eq:s100}. We rewrite it as
\begin{equation}\label{eq:s112}
\frac{\lambda}{2}=\Lambda\,z_\ast\,I_1[h_\Lambda(x)]\,,
\end{equation}
with $h_\Lambda(1)\simeq 1/(8\sqrt{6})+{\cal O}(\epsilon)$ 
where $\simeq$ denotes equality up to terms of irrelevant order in $\lambda$. Using \eqref{eq:int1} for small $\epsilon$ the integral \eqref{eq:s112} yields
\begin{equation}\label{eq:s110}
\Lambda \simeq  2\sqrt{6}\epsilon\lambda\hor + {\cal O}(\lambda\epsilon^2\ln\epsilon) \,.
\end{equation}

The next integral we consider is \eqref{eq:s38}, which determines $\epsilon$ defined in \eqref{eq:s14} in terms of $\ell$ and $\lambda$. Again we slightly rewrite the integral
\begin{equation}\label{eq:s116}
\frac{\ell+\lambda}{2z_\ast}\simeq I_2[h_z(x)]
+\lambda^2\hor^2 \epsilon\,I_1[h_\lambda(x)]\,,
\end{equation}
which for small $\epsilon$ by virtue of \eqref{eq:int1} and \eqref{eq:int2} expands as
\begin{equation}\label{eq:s119}
\frac{\ell+\lambda}{2z_\ast}\simeq-h_z(1)\ln\frac\epsilon4 + h_z^0 + 2\lambda^2 h_\lambda(1) + {\cal O}(\epsilon\ln\epsilon)\,,
\end{equation}
with $h_z(1)=2h_\lambda(1)=1/(2\sqrt{6})+{\cal O}(\epsilon)$ and
$h_z^0\approx -0.25032$~\footnote{
The explicit expression for $h_z^0$ follows from the integral formula \eqref{eq:int2} and reads $h_z^0 = \int_0^1d x \,[h_z(x)-h_z(1)]/(1-x)$  with $h_z(x)=x^3/[(1+x)W(x)]$ where $W(x)=\sqrt{(1 + x^2) (1 - x + x^2) (1 + x + x^2)}$.
}, 
yielding
\begin{equation}\label{eq:s111}
\epsilon \simeq \epsilon_0 \exp\big[-\sqrt{6}(\ell+\lambda)\hor+\lambda^2\hor^2\big] + \dots\,,
\end{equation}
where the ellipsis refers to terms that are exponentially suppressed as compared to the one displayed. Numerically, $\epsilon_0=4\exp[h_z^0/h_z(1)]\approx1.173487$.

Finally, we evaluate the area integral \eqref{eq:s37}. We split it into $\lambda$-independent and $\lambda$-dependent terms
\begin{eqnarray}
I_{\cal A} &\simeq& I_2[h_z(x)+ \epsilon\,k_z(x)] \lambda^2\hor^2\epsilon\,I_1[h_\lambda(x) + \epsilon\,k_\lambda(x)]\nonumber \\ 
&\simeq& \frac{\ell+\lambda}{2z_\ast} + \epsilon\,\big(I_2[k_z(x)] + \lambda^2\hor^2\epsilon\,I_1[k_\lambda(x)] \big)\,,
\label{eq:s121}
\end{eqnarray}
with the same functions $h_z$ and $h_\lambda$ as in \eqref{eq:s116}, $k_z(1)=-1/2$ and $k_\lambda(1)=-\sqrt{6}/4$. Physically, the reason why the split of the integrals in \eqref{eq:s121} into $h$ and $k$ is useful is related to the fact that for large $T\ell$ entanglement entropy scales linearly with $\ell$. 

The integration formulas \eqref{eq:int1} and \eqref{eq:int2} together with the results above yield for the area \eqref{eq:s36}
\begin{equation}
{\cal A} \simeq \frac{1}{z_{\textrm{cut}}^2} + \frac{\ell+\lambda}{z_\ast^3} + \frac{1}{z_\ast^2}\,\big(b_0 + b_1\epsilon + b_{\textrm{\tiny log}}\epsilon\ln\epsilon\big)+ \lambda^2\hor^4\,b_2\epsilon + {\cal O}(z_{\textrm{cut}}^2) + {\cal O}(\epsilon^2\ln\epsilon)\,,
\label{eq:s42}
\end{equation}
with $b_0\approx -0.66589$, $b_1\approx  -0.08889$~\footnote{%
To evaluate $b_1$ also the first subleading term not displayed in the integral formula \eqref{eq:int2} is needed. The explicit expression for $b_1$ reads
 $b_1 = 1+2k_0(1)-b_{\textrm{\tiny log}}\ln4  + 2\int_0^1d x\,[k_1(x)+\frac12]/(1-x) 
 -\int_0^1d x\,\int_x^1d y\,[k_0(y)-k_0(1)]/(1-y)$
with the functions $k_0(x)=[x^4+x^2+1-W(x)]/[x^3W(x)]$, $k_1(x)=(1-x)(3x^6 + 2 x^5  + 4x^4 + 2x^3 + 4x^2  + 2x + 1)/[2 x^2 (1 + x) (1 + x^2) W(x)]-1/(2x^2)$ and $W(x)=\sqrt{(1 + x^2) (1 - x + x^2) (1 + x + x^2)}$. The explicit expression for $b_0$ follows from the integral formula \eqref{eq:int2},  $b_0 =-1+2\int_0^1d x\,k_0(x)$.
}, $b_2=-\sqrt{6}$ and $b_{\textrm{\tiny log}}=\sqrt{6}/2$.
For $\lambda=0$ the area \eqref{eq:s42} establishes a result for entanglement entropy,
\begin{equation}
\mathcal{S}_{\textrm{EE}} = \frac{1}{2\pi}\,\Big[\frac{1}{z_{\textrm{cut}}^2} + \ell \hor^3 + \hor^2 \,b_0+ e^{-\sqrt{6}\ell\hor} \hor^2\,\epsilon_0\, \big(2b_0+b_1+b_{\textrm{\tiny log}}\ln\epsilon_0 \big)\Big]\,,
\label{eq:s23}
\end{equation}
where we neglected terms that vanish as the cutoff is removed, $z_{\textrm{cut}}\to 0$, and terms that are exponentially suppressed like $\ell\exp[-2\sqrt{6}\ell\hor]$. Note that all terms of the form $\ell \exp[-\sqrt{6}\ell\hor]$ cancel. Numerically, the cutoff-independent terms read (setting $\pi T=1$)
\begin{equation}\label{eq:s29}
2\pi\mathcal{S}_{\textrm{fin}} \approx \ell-0.666 - 1.437 \,e^{-\sqrt{6}\ell} + {\cal O}(\ell\,e^{-2\sqrt{6}\ell})\,.
\end{equation}
The result above agrees with (5.27) and (B.26) in \cite{Fischler:2012ca}.

The second derivative of the area \eqref{eq:s42} with respect to $\pm\lambda$ evaluated at $\lambda=0$ yields again the QNEC quantity $\mathcal{S}''_\pm$ used in the main text 
\begin{equation}\label{eq:s126}
\frac{1}{2\pi}\,\mathcal{S}_\pm'' = - \frac{5\sqrt{6}\,\epsilon_0}{4\pi^2}\,\hor^4\,e^{-\sqrt{6}\ell\hor} + \dots\,,
\end{equation}
where we neglected terms that are suppressed like $\ell\,\exp[-2\sqrt{6}\ell\hor]$ and used the numerical identity $b_{\textrm{\tiny log}}\ln\epsilon_0 = -2b_0-b_1-b_{\textrm{\tiny log}}$. Note that again all terms of the form $\ell \exp[-\sqrt{6}\ell\hor]$ cancel. Inserting numbers into our large width result \eqref{eq:s126} yields (setting $\pi T=1$)
\begin{equation}\label{eq:s125}
\frac{1}{2\pi}\,\mathcal{S}_\pm'' \approx -0.364053\,e^{-2.44949\ell} \,.
\end{equation}
The exponential behavior in \eqref{eq:s125} agrees rather precisely with the numerical data displayed in Figure~\ref{fig:QNECblackbrane}. 

\chapter[Shooting in Mathematica]{Mathematica Implementation of the Shooting Method}\label{App:Shoot}
In this appendix we give a Mathematica implementation of a simple shooting method that solves the auxiliary geodesic equations for minimal surfaces with infinite stripe boundary conditions in a homogeneous and isotropic AdS$_5$ Schwarzschild black brane geometry.
 Furthermore it computes the regularized (=vacuum subtracted) surface area which corresponds to the entanglement entropy of the of a stripe region in the boundary CFT which has finite temperature given by the Hawking temperature of the black brane. 
The reader can simply copy and paste the code into a Mathematica notebook or download the file\footnote{[\href{http://christianecker.com/wp-content/uploads/2018/08/AppendixD.nb\_.tar.gz}{{\tt http://christianecker.com/wp-content/uploads/2018/08/AppendixD.nb\_.tar.gz}}]} and run the simulation.\footnote{The code is implemented and tested in Mathematica 11.1.1.0 but since only very basic features are used it should run in any reasonable Mathematica version.}

\begin{mmaCell}[moredefined={xmu, d, metric, dim, M, metricInv}]{Input}
  (* coordinates and metric *)
  xmu=\{z,t,y\};
  d=Length[xmu];
  metric=1/z^(2(dim-2))\{\{0,-1,0\},\{-1,-(1-M z^(dim-1)),0\},\{0,0,1\}\};
  metricInv=Inverse[metric];
\end{mmaCell}

\begin{mmaCell}[moredefined={Gudd, metricInv, metric, xmu, \
d},morefunctionlocal={i, l, k, j}]{Input}
  (* Christoffel symbols *)
  Gudd=Table[Sum[1/2metricInv[[i,l]](D[metric[[k,l]],xmu[[j]]]
  +D[metric[[j,l]],xmu[[k]]]-D[metric[[j,k]],xmu[[l]]]),\{l,1,d\}],
  \{i,1,d\},\{j,1,d\},\{k,1,d\}];
\end{mmaCell}

\begin{mmaCell}[moredefined={Xmu, replXmu}]{Input}
  (* embedding functions *)
  Xmu=\{Z[s],T[s],Y[s]\};
  replXmu=\{z->Z[s],t->T[s],y->Y[s]\};
\end{mmaCell}

\begin{mmaCell}[moredefined={geoEQ, Xmu, Gudd, d, \
replXmu},morefunctionlocal={i, j, k}]{Input}
  (* geodesic equation *)
  geoEQ=Table[0==D[Xmu[[i]],\{s,2\}]+Sum[Gudd[[i,j,k]]
  D[Xmu[[j]],s]D[Xmu[[k]],s],\{j,1,d\},\{k,1,d\}],\{i,1,d\}]/.replXmu;
\end{mmaCell}
\newpage
\begin{mmaCell}[moredefined={ShootGeo, zIni, ErrMax, ItMax, geoEQ, \
Xmu, sCut, zCut, lCut, ICnew, dz, resGeoNew, sCutNew, lCutNew, delta, \
zNew, metric,
replXmu},morepattern={l_, zCut_, zIni_, dz_, ErrMax_, ItMax_, \
l},morelocal={Error, zNow, IC, resGeo, i},morefunctionlocal={s}]{Input}
  (* shooting method that computes embedding functions 
     and geodesics length *)
  ShootGeo[l_,zCut_,zIni_,dz_,ErrMax_,ItMax_]:=
  Module[\{Error,zNow,IC,GeoLength,resGeo,i\},
  zNow=zIni;
  i=0;Error=10^5;
  While[Error>ErrMax&&i<ItMax,
  (* boundary conditions *)
  IC=\{Z[0]==zNow,T[0]==1,Y[0]==0,Derivative[1][Z][0]==0,
   Derivative[1][T][0]==0,Derivative[1][Y][0]==1\};
  resGeo=NDSolve[geoEQ~Join~IC,Xmu,\{s,-10^8,10^8\},AccuracyGoal->10,
  PrecisionGoal->10, Method->\{"ExplicitRungeKutta",
  "DifferenceOrder"->8\},InterpolationOrder->8][[1]];
  (* computing parameter value at cutoff *)
  sCut=Solve[(Z[s]/.resGeo)==zCut,s][[1,1]]//Quiet;
  lCut=2Y[\mmaUnd{s}]/.resGeo/.sCut;
  Error=Abs[lCut-l];
  Print["i=",i," \mmaSub{z}{*}=",zNow,", Error=",Error,", l=",lCut];
  (* Newton step *)
  ICnew=\{Z[0]==zNow+dz,T[0]==1,Y[0]==0,Derivative[1][Z][0]==0,
   Derivative[1][T][0]==0,Derivative[1][Y][0]==1\};
  resGeoNew=NDSolve[geoEQ~Join~ICnew,Xmu,\{s,-10^8,10^8\},
  AccuracyGoal->10,PrecisionGoal->10, Method->\{"ExplicitRungeKutta",
  "DifferenceOrder"->8\},InterpolationOrder->8][[1]];
  sCutNew=Solve[(Z[s]/.resGeoNew)==zCut,s][[1,1]]//Quiet;
  lCutNew=2Y[\mmaUnd{s}]/.resGeoNew/.sCutNew;
  (* updating previous guess *)
  delta=If[Error>10^-1,1/10,1];
  zNew=zNow-delta dz (lCut-l)/(lCutNew-lCut);
  zNow=zNew;
  i++;
  ];
  (* geodesic length *)
  GeoLength=NIntegrate[Sqrt[D[Xmu/.resGeo,s].metric.
   D[Xmu/.resGeo,s]/.replXmu/.resGeo],\{s,-s/.sCut,s/.sCut\}];
  \{resGeo,GeoLength\}
  ]
\end{mmaCell}

\begin{mmaCell}[moredefined={zIni, zCut, M, dim, ItMax, ErrMax, dz, \
lList, zIniNow, solVac, ShootGeo},morefunctionlocal={i, j}]{Input}
  (* example: AdS5 vacuum *)
  zIni=2*10^-1;zCut=5*10^-2;M=0;dim=5;
  ItMax=100;ErrMax=10^-10;dz=10^-9;
  (* computing embedding functions and geodesic length *)
  lList=Table[i,\{i,1/10,5,1/10\}];
  zIniNow=zIni;
  Do[(*Print["l=",lList[[j]]];*)
  solVac[j]=ShootGeo[lList[[j]],zCut,zIniNow,dz,ErrMax,ItMax];
  (* using the previous result as new initial guess *)
  zIniNow=Z[s]/.solVac[j][[1]]/.s->0;
  ,\{j,1,Length[lList]\}]
\end{mmaCell}

\begin{mmaCell}[moredefined={GeoPlot2D, solVac, sCut, \
lList},morepattern={i_},morefunctionlocal={s, i}]{Input}
  (* plotting embedding functions and geodesic length *)
  GeoPlot2D[i_]:=ParametricPlot[\{Y[s],Z[s]\}/.solVac[i][[1]],
  \{s,-s/.sCut,s/.sCut\},PlotRange->\{All,\{zCut,6\}\},
  BaseStyle->Thick,AxesLabel->\{"y","z"\},LabelStyle->\{Black,20\},
  ImageSize->400,PlotStyle->ColorData["Rainbow",i/Length[lList]],
  MaxRecursion->15];
  \{Show[Table[GeoPlot2D[i],\{i,1,Length[lList]\}]],
  ListPlot[Table[\{lList[[i]],solVac[i][[2]]\},\{i,1,Length[lList]\}],
  BaseStyle->Thick,AxesLabel->\{"l","S"\},LabelStyle->\{Black,20\},
  ImageSize->500,Joined->True,PlotRange->All]\}
\end{mmaCell}

 \begin{mmaCell}[moregraphics={moreig={scale=.7}}]{Output}
   \mmaGraphics{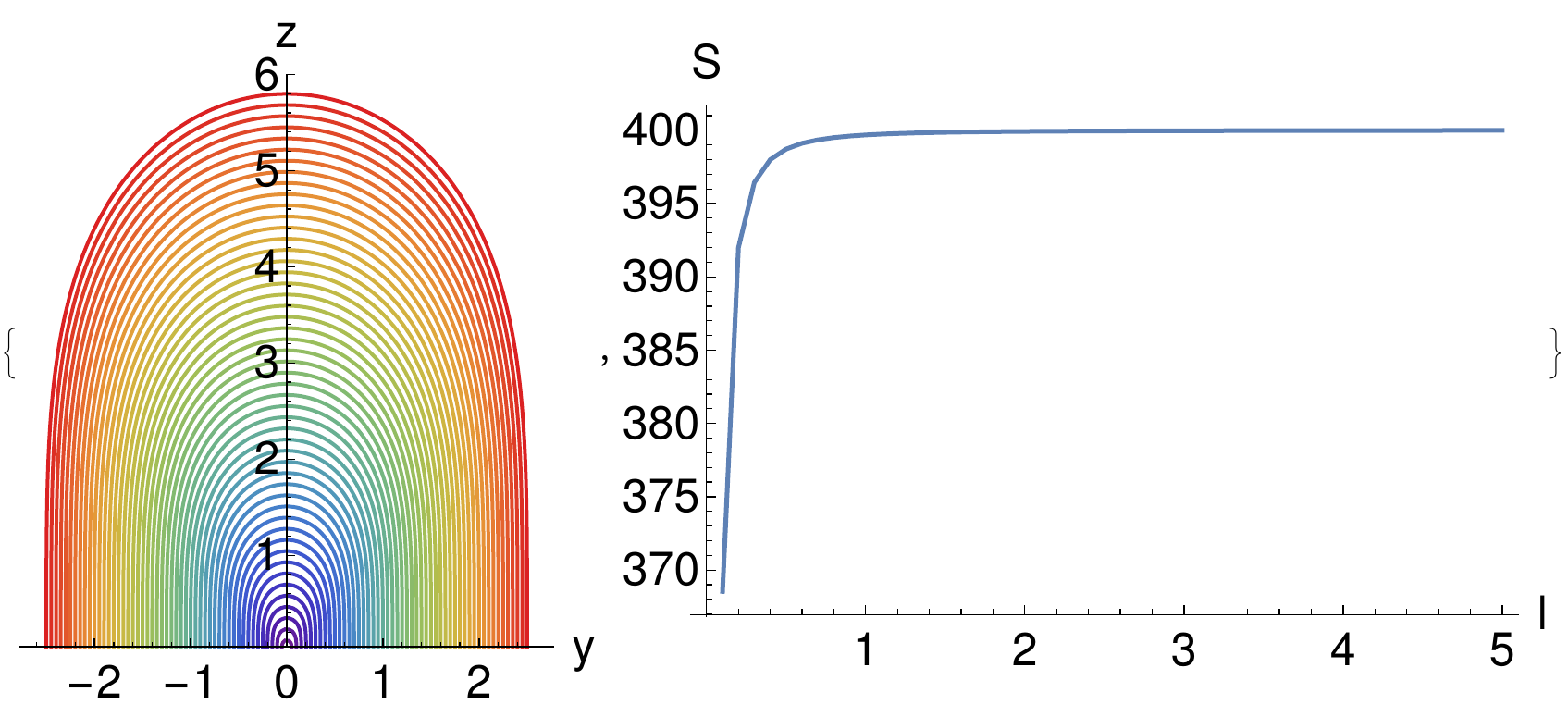}
 \end{mmaCell}

\begin{mmaCell}[moredefined={zIni, zCut, M, dim, ItMax, ErrMax, dz, \
lList, zIniNow, sol, ShootGeo},morefunctionlocal={i, j}]{Input}
  (* example: AdS5 black brane *)
  \
zIni=2*10^-1;zCut=5*10^-2;M=1;dim=5;
  ItMax=100;ErrMax=10^-10;dz=10^-9;
  (* computing embedding functions and geodesic length *)
  lList=Table[i,\{i,1/10,5,1/10\}];
  zIniNow=zIni;
  Do[
  (*Print["l=",lList[[j]]];*)
  sol[j]=ShootGeo[lList[[j]],zCut,zIniNow,dz,ErrMax,ItMax];
  (* using the previous result as new initial guess *)
  zIniNow=Z[s]/.sol[j][[1]]/.s->0;
  ,\{j,1,Length[lList]\}]
\end{mmaCell}

\begin{mmaCell}[moredefined={GeoPlot2D, sol, sCut, \
lList},morepattern={i_},morefunctionlocal={s, i, x}]{Input}
  (* plotting embedding functions and geodesic length *)
  GeoPlot2D[i_]:=ParametricPlot[\{Y[s],Z[s]\}/.sol[i][[1]]
  ,\{s,-s/.sCut,s/.sCut\},PlotRange->\{All,\{zCut,1.1\}\},
  BaseStyle->Thick,AxesLabel->\{"y","z"\},LabelStyle->\{Black,20\},
  ImageSize->550,PlotStyle->ColorData["Rainbow",
  \mmaPat{i}/Length[lList]],MaxRecursion->15,AspectRatio->0.4];
  \{Show[Table[GeoPlot2D[i],\{i,1,Length[lList]\}],
  Plot[1,\{x,-2.5,2.5\},PlotStyle->\{Black,Dashed,Thick\}]],
  ListPlot[Table[\{lList[[i]],sol[i][[2]]\},\{i,1,Length[lList]\}],
  BaseStyle->Thick,AxesLabel->\{"l","S"\},
  LabelStyle->\{Black,20\},ImageSize->400,Joined->True,
  PlotRange->All]\}
\end{mmaCell}

 \begin{mmaCell}[moregraphics={moreig={scale=.7}}]{Output}
   \mmaGraphics{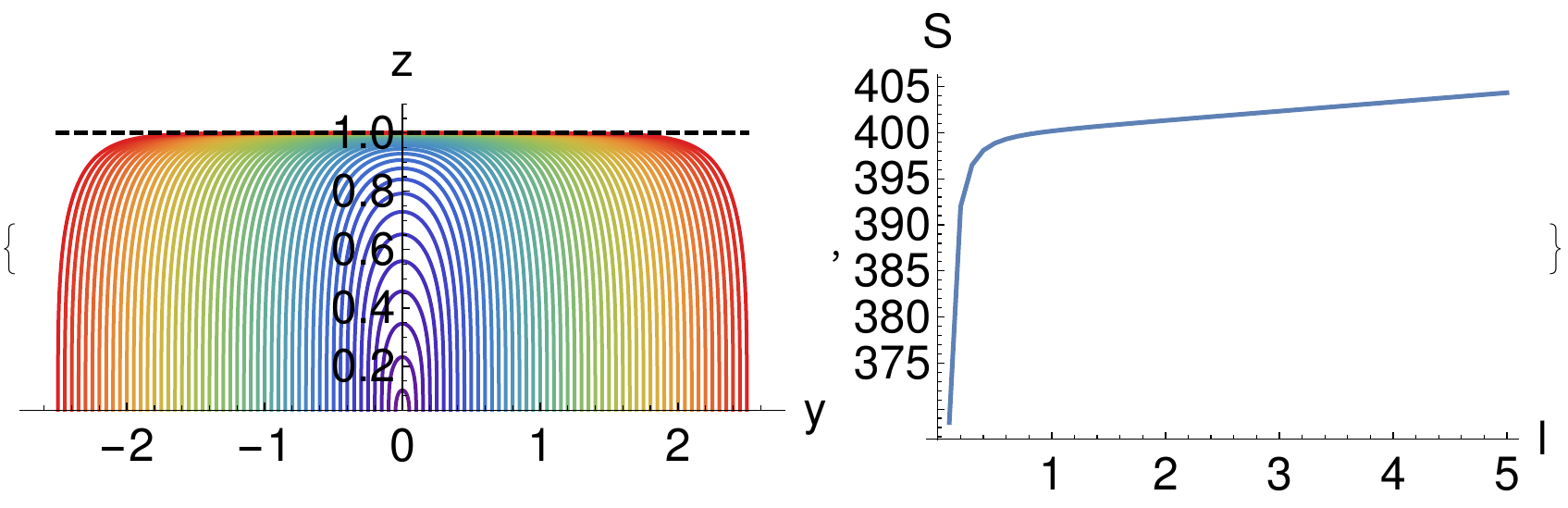}
 \end{mmaCell}

\chapter[Relaxation in Mathematica]{Mathematica Implementation of the Relaxation Method}\label{App:Relax}
In this appendix we give a Mathematica implementation of the relaxation method which solves the two-point boundary value problem for the geodesics equation.
The example provided here is given for the analytically available Vaidya spacetime, but generalizing to numeric backgrounds, as it was done in the examples in the main text is straight forward.
The Mathematica file is also available from the link\footnote{[\href{http://christianecker.com/wp-content/uploads/2018/08/AppendixE.nb\_.tar.gz}{{\tt http://christianecker.com/wp-content/uploads/2018/08/AppendixE.nb\_.tar.gz}}]}.
\begin{mmaCell}[moredefined={xmu, d, M, m, a, omega, dim, metric, metricinv, g, Gudd, i, k, j, gamma, Dgamma},morepattern={t_, z_, y_},morefunctionlocal={l}]{Input}
  (* Coordinates and metric for the relevant 3dim. subspace 
     in the Poincare patch *)
  xmu=\{z,t,y\};
  d=Length[xmu];
  (* smooth step function determining the location of the shell *)
  M[t_]:=m (1.+Tanh[a \mmaPat{t}])/2;
  (* conformal factor *)
  omega=z^-(dim-2);
  (* effective metric for the extremal surfaces *)
  metric=omega^2\{\{0,-1/z^2,0\},\{-1/z^2,-(1-M[t] z^dim)/z^2,0\},
  \{0,0,1/z^2\}\};
  metricinv=Inverse[metric];
  g=Det[metric];
  (* Christoffel symbol and contraction *)
  Gudd=Table[Sum[1/2metricinv[[i,l]](D[metric[[k,l]],xmu[[j]]]
  +D[metric[[j,l]],xmu[[k]]]-D[metric[[j,k]],xmu[[l]]])
  ,\{l,1,d\}],\{i,1,d\},\{j,1,d\},\{k,1,d\}];
  (* building the Christoffels used in the relaxation code *)
  gamma[z_,t_,y_]=Table[Gudd[[i,j,k]],\{i,d\},\{j,d\},\{k,d\}];
  Dgamma[z_,t_,y_]=Table[Table[D[Gudd[[i,j,k]],xmu[[l]]]
  ,\{i,d\},\{j,d\},\{k,d\}],\{l,d\}]//Simplify;
\end{mmaCell}
\newpage
\begin{mmaCell}[moredefined={makeGeoAnsatz, l0, dim, gridsize, i, t0},morepattern={gridsize_, zCUT_, l0_, t0_, zCUT},morelocal={tt, tav, f, tau,
Jf, tte, tts, tCUTOFF, lambda, dlambda, h, Jav, Y, Z, T, dY, dZ, dT, x, Zmax, lambdap, dlambdap, Yp, Ym, dYp, dYm, x_}]{Input}
  (* initial guess for the geodesic *)
  makeGeoAnsatz[gridsize_,zCUT_,l0_,t0_]:=
  Module[\{tt,tav,f,tau,Jf,tte,tts,tCUTOFF,lambda,dlambda,h,Jav,
  Y,Z,T,dY,dZ,dT,x,Zmax,lambdap,dlambdap,Yp,Ym,dYp,dYm\},
  (* turning point in the bulk *)
  Zmax=l0/2Gamma[1/(2(dim-1))]/(Sqrt[Pi]*Gamma[dim/(2(dim-1))]);
  (* parameter value at the z-cutoff *)
  tCUTOFF=1-Sqrt[1-zCUT/Zmax];
  tts=-1+tCUTOFF;
  tte=1-tCUTOFF;
  h=(tte-tts)/(gridsize-1);
  tt=Table[i, \{i, tts, tte, h\}];
  tav=Table[(i+i+h)/2, \{i,tts,tte-h,h\}];
  f[x_]=x;
  tau[x_]=ArcTanh[f[x]*Sqrt[2 - f[x]^2]];
  Jf[x_]=Which[
  dim==2,-Derivative[2][tau][x]/Derivative[1][tau][x],
  dim==3,-1*(Derivative[2][f][x]/Derivative[1][f][x])
   +(Derivative[1][f][x]*f[x]*(-22+38*f[x]^2-27*f[x]^4+7*f[x]^6))
  /(4-10f[x]^2+10*f[x]^4-5*f[x]^6+f[x]^8),
  dim==4,-1*(Derivative[2][f][x]/Derivative[1][f][x])
   +(Derivative[1][f][x]*f[x]*(-51+145*f[x]^2-205*f[x]^4+
   159*f[x]^6-65*f[x]^8+11*f[x]^10))
  /((-2+f[x]^2)*(-1+f[x]^2)*(3-3*f[x]^2+f[x]^4)*(1-f[x]^2+f[x]^4))
  ];
  Jav=Jf[tav];
  lambda=f[tt];
  dlambda=Derivative[1][f][tt];
  lambdap=lambda[[1;;Length[tt]/2]];
  dlambdap=dlambda;
  Yp=-l0/2+(Zmax*(1-lambdap^2))^dim/(dim*Zmax^(dim-1))*
  Hypergeometric2F1[1/2,dim/(2(dim -1)),(3dim-2)/(2dim-2),
  (-1+ lambdap^2)^(2(dim-1))];
  Ym=Reverse[-Yp];
  Y=Join[Yp,Ym];
  dYp=dlambdap l0 lambdap (1-lambdap^2)^dim Gamma[1/(2(dim-1))]
  /((lambdap^2-1)Sqrt[Pi]Sqrt[1-(lambdap^2-1)^(2(dim-1))]*
   Gamma[dim/(2(dim-1))]);
  dYm=Reverse[dYp];
  dY=Join[dYp,dYm];
  Z=Zmax*(1 - lambda^2);
  dZ=-2*Zmax*dlambda*lambda;
  T=t0-Z;
  dT=-dZ;
  \{\{Z,T,Y\},\{dZ,dT,dY\},Jav,lambda,dlambda,h\}
  ];
\end{mmaCell}
\newpage
\begin{mmaCell}[moredefined={evaluateChristoffels, i, j, gridsize, gamma, Dgamma, k},morepattern={gridsize_, X_, dX_, X, dX},morelocal={Xav, dXav,
gammaALL, DgammaALL},morefunctionlocal={l}]{Input}
  (* evaluating the numerical values of the Christoffel symbols *)
  evaluateChristoffels[gridsize_,X_,dX_]:=Module[
  \{Xav,dXav,gammaALL,DgammaALL\},
  Xav=Table[1/2(X[[i,j+1]]+X[[i,j]]),\{i,Length[X]\},\{j,1,gridsize-1\}];
  dXav=Table[1/2(dX[[i,j+1]]+dX[[i,j]]),\{i,Length[X]\},\{j,1,gridsize-1\}];
  gammaALL=gamma[Xav[[1]],Xav[[2]],Xav[[3]]];
  DgammaALL=Table[Dgamma[Xav[[1]],Xav[[2]],Xav[[3]]][[i]],\{i,Length[X]\}];
  For[i=1,i<4,i++,For[j=1,j<4,j++,For[k=1,k<4,k++,
  If[Length[gammaALL[[i,j,k]]]==0,
  gammaALL[[i,j,k]]=Table[0,\{gridsize-1\}]];
  Do[If[Length[DgammaALL[[l,i,j,k]]]==0,
  DgammaALL[[l,i,j,k]]=Table[0,\{gridsize-1\}]],\{l,Length[X]\}];]]];
  \{Xav,dXav,gammaALL,DgammaALL\}];
\end{mmaCell}

\begin{mmaCell}[moredefined={writeFDEs, i, j, gridsize, m, k},morepattern={gridsize_, X_, dX_, dXav_, Jav_, h_, gammaALL_, X, h, dXav, dX, Jav, gammaALL},morelocal={FDEs},morefunctionlocal={l}]{Input}
  (* building the finite differences equations *)
  writeFDEs[gridsize_,X_,dX_,dXav_,Jav_,h_,gammaALL_]:=Module[
  \{i,j,FDEs\},
  FDEs=Table[0,\{6*gridsize\}];
  j=1;
  For[i=1,i<6*(gridsize-1),i+=6,
  Do[
  FDEs[[i+2m+1]]=X[[m,j+1]]-X[[m,j]]-h*dXav[[m,j]];
  FDEs[[i+2m+2]]=dX[[m,j+1]]-dX[[m,j]]+h*Jav[[j]]*dXav[[m,j]]
  +h*Sum[gammaALL[[m,k,l,j]]*dXav[[k,j]]*dXav[[l,j]],
  \{k,Length[X]\},\{l,Length[X]\}];
  ,\{m,1,3\}]; 
  j+=1;
  ];
  FDEs];
\end{mmaCell}
\newpage
\begin{mmaCell}[moredefined={findSElements, j, gridsize, m, k},morepattern={gridsize_, dXav_, Jav_, h_, gammaALL_, DgammaALL_, h, Jav, dXav, gammaALL,
DgammaALL},morelocal={Sp, Sm},morefunctionlocal={n, l}]{Input}
  (* computing the numerical entries of the S-matrix *)
  findSElements[gridsize_,dXav_,Jav_,h_,gammaALL_,DgammaALL_]:=
  Module[\{j,Sp,Sm\},
  Sp=Sm=Table[Table[Table[0,\{gridsize-1\}],\{6\}],\{6\}];
  For[j=1,j<(gridsize),j+=1,
  Do[
  If[m==n,
  Sp[[2m-1,2n-1,j]]=1;
  Sm[[2m-1,2n-1,j]]=-1;
  Sp[[2m-1,2n,j]]=-h/2;
  Sm[[2m-1,2n,j]]=-h/2;
  Sp[[2m,2n,j]]=1+h/2Jav[[j]]
  +h Sum[dXav[[k,j]]*gammaALL[[m,n,k,j]],\{k,3\}];
  Sm[[2m,2n,j]]=-1+h/2Jav[[j]]
  +h Sum[dXav[[k,j]]*gammaALL[[m,n,k,j]],\{k,3\}];
  ,
  Sp[[2m,2n,j]]=h Sum[dXav[[k,j]]*gammaALL[[m,n,k,j]],\{k,3\}];
  Sm[[2m,2n,j]]=Sp[[2m,2n,j]];
  ];
  Sp[[2m,2n-1,j]]=h/2Sum[dXav[[k,j]]*dXav[[l,j]]
  *DgammaALL[[n,m,k,l,j]] ,\{k,3\},\{l,3\}];
  Sm[[2m,2n-1,j]]=Sp[[2m,2n-1,j]];
  ,\{m,3\},\{n,3\}];
  ];
  \{Sp,Sm\}];
\end{mmaCell}
\newpage
\begin{mmaCell}[moredefined={createSmatrix, i, j, k, gridsize},morepattern={gridsize_, Sp_, Sm_, h_, Sp, Sm},morelocal={dvS, dvSh, dvM, rvS, rvM,
cvS, cvSh, cvM, sMatrix, n},morefunctionlocal={l}]{Input}
  (* building the sparse matrix *)
  createSmatrix[gridsize_,Sp_,Sm_,h_]:=
   Module[\{i,j,dvS,dvSh,dvM,rvS,rvM,cvS,cvSh,cvM,sMatrix,k,n,\mmaLoc{l}\},
  \{rvS,cvSh,dvS,dvSh\}=Table[Table[0,\{24*(gridsize-1)\}],\{4\}];
  j=0;
  For[i=1,i<24*(gridsize-1),i+=24,
  k=0;n=1;
  While[k<23,
  Do[
  rvS[[i+k]]=If[l<3,2(n+1),2n+3]+6j;
  cvSh[[i+k]]=If[l<3,2n-2+l,l-2]+6j;
  dvS[[i+k]]=Sp[[If[l<3,2n-1,2n],If[l<3,2n-2+l,l-2],j+1]];
  dvSh[[i+k]]=Sm[[If[l<3,2n-1,2n],If[l<3,2n-2+l,l-2],j+1]];
  k++;
  ,\{l,1,8\}];
  n++;
  ];
  j+=1;];
  cvS=cvSh+6;
  rvM=Flatten[\{1,2,3,rvS,rvS,(6*gridsize-2),(6*gridsize-1),
  (6*gridsize)\}];
  cvM=Flatten[\{1,3, 5,cvSh,cvS,(6*gridsize-5),(6*gridsize-3),
  (6*gridsize-1)\}];
  dvM=Flatten[\{1,1,1,dvSh,dvS,1,1,1\}];
  sMatrix=SparseArray[Table[\{rvM[[k]],cvM[[k]]\}->dvM[[k]],
  \{k,1, Length[cvM]\}]];
  sMatrix];
\end{mmaCell}

\begin{mmaCell}[moredefined={NewtonStep, i, j, gridsize, k},morepattern={gridsize_, errorFDE_, deltaX_, X_, dX_, X, errorFDE, deltaX, dX},morelocal={eps,
Xnew, dXnew}]{Input}
  (* Newton iteration *)
  NewtonStep[gridsize_,errorFDE_,deltaX_,X_,dX_]:=Module[
  \{i,j,eps,Xnew,dXnew\},
  Xnew=Table[0,\{Length[X]\},\{gridsize\}];
  dXnew=Table[0,\{Length[X]\},\{gridsize\}];
  If[errorFDE<1*10^-3,eps=1.,eps=5*10^-1];
  j=1;
  For[i=1,i<6*gridsize,i+=6,
  Do[
  Xnew[[k,j]]=X[[k,j]]+eps*deltaX[[i+2k-2]];
  dXnew[[k,j]]=dX[[k,j]]+eps*deltaX[[i+2k-1]];
  ,\{k,1,3\}];
  j+=1;
  ];
  \{Xnew,dXnew\}];
\end{mmaCell}
\newpage
\begin{mmaCell}[moredefined={relax, evaluateChristoffels, gridsize, writeFDEs, i, findSElements, createSmatrix, NewtonStep},morepattern={gridsize_,
X_, dX_, h_, Jav_, X, dX, Jav, h},morelocal={Xav, dXav, gammaALL, DgammaALL, FDEs, errorFDE, Sp, Sm, sMatrix, deltaX, Xnew, dXnew}]{Input}
  (* relaxation routine *)
  relax[gridsize_,X_,dX_,h_,Jav_]:=
  Module[\{Xav,dXav,gammaALL,DgammaALL,FDEs,errorFDE,
  Sp,Sm,sMatrix,deltaX,Xnew,dXnew\},
  \{Xav,dXav,gammaALL,DgammaALL\}=evaluateChristoffels[gridsize,X,dX];
  FDEs=writeFDEs[gridsize,X,dX,dXav,Jav,h,gammaALL];
  errorFDE=Sum[Abs[FDEs[[i]]],\{i,Length[FDEs]\}]/(6*gridsize);
  \{Sp,Sm\}=findSElements[gridsize,dXav,Jav,h,gammaALL,DgammaALL];
  sMatrix=createSmatrix[gridsize,Sp,Sm,h];
  deltaX=-LinearSolve[sMatrix,FDEs];
  \{Xnew,dXnew\}=NewtonStep[gridsize,errorFDE,deltaX,X,dX];
  \{Xnew,dXnew,errorFDE\}];
\end{mmaCell}

\begin{mmaCell}[moredefined={findGeoLength, i, dim, M},morepattern={X0_, dX0_, X_, dX_, lambda_, zCUT_, lambda, X, X0},morelocal={Xint, X0int, dsint,
ds0int, L, L0}]{Input}
  (* computing the geodesic length *)
  findGeoLength[X0_,dX0_,X_,dX_,lambda_,zCUT_]:=
  Module[\{Xint,X0int,dsint,ds0int,L,L0\},
  Xint=Table[Interpolation[Thread[\{lambda,X[[i,;;]]\}]],
  \{i,Length[X]\}];
  X0int=Table[Interpolation[Thread[\{lambda,X0[[i,;;]]\}]],
  \{i,Length[X0]\}];
  dsint=Xint[[1]][x]^-(dim-1)*
  Sqrt[-(1-M[ Xint[[2]][x]]Xint[[1]][x]^dim)D[Xint[[2]][x],x]^2
  -2 D[Xint[[2]][x],x] D[Xint[[1]][x],x] + D[Xint[[3]][x],x]^2];
  ds0int=X0int[[1]][x]^-(dim-1) Sqrt[-D[X0int[[2]][x],x]^2 
  -2 D[X0int[[2]][x],x] D[X0int[[1]][x],x] + D[X0int[[3]][x],x]^2];
  L=NIntegrate[dsint, \{\mmaFnc{x}, First[lambda],Last[lambda]\}];
  L0=NIntegrate[ds0int, \{\mmaFnc{x}, First[lambda],Last[lambda]\}];
  \{L,L0\}];
\end{mmaCell}

\begin{mmaCell}[moredefined={relaxOneGeodesic, ansatz, geodesic, i, makeGeoAnsatz, gridsize, l0, t0, errMAX, relax, itMAX, findGeoLength},morepattern={gridsize_,
zCUT_, l0_, t0_, errMAX_, itMAX_, zCUT},morelocal={L, L0, X0, dX0, Jav, lambda, dlambda, h, X, dX, errorFDE}]{Input}
  (* routine that relaxes a single geodesic  *)
  relaxOneGeodesic[gridsize_,zCUT_,l0_,t0_,errMAX_,itMAX_]:=
  Module[\{L,L0,ansatz,geodesic,i,X0,dX0,Jav,lambda,dlambda,
  h,X,dX,errorFDE\},
  \{X0,dX0,Jav,lambda,dlambda,h\}=makeGeoAnsatz[gridsize,zCUT,l0,t0];
  \{X,dX\}=\{X0,dX0\};
  \{X[[2,1]],X[[2,-1]],X[[3,1]],X[[3,-1]]\}=
  \{t0-zCUT,t0-zCUT,-l0/2,l0/2\};
  errorFDE=10.^5;
  i=1;
  While[errorFDE>errMAX,
  \{X,dX,errorFDE\}=relax[gridsize,X,dX,h,Jav];
  If[i==itMAX,Break[]];
  i+=1;
  ];
  Print["Iteration: ",i," with errorFDE: ",errorFDE];
  \{L,L0\}=findGeoLength[X0,dX0,X,dX,lambda,zCUT];
  ansatz=\{t0,l0,gridsize,X0,dX0,lambda,0,0,L0\};
  geodesic=\{t0,l0,gridsize,X,dX,lambda,errorFDE,i-1,L\};
  \{ansatz,geodesic\}
  ];
\end{mmaCell}

\begin{mmaCell}[moredefined={relaxTevolution, i, j, ansatz, geodesic, tMax, t0, Nt, makeGeoAnsatz, gridsize, l0, errMAX, relax, itMAX, findGeoLength},morepattern={gridsize_,
zCUT_, l0_, t0_, tMax_, Nt_, errMAX_, itMAX_, zCUT},morelocal={t, deltat, X0, dX0, Jav, lambda, dlambda, h, X, dX, errorFDE, L, L0}]{Input}
  (* Compute Entanglement entropy as function of time *)
  relaxTevolution[gridsize_,zCUT_,l0_,t0_,tMax_,Nt_,errMAX_,itMAX_]:=
  Module[\{i,j,t,deltat,X0,dX0,Jav,lambda,dlambda,
  h,X,dX,errorFDE,L,L0,ansatz,geodesic\},
  deltat=(tMax-t0)/(Nt-1);
  t=t0;ansatz=\{\};geodesic=\{\};
  \{X0,dX0,Jav,lambda,dlambda,h\}=makeGeoAnsatz[gridsize,zCUT,l0,t0];
  For[j=1,j<=Nt,j++,
   Print["step: ",j," @ t = ",t];
   If[j==1,\{X,dX\}=\{X0,dX0\}];
  \{X[[2,1]],X[[2,-1]],X[[3,1]],X[[3,-1]]\}=\{t-zCUT,t-zCUT,-l0/2,l0/2\};
  errorFDE=10^5;i=1;
  While[errorFDE>errMAX,
  \{X,dX,errorFDE\}=relax[gridsize,X,dX,h,Jav];
   (*Print["Iteration: ",i," with errorFDE: ",errorFDE];*)
  If[i==itMAX,Break[]];
  i+=1;
  ];
  Print[i-1," iterations with error= ",errorFDE];
  \{L,L0\}=findGeoLength[X0,dX0,X,dX,lambda,zCUT];
  AppendTo[ansatz,\{t,l0,gridsize,X0,dX0,lambda,0,0,L0\}];
  AppendTo[geodesic,\{t,l0,gridsize,X,dX,lambda,errorFDE,i-1,L\}];
  t=t+deltat;
  ];
  \{ansatz,geodesic\}];
\end{mmaCell}
\newpage
\begin{mmaCell}[moredefined={relaxLevolution, i, j, ansatz, geodesic, lMax, l0, Nl, makeGeoAnsatz, gridsize, t0, errMAX, relax, itMAX, findGeoLength},morepattern={gridsize_,
zCUT_, l0_, t0_, lMax_, Nl_, errMAX_, itMAX_, zCUT},morelocal={l, deltal, X0, dX0, Jav, lambda, dlambda, h, X, dX, errorFDE, L, L0}]{Input}
  (* compute entanglement entropy as function of the separation *)
  relaxLevolution[gridsize_,zCUT_,l0_,t0_,lMax_,Nl_,errMAX_,itMAX_]:=
  Module[\{i,j,l,deltal,X0,dX0,Jav,lambda,dlambda,
  h,X,dX,errorFDE,L,L0,ansatz,geodesic\},
  deltal=(lMax-l0)/(Nl-1);
  l=l0;ansatz=\{\};geodesic=\{\};
  For[j=1,j<=Nl,j++,
   Print["step: ",j," @ L = ",l];
  \{X0,dX0,Jav,lambda,dlambda,h\}=makeGeoAnsatz[gridsize,zCUT,l,t0];
   If[j==1,\{X,dX\}=\{X0,dX0\}];
  \{X[[2,1]],X[[2,-1]],X[[3,1]],X[[3,-1]]\}=\{t0-zCUT,t0-zCUT,-l/2,l/2\};
  errorFDE=10^5;i=1;
  While[errorFDE>errMAX,
  \{X,dX,errorFDE\}=relax[gridsize,X,dX,h,Jav];
   (*Print["Iteration: ",i," with errorFDE: ",errorFDE];*)
  If[i==itMAX,Break[]];
  i+=1;
  ];
  Print[i-1," iterations with error= ",errorFDE];
  \{L,L0\}=findGeoLength[X0,dX0,X,dX,lambda,zCUT];
  AppendTo[ansatz,\{t0,l,gridsize,X0,dX0,lambda,0,0,L0\}];
  AppendTo[geodesic,\{t0,l,gridsize,X,dX,lambda,errorFDE,i-1,L\}];
  l=l+deltal;
  ];
  \{ansatz,geodesic\}];
\end{mmaCell}

\begin{mmaCell}[moredefined={gridsize, zCut, l0, t0, dim, a, m, itMAX, errMAX, tMax, Nt, Nl, lMax, ansatz, geodesic, relaxTevolution}]{Input}
  (* time evolution of entanglement entropy *)
  gridsize=300;zCut=0.01;l0=6;t0=-0.5;dim=2;a=30;m=1;
  itMAX=20;errMAX=10^-10;tMax=4.;Nt=30;Nl=10;lMax=2;
  \{ansatz,geodesic\}=relaxTevolution[gridsize,zCut,l0,t0,
   tMax,Nt,errMAX,itMAX];
\end{mmaCell}

\begin{mmaCell}[moredefined={geodesic, i, ansatz}]{Input}
  (* plotting the result *)
  \{ListPlot[Table[Thread[
  \{geodesic[[i,4,2,;;]],geodesic[[i,4,1,;;]]\}],
  \{i,1,Length[geodesic]\}],
  PlotStyle->Table[ColorData["Rainbow",i/(Length[geodesic]-1)],
  \{i,0,(Length[geodesic]-1)\}],PlotStyle->Thick,PlotRange->All,
  Joined->True,ImageSize->400,LabelStyle->\{Black,20\},
  AxesLabel->\{"t","z"\}],
  ListPlot[Thread[\{geodesic[[;;,1]],
  geodesic[[;;,-1]]-ansatz[[;;,-1]]\}],
  Joined->True,ImageSize->400,LabelStyle->\{Black,20\},
  AxesLabel->\{"t","S"\}]\}
\end{mmaCell}

\begin{mmaCell}[moregraphics={moreig={scale=0.7}}]{Output}
 \mmaGraphics{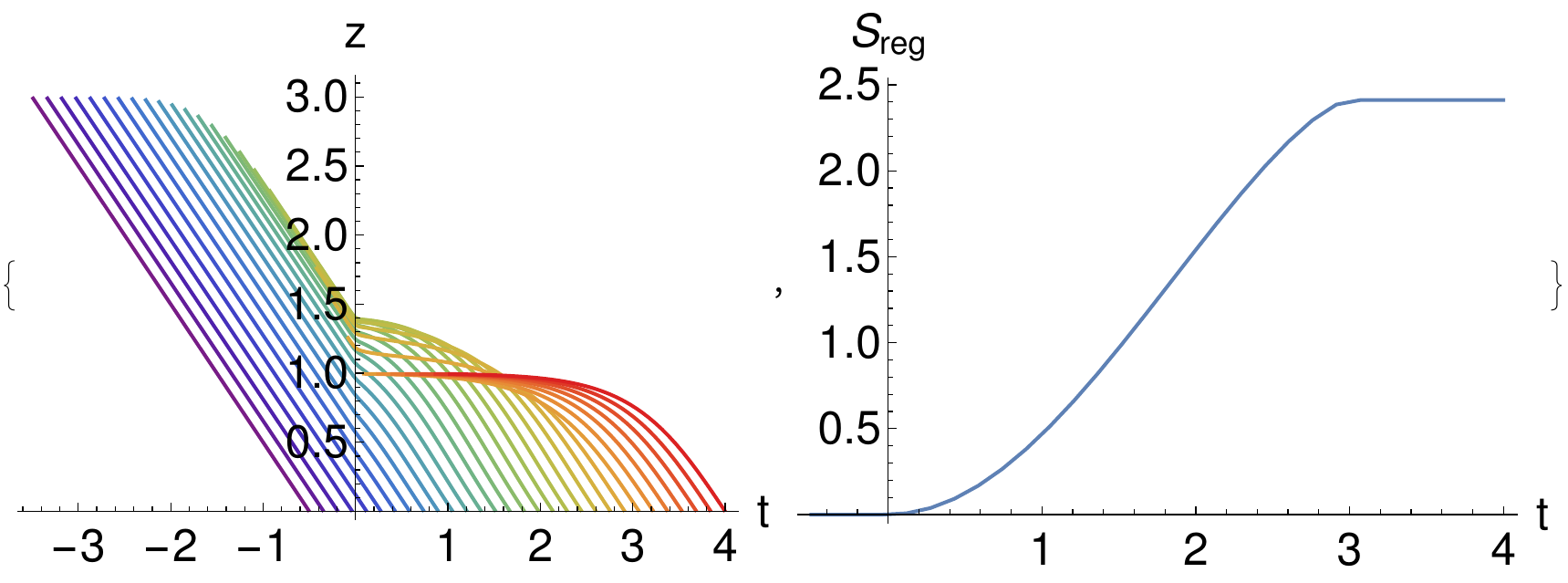}
\end{mmaCell}

\begin{mmaCell}[moredefined={gridsize, zCut, l0, t0, dim, a, m, itMAX, errMAX, tMax, Nt, Nl, lMax, ansatz, geodesic, relaxLevolution}]{Input}
  (* entanglement entropy as function of the separation *)\\
  gridsize=300;zCut=0.01;l0=0.5;t0=2.;dim=2;a=30;m=1;
  itMAX=20;errMAX=10^-10;tMax=5;Nt=30;Nl=30;lMax=8;
  \{ansatz,geodesic\}=relaxLevolution[gridsize,zCut,l0,t0,
  lMax,Nl,errMAX,itMAX];
\end{mmaCell}

\begin{mmaCell}[moredefined={geodesic, i, ansatz}]{Input}
  (* plotting results *)
  \{ListPlot[Table[Thread[
  \{geodesic[[i,4,3,;;]],geodesic[[i,4,1,;;]]\}],
  \{i,1,Length[geodesic]\}],
  PlotStyle->Table[ColorData["Rainbow",i/(Length[geodesic]-1)],
  \{i,0,(Length[geodesic]-1)\}],PlotStyle->Thick,PlotRange->All,
  Joined->True,ImageSize->400,LabelStyle->\{Black,20\}
  ,AxesLabel->\{"y","z"\}],
  ListPlot[Thread[\{geodesic[[;;,2]],
  geodesic[[;;,-1]]-ansatz[[;;,-1]]\}],
  Joined->True,ImageSize->400,LabelStyle->\{Black,20\},
  AxesLabel->\{"l","S"\}]\}
\end{mmaCell}

\begin{mmaCell}[moregraphics={moreig={scale=0.7}}]{Output}
 \mmaGraphics{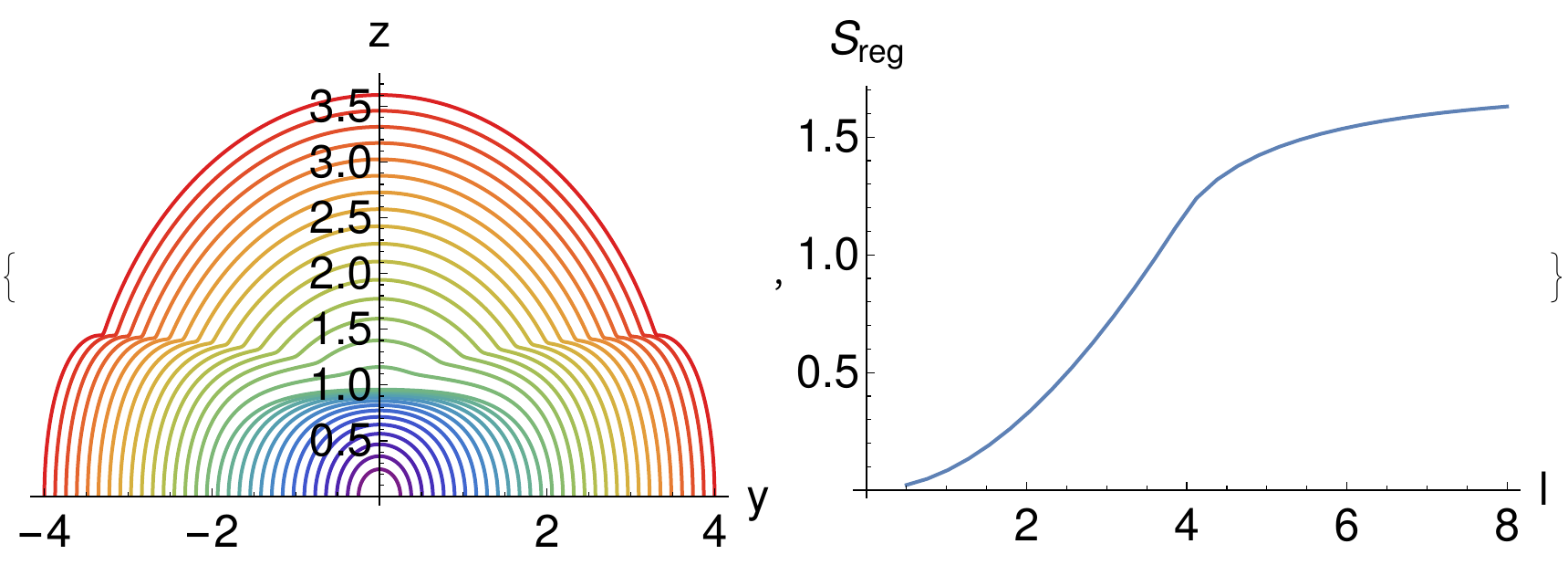}
\end{mmaCell}

\chapter{Mathematica Routine for QNEC}\label{App:MathematicaQNEC}
In this appendix we give a Mathematica routine that computes QNEC using the relaxation code provided in the previous appendix.
Again, the routine can be copied, together with the code in Appendix \ref{App:Relax}, into an empty Mathematica notebook and executed or downloaded from\footnote{[\href{http://christianecker.com/wp-content/uploads/2018/08/AppendixF.nb\_.tar.gz}{{\tt http://christianecker.com/wp-content/uploads/2018/08/AppendixF.nb\_.tar.gz}}]}.
\newpage
\begin{mmaCell}[moredefined={relaxQNECLevo, k, ansatz, geodesic, lMax, l0, Nl, makeGeoAnsatz, gridsize, zCUT, t0, side, epsilon, errMAX, relax, itMAX,
findGeoLength, Nk},morepattern={gridsize_, zCUT_, l0_, t0_, lMax_, Nl_, errMAX_, itMAX_, k0_, k1_, Nk_, epsilon_, side_:, k0, k1},morelocal={i, j,
deltal, l, X0, dX0, X, dX, Xk, dXk, Jav, lambda, dlambda, h, errorFDE, geodesick, L, L0},morefunctionlocal={kk}]{Input}
  (* compute QNEC as function of the separation *)
  relaxQNECLevo[gridsize_,zCUT_,l0_,t0_,lMax_,Nl_,errMAX_,itMAX_,
  \{k0_,k1_\},Nk_,epsilon_,side_:-1]:=Module[\{i,j,k,deltal,kk,
  l,X0,dX0,X,dX,Xk,dXk,Jav,lambda,dlambda,h,errorFDE,
  ansatz,geodesic,geodesick,L,L0\},
  deltal=(lMax-l0)/(Nl-1);
  l=l0;ansatz=\{\};geodesic=\{\};
  For[i=1,i<=Nl,i++,
  Print["step:",i," l=",l];
  \{X0,dX0,Jav,lambda,dlambda,h\}=
  makeGeoAnsatz[gridsize,zCUT,l,t0];
  If[i==1,
  \{X,dX\}=\{X0,dX0\},\{X[[2,1]],X[[2,-1]],
   X[[3,1]],X[[3,-1]]\}=\{t0-zCUT,t0-zCUT,-l/2,l/2\};
  ];
  geodesick=\{\};
  Do[
  (*Print["k*eps=",kk*epsilon*\{k0,k1\}];*)
  \{Xk,dXk\}=\{X,dX\};
  If[side==1,
  \{Xk[[2,1]],Xk[[3,1]]\}=\{X[[2,1]],X[[3,1]]\}+
  kk*epsilon*\{k0,k1\};,
  \{Xk[[2,-1]],Xk[[3,-1]]\}=\{X[[2,-1]],X[[3,-1]]\}+
  kk*epsilon*\{k0,k1\};
  ];
  errorFDE=10^5;j=1;
  While[errorFDE>errMAX,
  \{Xk,dXk,errorFDE\}=relax[gridsize,Xk,dXk,h,Jav];
   (*Print["Iteration: ",i," with errorFDE: ",errorFDE];*)
  If[j==itMAX,Break[]];
  j+=1;
  ];
  (*Print[j-1," iterations with errorFDE= ",errorFDE];*)
  \{L,L0\}=findGeoLength[X0,dX0,Xk,dXk,lambda,zCUT];
  AppendTo[geodesick,\{t0,l,gridsize,Xk,dXk,epsilon,
  lambda,errorFDE,j-1,L\}];
  If[kk==0,\{X,dX\}=\{Xk,dXk\}];
  ,\{kk,(-Nk+Mod[Nk,2])/2,(Nk-Mod[Nk,2])/2\}];
  AppendTo[geodesic,geodesick];
  AppendTo[ansatz,\{t0,l,gridsize,X0,dX0,lambda,0,0,L0\}];
  l=l+deltal;
  ];
  \{ansatz,geodesic\}];
\end{mmaCell}
\newpage
\begin{mmaCell}[moredefined={relaxQNECTevo, k, ansatz, geodesic, tMax, t0, Nt, makeGeoAnsatz, gridsize, zCUT, l0, side, epsilon, errMAX, relax, itMAX,
findGeoLength, Nk},morepattern={gridsize_, zCUT_, l0_, t0_, tMax_, Nt_, errMAX_, itMAX_, k0_, k1_, Nk_, epsilon_, side_, k0, k1},morelocal={i, j,
deltat, t, X0, dX0, X, dX, Xk, dXk, Jav, lambda, dlambda, h, errorFDE, geodesick, L, L0},morefunctionlocal={kk}]{Input}
  (* compute QNEC as function of time *)
  relaxQNECTevo[gridsize_,zCUT_,l0_,t0_,tMax_,Nt_,errMAX_,itMAX_,
  \{k0_,k1_\},Nk_,epsilon_,side_]:=Module[\{i,j,k,deltat,
  kk,t,X0,dX0,X,dX,Xk,dXk,Jav,lambda,dlambda,h,errorFDE,
  ansatz,geodesic,geodesick,L,L0\},
  deltat=(tMax-t0)/(Nt-1);
  t=t0;ansatz=\{\};geodesic=\{\};
  \{X0,dX0,Jav,lambda,dlambda,h\}=makeGeoAnsatz[gridsize,zCUT,l0,t];
  For[j=1,j<=Nt,j++,
  Print["step:",j,", @ t=",t];
  If[j==1,\{X,dX\}=\{X0,dX0\}];
  \{X[[2,1]],X[[2,-1]],X[[3,1]],X[[3,-1]]\}=
  \{t-zCUT,t-zCUT,-l0/2,l0/2\};
  geodesick=\{\};
  \{Xk,dXk\}=\{X,dX\};
  Do[
  (*Print["k*eps=",kk*epsilon*\{k0,k1\}];*)
  \{Xk,dXk\}=\{X,dX\};
  If[side==1,
  \{Xk[[2,1]],Xk[[3,1]]\}=\{X[[2,1]],X[[3,1]]\}+
  kk*epsilon*\{k0,k1\};,
  \{Xk[[2,-1]],Xk[[3,-1]]\}=\{X[[2,-1]],X[[3,-1]]\}+
  kk*epsilon*\{k0,k1\};
  ];
  errorFDE=10^5;i=1;
  While[errorFDE>errMAX,
  \{Xk,dXk,errorFDE\}=relax[gridsize,Xk,dXk,h,Jav];
   (*Print["Iteration: ",i," with errorFDE: ",errorFDE];*)
  If[i==itMAX,Break[]];
  i+=1;
  ];
  (*Print[i-1," iterations with errorFDE= ",errorFDE];*)
  \{L,L0\}=findGeoLength[X0,dX0,Xk,dXk,lambda,zCUT];
  AppendTo[geodesick,\{t,l0,gridsize,Xk,dXk,epsilon,
  lambda,errorFDE,i-1,L\}];
  If[kk==0,\{X,dX\}=\{Xk,dXk\}];
  ,\{kk,(-Nk+Mod[Nk,2])/2,(Nk-Mod[Nk,2])/2\}];
  AppendTo[geodesic,geodesick];
  AppendTo[ansatz,\{t,l0,gridsize,X0,dX0,lambda,0,0,L0\}];
  t=t+deltat;
  ];
  \{ansatz,geodesic\}];
\end{mmaCell}
\newpage
\begin{mmaCell}[moredefined={epsilon, gridsize, zCUT, errMAX, itMAX, Nk}]{Input}
  (* numerical parameters *)
  epsilon=0.001;gridsize=300;zCUT=0.001;
  errMAX=10^-15;itMAX=20;Nk=7;
\end{mmaCell}

\begin{mmaCell}[moredefined={dim, m, a, k, side, l0, t0, tMax, Nt, ansatz, geodesic, relaxQNECTevo, gridsize, zCUT, errMAX, itMAX, Nk, epsilon}]{Input}
  (* compute QNEC surfaces and areas as function of time *)
  dim=2;m=1;a=30;k=\{1,1\};side=1;l0=5;t0=-2.0;tMax=5.0;Nt=70;
  \{ansatz,geodesic\}=relaxQNECTevo[gridsize,zCUT,l0,t0,tMax,Nt,
   errMAX,itMAX,k,Nk,epsilon,side];
\end{mmaCell}

\begin{mmaCell}[moredefined={S, epsilon, geodesic, Sfit, Sd, Sdd, c, QNEClist},morefunctionlocal={i, j}]{Input}
  (* computing QNEC *)
  S=Table[Table[\{i*epsilon,geodesic[[j,i+4,-1]]/4\},\{i,-3,3\}],
   \{j,1,Length[geodesic]\}];
  Sfit=Table[Fit[S[[i]],\{1,eps,eps^2,eps^3\},eps],\{i,1,Length[S]\}];
  Sd=Table[D[Sfit[[i]],\{eps,1\}]/.eps->0,\{i,1,Length[Sfit]\}];
  Sdd=Table[D[Sfit[[i]],\{eps,2\}]/.eps->0,\{i,1,Length[Sfit]\}];
  c=3/2;
  QNEClist=Table[\{geodesic[[i,1,1]],
  (Sdd[[i]]+6/c Sd[[i]]^2)/(2 Pi)\},\{i,1,Length[geodesic]\}];
\end{mmaCell}

\begin{mmaCell}[moredefined={plotSurf, geodesic, Nt, plotHorizon, M, dim, Tkkplot, plotQNEC, QNEClist},morefunctionlocal={j, t}]{Input}
  (* plotting surfaces *)
  plotSurf=ListPlot[Table[Thread[\{geodesic[[j,4,4,2,;;]],
   geodesic[[j,4,4,1,;;]]\}],\{j,1,Nt\}],Joined->True,ImageSize->500,
   PlotStyle->Table[ColorData["Rainbow",j/(Nt-1)],\{j,0,Nt-1\}],
   LabelStyle->\{Black,20\},AxesLabel->\{"t","z"\}];
  (* plotting horizon *)
  plotHorizon=Plot[1/M[t]^(1/dim),\{t,-0.1,5\},AxesOrigin->\{0,0\},
   PlotStyle->\{Thickness[0.005], Dashed,Black\}];
  (* null-projection of the EMT *)
  Tkkplot=Plot[1/(8Pi)M[t],\{t,-2,5\},PlotRange->All,
   PlotStyle->\{Black,Thick,Dashed\},
   PlotLegends->Placed[\{"Tkk"\},Above],LabelStyle->\{Black,20\}];
  (* plotting QNEC *)
  plotQNEC=ListPlot[QNEClist,Joined->True,PlotStyle->\{Red,Thick\},
   LabelStyle->\{Black,20\},AxesLabel->\{"t"\},
   ImageSize->500,PlotRange->All,
   PlotLegends->Placed[\{"(S''+6/c(S')^2)/(2Pi)"\},Above]];
  \{Show[plotSurf,plotHorizon],Show[plotQNEC,Tkkplot]\}
\end{mmaCell}

\begin{mmaCell}[moregraphics={moreig={scale=0.8}}]{Output}
 \mmaGraphics{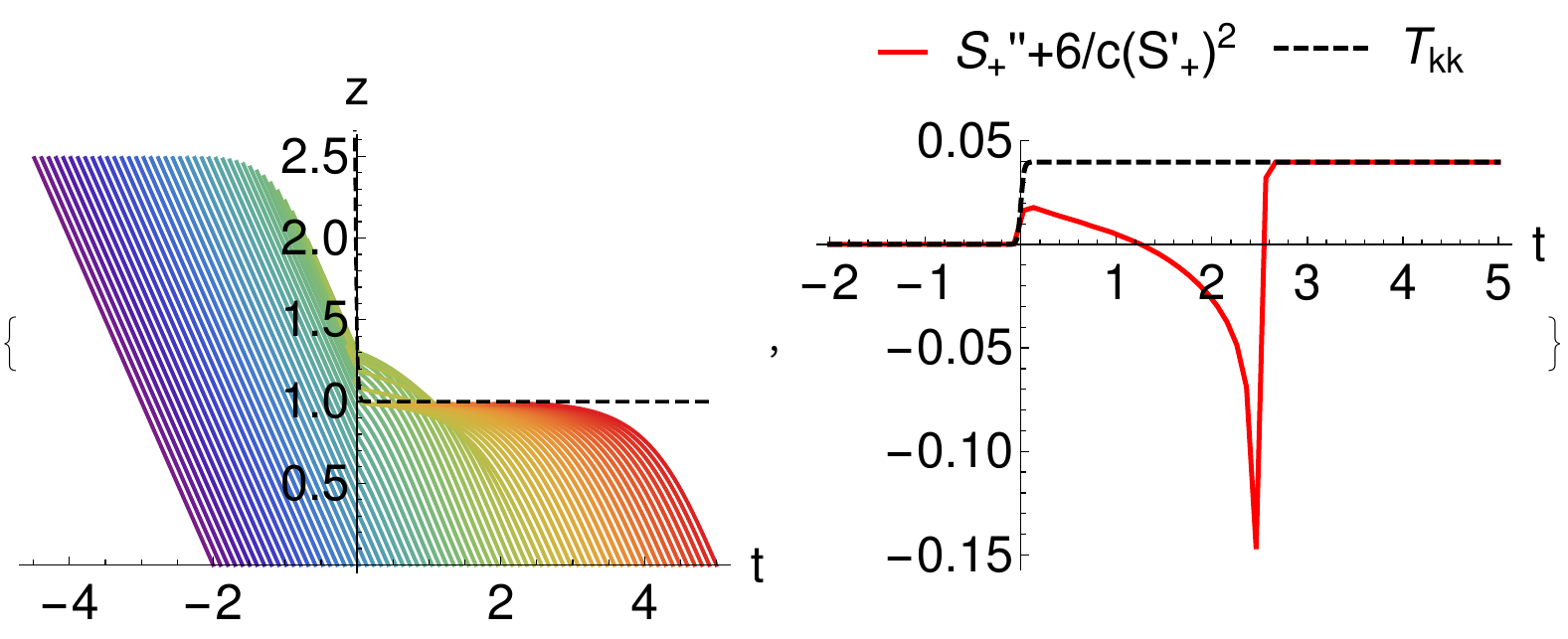}
\end{mmaCell}

\begin{mmaCell}[moredefined={dim, m, a, k, side, t0, l0, lMax, Nl, ansatz, geodesic, relaxQNECLevo, gridsize, zCUT, errMAX, itMAX, Nk, epsilon}]{Input}
  (* compute QNEC surfaces and areas as function of separation *)
  dim=2;m=1;a=30;k=\{1,1\};side=1;t0=1.0;l0=0.1;lMax=5.0;Nl=30;
  \{ansatz,geodesic\}=relaxQNECLevo[gridsize,zCUT,l0,t0,lMax,Nl,
  errMAX,itMAX,k,Nk,epsilon,side];
\end{mmaCell}

\begin{mmaCell}[moredefined={S, epsilon, geodesic, Sfit, Sd, Sdd, c, QNEClist},morefunctionlocal={i, j}]{Input}
  (* computing QNEC *)
  S=Table[Table[\{i*epsilon,geodesic[[j,i+4,-1]]/4\},\{i,-3,3\}],
   \{j,1,Length[geodesic]\}];
  Sfit=Table[Fit[S[[i]],\{1,eps,eps^2,eps^3\},eps],\{i,1,Length[S]\}];
  Sd=Table[D[Sfit[[i]],\{eps,1\}]/.eps->0,\{i,1,Length[Sfit]\}];
  Sdd=Table[D[Sfit[[i]],\{eps,2\}]/.eps->0,\{i,1,Length[Sfit]\}];
  c=3/2;
  QNEClist=Table[\{geodesic[[i,1,2]],(Sdd[[i]]+6/c Sd[[i]]^2)/(2 Pi)\},
  \{i,1,Length[geodesic]\}];
\end{mmaCell}

\begin{mmaCell}[moredefined={plotSurf, geodesic, Nl, Tkkplot, M, t0, l0, lMax, plotQNEC, QNEClist},morefunctionlocal={j, l}]{Input}
  (* plotting surfaces *)
  plotSurf=ListPlot[Table[Thread[\{geodesic[[j,4,4,3,;;]],
   geodesic[[j,4,4,1,;;]]\}],\{j,1,Nl\}],Joined->True,
  ImageSize->500,LabelStyle->\{Black,20\},AxesLabel->\{"x","z"\},
  PlotStyle->Table[ColorData["Rainbow",j/(Nl-1)],\{j,0,Nl-1\}]];
  (* null-projection of the EMT *)
  Tkkplot=Plot[1/(8Pi)M[t0],\{l,l0,lMax\},PlotRange->All,
  PlotStyle->\{Black,Thick,Dashed\},
  PlotLegends->Placed[\{"Tkk"\},Above],LabelStyle->\{Black,20\}];
  (* plotting QNEC *)
  plotQNEC=ListPlot[QNEClist[[2;;]],Joined->True,
  PlotStyle->\{Red,Thick\},LabelStyle->\{Black,20\},AxesLabel->\{"l"\},
  PlotLegends->Placed[\{"(S''+6/c(S')^2)/(2Pi)"\},Above],ImageSize->500];
  \{Show[plotSurf],Show[plotQNEC,Tkkplot]\}
\end{mmaCell}

\begin{mmaCell}[moregraphics={moreig={scale=0.8}}]{Output}
 \mmaGraphics{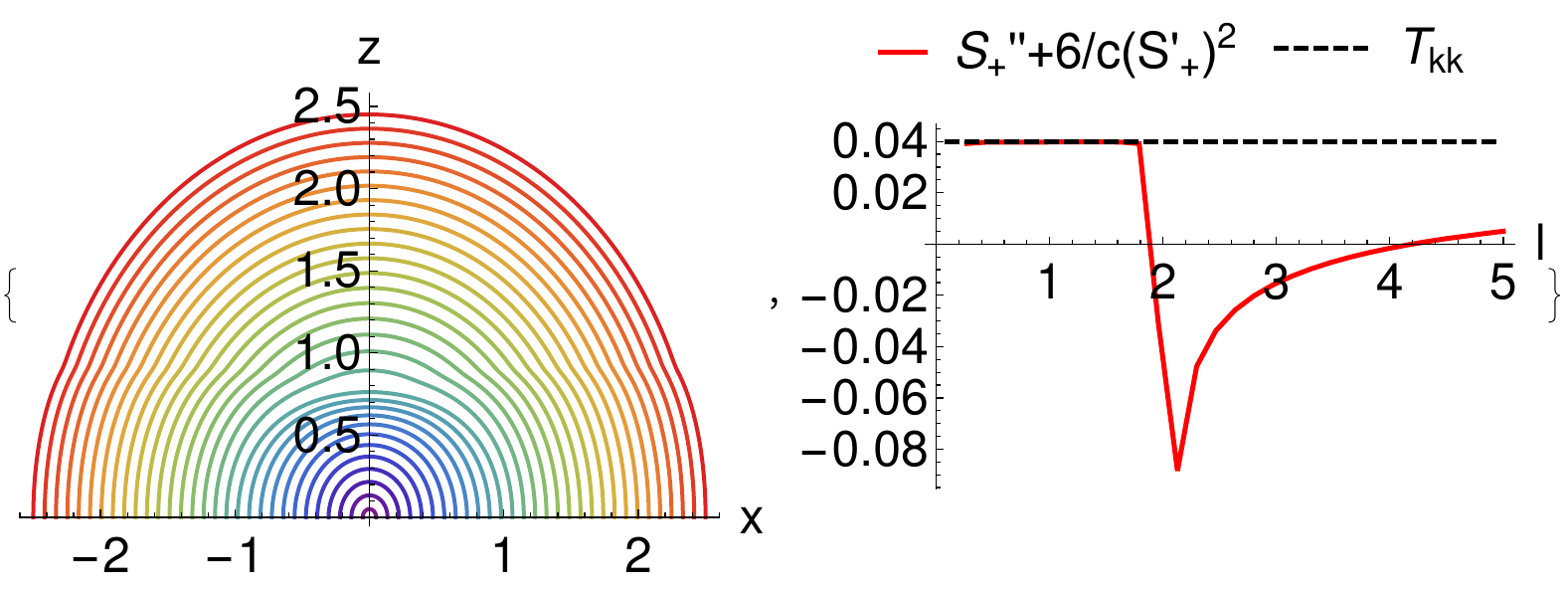}
\end{mmaCell}

\end{appendices}
\backmatter

\bibliographystyle{JHEP-2} 
\bibliography{thesis}

\printindex

\end{document}